\documentclass[usenatbib]{mn2e}
\usepackage{natbib}
\usepackage{epsfig}
\usepackage{amsmath}
\usepackage{color}

\title[Water towards methanol: 341$^{\circ}$ to 6$^{\circ}$]{A search for water masers associated with class II methanol masers - II. Longitude range 341$^{\circ}$ to 6$^{\circ}$}
\author[A. M. Titmarsh, S. P. Ellingsen, S. L. Breen, J. L. Caswell, M. A. Voronkov]{A. M. Titmarsh$^{1,2}$\thanks{E-mail:
Anita.Titmarsh@utas.edu.au}, S. P. Ellingsen$^{1}$, S. L. Breen$^{2}$, J. L. Caswell$^{2}$\thanks{Deceased 2015 January 14}, M. A. Voronkov$^{2}$ \\
$^{1}$School of Mathematics and Physics, University of Tasmania, Private Bag 37, Hobart, Tasmania 7001, Australia \\
$^{2}$CSIRO Astronomy and Space Science, Australia Telescope National Facility, PO Box 76, Epping, NSW 1710}
\begin{document}

\date{}

\pagerange{\pageref{firstpage}--\pageref{lastpage}} \pubyear{2016}

\maketitle

\label{firstpage}

\begin{abstract}

This is the second paper in a series of catalogues of 22-GHz water maser observations towards the 6.7-GHz methanol masers from the Methanol Multibeam (MMB) Survey. In this paper we present our water maser observations made with the Australia Telescope Compact Array towards the masers from the MMB survey between $l = 341^{\circ}$ through the Galactic centre to $l = 6^{\circ}$. Of the 204 6.7-GHz methanol masers in this longitude range we found 101 to have associated water maser emission ($\sim$ 50 per cent). We found no difference in the 6.7-GHz methanol maser luminosities of those with and without water masers. In sources where both maser species are observed, the luminosities of the methanol and water masers are weakly correlated. Studying the mid-infrared colours from GLIMPSE we found no differences between the colours of those sources associated with both methanol and water masers and those associated with just methanol. Comparing the column density and dust mass calculated from the 870-$\mu$m thermal dust emission observed by ATLASGAL, we found no differences between those sources associated with both water and methanol masers and those with methanol only. Since water masers are collisionally pumped and often show emission further away from their accompanying YSO than the radiatively pumped 6.7-GHz methanol masers, it is likely water masers are not as tightly correlated to the evolution of the parent YSO and so do not trace such a well defined evolutionary state as 6.7-GHz methanol masers.

\end{abstract}

\begin{keywords}
masers -- surveys -- stars: formation -- ISM: molecules
\end{keywords}

\section{Introduction}

Masers are useful tools for studying star formation as they are bright, intense and observable at radio frequencies, thus allowing us to see into the dusty environments of young stellar objects. 

Water masers were first discovered by \cite{cheung69} towards Sgr B2, Orion A and W49 and are now known to be the most widespread of the known maser species \citep[e.g.][]{walsh11}. They are found within the envelopes of evolved asymptotic giant branch stars and around young stellar objects. For water masers accompanying high-mass young stellar objects (YSOs), VLBI observations have shown that they are found the disks surrounding them \citep{torrelles98}, in their high velocity bipolar outflows \citep[e.g.][]{genzel81,hofner96} and in bow shocks from outflows and expanding shells \citep{hofner96,carrasco15}. The 22-GHz water maser $6_{1,6} \rightarrow 5_{2,3}$ rotational transition is the brightest spectral line at radio frequencies and shows much greater temporal variability than those of other species \citep{brand03}. Other transitions of water have also been observed, especially the higher frequency vibrational transitions towards evolved stars \citep{menten89}. Water masers are pumped by collisions with H$_2$ molecules within shocks associated with outflows or accretion \citep{elitzur89,garay99} and may reside further away from their parent YSO than 6.7-GHz methanol masers (see \cite{forster89} c.f. \cite{caswell97}).

Methanol masers have been empirically divided into class I and class II types. Class I methanol masers appear spread around a star forming region, distributed over linear scales of up to 1~pc \citep[e.g.][]{kurtz04,voronkov06,cyganowski09,voronkov14,jordan15}, whereas, class II methanol masers reside close to their parent YSO \citep[e.g.][]{caswell97}. Modelling of maser pumping mechanisms has shown that class I methanol masers are pumped by collisions with molecular hydrogen, whereas the class II methanol masers are pumped by infrared radiation \citep[e.g.][]{cragg92,cragg02,voronkov05m}.

6.7-GHz methanol masers make ideal candidates to study the conditions at the sites of young, high-mass stars as they are found exclusively at sites of high-mass star formation \citep{pestalozzi02,minier03,xu08,breen13}. The Methanol Multibeam (MMB) survey was conducted to make an unbiased survey of the Galactic plane for 6.7-GHz methanol masers. The MMB covered the entire Galactic Plane observable with the Parkes 64-m dish in Australia (longitudes 186$^{\circ}$, through 0$^{\circ}$, to 60$^{\circ}$) with latitude coverage of $|b| \leq 2^{\circ}$. For sources without interferometric positions already available, the detections were followed-up with the Australia Telescope Compact Array (ATCA) and the Multi-Element Radio Linked Interferometer Network (MERLIN) to obtain precise positions. Catalogues covering the entire MMB region have now been published \citep{green09,caswell10,green10,caswell11,green12,breen15}.

Another large-scale, unbiased survey for masers is the H$_2$O southern Galactic Plane Survey \citep[HOPS;][]{walsh11,walsh14}. HOPS surveyed 100~deg$^2$ of our Galaxy for water masers and many other molecular lines with the Mopra Radio Telescope and the water maser detections were subsequently followed-up with the ATCA. HOPS is ideal for studying water masers in all the different environments that they form, however, for comparison with the MMB it lacks sensitivity (HOPS is estimated to be 98 per cent complete down to 8.4~Jy compared to the the MMB which is estimated to be approaching 100 per cent completeness at 1~Jy).

It has been proposed \citep[e.g.][]{ellingsen13,breen10ev,ellingsen07} that the presence and/or absence of different maser transitions correspond to different evolutionary stages of the parent YSO. Masers can be associated with ultra-compact H{\small{II}} regions \citep{phillips98,walsh98} and at younger stages embedded within infrared dark clouds \citep{ellingsen06}. \cite{breen10ev} have shown that YSO with only 6.7-GHz methanol masers are at an earlier stage of evolution than those with 12.2-GHz methanol and OH masers. How water masers fit into this evolutionary sequence remains unclear. There is now some evidence that water masers may not fit such a well defined evolutionary stage as other types of masers. Water masers can be found at a wider range of projected physical separations from their associated YSOs, and so may not always be closely tied to the physical conditions there \citep{breen14,titmarsh14}.

The conditions under which methanol and water masers form are also very different. 6.7-GHz methanol masers benefit from strong background radiation, however, this is not needed for 22-GHz water masers to exist and modelling suggests that this may even suppress them. In addition, after shock destruction, water masers re-form quickly \citep[e.g.][]{hollenbach13} whereas, methanol forms slowly on dust grains \citep{dartois99} and once it has been destroyed it is lost in that region during that star formation episode. Water and methanol masers also have very different excitation temperatures ($\sim$49~K and $\sim$646~K respectively) and conditions for strong inversion. The 6.7-GHz methanol masers are found in dust temperatures of $<$~500~K and their ideal number densities are 10$^{12}$~m$^{-3}$ -- 10$^{13}$~m$^{-3}$ and are quenched at $>$~10$^{14}$~m$^{-3}$ \citep{sobolev97}. These temperatures and densities are at the bottom of the range of conditions water masers can be found in. They can be found under a wide range of conditions, but they most favour temperatures of $\sim$800~K and number densities of 10$^{15}$~m$^{-3}$ \citep{gray15}.

Since masers are found within the dusty envelopes of high-mass star formation, is it useful to compare maser data to the thermal dust emission. Dust emission is optically thin at submillimetre wavelengths and is useful for tracing column densities and clump masses. Submillimetre emission allows us to study the coldest, densest regions where stars are forming. Recently, there have been two large, unbiased surveys of the Galactic Plane in the submillimetre regime. The Bolocam Galactic Plane Survey \citep[BGPS;][]{rosolowsky10} at 1.1~mm and the APEX Telescope Large Area Survey of the GALaxy \citep[ATLASGAL;][]{schuller09,contreras13,csengeri14} at 870~$\mu$m. 

In this paper we report the 22-GHz water maser follow-up of the MMB in the longitude range $l = 341^{\circ}$ through the Galactic centre to $l = 6^{\circ}$. All the MMB sources in the range $l = 310^{\circ} - 341^{\circ}$ have also been observed for water masers. In addition, during all our MMB follow-up observations we also observed the ammonia (1,1) and (2,2) transitions and, when available, an additional 2~x~2~GHz continuum bands with 32~x~64~MHz channels. The rest of the water maser data ammonia and continuum observations will be reported in future publications.

\section{Observations and data reduction}

The observing strategy for the water maser follow-up of the MMB survey is described in detail in \cite{titmarsh14}. Here we only repeat the essential details.

We targeted all the MMB sources for water maser emission in the range $l = 341^{\circ}$ through the Galactic centre to $l = 6^{\circ}$, including those that had been observed before, with the ATCA. We re-observed those that had been targeted before to ensure that we had a statistically complete sample of water masers at one epoch. The sources in the range $l = 341^{\circ} - 2^{\circ}$ were observed with the H214 array configuration on 2011 June 3-5 and $l = 2^{\circ} - 6^{\circ}$ were observed using the H168 configuration on 2011 August 8. At 22~GHz the primary beam of the ATCA is 2.1~arcminutes and the synthesised beam of the H214 and H168 configurations are $\sim$~9.6 and $\sim$~12.4 arcseconds respectively. 

For calibration we used PKS~B1934-638 for the primary flux density, PKS~B1253-055 for the bandpass and PKS~B1646-50 for the phase calibrator on the 2011 June 3 observations and PKS~B1714-336 for the 2011 June 4-5 and August 8 observations. We spent 1.5~minutes observing the phase calibrator every $\sim$~10 minutes and in between we observed groups of six nearby MMB targets for 1.5~minutes each. Each target was observed at least 4 times, giving a total on-source time of $\geq$~6~minutes, over an hour angle range of six hours or more to ensure sufficient \textit{uv}-coverage for imaging. 

We used the Compact Array Broadband Backend \citep[CABB;][]{wilson11} with two zoom bands of 64 MHz bandwidth with 2048 spectral channels. We centred the first band at the 22.235-GHz water maser transition and the second half-way between the ammonia (1,1) and (2,2) transitions at 23.708 GHz. This corresponded to a channel width in velocity of 0.42~km~s$^{-1}$ and 0.39~km~s$^{-1}$ for the water and ammonia bands respectively and velocity resolutions for uniform weighting of the spectral channels of 0.50~km~s$^{-1}$ and 0.47~km~s$^{-1}$ and velocity coverages of $>$~800~km~s$^{-1}$. The 5~$\sigma$ detection limit for these observations ranged from $\sim$125~mJy - $\sim$250~mJy depending on weather conditions and total time on-source. In addition 2 x 2 GHz continuum bands with 32 x 64 MHz channels were available for the 2011 August observations. 

The water maser data were reduced in {\sc miriad} \cite[]{sault95} using standard techniques for ATCA spectral line data and all the velocities were corrected to the Local Standard of Rest. We inspected the image cubes within an arcminute radius of each methanol maser position to find the associated water masers and the spectra were produced from these image cubes by integrating the emission at each maser site.  

For masers previously observed with the ATCA by  \cite[][]{breen10oh}, our positions agreed to within $\leq$~2 arcseconds. The astrometric accuracy of ATCA measurements of water masers is, under ideal conditions, around 0.4 arcseconds \citep{caswell97} and is set by the precision to which the coordinates of the phase calibrators have been measured and other systematics such as the antenna positions. For observations at 22 GHz, varying water vapour content in the atmosphere degrades the phase calibration and the typical astrometric accuracy for the current observations is estimated to be within an arcsecond in good weather conditions.

Many maser sites consist of a collection of emission "spots". The spread of the maser spots also limits the positional accuracy we can achieve for each maser site since there is some physical scatter of these spots. Hence, the measured position of the maser site can change as water masers are variable and a different spot may now be the strongest. However, from epoch to epoch we expect the change due to scatter in the strongest maser spots to be of the same order as the astrometric accuracy or lower.

\section{Results}
\label{sec:results}

\begin{figure}
\includegraphics[width=3.4in]{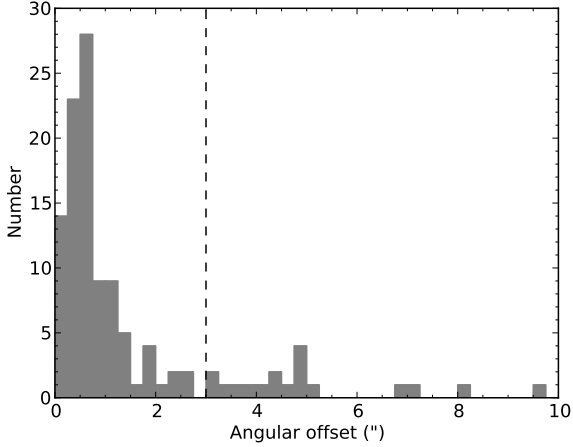}
\caption{Separations between methanol maser targets and our detected water masers. We found masers out to offsets of over 50 arcseconds, so we have zoomed in on the x-axis for clarity.}
\label{fig:offset_hist}
\end{figure}

Here we report the 22-GHz water masers detected towards the 6.7-GHz methanol masers in the Galactic longitude range 341$^{\circ}$ to 6$^{\circ}$. We found 101 of the 204 MMB sources in this range to have associated water maser emission ($\sim$ 50 per cent). Figure \ref{fig:offset_hist} shows a histogram of the offsets between the target methanol masers and the water masers detected around them. We have used an offset of $\leq$ 3.0 arcseconds from the 6.7-GHz methanol maser to establish association with the exception of G05.885--0.393 (for details see Section \ref{sec:comments}). Note: 3 arcseconds is not the positional uncertainty of our maser sites, rather the angular separation we deem reasonable to determine if masers are associated with the same YSO. We have used a 3~arcsecond criteria in this paper and in \cite{titmarsh14} as it will include all the real associations and is consistent with what has been used in other large, high resolution surveys of water masers with the ATCA such as \cite{breen10oh} and \cite{caswell10}. The spectra of the associated water masers taken with the  ATCA are presented in Figure \ref{fig:spectra} and comments on individual sources of interest are in Section \ref{sec:comments}. We have only determined maser associations based on their angular separation. We did not use their velocities as water masers are well known for showing emission a long way from the systemic velocity of the region often spanning many tens to hundreds of km~s$^{-1}$.

We have looked at the distributions of the physical separations in the plane of the sky as we have good distance estimates to these masers. The distances were taken from \cite{green11} or \cite{motogi11} where available, of which a few are astrometric distances but the majority are the kinematic distances with the near-far ambiguity resolved using H{\small{I}} self-absorption. For the remaining sources we used the 6.7-GHz methanol peak velocities and the \cite{reid09} rotation curve to estimate their distances. For 15 sources the \cite{reid09} rotation curve produced very unrealistic distance estimates, particularly near the Galactic centre where non-circular motions are large. For these sources we used estimates which utilise a Baysian approach which takes into account a range of distance estimations including any trigonometric parallax distances of nearby sources, CO and kinematic distances (Mark Reid, private communication).

\begin{figure}
\includegraphics[width=3.4in]{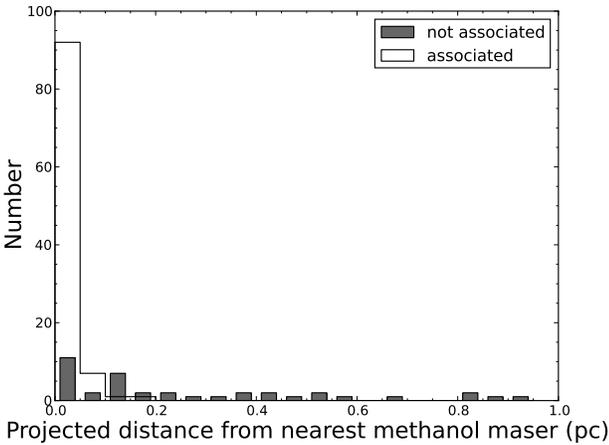}
\caption{Linear separations between methanol maser targets and our detected water masers. We only show up to 1~pc for clarity.}
\label{fig:linsep}
\end{figure}

The distribution of projected distances between the methanol and water masers is given in Figure \ref{fig:linsep}. Given the varying distances to the masers, determining association based on linear separations is desirable, however for ease of comparison with previous work and consistency with the previous instalment of this catalogue, we have used the 3 arcsecond angular separation criteria. This will make little difference to our results as most of the water masers we have classified as associated are within 0.05~pc of their methanol maser and all are within 0.2~pc. There are a small number of masers that we have classified as not associated within 0.05~pc. A couple are separated by less than 3 arcseconds, however, these were close pairs of methanol masers with only one water maser detected (within 3 arcseconds of both of them) and the water masers have been assigned to the closest methanol maser. The rest have separations greater than 3 arcseconds. Further analysis of linear separations has been left to future publications.

The results of the search are summarised in Table \ref{table:assoc_sources}: column 1 gives the source name (Galactic Longitude and Latitude) of the target 6.7-GHz methanol maser; columns 2 and 3 give the position of the water maser in Right Ascension and Declination (J2000); column 4 the peak velocity; columns 5 and 6 the minimum and maximum velocities of the emission; column 7 the peak flux densities; column 8 the integrated flux densities; columns 9, 10 and 11 the peak, minimum and maximum velocities of the associated 6.7-GHz methanol masers; column 12 the angular offsets; column 13 the epoch of the water observations and column 14 lists the distances to the MMB sources. 

Column 15 of Table \ref{table:assoc_sources} lists the associations with 12.2-GHz methanol and OH masers from \cite{breen12,caswell98} and \cite{caswell13}. Note that the 12.2-GHz associations are complete, coming from targeted observations towards the MMB sources. The OH observations are not complete towards all the MMB sources. \cite{caswell98} surveyed the Galactic Plane between $l = 312^{\circ} - 356^{\circ}$ and $|b| < 0.6^{\circ}$ and will have detected most of the masers above $\sim$3~Jy and we have indicated MMB sources that lie within the area of this survey, but no OH emission was detected. Also indicated in this column, are sources that were observed for water maser emission in \cite{breen10oh}.

MMB targets for which no associated water maser emission was detected are presented in Table \ref{table:notass_sources}. Column 1 is the source name of the target MMB maser; column 2 is the 5$\sigma$ detection limit; columns 3, 4 and 5 are the epochs, associations and distances respectively, the same as for Table \ref{table:assoc_sources}. Since water masers are variable, some of these methanol maser targets have been found to have water maser emission in past studies. The 15 sources where this has occurred were observed by \cite{breen10oh} to be associated with the target methanol masers and are discussed in Section \ref{sec:comments} and in Section \ref{sec:var}. Extending to a multi-epoch search by including these sources with the ones in Table \ref{table:assoc_sources}, we can partially account for variability over several years and increase to a detection rate of $\sim$~56 per cent. 

\begin{table*}
\caption{Positions and parameters of 101 water masers associated with 6.7-GHz methanol masers. RA and DEC are in J200 coordinates. Epochs are coded 1, 2, 3 and 4 for 2011 June 3, 2011 June 4, 2011 June 5 and 2011 August 8 respectively. Associations with 12.2-GHz methanol masers (Breen et al., 2012) are indicated with an `m', associations with the 1665~MHz transition of OH masers (Caswell, 1998; Caswell et al. 2013) are indicated with an `o' and sources marked with a '*' are within the survey region for OH by Caswell (1998) and no emission was detected. Sources observed for water masers in 2003 and/or 2004 by Breen et al. (2010a) are indicated with a `w'. Distances estimates are from Green et al. (2011) or Motogi et al. (2011) where available, others are the near kinematic distances (these are in italics) and the remainder are from Mark Reid (private communication) (these are in square brackets).}
\begin{tabular}{l l l l l l l l l l l l l l l}
\hline
MMB Target & \multicolumn{2}{l}{Equatorial Coordinates} & & & & & S$_{int}$ & \multicolumn{3}{l}{6.7-GHz methanol} & & & & \\
Name & RA & DEC & V$_{pk}$ & V$_{l}$ & V$_{h}$ & S$_{pk}$ & (Jy km & V$_{pk}$ & V$_{l}$ & V$_{h}$ & Offsets & Epoch & Dist. & Assoc. \\
(l, b) & (h m s) & ($^{\circ}$ ' ``) & \multicolumn{3}{c}{(km s$^{-1}$)} & (Jy) & s$^{-1}$) & \multicolumn{3}{c}{(km s$^{-1}$)} & (``) & & (kpc) & \\
\hline
341.218$-$0.212  & 16 52 17.88 & --44 26 52.8 & --39.5    & --51.1  & --34.8 & 106.   & 505.   & --37.9  & --50.0  & --35.0  & 0.6 & 1 & 3.1   & mow  \\
341.276$+$0.062  & 16 51 19.38 & --44 13 44.9 & --74.9    & --77.1  & --70.8 & 1.6    & 6.9    & --70.5  & --77.5  & --65.5  & 0.5 & 1 & 11.5  & mow  \\
341.973$+$0.233  & 16 53 02.96 & --43 34 55.1 & --13.9    & --17.1  & --8.7  & 1.2    & 6.1    & --11.5  & --12.5  & --10.5  & 0.4 & 1 & 1.0   & *   \\
342.446$-$0.072  & 16 55 59.91 & --43 24 23.1 & --20.2    & --28.2  & --18.6 & 1.0    & 4.3    & --30.0  & --32.5  & --14.0  & 0.7 & 1 & 2.0   & m*  \\
342.484$+$0.183  & 16 55 02.30 & --43 13 00.1 & --49.4    & --53.2  & --31.5 & 14.9   & 108.   & --41.9  & --44.5  & --38.5  & 0.3 & 1 & 12.9  & m*w  \\
342.954$-$0.019  & 16 57 30.63 & --42 58 34.6 & --6.1     & --15.8  & 2.7    & 3.3    & 20.9   & --4.1   & --14.0  & --2.0   & 0.3 & 1 & 0.7   & *   \\ 
343.502$-$0.472  & 17 01 18.46 & --42 49 37.3 & --48.6    & --50.7  & --25.5 & 3.6    & 26.0   & --42.0  & --43.0  & --32.0  & 0.9 & 1 & 3.0   & m*  \\
343.756$-$0.163  & 17 00 49.88 & --42 26 08.8 & --38.8    & --45.0  & --5.3  & 34.8   & 268.   & --30.8  & --32.5  & --24.0  & 0.7 & 1 & 2.5   & *  \\
344.227$-$0.569  & 17 04 07.90 & --42 18 39.3 & --1.4     & --52.1  & 10.1   & 11.0   & 230.   & --19.8  & --33.0  & --10.5  & 1.4 & 1 & 2.1   & mow  \\ 
344.581$-$0.024  & 17 02 57.71 & --41 41 53.9 & --3.4     & --35.3  & 19.1   & 366.   & 2840   & 1.6     & --5.0   & 2.5     & 0.2 & 1 & 16.2  & ow  \\ 
345.003$-$0.223  & 17 05 10.92 & --41 29 06.6 & --83.6    & --86.8  & 21.8   & 11.3   & 62.3   & --23.1  & --25.0  & --20.1  & 0.6 & 1 & 2.2   & mow  \\ 
345.012$+$1.797  & 16 56 46.88 & --40 14 08.2 & --11.3    & --31.2  & 9.6    & 88.7   & 534.   & --12.2  & --16.0  & --10.0  & 0.9 & 1 & 1.3   & mw   \\
345.131$-$0.174  & 17 05 23.23 & --41 21 10.6 & --27.1    & --32.3  & --20.9 & 4.9    & 40.8   & --28.9  & --31.0  & --28.0  & 0.2 & 1 & 2.7   & *   \\
345.407$-$0.952  & 17 09 35.43 & --41 35 56.3 & --16.2    & --25.4  & --11.8 & 0.4    & 3.3    & --14.3  & --15.5  & --14.0  & 0.8 & 1 & 1.5   & ow  \\
345.424$-$0.951  & 17 09 38.55 & --41 35 04.6 & --14.1    & --26.3  & --9.0  & 0.9    & 7.7    & --13.2  & --21.0  & --5.0   & 0.1 & 1 & 1.4   & w  \\
345.441$+$0.205  & 17 04 46.87 & --40 52 38.1 & --20.3    & --37.5  & 3.6    & 2.7    & 42.4   & 0.9     & --13.0  & 2.0     & 0.1 & 1 & 10.8  & *  \\
345.487$+$0.314  & 17 04 28.37 & --40 46 29.8 & 5.7       & --58.4  & 7.6    & 1.7    & 12.9   & --22.6  & --24.0  & --21.5  & 1.9 & 1 & 2.3   & *w   \\
345.505$+$0.348  & 17 04 22.95 & --40 44 24.6 & --22.4    & --26.2  & --9.6  & 4.8    & 45.4   & --17.8  & --23.1  & --10.5  & 3.0 & 1 & 10.8  & mow  \\
345.807$-$0.044  & 17 06 59.84 & --40 44 08.2 & 2.7       & --3.8   & 3.6    & 1.0    & 4.1    & --2.0   & --3.0   & --0.5   & 0.0 & 1 & 10.8  & m*  \\
345.824$+$0.044  & 17 06 40.70 & --40 40 09.8 & --20.4    & --24.1  & --1.2  & 8.1    & 44.5   & --10.3  & --12.0  & --9.0   & 0.1 & 1 & 10.9  & *  \\
345.949$-$0.268  & 17 08 23.61 & --40 45 21.3 & --8.1     & --11.8  & --3.8  & 1.2    & 7.8    & --21.9  & --22.5  & --21.4  & 0.3 & 1 & 14.1  & *   \\
345.985$-$0.020  & 17 07 27.57 & --40 34 43.4 & --77.6    & --90.5  & --69.9 & 11.3   & 77.7   & --84.1  & --85.5  & --81.7  & 0.2 & 1 & 11.0  & *   \\
346.036$+$0.048  & 17 07 20.01 & --40 29 48.9 & --12.3    & --20.5  & --7.6  & 5.0    & 19.7   & --6.4   & --14.5  & --3.9   & 0.1 & 1 & 10.9  & *  \\
346.231$+$0.119  & 17 07 39.05 & --40 17 52.6 & --91.9    & --109.4 & --90.2 & 3.1    & 10.0   & --95.0  & --96.6  & --92.6  & 0.6 & 1 & 10.7  & *  \\
346.517$+$0.117  & 17 08 33.07 & --40 04 14.7 & 2.2       & --4.3   & 3.5    & 0.6    & 3.8    & --1.7   & --3.0   & 1.0     & 1.4 & 1 & 10.9  & *  \\
346.522$+$0.085  & 17 08 42.21 & --40 05 08.4 & 8.1       & --5.9   & 10.2   & 2.3    & 16.8   & 5.7     & 4.7     & 6.1     & 1.1 & 1 & 10.9  & *w  \\
347.230$+$0.016  & 17 11 11.14 & --39 33 27.2 & --74.8    & --78.8  & --68.8 & 9.6    & 33.1   & --68.9  & --69.9  & --68.0  & 0.4 & 1 & 11.5  & *  \\
347.583$+$0.213  & 17 11 26.74 & --39 09 22.4 & --100.4   & --104.5 & --93.1 & 0.5    & 6.0    & --102.5 & --103.8 & --96.0  & 0.3 & 1 & 5.3  & m*  \\
347.628$+$0.149  & 17 11 50.97 & --39 09 29.8 & --92.3    & --95.0  & --90.4 & 6.7    & 20.3   & --96.5  & --98.9  & --95.0  & 0.9 & 1 & 5.3  & ow  \\
347.631$+$0.211  & 17 11 36.13 & --39 07 06.9 & --89.8    & --97.1  & --85.4 & 24.5   & 97.3   & --91.9  & --94.0  & --89.0  & 1.0 & 1 & 5.7  & *w   \\
348.550$-$0.979  & 17 19 20.30 & --39 03 51.8 & --24.1    & --25.4  & --15.4 & 0.8    & 4.1    & --10.6  & --19.0  & --7.0   & 1.3 & 1 & 1.7  & mo  \\
348.579$-$0.920  & 17 19 10.61 & --39 00 24.5 & --10.6    & --16.9  & --8.5  & 8.0    & 30.4   & --15.0  & --16.0  & --14.0  & 0.4 & 1 & 1.9  & o  \\
348.617$-$1.162  & 17 20 18.64 & --39 06 50.7 & --9.3     & --23.5  & --4.2  & 1.5    & 8.0    & --11.4  & --21.5  & --8.5   & 0.1 & 1 & 1.9  & m  \\
348.654$+$0.244  & 17 14 32.41 & --38 16 16.7 & 18.1      & 13.3    & 20.9   & 17.0   & 63.3   & 16.9    & 16.5    & 17.5    & 0.5 & 1 & 11.2  & *   \\
348.884$+$0.096  & 17 15 50.08 & --38 10 12.7 & --78.5    & --90.3  & --73.1 & 5.3    & 46.5   & --74.5  & --79.0  & --73.0  & 0.6 & 2 & 11.1  & mow  \\
348.892$-$0.180  & 17 17 00.17 & --38 19 28.2 & 7.1       & --18.3  & 13.3   & 1.5    & 14.0   & 1.5     & 1.0     & 2.0     & 0.9 & 2 & 11.2  & ow  \\
349.067$-$0.017  & 17 16 50.70 & --38 05 14.3 & 13.0      & --9.0   & 16.9   & 0.9    & 3.8    & 11.6    & 6.0     & 16.0    & 0.4 & 2 & 11.3  & ow  \\
349.092$+$0.106  & 17 16 24.54 & --37 59 45.8 & --80.5    & --87.3  & --54.6 & 15.8   & 115.   & --81.5  & --83.0  & --78.0  & 0.6 & 2 & \textit{5.6}  & mow  \\
349.151$+$0.021  & 17 16 55.86 & --37 59 47.7 & 15.2      & 12.5    & 22.9   & 0.8    & 6.2    & 14.6    & 14.1    & 25.0    & 0.3 & 2 & 11.3  & *   \\
349.799$+$0.108  & 17 18 27.70 & --37 25 03.5 & --60.7    & --73.5  & --55.0 & 4.1    & 18.2   & --62.4  & --65.5  & --57.4  & 0.5 & 2 & \textit{5.1}  & m*  \\
350.015$+$0.433  & 17 17 45.36 & --37 03 11.5 & --25.3    & --41.6  & --8.5  & 1.8    & 17.7   & --30.4  & --37.0  & --29.0  & 1.0 & 2 & 12.9  & ow  \\
350.104$+$0.084  & 17 19 26.64 & --37 10 53.0 & --74.0    & --83.3  & --62.1 & 15.2   & 140.   & --68.1  & --69.0  & --67.5  & 0.4 & 2 & \textit{5.3}  & *  \\
350.189$+$0.003  & 17 20 01.38 & --37 09 30.3 & --65.5    & --67.6  & --61.3 & 0.6    & 2.3    & --62.4  & --65.0  & --62.0  & 0.4 & 2 & \textit{5.1}  & *  \\
350.340$+$0.141  & 17 19 53.39 & --36 57 21.2 & --54.9    & --90.1  & --51.4 & 0.8    & 14.2   & --58.4  & --60.0  & --57.5  & 2.5 & 2 & 11.5  & m*  \\
350.356$-$0.068  & 17 20 47.54 & --37 03 41.6 & --71.3    & --74.4  & --63.1 & 0.4    & 1.6    & --67.6  & --68.5  & --66.0  & 0.3 & 2 & 11.2  & *  \\
350.520$-$0.350  & 17 22 25.40 & --37 05 13.4 & --26.9    & --38.4  & --9.2  & 3.3    & 30.8   & --24.6  & --25.0  & --22.0  & 1.0 & 2 & 3.0  & *  \\
350.686$-$0.491  & 17 23 28.59 & --37 01 48.6 & --14.1    & --17.1  & --11.5 & 22.0   & 57.5   & --13.8  & --15.0  & --13.0  & 0.5 & 2 & 2.1  & mow  \\
351.161$+$0.697  & 17 19 57.44 & --35 57 53.2 & --3.7     & --17.4  & 2.8    & 45.1   & 341.   & --5.2   & --7.0   & --2.0   & 0.8 & 2 & 1.8  & ow  \\
351.417$+$0.645  & 17 20 53.36 & --35 47 00.0 & --7.3     & --43.4  & 13.7   & 156.   & 2800   & --10.4  & --12.0  & --6.0   & 1.1 & 2 & \textit{1.7}  & mow  \\
351.581$-$0.353  & 17 25 25.20 & --36 12 44.0 & --91.6    & --112.8 & --58.8 & 110.   & 864.   & --94.2  & --100.0 & --88.0  & 2.3 & 2 & 5.1  & ow  \\
351.611$+$0.172  & 17 23 21.23 & --35 53 32.3 & --20.1    & --88.7  & --17.2 & 1.1    & 27.2   & --43.7  & --46.0  & --31.5  & 0.3 & 2 & \textit{4.6}  & m*  \\
351.775$-$0.536  & 17 26 42.57 & --36 09 19.1 & --0.4     & --28.7  & 28.1   & 61.9   & 769.   & 1.3     & --9.0   & 3.0     & 1.6 & 2 & 0.4  & mo  \\ 
352.133$-$0.944  & 17 29 22.26 & --36 05 00.5 & --12.4    & --13.6  & 2.6    & 2.2    & 12.6   & --7.8   & --18.8  & --5.6   & 0.7 & 2 & 2.1  & w   \\
352.517$-$0.155  & 17 27 11.29 & --35 19 32.0 & --48.6    & --50.4  & --44.3 & 2.2    & 10.0   & --51.3  & --52.0  & --49.0  & 0.7 & 2 & 11.5  & ow  \\
352.584$-$0.185  & 17 27 29.51 & --35 17 14.3 & --77.2    & --79.7  & --76.4 & 1.2    & 2.6    & --85.6  & --92.6  & --79.7  & 0.8 & 2 & 5.1  & *   \\
352.630$-$1.067  & 17 31 13.82 & --35 44 08.5 & 10.6      & --18.2  & 18.1   & 14.1   & 126.   & --3.0   & --8.0   & --2.0   & 1.1 & 2 & 0.9  & ow  \\ 
352.855$-$0.201  & 17 28 17.61 & --35 04 12.5 & --60.2    & --62.7  & --5.6  & 0.8    & 9.6    & --51.4  & --54.1  & --50.1  & 0.5 & 2 & 11.3  & *   \\
\end{tabular}
\label{table:assoc_sources}
\end{table*}

\begin{table*}\addtocounter{table}{-1}
  \caption{-- {\emph {continued}}}
\begin{tabular}{l l l l l l l l l l l l l l l}
\hline
MMB Target & \multicolumn{2}{l}{Equatorial Coordinates} & & & & & S$_{int}$ & \multicolumn{3}{l}{6.7-GHz methanol} & & & & \\
Name & RA & DEC & V$_{pk}$ & V$_{l}$ & V$_{h}$ & S$_{pk}$ & (Jy km & V$_{pk}$ & V$_{l}$ & V$_{h}$ & Offsets & Epoch & Dist. & Assoc. \\
(l, b) & (h m s) & ($^{\circ}$ ' ``) & \multicolumn{3}{c}{(km s$^{-1}$)} & (Jy) & s$^{-1}$) & \multicolumn{3}{c}{(km s$^{-1}$)} & (``) & & (kpc) & \\
\hline
353.216$-$0.249  & 17 29 27.68 & --34 47 48.3 & --14.5    & --51.4  & 10.5   & 4.8    & 30.6   & --23.0  & --25.0 & --15.0  & 1.8 & 2 & 3.3  & *  \\
353.273$+$0.641  & 17 26 01.55 & --34 15 15.1 & --52.2    & --117.8 & 6.3    & 182.   & 954.   & --4.4   & --7.0  & --3.0   & 0.4 & 2 & \textit{0.8}  & w  \\
353.537$-$0.091  & 17 29 41.25 & --34 26 28.6 & --60.3    & --72.1  & --55.6 & 0.6    & 3.7    & --56.6  & --59.0 & --54.0  & 0.2 & 2 & 11.0  & m*  \\
354.308$-$0.110  & 17 31 48.52 & --33 48 29.2 & 9.1       & 4.0     & 20.9   & 6.1    & 39.7   & 18.7    & 11.0   & 19.5    & 0.5 & 2 & [11.6]  & *  \\  
355.344$+$0.147  & 17 33 29.02 & --32 47 58.8 & 10.0      & 5.7     & 124.0  & 12.7   & 38.0   & 19.9    & 19.0   & 21.0    & 0.4 & 2 & [11.5]  & o  \\  
355.346$+$0.149  & 17 33 28.89 & --32 47 49.5 & 9.2       & 6.8     & 21.3   & 4.9    & 16.7   & 9.9     & 9.0    & 12.5    & 0.3 & 2 & [11.5]  & *w   \\  
355.538$-$0.105  & 17 34 59.58 & --32 46 23.2 & --1.5     & --4.7   & 0.9    & 2.3    & 7.1    & 3.8     & --3.5  & 5.0     & 0.6 & 2 & 17.7  & *  \\ 
355.666$+$0.374  & 17 33 24.89 & --32 24 21.0 & --3.3     & --5.6   & 3.1    & 0.4    & 2.0    & --3.4   & --4.5  & 0.6     & 0.3 & 2 & 16.4  & *  \\  
357.558$-$0.321  & 17 40 57.15 & --31 10 59.6 & --4.6     & --56.6  & 38.8   & 28.7   & 455.   & --3.9   & --5.5  & 0.0     & 0.6 & 2 & \textit{1.8}  & --  \\
357.922$-$0.337  & 17 41 55.00 & --30 52 54.8 & 4.0       & --14.1  & 7.9    & 1.4    & 8.0    & --4.9   & --5.5  & --4.0   & 0.8 & 3 & 2.5  & --  \\  
357.965$-$0.164  & 17 41 20.10 & --30 45 16.7 & --5.2     & --61.1  & 37.3   & 3.3    & 34.1   & --8.8   & --9.0  & 3.0     & 2.3 & 3 & 3.9 & w \\
357.967$-$0.163  & 17 41 20.27 & --30 45 06.2 & 0.6       & --74.2  & 103.1  & 57.8   & 2220   & --4.2   & --6.0  & 0.0     & 0.6 & 3 & 2.2  & mow  \\  
358.386$-$0.483  & 17 43 37.69 & --30 33 50.2 & --1.9     & --6.8   & 4.2    & 7.1    & 28.3   & --6.0   & --7.0  & --5.0   & 1.9 & 3 & \textit{3.5}  & ow  \\
358.460$-$0.391  & 17 43 26.72 & --30 27 11.8 & --4.5     & --19.5  & --1.8  & 3.2    & 28.0   & 1.2     & --0.5  & 4.0     & 0.7 & 3 & [3.5]  & --  \\   
358.931$-$0.030  & 17 43 09.96 & --29 51 46.0 & --21.6    & --25.3  & --12.5 & 0.7    & 7.2    & --15.9  & --22.0 & --14.5  & 0.7 & 3 & \textit{6.4}  & --  \\
358.980$+$0.084  & 17 42 50.32 & --29 45 41.4 & --8.1     & --10.3  & 6.1    & 0.4    & 3.0    & 6.2     & 5.0    & 7.0     & 1.9 & 3 & [3.8]  & --  \\  
359.138$+$0.031  & 17 43 25.62 & --29 39 17.3 & --1.3     & --121.4 & 49.7   & 304.   & 2300   & --3.9   & --7.0  & 1.0     & 0.6 & 3 & \textit{3.7}  & ow  \\
359.436$-$0.104  & 17 44 40.56 & --29 28 15.8 & --56.7    & --61.1  & --38.8 & 13.4   & 70.4   & --46.7  & --53.0 & --45.0  & 0.5 & 3 & \textit{7.9}  & mow  \\
359.436$-$0.102  & 17 44 40.05 & --29 28 12.4 & --57.1    & --61.2  & --54.1 & 13.2   & 80.3   & --54.0  & --58.0 & --53.6  & 2.1 & 3 & \textit{8.0}  & w   \\
359.615$-$0.243  & 17 45 39.08 & --29 23 30.3 & 23.4      & --6.1   & 65.7   & 80.2   & 457.   & 22.6    & 14.0   & 27.0    & 0.4 & 3 & \textit{7.7}  & mow  \\
359.970$-$0.457  & 17 47 20.18 & --29 11 59.0 & 4.7       & 2.7     & 12.8   & 1.1    & 6.1    & 23.8    & 20.0   & 24.1    & 0.4 & 3 & \textit{8.3}  & ow  \\
0.167$-$0.446    & 17 47 45.46 & --29 01 29.4 & 14.0      & 7.6     & 18.1   & 2.8    & 22.5   & 13.8    & 9.5    & 17.0    & 0.2 & 3 & \textit{8.0}  & --   \\
0.315$-$0.201    & 17 47 09.10 & --28 46 16.0 & 19.9      & 13.3    & 33.8   & 2.9    & 18.7   & 19.4    & 14.0   & 27.0    & 0.5 & 3 & \textit{7.8}  & mw  \\
0.376$+$0.040    & 17 46 21.38 & --28 35 39.9 & 38.8      & --15.6  & 71.5   & 119.   & 483.   & 37.0    & 35.0   & 40.0    & 0.4 & 3 & \textit{8.0}  & ow  \\
0.496$+$0.188    & 17 46 03.95 & --28 24 51.8 & 0.9       & --9.9   & 20.8   & 1.5    & 12.4   & 0.8     & --12.0 & 2.0     & 1.0 & 3 & \textit{3.6}  & mow  \\
0.546$-$0.852    & 17 50 14.41 & --28 54 30.1 & 17.3      & --95.0  & 123.7  & 204.   & 3250   & 11.8    & 8.0    & 20.0    & 1.3 & 3 & \textit{7.0}  & mow  \\
0.836$+$0.184    & 17 46 52.80 & --28 07 36.0 & --52.8    & --58.7  & 16.1   & 0.9    & 20.3   & 3.6     & 2.0    & 5.0     & 1.4 & 3 & \textit{4.6}  & m  \\
1.008$-$0.237    & 17 48 55.28 & --28 11 48.0 & 0.8       & --0.5   & 3.2    & 1.7    & 4.5    & 1.6     & 1.0    & 7.0     & 0.2 & 3 & \textit{2.9}  & --   \\
1.147$-$0.124    & 17 48 48.50 & --28 01 11.1 & --22.2    & --25.4  & 0.8    & 4.3    & 51.6   & --15.3  & --20.5 & --14.0  & 0.3 & 3 & \textit{4.1}  & --   \\
2.143$+$0.009    & 17 50 36.11 & --27 05 47.1 & 60.0      & 54.9    & 79.4   & 1.7    & 13.0   & 62.7    & 54.0   & 65.0    & 0.7 & 4 & \textit{7.3}  & ow  \\
2.521$-$0.220    & 17 52 21.15 & --26 53 20.3 & 2.4       & --0.8   & 4.1    & 3.8    & 18.6   & 4.2     & --7.5  & 5.0     & 0.8 & 4 & \textit{2.6}  & --   \\
2.536$+$0.198    & 17 50 46.49 & --26 39 45.1 & 4.9       & --0.9   & 12.5   & 0.5    & 5.4    & 3.2     & 2.0    & 20.5    & 0.4 & 4 & \textit{2.2}  & mw  \\
2.591$-$0.029    & 17 51 46.70 & --26 43 50.5 & --12.2    & --15.2  & --8.9  & 2.8    & 12.4   & --8.2   & --9.5  & --4.0   & 0.7 & 4 & [4.2]  & --   \\  
2.615$+$0.134    & 17 51 12.28 & --26 37 36.7 & 97.6      & 95.7    & 103.2  & 0.5    & 3.2    & 94.5    & 93.5   & 104.0   & 0.5 & 4 & \textit{7.5}  & m  \\
3.312$-$0.399    & 17 54 50.02 & --26 17 48.6 & 7.4       & 0.3     & 10.3   & 3.4    & 12.9   & 0.5     & 0.0    & 10.0    & 3.0 & 4 & \textit{0.6}  & --   \\
3.502$-$0.200    & 17 54 30.08 & --26 02 00.3 & 47.5      & 45.3    & 48.5   & 0.6    & 1.5    & 43.9    & 43.0   & 45.5    & 1.0 & 4 & \textit{6.3}  & m  \\
4.434$+$0.129    & 17 55 19.73 & --25 03 44.6 & --6.0     & --28.0  & 53.8   & 2.0    & 26.9   & --0.9   & --1.5  & 8.0     & 0.2 & 4 & [4.3]  & --  \\  
4.676$+$0.276    & 17 55 18.32 & --24 46 45.3 & --0.9     & --6.9   & 1.7    & 0.2    & 1.5    & 4.4     & --5.5  & 6.0     & 0.2 & 4 & \textit{1.6}  & --  \\
5.618$-$0.082    & 17 58 44.90 & --24 08 38.4 & --27.4    & --33.0  & --15.6 & 0.3    & 3.5    & --27.0  & --28.0 & --18.5  & 2.4 & 4 & 5.1  & m  \\
5.630$-$0.294    & 17 59 34.52 & --24 14 23.6 & 18.0      & 16.1    & 20.9   & 1.8    & 6.9    & 10.6    & 9.0    & 22.0    & 1.1 & 4 & 3.4  & m  \\
5.657$+$0.416    & 17 56 56.53 & --23 51 41.3 & 18.8      & 12.6    & 21.2   & 4.6    & 30.4   & 20.1    & 13.0   & 22.0    & 0.7 & 4 & 13.1  & --  \\
5.677$-$0.027    & 17 58 39.94 & --24 03 56.7 & --9.3     & --14.4  & 0.1    & 1.4    & 9.8    & --11.5  & --14.5 & --11.0  & 0.6 & 4 & [4.5]  & --  \\  
5.885$-$0.393    & 18 00 30.44 & --24 04 00.8 & 11.8      & --59.7  & 46.0   & 46.3   & 269.   & 6.7     & 6.0    & 7.5     & 3.8 & 4 & 1.2   & ow  \\  
5.900$-$0.430    & 18 00 40.72 & --24 04 18.9 & 12.3      & --13.4  & 30.4   & 2.2    & 28.2   & 10.4    & 0.0    & 10.6    & 2.6 & 4 & 1.6   & w   \\  
\end{tabular}
\end{table*}

\begin{table*}
\caption{6.7-GHz methanol masers with no associated water maser emission. Column 1 is the name of the target methanol maser given in Galactic coordinates; column 2 is the 5 sigma detection limit; column 3 is the epoch of observation coded 1, 2, 3 and 4 for 2011 June 3, 2011 June 4, 2011 June 5 and 2011 August 8 respectively. Column 4 gives associations with water masers detected in Pillai et al. (2006b) or the Breen et al. (2010a) 2003 or 2004 observations but not in ours are indicated with a `w', associations with the 1665~MHz transition of OH masers (Caswell, 1998; Caswell et al. 2013) are indicated with an `o' and sources marked with a '*' are within the survey region for OH by Caswell (1998) and no emission was detected. Column 5 is the distance to the methanol maser. Distances estimates are from Green et al. (2011) where available, others are the near kinematic distance (these are in italics) and the remainder are from Mark Reid (private communication) (these are in square brackets).}
\begin{tabular}{ccccc @{\hskip 2cm} ccccc}
\hline
MMB Target & Det. & Epoch & Assoc. & & MMB Target & Det. & Epoch & Assoc. &  \\
Source Name &  lim. & & & Distance & Source Name &  lim. & & & Distance \\
(l, b) & (mJy) & & & (kpc) & (l, b) & (mJy) & & & (kpc) \\
\hline
341.124$-$0.361    & 250 & 1 & *   & 3.0            &    353.378$+$0.438         & 200 & 2 & * & 13.9 \\
341.238$-$0.270    & 250 & 1 & m*  & 3.5            &    353.410$-$0.360         & 200 & 2 & mo & \textit{3.4} \\
341.367$+$0.336    & 250 & 1 & *   & 11.2           &    353.429$-$0.090         & 200 & 2 & * & 11.1 \\
341.990$-$0.103    & 250 & 1 & *   & 3.0            &    353.464$+$0.562         & 200 & 2 & wo & 11.2 \\
342.251$+$0.308    & 250 & 1 & *   & 9.9            &    354.206$-$0.038         & 200 & 2 & * & \textit{5.0} \\
342.338$+$0.305    & 250 & 1 & m*  & 10.3           &    354.496$+$0.083         & 200 & 2 & m* & 11.7 \\
342.368$+$0.140    & 250 & 1 & *   & 0.6            &    354.615$+$0.472         & 200 & 2 & wmo & 3.8 \\  
343.354$-$0.067    & 250 & 1 & m*  & 9.9            &    354.701$+$0.299         & 200 & 2 & * & 6.1 \\
343.929$+$0.125    & 250 & 1 & m*  & 18.6           &    354.724$+$0.300         & 200 & 2 & mo & 5.8 \\
344.419$+$0.044    & 250 & 1 & *   & 4.4            &    355.184$-$0.419         & 210 & 2 & * & 0.1  \\  
344.421$+$0.045    & 250 & 1 & wm* & 4.7            &    355.343$+$0.148         & 215 & 2 & wmo & [1.6]  \\  
345.003$-$0.224    & 250 & 1 & wm* & 2.7            &    355.545$-$0.103         & 225 & 2 & * & 11.7  \\  
345.010$+$1.792    & 250 & 1 & wm  & 2.0            &    355.642$+$0.398         & 225 & 2 & * & 14.5  \\  
345.198$-$0.030    & 250 & 1 & m*  & 10.8           &    356.054$-$0.095         & 225 & 2 & * & [11.4] \\  
345.205$+$0.317    & 250 & 1 & *   & 11.8           &    356.662$-$0.263         & 225 & 2 & o & 6.6 \\
345.498$+$1.467    & 250 & 1 & --  & 1.5            &    357.559$-$0.321         & 225 & 3 & m & [4.1] \\  
345.576$-$0.225    & 250 & 1 & *   & 5.5            &    357.924$-$0.337         & 210 & 3 & m & 16.9  \\  
346.480$+$0.221    & 250 & 1 & m*  & 14.4           &    358.263$-$2.061         & 200 & 3 & m & [2.2] \\              
346.481$+$0.132    & 250 & 1 & w*  & 10.9           &    358.371$-$0.468         & 200 & 3 & wm & [3.4] \\             
347.817$+$0.018    & 250 & 1 & *   & 13.8           &    358.460$-$0.393         & 200 & 3 & -- & \textit{4.1} \\      
347.863$+$0.019    & 250 & 1 & m*  & 13.1           &    358.721$-$0.126         & 200 & 3 & m & \textit{0.6} \\       
347.902$+$0.052    & 250 & 1 & m*  & \textit{2.9}   &    358.809$-$0.085         & 200 & 3 & m & \textit{7.7} \\       
348.027$+$0.106    & 250 & 1 & *   & \textit{6.4}   &    358.841$-$0.737         & 180 & 3 & m & \textit{6.7} \\       
348.195$+$0.768    & 250 & 1 & m   & \textit{16.5}  &    358.906$+$0.106         & 175 & 3 & m & \textit{6.6} \\        
348.550$-$0.979n   & 250 & 1 & m   & 2.2            &    359.938$+$0.170         & 170 & 3 & -- & [7.2] \\              
348.723$-$0.078    & 250 & 1 & m*  & 11.2           &    0.092$+$0.663           & 150 & 3 & m & \textit{8.2} \\       
348.703$-$1.043    & 250 & 1 & m   & 1.3            &    0.212$-$0.001           & 160 & 3 & wm & \textit{8.2} \\      
348.727$-$1.037    & 125 & 2 & m   & 1.2            &    0.316$-$0.201           & 200 & 3 & w & \textit{7.9} \\       
349.092$+$0.105    & 125 & 2 & wm* & 11.1           &    0.409$-$0.504           & 225 & 3 & -- & \textit{7.8} \\      
349.579$-$0.679    & 150 & 2 & --  & 13.5           &    0.475$-$0.010           & 225 & 3 & -- & \textit{7.8} \\      
349.884$+$0.231    & 160 & 2 & *   & 11.3           &    0.645$-$0.042           & 250 & 3 & -- & \textit{7.9} \\      
350.011$-$1.342    & 170 & 2 & o   & \textit{3.1}   &    0.647$-$0.055           & 250 & 3 & -- & \textit{7.9} \\      
350.105$+$0.083    & 175 & 2 & wm* & \textit{5.5}   &    0.651$-$0.049           & 250 & 3 & m & \textit{7.9} \\       
350.116$+$0.084    & 175 & 2 & *   & 11.4           &    0.657$-$0.041           & 250 & 3 & wo & \textit{7.9} \\      
350.116$+$0.220    & 175 & 2 & m*  & 17.8           &    0.665$-$0.036           & 250 & 3 & -- & \textit{8.0} \\      
350.299$+$0.122    & 175 & 2 & wm* & 11.3           &    0.666$-$0.029           & 250 & 3 & m & \textit{8.0} \\       
350.344$+$0.116    & 175 & 2 & m*  & 11.4           &    0.667$-$0.034           & 250 & 3 & m & \textit{8.0} \\       
350.470$+$0.029    & 175 & 2 & m*  & 1.3            &    0.672$-$0.031           & 250 & 3 & -- & \textit{8.0} \\      
350.776$+$0.138    & 175 & 2 & *   & 11.4           &    0.673$-$0.029           & 250 & 3 & -- & \textit{8.0} \\      
351.242$+$0.670    & 175 & 2 & --  & 1.8            &    0.677$-$0.025           & 240 & 3 & -- & \textit{8.1} \\      
351.251$+$0.652    & 175 & 2 & --  & 1.8            &    0.695$-$0.038           & 230 & 3 & -- & \textit{8.0} \\      
351.382$-$0.181    & 175 & 2 & m*  & 5.4            &    1.329$+$0.150           & 225 & 3 & -- & \textit{17.0} \\      
351.417$+$0.646    & 175 & 2 & wmo & 1.8            &    1.719$-$0.088           & 225 & 3 & m & [4.2] \\              
351.445$+$0.660    & 185 & 2 & m   & 1.17           &    2.703$+$0.040           & 125 & 4 & m & \textit{7.5} \\        
351.688$+$0.171    & 185 & 2 & m*  & 12.1           &    3.253$+$0.018           & 125 & 4 & m & \textit{1.4} \\        
352.083$+$0.167    & 185 & 2 & m*  & 11.0           &    3.442$-$0.348           & 125 & 4 & -- & \textit{2.0} \\      
352.111$+$0.176    & 190 & 2 & wm* & 5.3            &    3.910$+$0.001           & 125 & 4 & o & \textit{4.4} \\        
352.525$-$0.158    & 200 & 2 & w*  & 11.2           &    4.393$+$0.079           & 125 & 4 & m & \textit{1.0} \\        
352.604$-$0.225    & 200 & 2 & *   & 5.1            &    4.569$-$0.079           & 135 & 4 & -- & \textit{2.8} \\      
352.624$-$1.077    & 200 & 2 & --  & 17.8           &    4.586$+$0.028           & 140 & 4 & -- & \textit{4.8} \\       
353.363$-$0.166    & 200 & 2 & *   & 5.1            &    4.866$-$0.171           & 150 & 4 & -- & \textit{1.8} \\      
353.370$-$0.091    & 200 & 2 & m*  & \textit{5.2}   &    & & & & \\
\end{tabular}
\label{table:notass_sources}
\end{table*}

\subsection{Comments on individual sites of maser emission}
\label{sec:comments}

\textit{341.218--0.212}. This water maser showed strong variation in its flux density, with a peak flux density of 120~Jy in 2003 and 33~Jy in the 2004 observations of \cite{breen10oh}. In our observations in 2011 it had increased again to 106.~Jy. Although the intensity varied greatly, the velocities of the emission remained similar over these three epochs. 

\textit{345.003--0.223 and 345.003--0.224}. This pair of methanol masers has been found to show variability in both sites \citep{caswell95}. \cite{goedhart04} monitored these methanol masers over more than four years and found no periodicities on these timescales. The water maser associated with 345.003--0.223 was also detected by \cite{breen10oh} in 2003 to have a peak flux density of 3.0~Jy, in 2004 at 3.7~Jy. It had two main features and \cite{breen10oh} found the peak to be the feature at 15~kms$^{-1}$, but in our observations in 2011 the other feature at -83.6~kms$^{-1}$ had flared to 11.3~Jy to become the peak.

\textit{345.010+1.792 and 345.012+1.797}. 345.010+1.792 is associated with an UCH{\small{II}} region \citep{caswell97}, and many other class II methanol maser transitions \citep{ellingsen12}. \cite{breen10oh} detected water maser emission in 2003 with a peak flux density of 2.0~Jy, but it was not detected in their 2004 observations (detection limit of 0.2~Jy) or in our observations. 345.012+1.797 was observed by \cite{breen10oh} in 2003 and 2004 and in our observations to have peak flux densities of 50, 29 and 88.7~Jy, respectively. 

\textit{349.092+0.106}. This water maser showed strong variation in its flux density, with a peak flux density of 34~Jy in 2003 and 154~Jy in the 2004 observations of \cite{breen10oh}. In our observations in 2011 it had decreased again to 15.8~Jy. Although the intensity varied greatly, the overall shape of the spectra remained similar with the peak velocity at $\sim$80~kms$^{-1}$ in all three epochs.

\textit{351.417+0.645}. In 2003 this water maser had peak flux density of 1400~Jy by \cite[][note it was not observed by them in 2004]{breen10oh}. In our observations its peak flux density had decreased to 156.~Jy, and the peak was a different feature. The spectra at both epochs had many features, but the spectra in 2003 had a larger total velocity range than in 2011 (-58 to 50~kms$^{-1}$ compared to -43.4 to 13.7~kms$^{-1}$).

\textit{351.581--0.353}. In 2003 this water maser had peak flux density of 1600~Jy observed by \cite{breen10oh} (note it was not observed by them in 2004). In our observations in 2011 the spectra maintained similar velocities although its peak flux density had decreased to 110.~Jy. 

\textit{352.630--1.067}. This water maser showed strong variation in its flux density, with a peak flux density of 35~Jy in 2003 and 700~Jy in the 2004 observations of \cite{breen10oh}. The peak feature was in the centre of the spectra $\sim$0~kms$^{-1}$ for the \cite{breen10oh} observations and only this feature showed showed strong variation with the features on either side remaining similar in each epoch. In our observations the peak feature was at 10.6~kms$^{-1}$ (peak flux density of 14.1~Jy) and the central feature at $\sim$0~kms$^{-1}$ had decreased to $\sim$10~Jy. 

\textit{353.273+0.641}. This water maser was identified by \cite{caswell08} to be associated with an unusual water maser dominated by a blue-shifted outflow. It was observed by \cite{breen10oh} in 2004 with peak flux density of 366~Jy, in 2007 by \cite{caswell08} with a peak of 45~Jy and in our observations in 2011 with a peak flux density of 182.~Jy. It remained dominated by the blue-shifted emission at all epochs with the strongest emission coming from the features clustered around $\sim$50~kms$^{-1}$.

\textit{357.965--0.164 and 357.967--0.163}. This is a pair of two distinct methanol maser sites with a small separation (7 arcseconds). We used the near kinematic distances of 3.9 and 2.2~kpc respectively (they are the same within errors) rather than the H{\small{I}} self-absorption distance from \cite{green11} of 15.2~kpc. We have chosen the near distances as 357.965--0.164 is also associated with a rare 23.4-GHz class I methanol maser and very strong 9.9-GHz class I methanol maser whose flux density is more than an order of magnitude stronger than in other known sources \citep{voronkov11} and the \cite{green11} distance was identified as unreliable in their paper. The systemic velocity of the region is $\sim$~--3.0~kms$^{-1}$ as that is the median velocity of the 6.7-GHz methanol maser emission for both sources. \textit{357.967--0.163} has the stronger methanol maser emission and is also accompanied by a 12.2-GHz methanol maser and an OH maser \citep{breen12,caswell13}. It's water maser emission is continuous over 177~kms$^{-1}$. In our observations it had a velocity range of --74.2 to 103.1~kms$^{-1}$ with a peak velocity of 0.6~kms$^{-1}$ and a peak flux density of 57.8~Jy. \cite{breen10oh} observed this source in 2003 and found it to have continuous emission over --80 to +100~kms$^{-1}$ with a peak velocity of 0~kms$^{-1}$ and a peak flux density of 40~Jy and in 2004 a velocity range of --81 to +87~kms$^{-1}$, peak velocity of --65~kms$^{-1}$ and peak flux density 57~Jy. \textit{357.965--0.164} is the weaker 6.7-GHz methanol maser of the two with the smaller velocity range with most of its emission around the systemic velocity (--6.0 to 0.0~kms$^{-1}$) and no OH counterpart. It's water maser is also the weaker of the two and has a velocity range of --61.1 to 37.3~kms$^{-1}$, but most of the strong emission is within a couple of~kms$^{-1}$ of the peak (--5.2~kms$^{-1}$), close to the systemic velocity of the region. The higher velocity features are very weak in comparison ($<$~0.5~Jy).

\textit{359.615--0.243}. A strong methanol maser identified to be variable by \cite{caswell95} and monitored by \cite{goedhart04}, and no periodicities in the variability were observed. The associated water maser emission is also quite variable, in 2003 it had a peak flux density of 7~Jy, in 2004 14~Jy \citep{breen10oh} and in our 2011 observations 80.2~Jy.

\textit{0.546--0.852}.  The water maser has velocity range of 218.7~kms$^{-1}$. Water maser emission was previously detected by \cite{forster99} who found the velocities of the maser site to range from --12.95 to 56.90~kms$^{-1}$, these were some of the features in the middle of the total velocity range we observed in our observations (see Figure \ref{fig:spectra}). \cite{breen10oh} found an even greater velocity range of --60 to 110~kms$^{-1}$ in 2003, and in our observations we found weaker high-velocity features out to --95.0 and +123.7~kms$^{-1}$. 

\textit{5.885--0.393}. This source has been studied extensively in water, OH and methanol and the latest water information from \cite{motogi11} gives an astrometric distance of 1.28$^{+0.09}_{-0.08}$~kpc, and confirms the very large angular extent of more than 4 arcseconds, which is not surprising at this quite low distance. Some of the \cite{motogi11} water features are within 2 arcseconds of MMB methanol maser, so even though we find the angular offset to be $>$~3 arcseconds we can establish that this water maser emission is associated with the target methanol maser.

\section{Discussion}

\cite{titmarsh14} found $\sim$~46 per cent of the 6.7-GHz methanol masers had associated water maser emission. This is consistent within uncertainties with this sample where we found approximately 50 per cent. To estimate the uncertainty in our number of detections, we have used $\sqrt{N_{det}}$, where $N_{det}$ is the number of detections, however, this assumes that the non-detections are due to purely stochastic effects. We know that source evolution and variability play a major role in the detection statistics, however, neither of these are well enough understood at present to be able to account for them explicitly. Our results are also consistent with previous searches for water masers by \cite{beuther02,szymczak05,xu08,breen10oh}. 

The previous searches most similar to this study are those by \cite{szymczak05} and \cite{breen10oh}. \cite{szymczak05} searched a statistically complete, but flux-limited (peak intensity greater than 1.6~Jy), sample of methanol masers with the Effelsberg 100~m telescope for water masers. They found 41 out of 79 methanol masers had associated water masers (a 52 per cent detection rate, rms noise of 0.45~Jy). Our detection rate is consistent within the combined uncertainties, although our sample has significantly higher sensitivity (both in target methanol maser sample and water maser observations) and higher angular resolution. 

\cite{breen10oh} searched a large sample of OH and 6.7-GHz methanol maser targets with the ATCA for water masers (5-$\sigma$ detection limit in a 1~km~s$^{-1}$ channel typically less than 0.2~Jy) in 2003 and 2004. 198 of their 270 methanol masers had an associated water maser in this two epoch search; a detection rate of 73 per cent. The main difference between our sample of methanol masers and the \cite{breen10oh} sample, is that our sample is statistically complete. The \cite{breen10oh} methanol masers were primarily those with associated OH masers or those without an OH maser which had an accurate position from \cite{caswell09}, which was biased towards sources with a higher 6.7-GHz peak flux density.

Another major survey of water masers is the H$_2$O Southern Galactic Plane Survey (HOPS) which conducted an unbiased search for water masers in Galactic longitudes 290$^{\circ}$ through zero to 30$^{\circ}$ and latitudes of $\pm 0.5^{\circ}$ \citep{walsh11,walsh14}. In the overlapping region between our survey and the HOPS survey, there are 172 MMB masers. Of those, \cite{walsh14} detected 57 to be associated with water maser emission ($\sim$~33 per cent). This is much lower than our survey, and is due the significantly lower sensitivity of the initial HOPS search which has an rms noise of 1 - 2~Jy over most of the survey region and hence is likely to have detected the majority of sources with peak flux density above 10~Jy compared to our survey which will have detected most of the sources above $\sim$0.25~Jy. 

Considering only the sources in our survey above 10~Jy we find a detection rate of $\sim$~25 per cent. In addition, \cite{walsh14} used a less strict offset criterion in assessing whether 6.7-GHz methanol and water masers were associated with the methanol and water masers being separated by no more than 0$^{\circ}$.001 in Galactic coordinates which is equivalent to an angular separation of $\sim$~5.1 arcseconds. From inspection of Figure \ref{fig:offset_hist}, it can be seen this would yield only a few more masers than using our 3~arcsecond criterion, but enough that if we extend our association radius for our masers above 10~Jy to 5.1 arcseconds we have a detection rate consistent with that of HOPS.

In future large surveys of maser associations, it may be desirable to determine maser associations based on a linear separation criteria now that better distance estimates are becoming available. However, we caution this only measures the component of the linear separation in the plane of the sky. Also, some sources have outflows and outflow sizes have dependencies on age, environment and the luminosity of the driving source, so there may not be a well defined linear scale for the water masers either.

\subsection{Luminosities}

There is evidence that the luminosities of 6.7-GHz methanol masers increase as they evolve \citep{breen10oh,breen11b}. Hence, we compared the luminosities of the methanol masers with and without associated water masers to see if there was any evidence for these being at different evolutionary phases. Previous work by \cite{szymczak05} using a statistically complete, but smaller sample (79 methanol masers) with a greater detection-limit (peak intensity greater than 1.6~Jy) found no statistically significant difference between the methanol maser luminosities of those with and without associated water masers. 

\cite{titmarsh14} undertook the same comparison with the MMB masers in the range $l = 6^{\circ} - 20^{\circ}$. We found a statistically significant difference in the luminosities after the removal of the outlying, extremely luminous source 09.621+0.196 (luminosity $\sim$~60000~Jy~kms$^{-1}$~kpc$^2$). The difference became statistically significant because of the large reduction in variance in the population that have associated water masers. 

In this sample, however, we found no statistically significant differences in the methanol maser integrated luminosities between those with and without associated water masers. The mean of the methanol luminosities with an associated water maser was higher than that of methanol only sources as this was effected by the extremely luminous source 345.505+0.348 (85500~Jy~kms$^{-1}$~kpc$^2$), but the median luminosity was lower. With the exception of this extremely luminous methanol maser, the two distributions of integrated luminosities are very similar. 

\begin{figure*}
\includegraphics[width=3.4in]{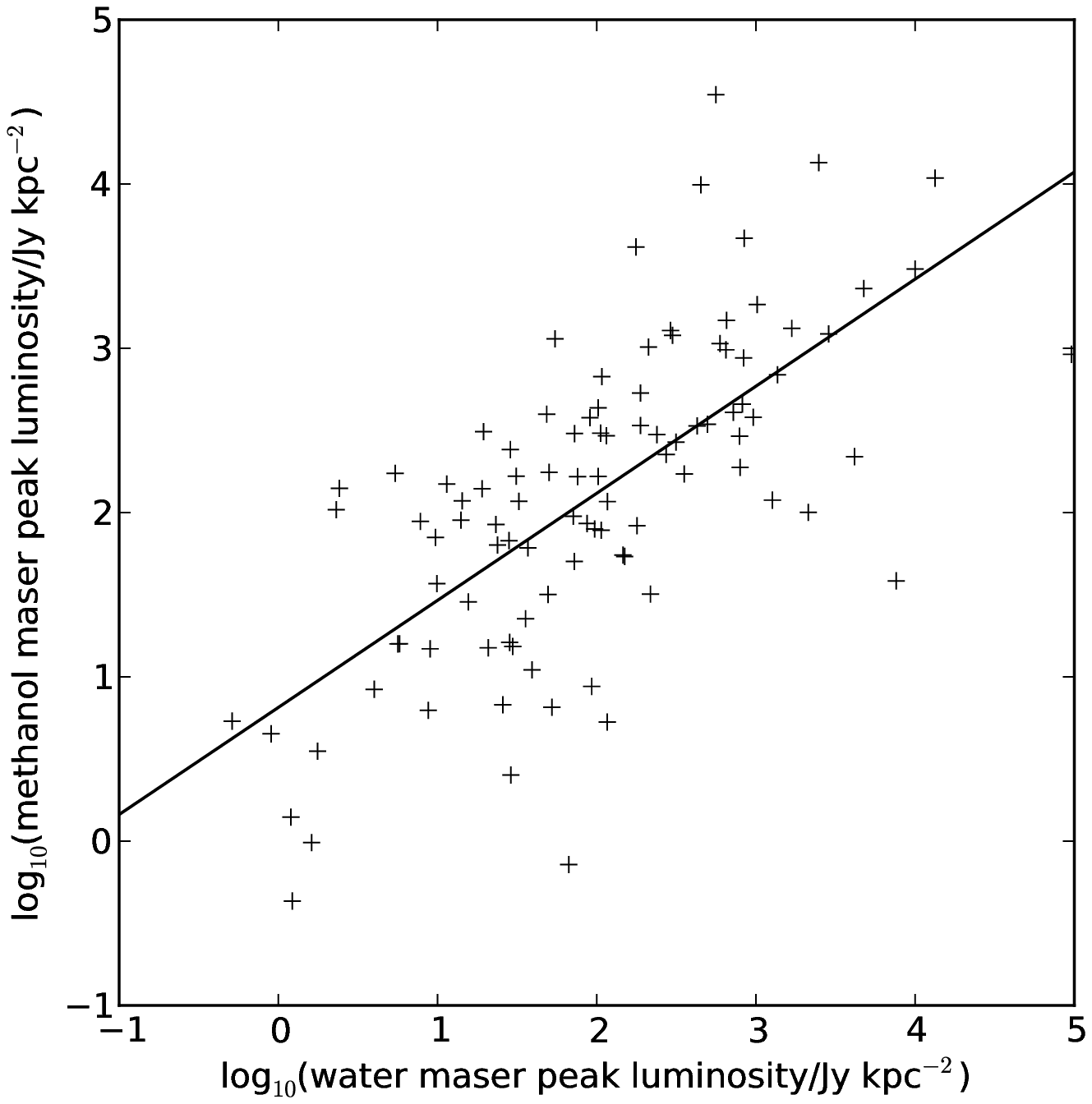}
\includegraphics[width=3.4in]{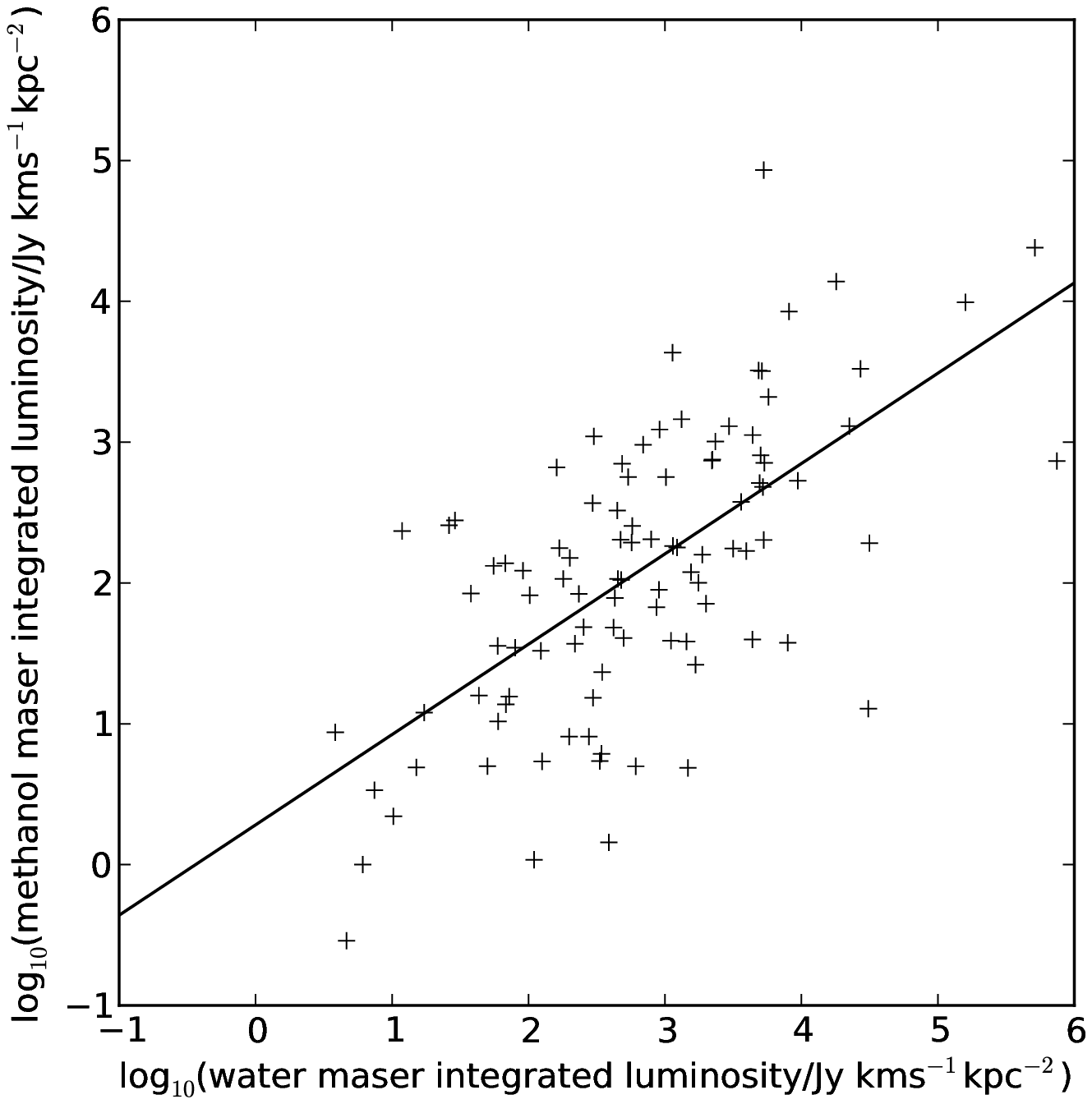}
\caption{Log 6.7-GHz methanol maser peak luminosities vs. log of associated water maser peak luminosities and log 6.7-GHz methanol maser integrated luminosities vs. log of associated water maser integrated luminosities. The lines are a linear least squares fits to the data.}
\label{fig:lums}
\end{figure*}

We previously found that there was a correlation between the water maser peak and integrated luminosities and those of their associated methanol masers \citep{titmarsh14}. Using the masers in this paper, we obtain a similar result. Figure \ref{fig:lums} shows water versus methanol peak and integrated luminosities respectively with linear least squares fits plotted. The peak luminosities have a Pearson's correlation coefficient of $R = 0.68$ and the fit has a slope of 0.65 (p-value of 5.8e-15). The integrated luminosities have a Pearson's correlation coefficient of $R = 0.64$ and the fit has a slope of 0.64 (p-value of 4.94e-13). 

These correlations may be reduced in any individual source as maser beaming means that the emission is not isotropic and the methanol and water masers may be beamed in different directions. However, over the whole sample this effect should average out. Also, we have used the emission integrated over the whole velocity range of the masers, rather than the peak, and this should average out the effects of beaming in an individual source.

We find the methanol and water maser luminosities are both correlated with distance, so we expect there to be a partial correlation between the maser luminosities due to this alone. Uncertainties in distance estimates for the masers don't vary significantly across the sample, with the exception of sources at LSR velocities near 0~kms$^{-1}$ which are at longitudes close to the Galactic Centre. Furthermore, we do not have any evidence to suggest that there are systematic differences in the intrinsic luminosity of star formation masers in different regions of the Milky Way.  However, flux density limited searches (such as the MMB and this survey) tend to produce correlations between distance and luminosity, since the less luminous sources are only detected nearby, whereas the more luminous sources can be detected over a greater volume.  This is believed to be the reason for the partial correlation we find between maser luminosity and distance.  We have used the procedure in \cite{darling02} to account for this. We found the correlation coefficients between the water and methanol maser peak and integrated luminosities and with distance (see Table \ref{table:mal}) and calculated the partial correlation coefficients using Equation \ref{eqn:mal}. Accounting for the partial correlation, we find the maser luminosities to still be loosely correlated. The water versus methanol maser peak and integrated luminosities have partial correlation coefficients of 0.44 and 0.43 respectively. Following the approach in \citet{Wall+03} we calculate the standard error in the partial correlation coefficients for a sample size of 100 and 3 degrees of freedom to be 0.082.  Applying a $t$-test to the partial correlation coefficient we find that for our sample they are significant at the 99.9 per cent confidence level.

\begin{table}
\caption{Correlation coefficients between the methanol and water luminosities and with distance used to account for bias due to distance.}
\begin{tabular}{l c c}
\hline
 & peak & integrated \\
 & luminosity & luminosity \\
 & \multicolumn{2}{c}{(Jy km s$^{-1}$)} \\
\hline
methanol - water ($R_{mw}$) & 0.68 & 0.64 \\
methanol - distance ($R_{md}$) & 0.65 & 0.59 \\
water - distance ($R_{wd}$) & 0.65 & 0.62 \\
\end{tabular}
\label{table:mal}
\end{table}

\begin{equation}
R_c  = \frac{R_{wm} - R_{wd}R_{md}}{\sqrt{(1 - R_{wd}^2)(1 - R_{md}^2)}}
\label{eqn:mal}
\end{equation}

Since there is evidence that 6.7-GHz methanol masers increase in luminosity as they age this may indicate that there is a trend for water masers to increase in luminosity as they age, although this may be over a long time scale as water masers are well known for their variability. This is consistent with \cite{breen11} who found, based on the 1.2-mm data from dust clumps associated with water masers, some evidence that water masers increase in luminosity as they evolve.

We also fitted a general linear model to the methanol maser luminosity, with the distance and water maser luminosity as the independent variables (we used the logarithms of each of these quantities in the modelling).  This modelling showed that while there is a significant dependence of the methanol maser luminosity on distance, after taking that into account a significant dependence on the water maser luminosity is still present.  This is consistent with the value calculated for the corrected PearsonÕs correlation coefficient which considers the partial correlation. 

\subsection{Velocities}

6.7-GHz methanol masers are useful probes of the velocities of the high-mass star formations they reside in. They typically have narrow total velocity ranges \citep[less than 16~km~s$^{-1}$;][]{caswell09} and central velocities within $\pm$~3~km~s$^{-1}$ of the systemic velocity of the region \citep{szymczak07,caswell09,pandian09}. 

Water masers on the other hand, are well known for their large total velocity ranges with highly red or blue-shifted spectral features commonly found. Within high-mass star formation regions, 10.472+0.027 is the most extreme, with features redshifted up to 250~km~s$^{-1}$ from the systemic velocity of the region and a total velocity range of nearly 300~km~s$^{-1}$ \citep{titmarsh13}.

\begin{figure}
\includegraphics[width=3.4in]{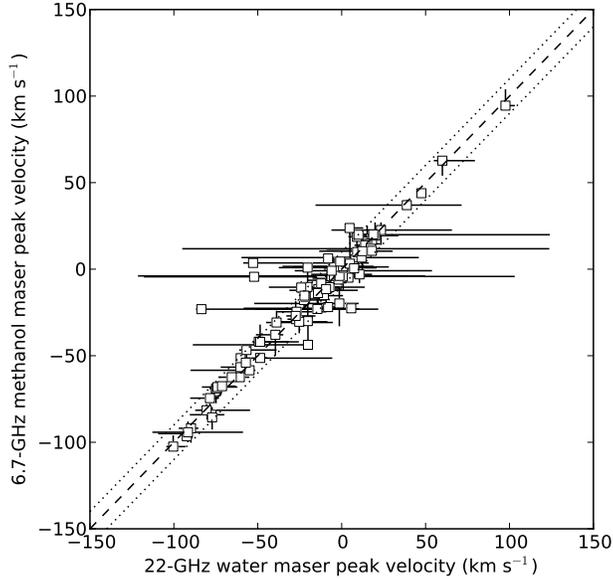}
\caption{Water maser peak velocities vs. associated 6.7-GHz methanol maser peak velocities are shown with squares. The horizontal and vertical bars represent the total velocity ranges of the water and methanol masers respectively. Also plotted is a dashed line with a slope of 1 and two dotted lines showing a deviation of $\pm$10~km~s$^{-1}$ from the dashed line.}
\label{fig:velpeaks}
\end{figure}

Figure \ref{fig:velpeaks} compares the velocity of the peak emission in the water and methanol masers. Consistent with \cite{titmarsh14} we found that most ($\sim$~88 per cent) of the water maser peak velocities lie within $\pm$~10~km~s$^{-1}$ of the peak velocity of the methanol masers. We note that in both \cite{titmarsh14} and this paper we found our peak velocities to be much more tightly correlated than \cite{breen10oh} who found 78 per cent of the water and methanol maser peak velocities to be within $\pm$~10~km~s$^{-1}$ of each other. This may be because their sample was biased towards sources with OH masers which have been shown to be more evolved \citep{breen10ev}. 

\begin{figure}
\includegraphics[width=3.4in]{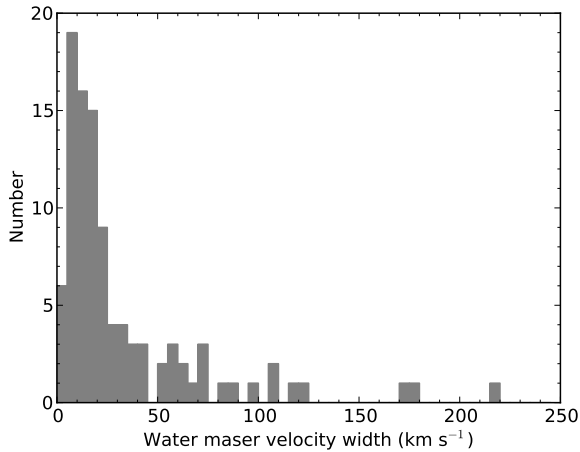}
\caption{Velocity ranges of the water masers associated with methanol masers.}
\label{fig:velwidths}
\end{figure}

\begin{figure}
\includegraphics[width=3.4in]{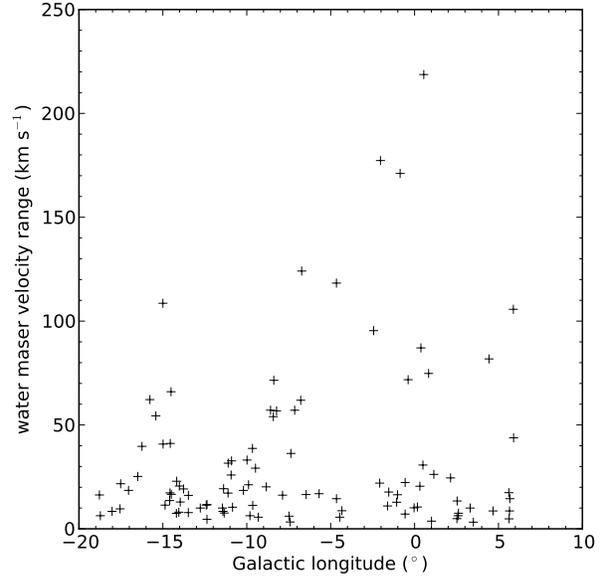}
\caption{Velocity ranges of the water masers vs. Galactic longitude.}
\label{fig:vellong}
\end{figure}

In Figure \ref{fig:velwidths} we show the total velocity ranges that the water maser emission spans in this sample. In \cite{titmarsh14}, where we performed the same survey over Galactic longitudes $l = 6^{\circ} - 20^{\circ}$, we found the median velocity range to be $\sim$~17~km~s$^{-1}$ and the mean to be 27~km~s$^{-1}$, consistent with this sample where we find the median and mean total velocity ranges to be $\sim$~17~km~s$^{-1}$ and 32~km~s$^{-1}$ respectively. In contrast to \cite{titmarsh14}, where only $\sim$~7 per cent of our of the masers had velocity ranges greater than 50~km~s$^{-1}$, this sample ($l = 341^{\circ} - 6^{\circ}$ through zero) has $\sim$~20 per cent with velocity ranges greater than 50~km~s$^{-1}$. We can rule out being unable to detect high velocity emission in the first paper as we detected 10.472+0.027 which has a total velocity range of nearly 300~km~s$^{-1}$ (the largest velocity range in this paper is 0.546--0.852 with a velocity range of 218.7~kms$^{-1}$). In Figure \ref{fig:vellong} we investigate the total velocity ranges versus Galactic longitude. The three sources with the largest velocity ranges (357.967--0.163, 359.138+0.031 and 0.546--0.852) are all close to the Galactic Centre, but otherwise the masers with velocity ranges greater than 50~km~s$^{-1}$ are spread across the longitudes surveyed.

We do however, find our velocity ranges to be similar to those of \cite{szymczak05} and \cite{caswell10}. \cite{szymczak05} found 33 per cent of their water masers to have velocity ranges greater than 20~km~s$^{-1}$ (we found 44 per cent) and 22 per cent greater than 40~km~s$^{-1}$ (we found 24 per cent). In the \cite{caswell10} sample of 32 masers from an unbiased survey for water masers, they found 34 per cent of water masers with both methanol and OH masers associated and 31 per cent of those with just methanol, had emission red or blue-shifted more than 30~km~s$^{-1}$ from the systemic velocity. For the current sample we find 26 per cent meet these criteria.

\subsection{Variability}
\label{sec:var}

Water masers are well known for showing variations in their spectra. Studying variability over long timescales is very time consuming, hence variability studies over many epochs tend to be with smaller sample sizes \citep{brand03,felli07}.

For comparison with our sample, we have focused on larger surveys. Two such surveys are \cite{breen10oh} and \cite{walsh11,walsh14} which have observed the same water maser sites at more than one epoch. In the \cite{breen10oh} survey, they observed 253 waters in 2003 and 2004 and 17 per cent of these were only detected in one epoch. These masers typically had simpler spectra with only a few velocity features and two-thirds had peak flux densities of less than 2~Jy when they were detected. 

\cite{walsh11} detected 540 water masers between 2008 - 2010 in an unbiased search for water masers in the Galactic Plane with the Mopra telescope. In 2011/2012 they followed up these detections with high resolution observations from the ATCA \citep{walsh14}. 31 sites of water maser emission were not detectable in the second epoch, and like \cite{breen10oh}, they found the most variable sources tended to be weaker with simpler spectra. Unlike \cite{breen10oh} they found a smaller fraction of the masers not detected in one epoch ($\sim$~6 per cent). This is likely to be because the \cite{breen10oh} survey was much more sensitive than the initial HOPS search (which is estimated to be 98 per cent complete down to 8.4~Jy) since there is a well established tendency for less luminous water masers to exhibit greater fractional variability.

\begin{figure*}
\includegraphics[width=3.4in]{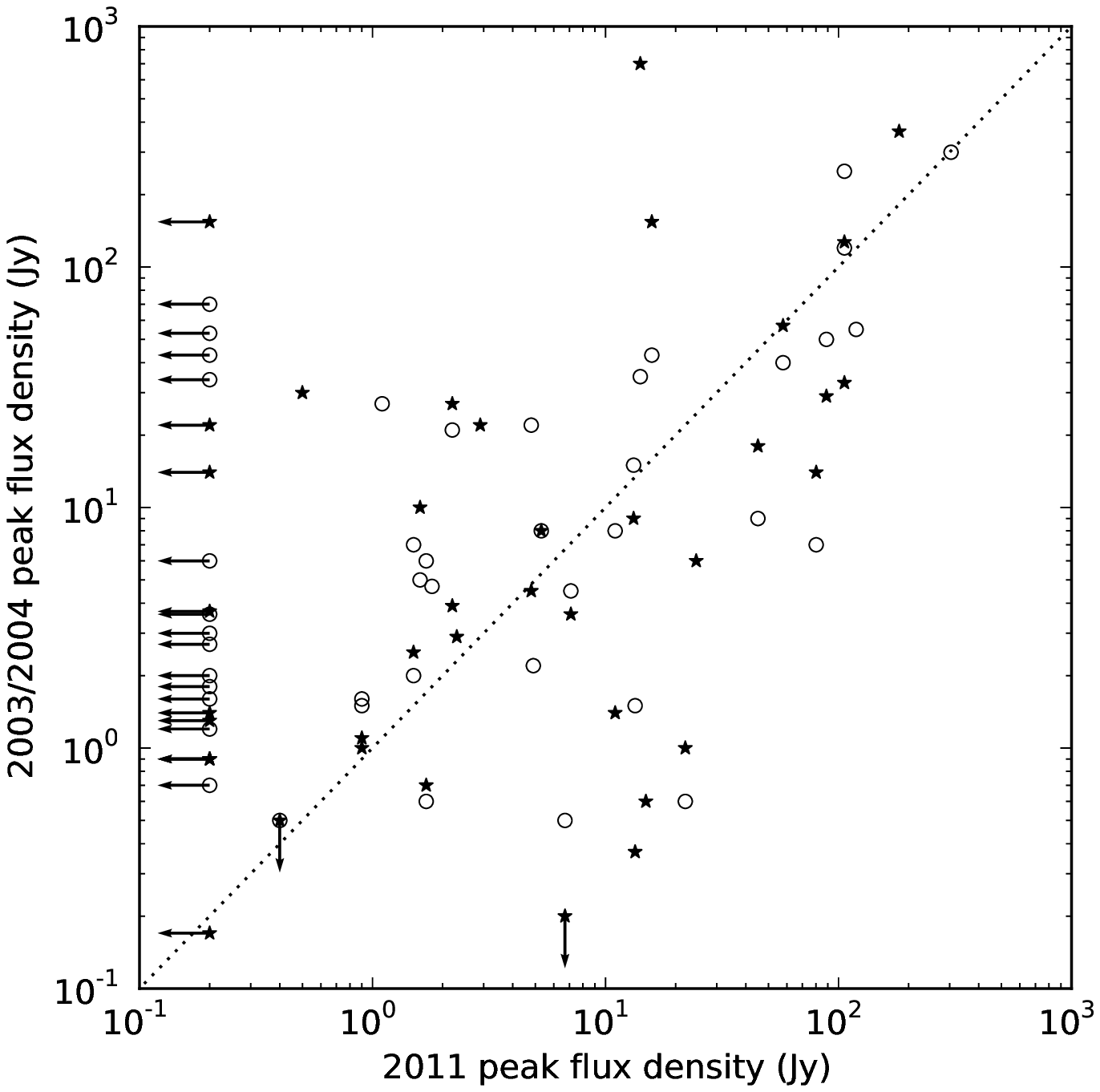}
\includegraphics[width=3.4in]{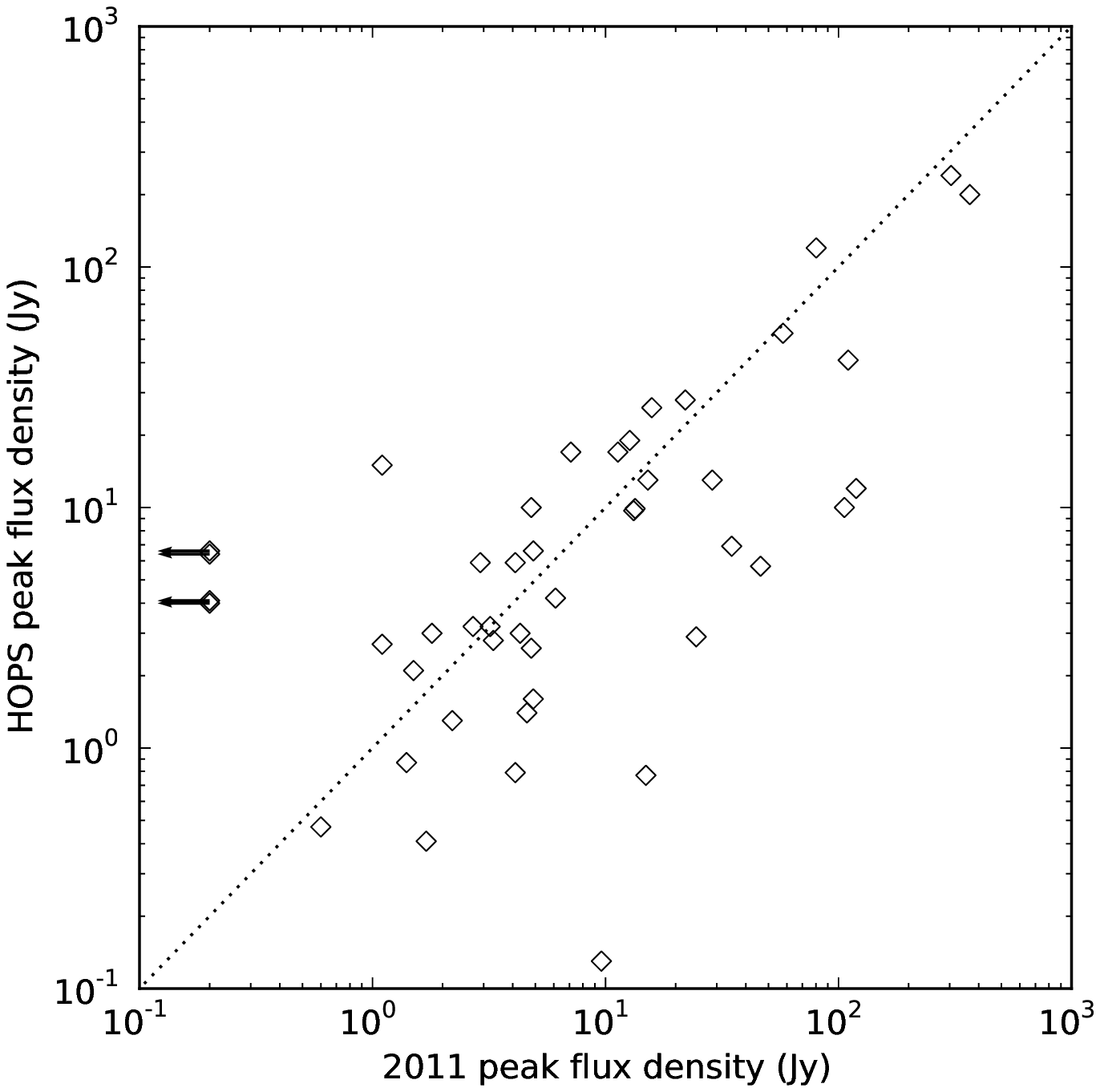}
\caption{The first panel compares the peak flux densities of the water masers that were observed in both the MMB follow-up (observed in 2011) and the Breen et al. (2010a) sample. The Breen et al. (2010a) observations that were made in 2003 are marked with open circles and those from 2004 are stars. The second panel shows peak flux densities of water masers detected in the MMB follow-up and the HOPS high resolution follow-up (observed in 2011 and 2012). There were many sources detected in our observations that were not reported in HOPS. We have not included these as we do not know if they were not found because the initial HOPS search was much less sensitive than ours, or because of variability. In both panels, where masers were detected in one epoch and not the other, 3~$\sigma$ upper limits on the flux densities are shown with arrows. Also plotted are dotted lines with slope of 1.}
\label{fig:var}
\end{figure*}

We have compared the properties of the water masers in the current sample that overlap with the \cite{breen10oh} and \cite{walsh14} observations. All of the masers for which we have data at more than one epoch showed some variations in their spectra. Figure \ref{fig:var} plots (on a log-log scale) the peak flux densities of our water maser observations in 2011 against those in 2003/2004 from \cite{breen10oh} and against the 2011/2012 observations from \cite{walsh14}. There are a handful of masers that have the same peak flux densities (within uncertainties) in two epochs, however, most show variation, some of which may be from different spectral features changing their relative intensities resulting in the peak flux density being measured from a different feature. Even though the masers in \cite{breen10oh} were observed 7 - 8 years apart from our observations, they showed no more scatter in Figure \ref{fig:var} than the HOPS masers that were observed within 12 months of ours.

\subsection{Associations with GLIMPSE sources}

The \textit{Spitzer} Galactic Legacy Infrared Midplane Survey Extraordinaire \citep[GLIMPSE;][]{benjamin03} surveyed the Galactic Plane in four infrared bands: 3.6, 4.5, 5.8 and 8.0~$\mu$m. \cite{ellingsen06} used point sources identified by GLIMPSE to compare the colours of the mid-infrared objects associated with 6.7-GHz methanol masers compared to the general population of point sources. They used 56 methanol masers, most of which came from a blind single dish survey for methanol masers in $l = 325^{\circ} - 335^{\circ}$ \citep{ellingsen96}. They found that the mid-IR point sources associated with masers tended to exhibit redder colours than sources without masers, which is consistent with them coming from high-mass star formation regions with similar SED properties to Class 0 low-mass stars. \cite{ellingsen06} also compared the colours of 6.7-GHz methanol masers with and without associated OH masers and found colours consistent with sources harbouring an OH maser to be more evolved than those without.

\cite{breen10ev} compared the GLIMPSE point source colours of a sample of 113 6.7-GHz methanol masers with and without associated 12.2-GHz methanol masers. They found no difference in the colours of the two groups and they proposed that the masers themselves are more sensitive than the mid-infrared data to evolutionary changes in the YSO.

Similar work by \cite{gallaway13} studied the mid-infrared colours of the mid-IR sources hosting the MMB masers. Since many of the mid-IR sources associated with methanol masers were more extended than the GLIMPSE point spread function, they used Adaptive Non-Circluar Aperture Photometry (ANCAP) to measure their extended flux densities in all four GLIMPSE bands. Not all of the maser counterparts were included, as some of the masers did not have interferometric positions yet, were outside the GLIMPSE survey range, GLIMPSE fluxes were not available in all four bands, or there was more than one possible counterpart. They found the infrared colours of the maser-associated sources to be very similar to those of \cite{ellingsen06}.

We used the \cite{gallaway13} ANCAP flux densities to compare the MMB sources with and without associated water masers in colour-colour and colour-magnitude diagrams. Any difference in the infrared colours could point to differences in evolutionary phase (like those with and without OH masers in \cite{ellingsen06}). However, we found no statistically significant difference in the mid-infrared colours between the infrared sources associated with both methanol and water masers and those with only methanol. 

\subsection{Associations with 870-$\mu$m emission from dust clumps}

The earliest stages of high-mass star formation occur embedded within clumps of gas and dust. These clumps are well traced by their continuum dust emission which is optically thin at submillimetre wavelengths allowing us to calculate their clump masses and column densities (the column density is the molecular hydrogen column inferred from the observed dust emission, an assumed dust-to-gas ratio of 1:100 and assuming that the dust emission is optically thin). The APEX Telescope Large Area Survey of the GALaxy \citep[ATLASGAL;][]{schuller09} is an unbiased survey of the Galactic Plane at 870~$\mu$m with the APEX telescope in Chile. ATLASGAL had a spatial resolution of 19.2 arcseconds and a 5~$\sigma$ sensitivity of 0.25~Jy~beam$^{-1}$.

About 95 per cent of the MMB masers lie within the region surveyed by ATLASGAL. \cite{urquhart13} used the ATLASGAL source catalogues of \cite{contreras13} and \cite{csengeri14} to match the methanol masers with an associated dust clump. They used a criteria of being within 120 arcseconds of a peak in the 870~$\mu$m emission to define an association as this was the largest clump radius \citep{contreras13}. If more than one clump was within that radius, they chose the clump with its peak emission closest. They then inspected the ATLASGAL images to determine if associations were genuine. $\sim$~94 per cent of MMB masers within the ATLASGAL survey range were found to have an associated 870-$\mu$m dust clump.

We used the ATLASGAL-MMB associations identified in \cite{urquhart13} to compare the dust emission of the 6.7-GHz methanol masers with and without associated water masers and also to compare the 870-$\mu$m emission with the results from \cite{titmarsh14} using the 1.1~mm emission from the Galactic Plane is the Bolocam Galactic Plane Survey \citep[BGPS;][]{rosolowsky10}. The BGPS is another unbiased submillimeter survey of the Galactic Plane which used the Bolocam instrument on the Caltech Submillimeter Observatory to make a continuum survey of the Galactic Plane at 1.1~mm  with an effective resolution of 33~arcseconds. Like ATLASGAL, the BGPS is ideal for tracing cold, dense gas and dust cores. 

\cite{chen12} used dust clumps from the BGPS for targeted class I methanol maser observations. When they compared the properties of the clumps with and without associated masers, they found BGPS sources with an associated class I methanol maser had higher BGPS flux densities and beam averaged column densities than those without. They also found a correlation between the intensity of the class I methanol masers and the beam averaged column density.

In \cite{titmarsh14} we found BGPS sources with an associated MMB maser to have higher integrated flux densities and higher column densities than the general population. However, we found no differences in these properties between the clumps associated with an MMB maser with and without a water maser present. Like class I methanol masers, water masers are also pumped by collisions, and like \cite{chen12} we found a weak correlation between the water maser integrated flux density and the beam averaged column density. 

To make our ATLASGAL comparisons, we recalculated the masses and column densities of the dust clumps, as we have distance estimates for all the MMB sources in our survey (\cite{urquhart13} did not have distances for many of the sources around the Galactic Centre) and we used $\mu$~=~2.3 for the mean molecular weight of the ISM to be consistent with \cite{chen12}. Column densities and clump masses were calculated according to equations 1 and 2 in \cite{chen12}. 

Unlike comparisons with the BGPS in \cite{titmarsh14}, we found no correlation between the integrated flux density of the water masers and the ATLASGAL column densities of their associated dust clumps. The correlation at 1.1~mm was weak, however, we expected to observe similar results at 870~$\mu$m as sources with higher integrated column densities tend to be more massive and there may be a larger volume of gas where the conditions are appropriate for maser emission in these clumps. Similar to our comparisons with the BGPS we also found no correlations between the water maser luminosities and the clump masses or between the methanol maser intensities and the column densities and masses. We also found no differences in the distributions of the ATLASGAL column densities and masses of the clumps with both methanol and water masers and those with only methanol.

\begin{figure*}
\includegraphics[width=3.4in]{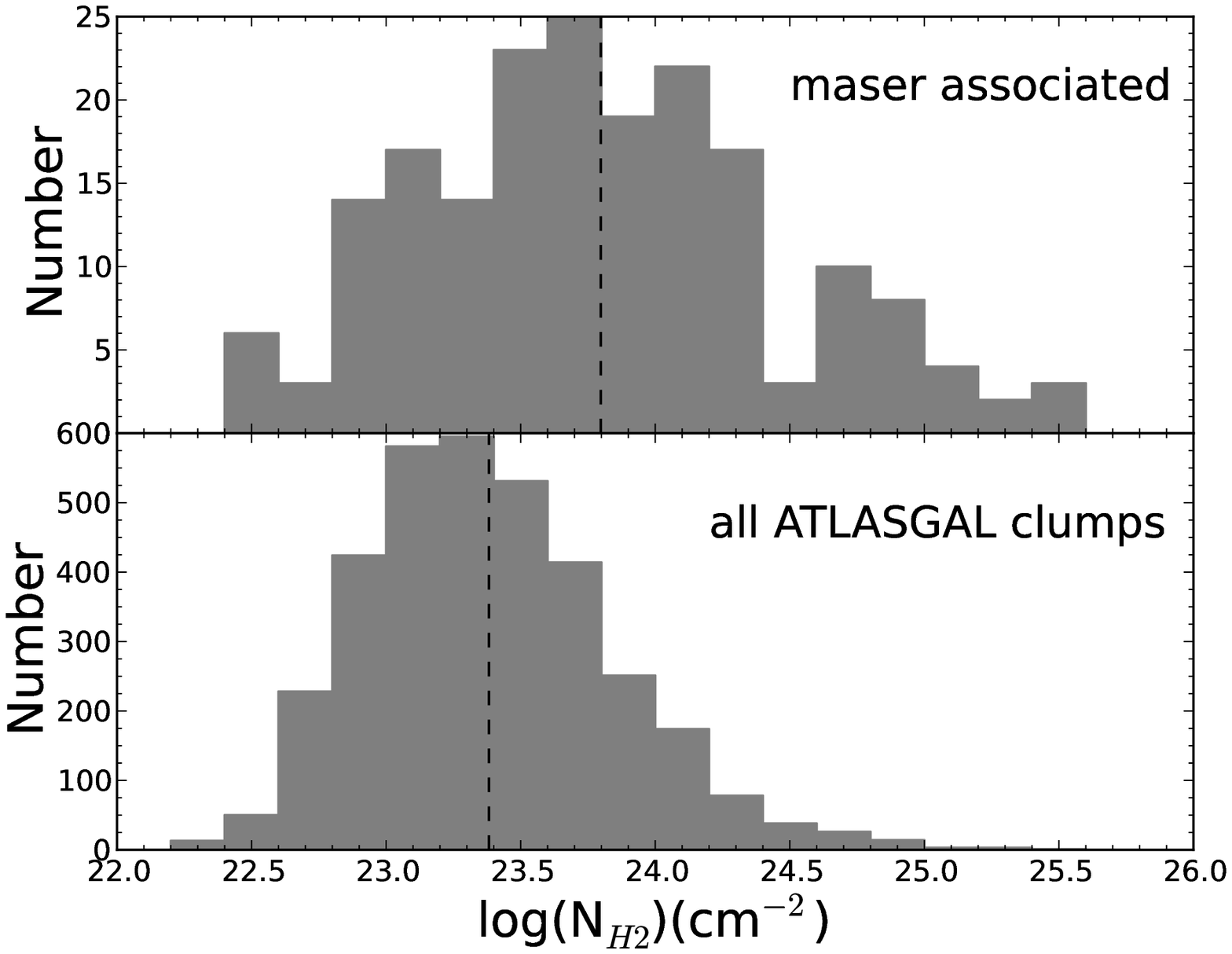}
\includegraphics[width=3.4in]{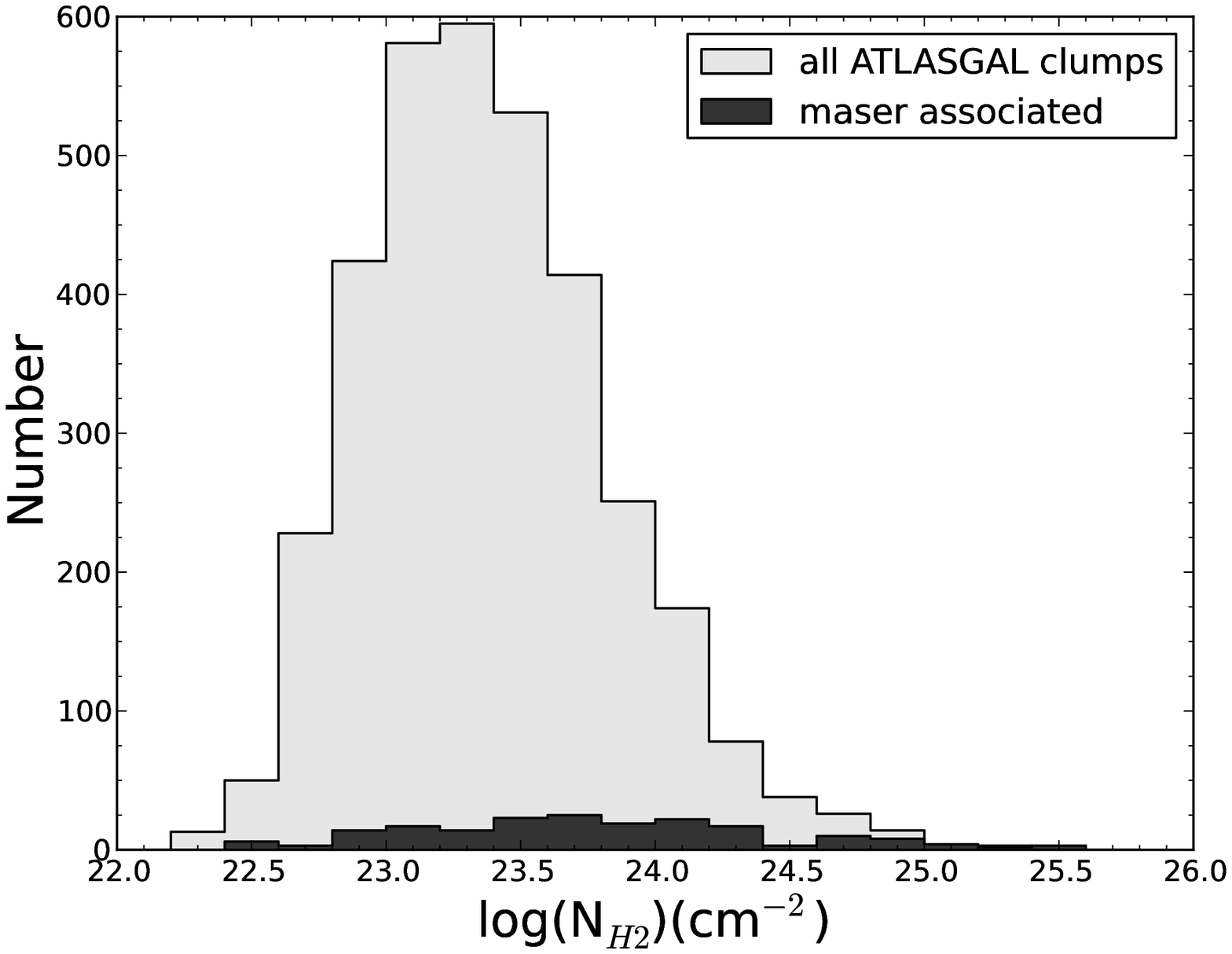}
\caption{Number of sources as a function of column density. In the left figure, the top panel shows the dust clumps detected in ATLASGAL associated with 6.7-GHz methanol masers and the bottom panel are all the clumps detected in the longitude range covered in this paper. The dashed lines represent the means of each sample. The figure on the right has both populations overlaid, showing that at the highest column densities (over $\sim 24.8$~cm$^{-2}$) almost all the dust clumps have an associated methanol maser.}
\label{fig:atlasgalhist}
\end{figure*}

Figure \ref{fig:atlasgalhist} shows the column densities of the clumps with an associated MMB maser compared to those of all the ATLASGAL clumps within our survey region. The left panel shows the distribution of the clumps with associated masers zoomed in compared to all the clumps within our survey region, and the right panel shows that the fraction of maser-associated dust clumps increases with increasing column density. At the very highest column densities, almost 100 per cent of clumps have an associated methanol maser. Like our comparisons with the BGPS, the maser associated clumps are skewed toward the higher mass clumps, unlike our BGPS comparisons, the dust clumps with an associated maser appear to cover the whole range of column densities of the general population. However, this may be a sensitivity issue as the 5~$\sigma$ sensitivity of ATLASGAL is 0.25~Jy~beam$^{-1}$, corresponding to a hydrogen column density of order 10$^{22}$~cm$^{-2}$ \citep{schuller10} whereas the BGPS was sensitive to column densities greater than 10$^{21}$~cm$^{-2}$.

\section{Conclusions}

We have observed all the known 6.7-GHz methanol masers in the Galactic Plane between $l = 341^{\circ}$ and $l = 6^{\circ}$ (through the Galactic centre) with the ATCA for water maser emission. We detected water masers towards 101 out of the 204 sources in this range ($\sim$ 50 per cent) consistent with previous studies within their flux density and angular resolution constraints.

In \cite{titmarsh14} we concluded that there may be a difference in the integrated luminosities between 6.7-GHz methanol masers with and without associated water masers, however, the current sample does not support this (which is consistent with \cite{szymczak05}). Like \cite{titmarsh14} we did find a correlation between the methanol and water peak and integrated luminosities, even after taking into account the partial correlation due to distance. Since methanol masers increase in brightness as they age, we interpret this to be evidence that water maser generally increase in brightness as they age.

Most of the peak velocities of the methanol and water masers are well correlated ($\sim$~88 per cent within $\pm$~10~km~s$^{-1}$ of each other), consistent with the previous paper in this series. However, in the longitude range of this paper ($l = 341^{\circ}$ through the Galactic centre to $l = 6^{\circ}$) there are many more sources with high total velocity ranges compared to the previous paper ($l = 6^{\circ} - 20^{\circ}$). In our current sample we had $\sim$~20 per cent with velocity ranges greater than 50~km~s$^{-1}$ compared to the previous paper where it was $\sim$~7 per cent. The total velocity ranges in our current sample are similar to those found in \cite{szymczak05} and \cite{caswell10}.

Using the GLIMPSE sources and their flux densities in the four \textit{Spitzer} bands extracted by \cite{gallaway13}, we found no difference in the colours of the mid-infrared sources associated with both methanol and water masers and those with methanol only. 

We used the ATLASGAL sources identified to have associated MMB masers in \cite{urquhart13}, and found that in the submillimetre emission there were no differences  between the column densities and masses of clumps with both water and methanol masers associated and those with just methanol. We did find that the clumps associated with methanol masers are skewed towards the higher column densities compared to the general population. This is consistent with our findings for 1.1-mm thermal dust continuum emission in \cite{titmarsh14}. 

Our findings support the conclusions of \cite{breen14} that unlike 6.7-GHz methanol masers, there is little evidence that water masers trace a well defined evolutionary stage of high-mass star formation. Nevertheless, water maser observations can still provide us with valuable information the physical conditions of the YSO.

\section*{Acknowledgments}

We would like to thank an anonymous referee for their constructive and helpful feedback. The Australia Telescope is funded by the Commonwealth of Australia for operation as a National Facility managed by CSIRO. This research has made use of NASA's Astrophysics Data System Abstract Service, the NASA/IPAC Infrared Science Archive (which is operated by the Jet Propulsion Laboratory, California Institute of Technology, under contract with the National Aeronautics and Space Administration) and data products from the GLIMPSE survey, which is a legacy science programme of the \textit{Spitzer Space Telescope}, funded by the National Aeronautics and Space Administration. The ATLASGAL project is a collaboration between the Max-Planck-Gesellschaft, the European Southern Observatory (ESO) and the Universidad de Chile. This research has made use of the SIMBAD data base operated at CDS, Strasbourg, France. SLB is the recipientÊ of an Australian Research Council DECRA FellowshipÊ (project number DE130101270) and a 2015 L'Or$\rm{\acute{e}}$al-UNESCO for Women in Science Fellowship.

\bibliography{bib_paper2}

\clearpage

\appendix

\section{Spectra}

\begin{figure*}
\includegraphics[width=2.2in]{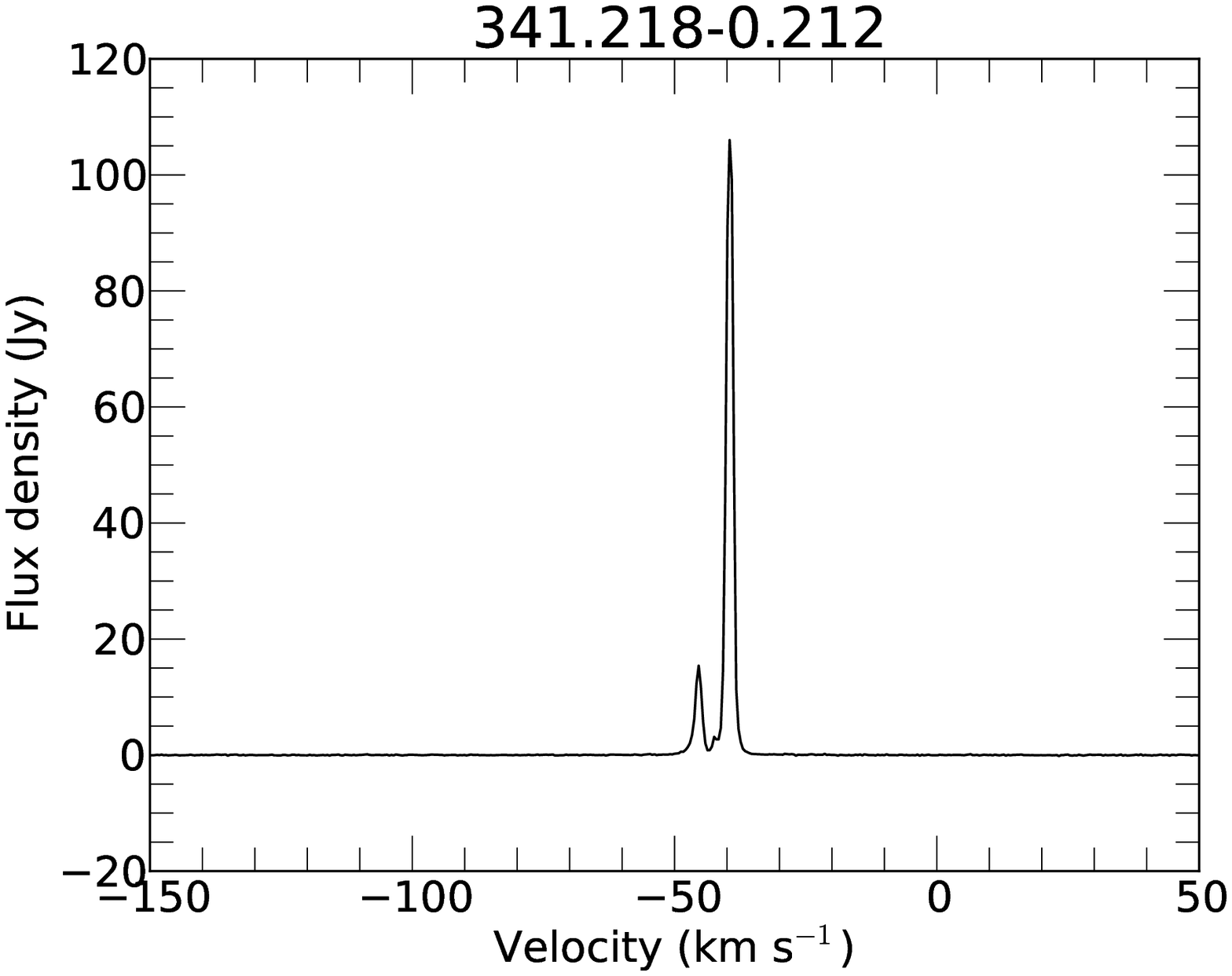}
\includegraphics[width=2.2in]{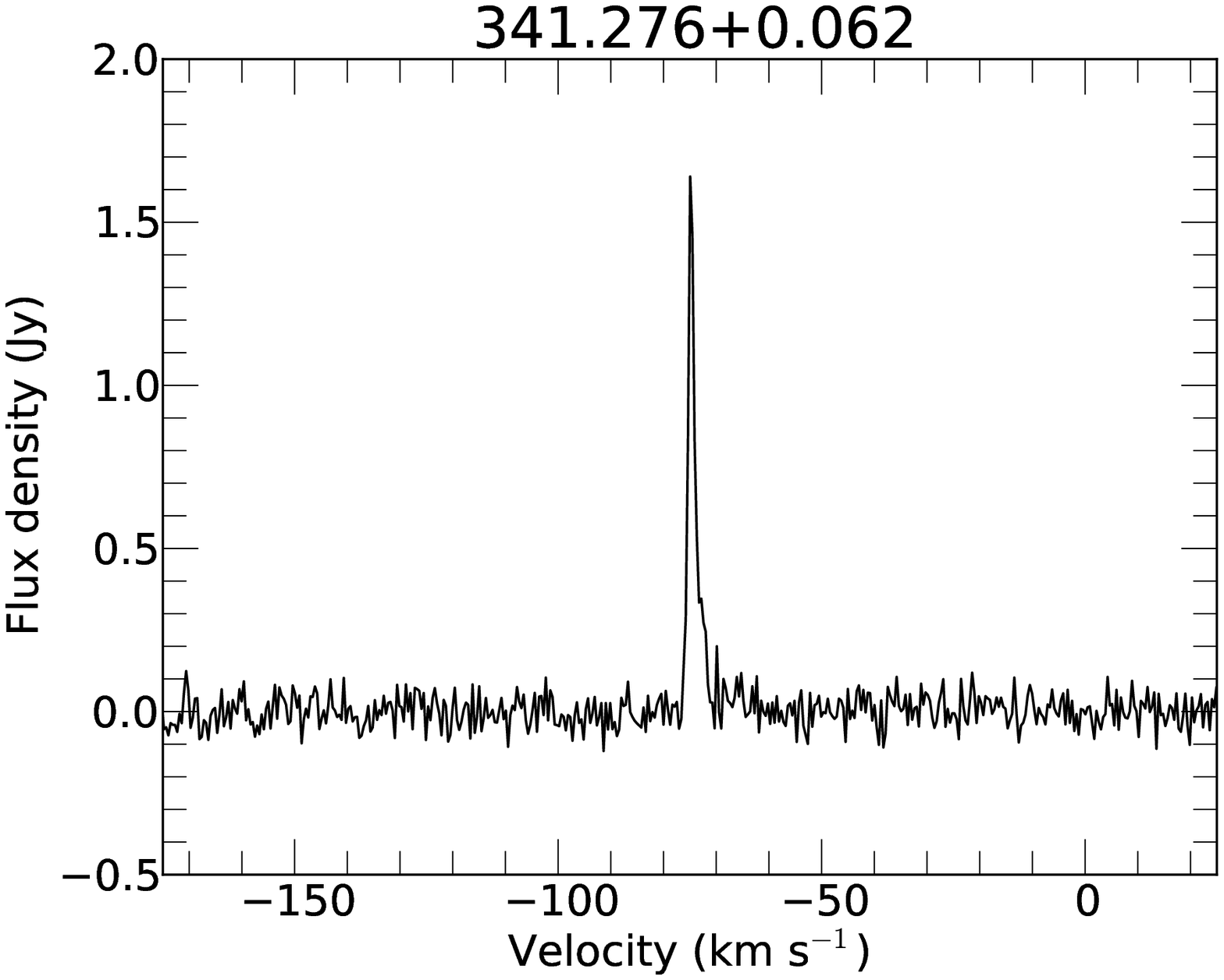}
\includegraphics[width=2.2in]{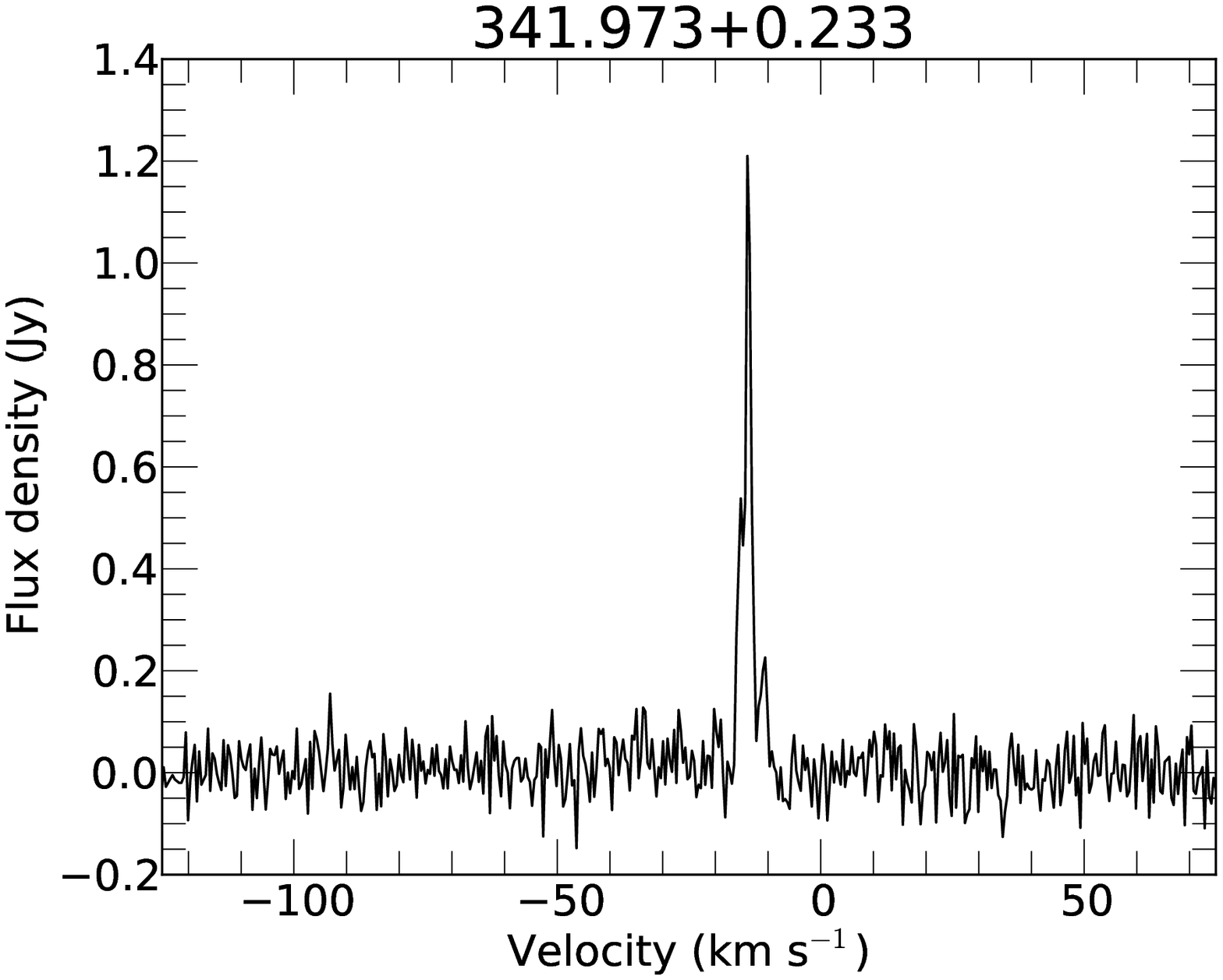}
\includegraphics[width=2.2in]{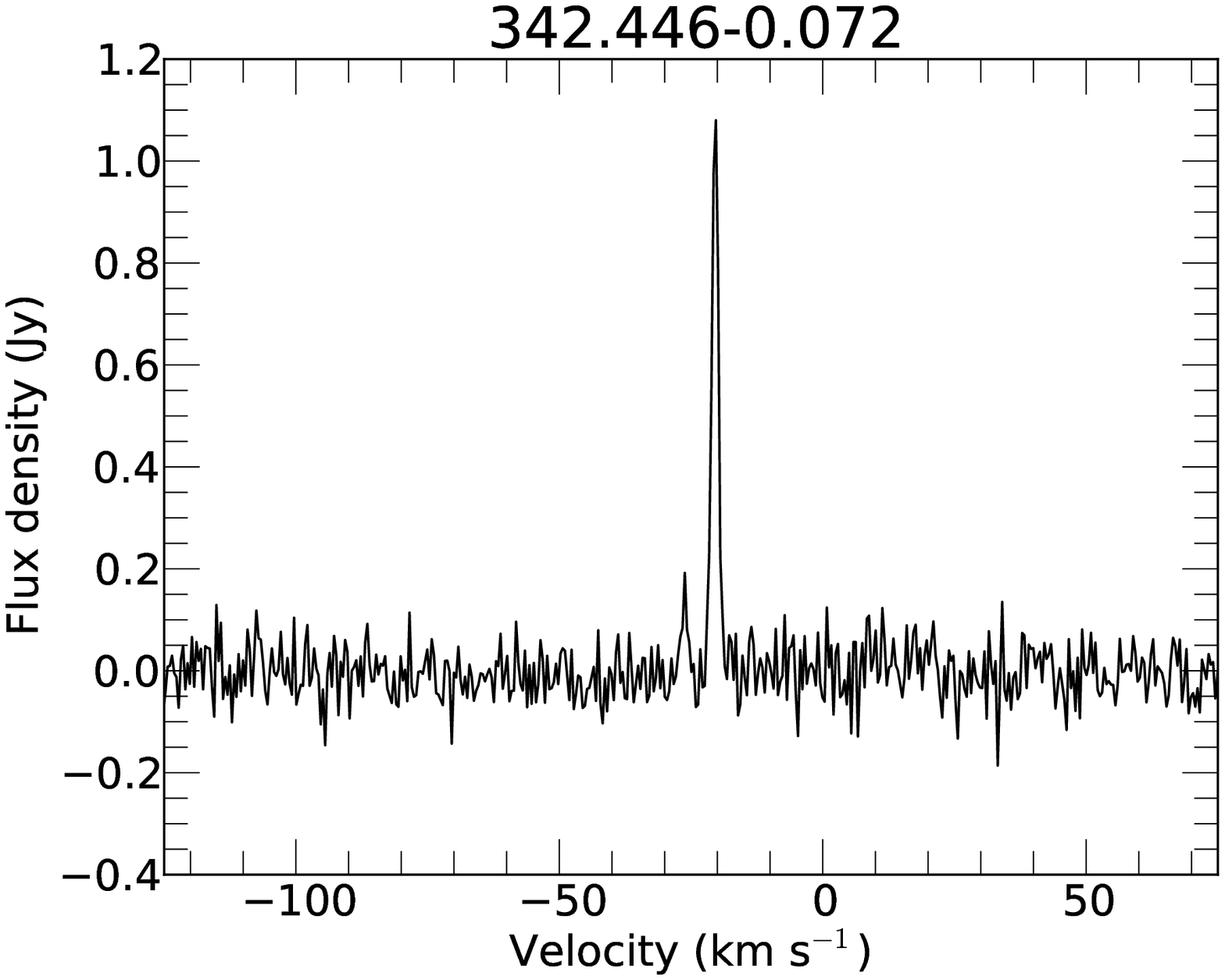}
\includegraphics[width=2.2in]{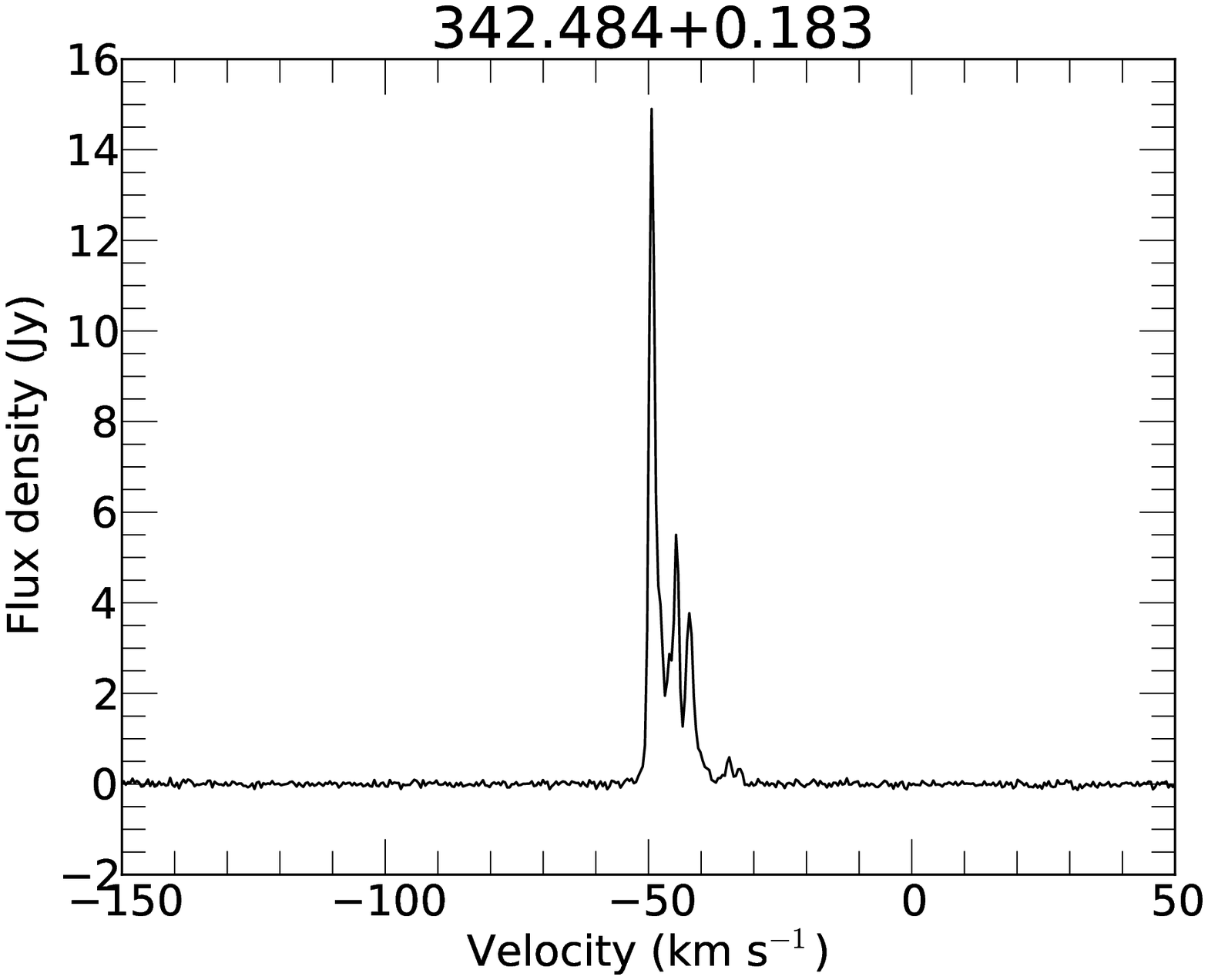}
\includegraphics[width=2.2in]{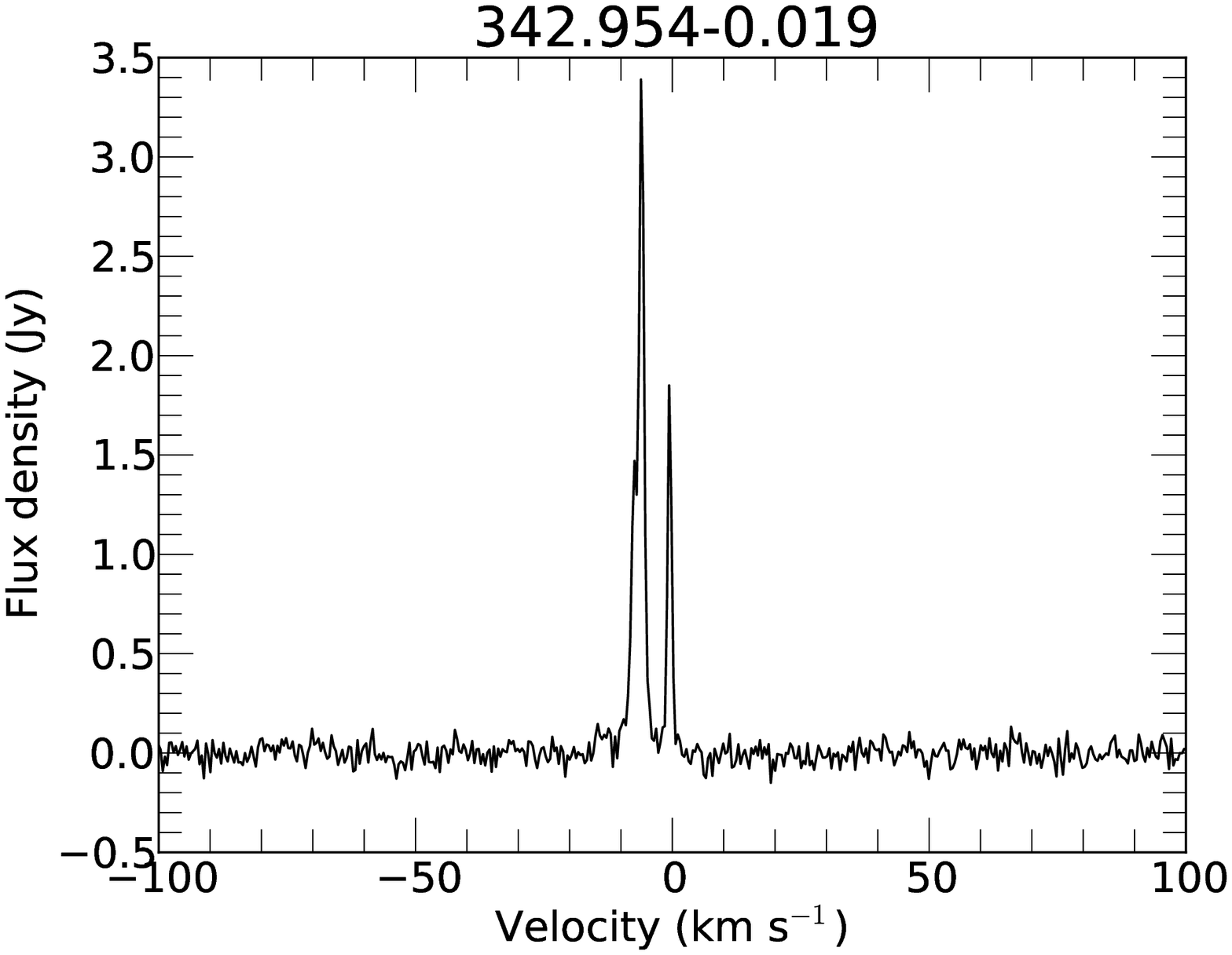}
\includegraphics[width=2.2in]{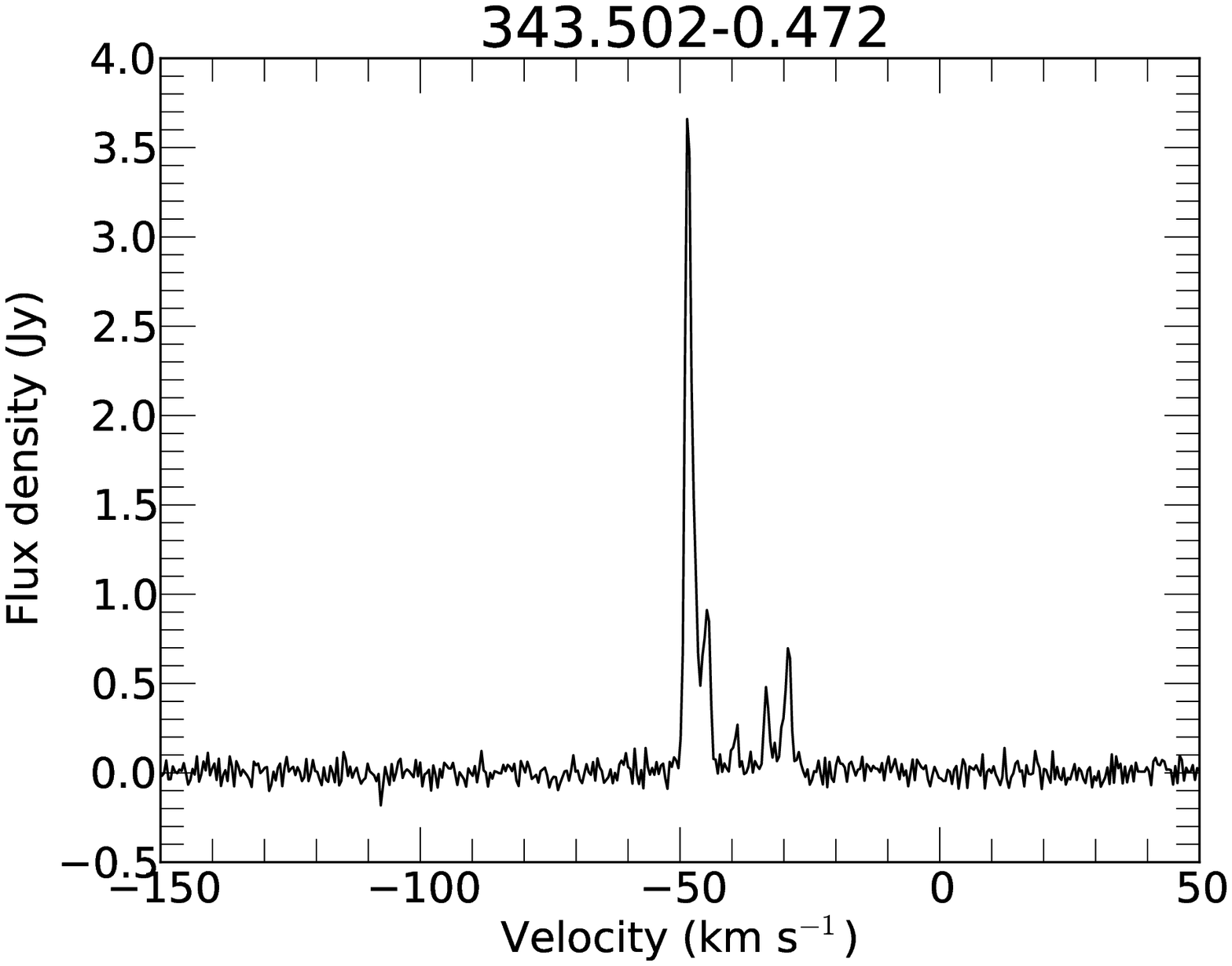}
\includegraphics[width=2.2in]{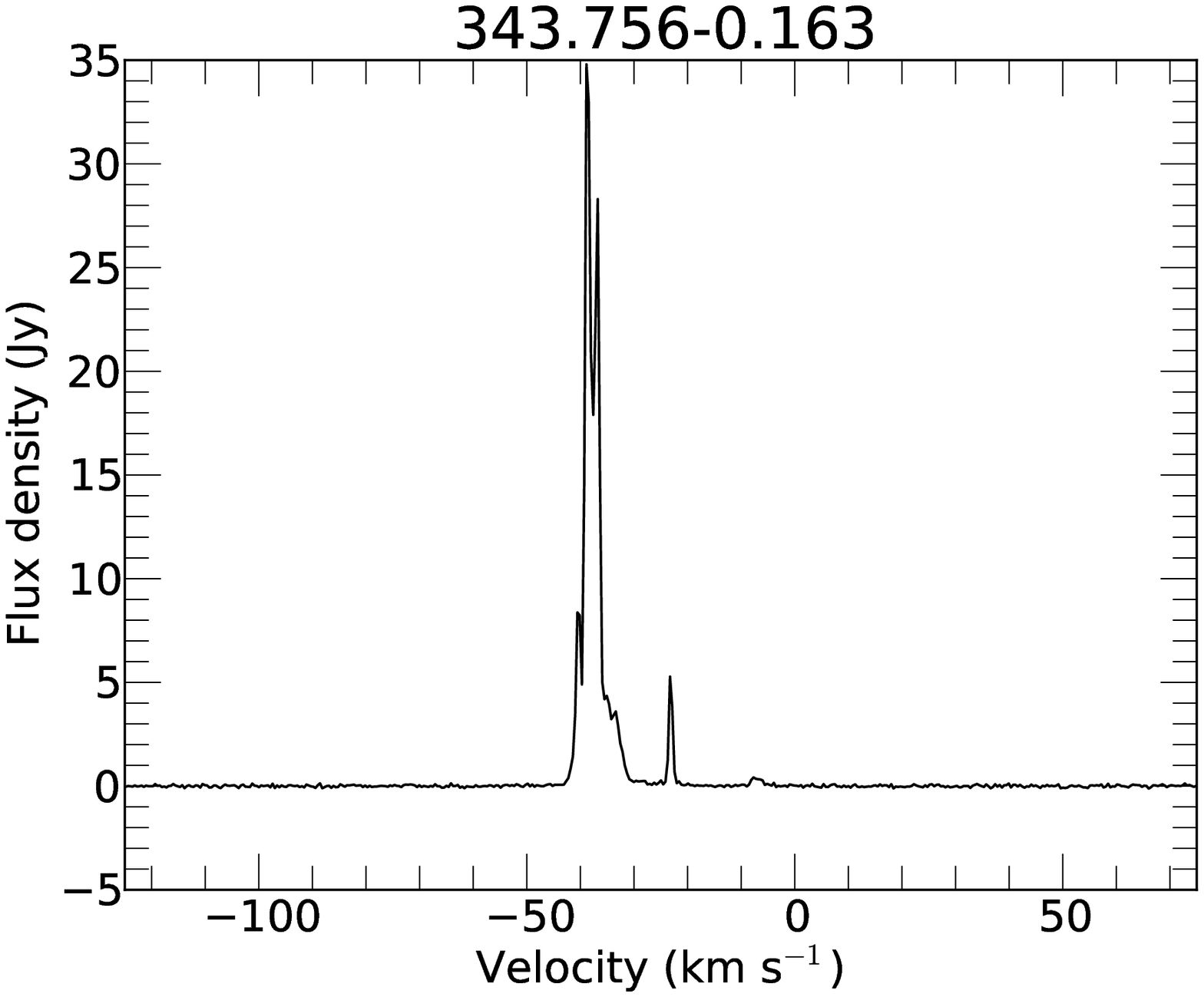}
\includegraphics[width=2.2in]{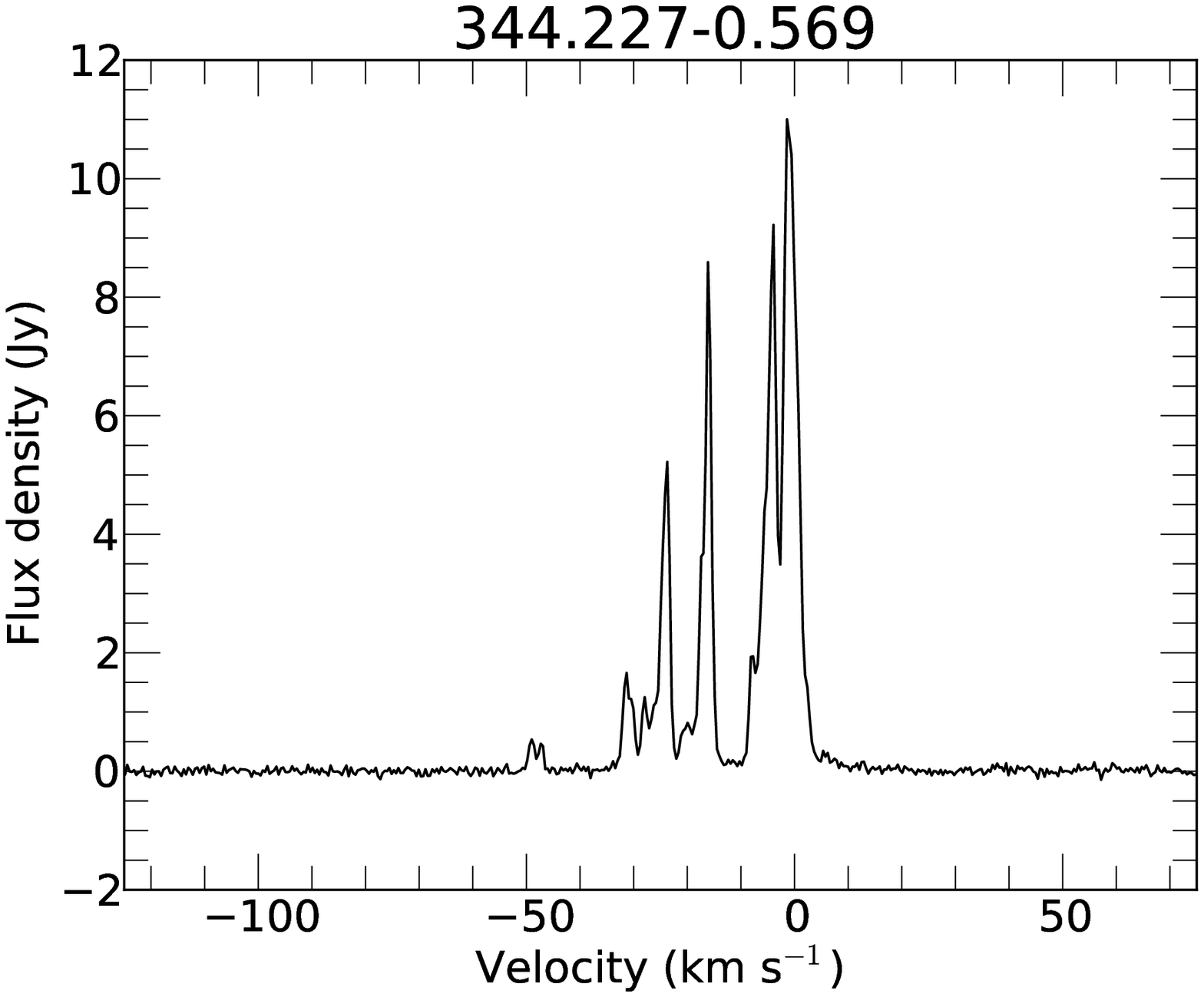}
\includegraphics[width=2.2in]{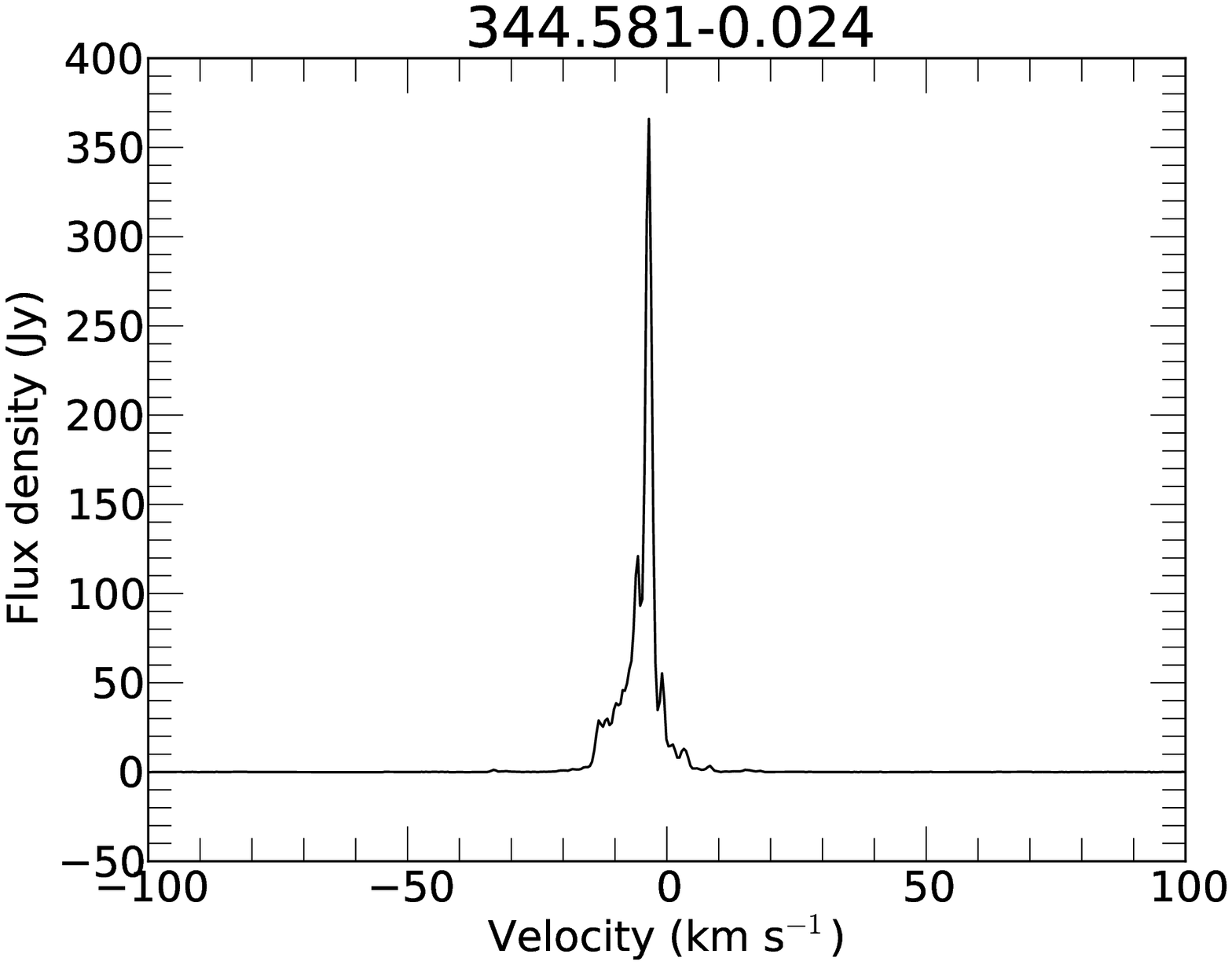}
\includegraphics[width=2.2in]{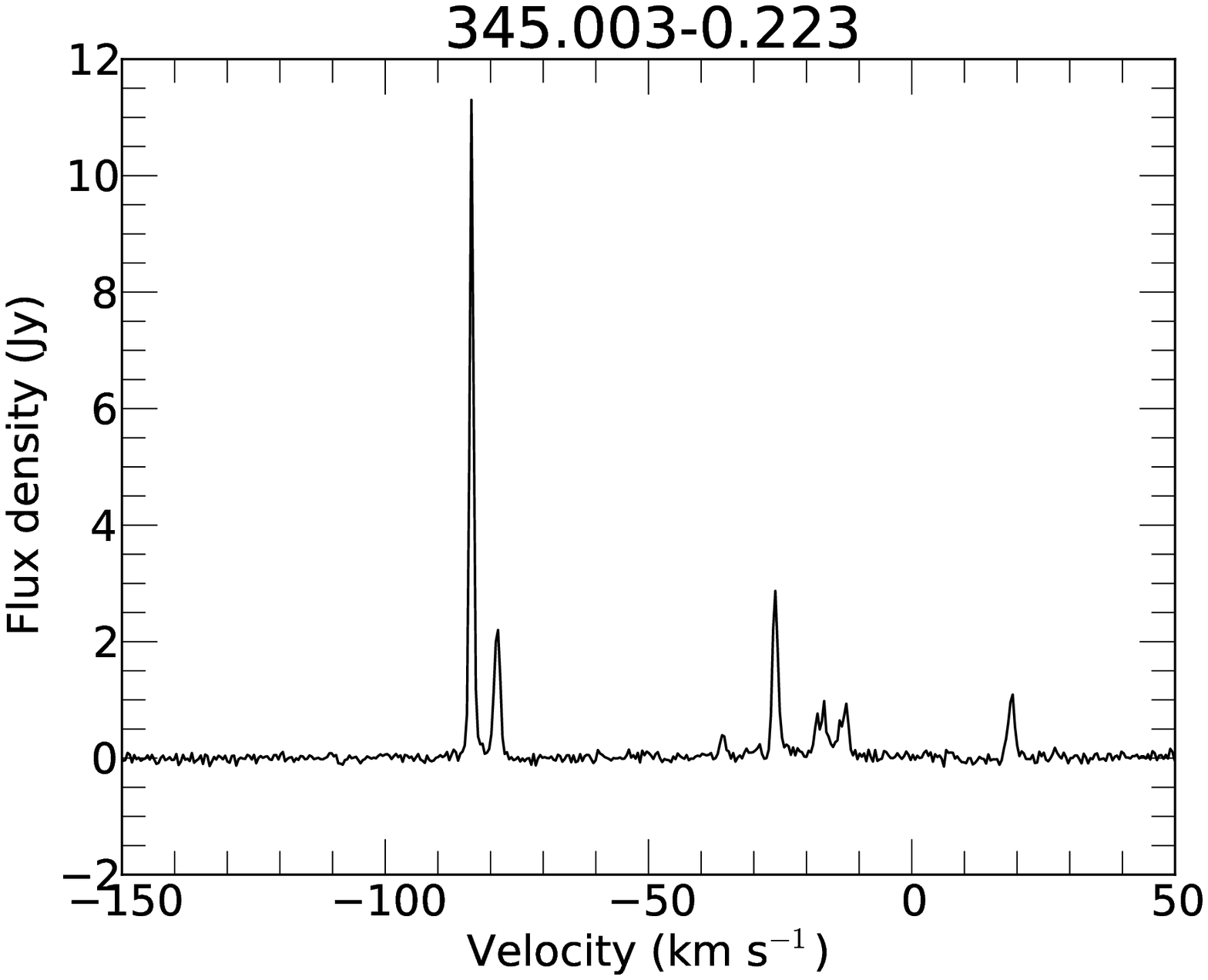}
\includegraphics[width=2.2in]{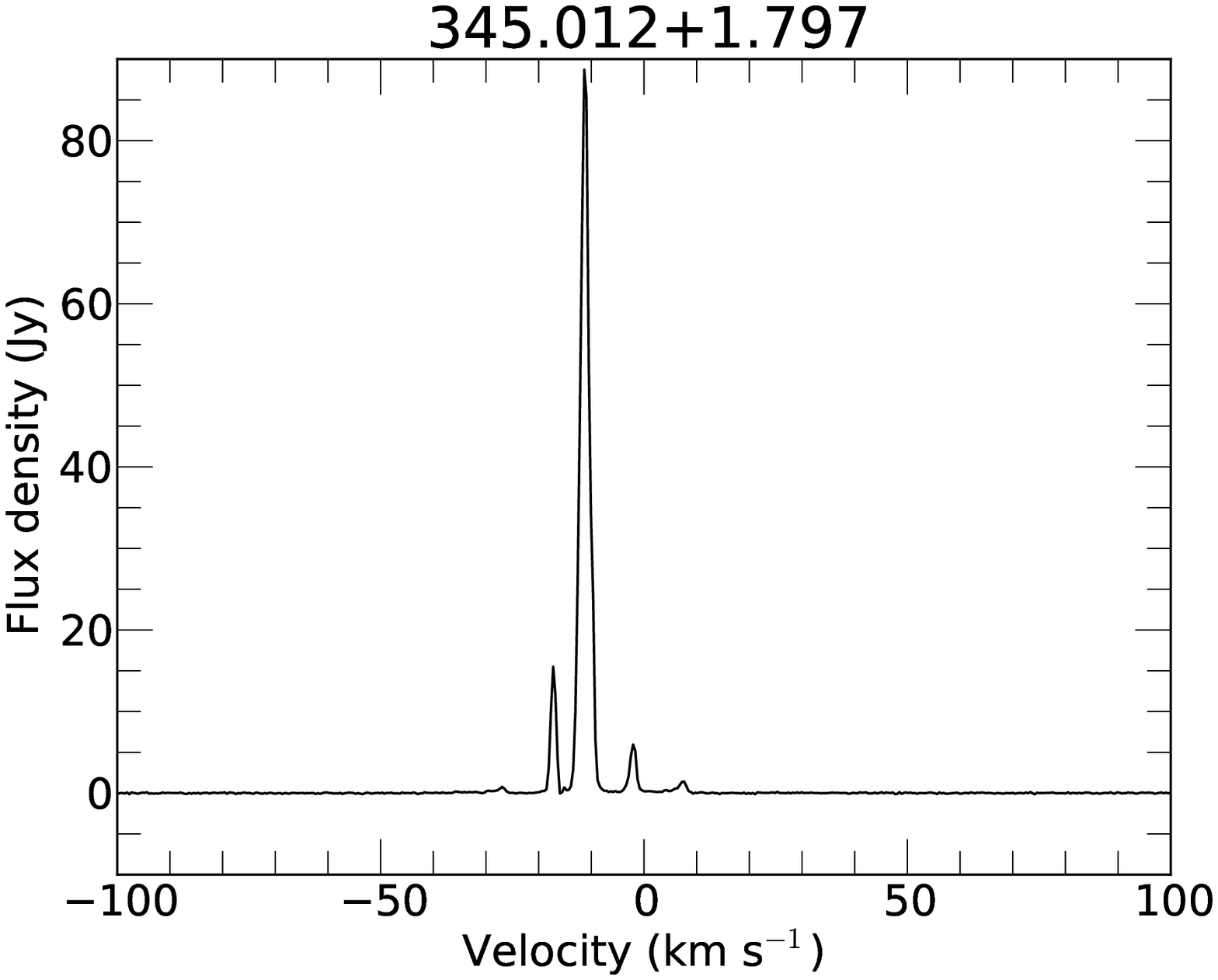}
\includegraphics[width=2.2in]{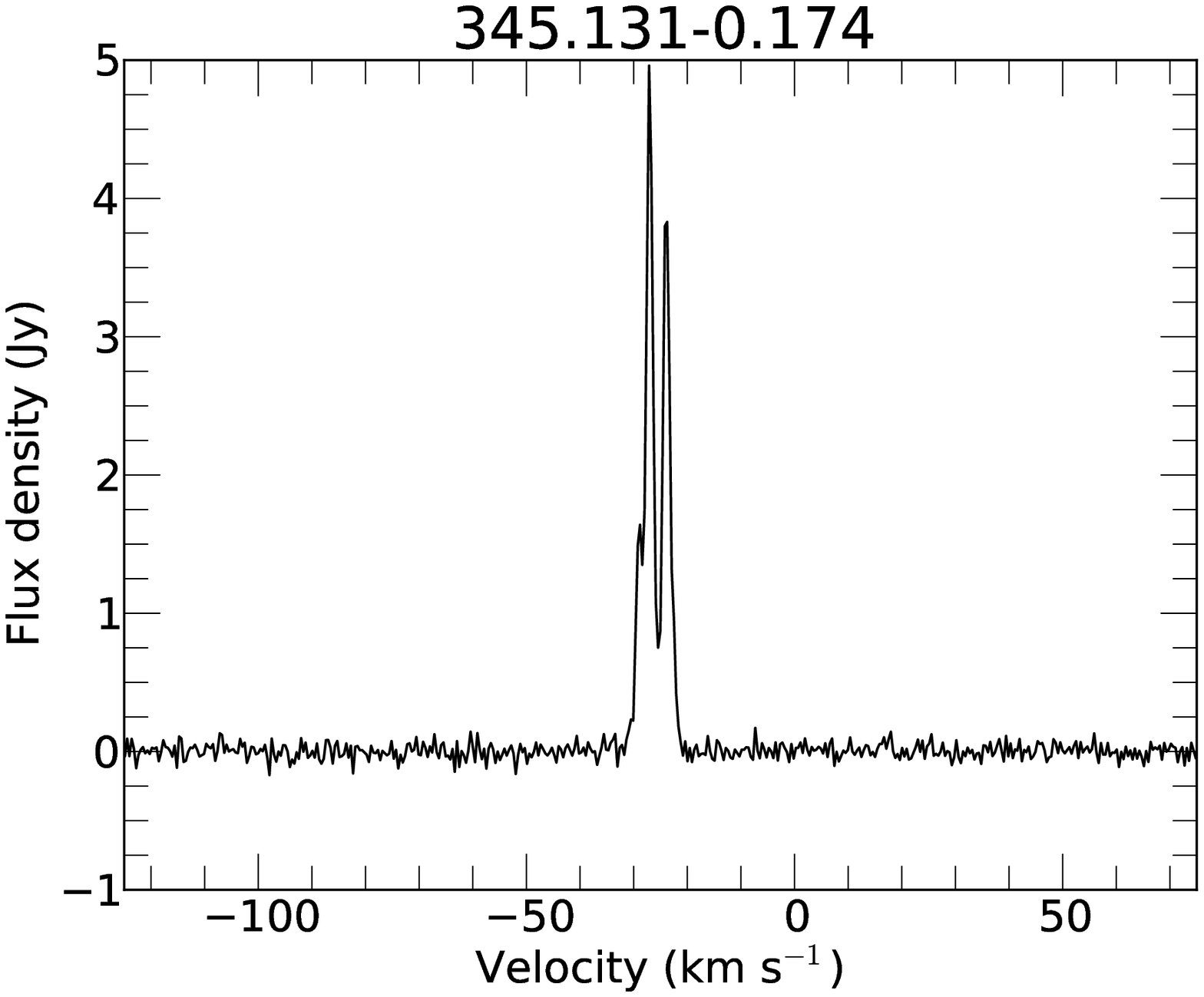}
\includegraphics[width=2.2in]{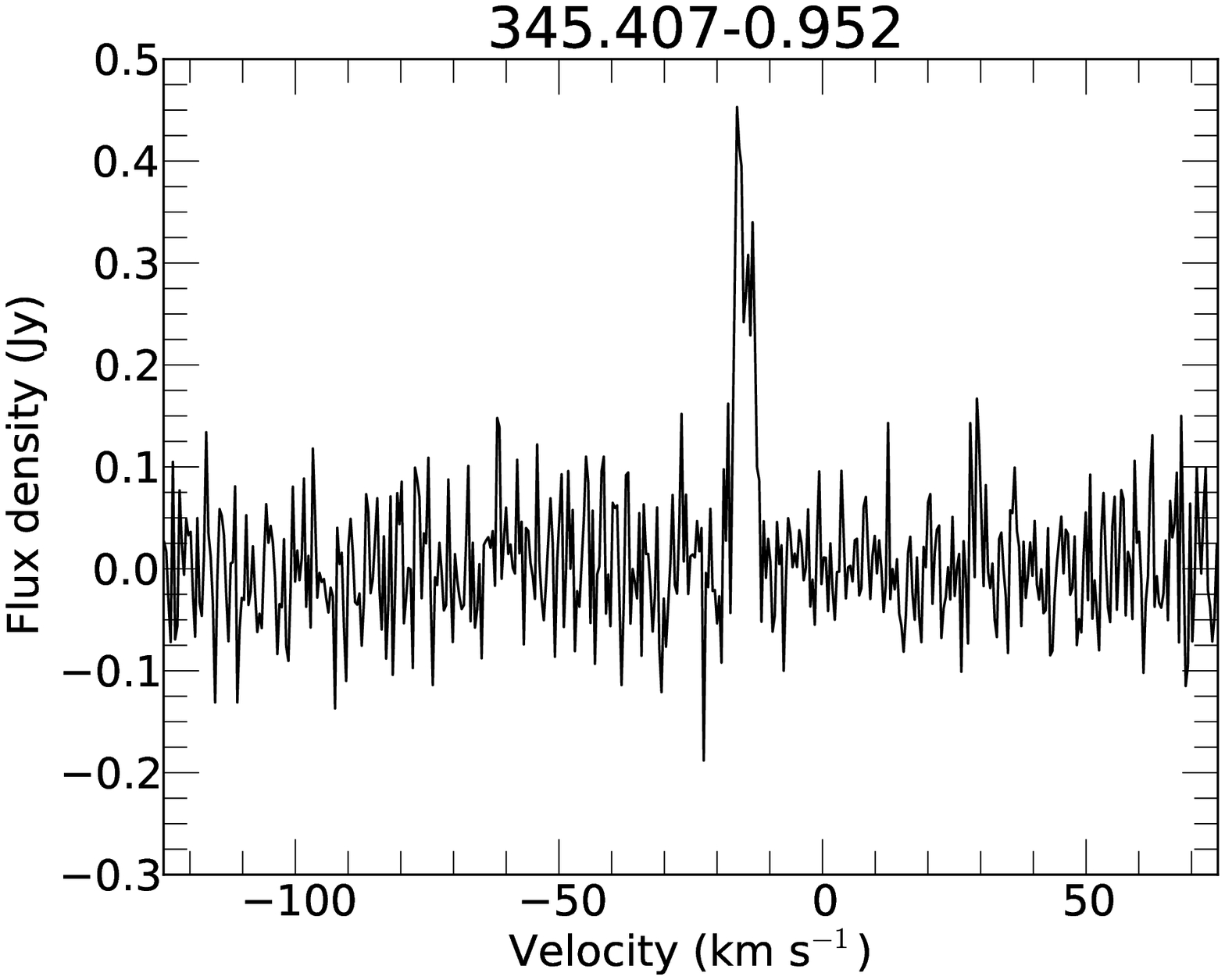}
\includegraphics[width=2.2in]{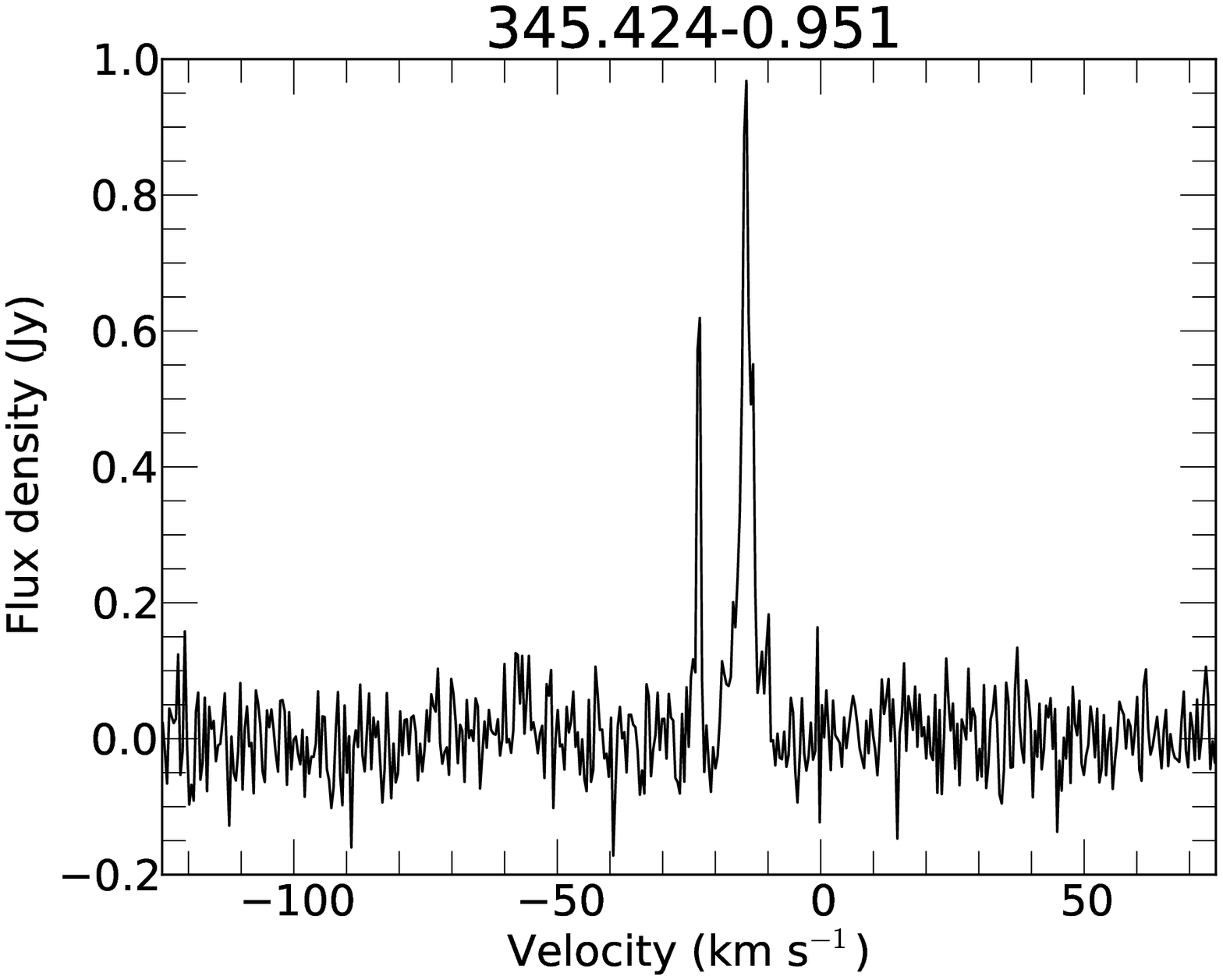}
\caption{Spectra obtained with the ATCA of water masers associated with 6.7-GHz methanol masers. The velocity is the velocity at the Local Standard of Rest.}
\label{fig:spectra}
\end{figure*}

\begin{figure*}

\includegraphics[width=2.2in]{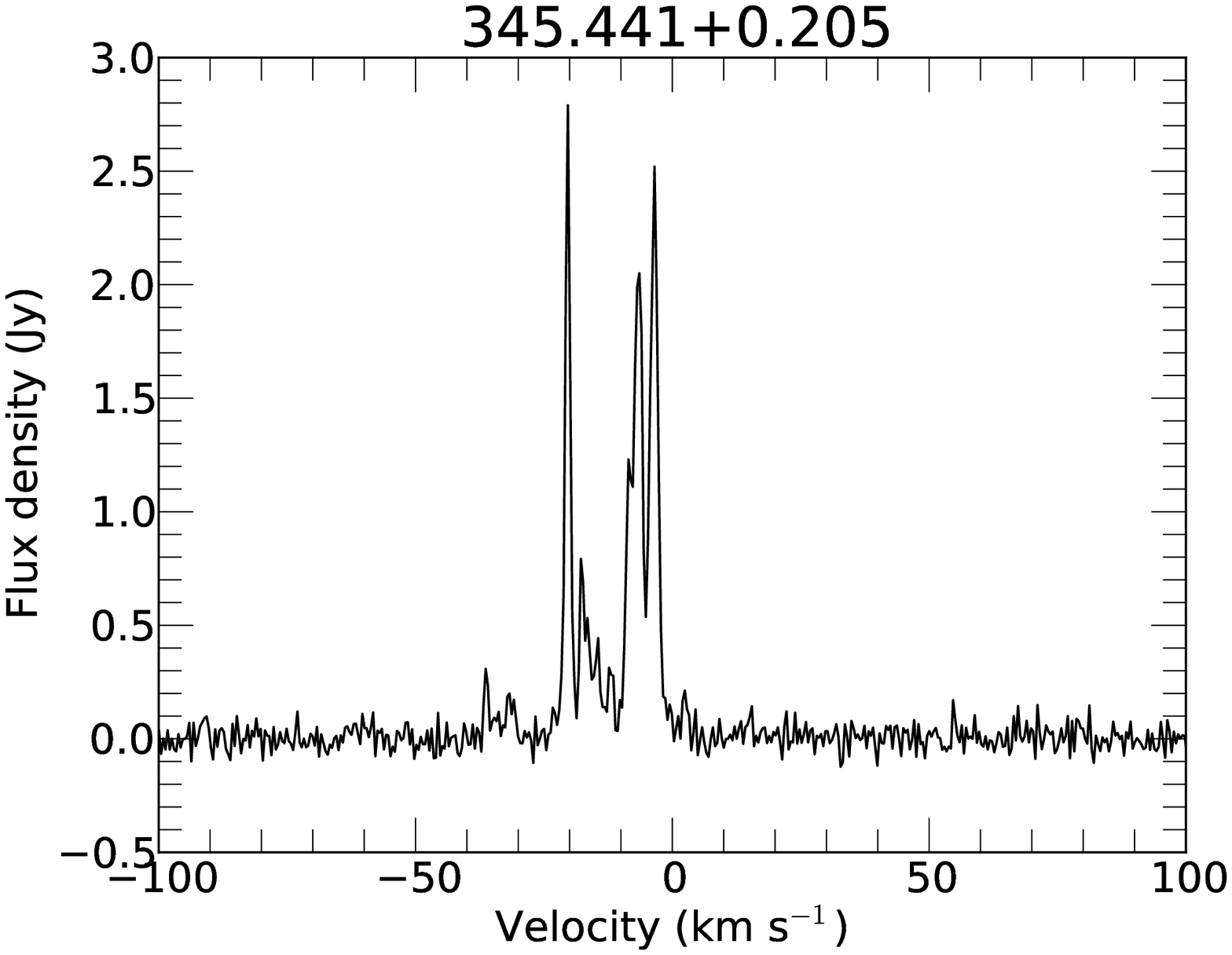}
\includegraphics[width=2.2in]{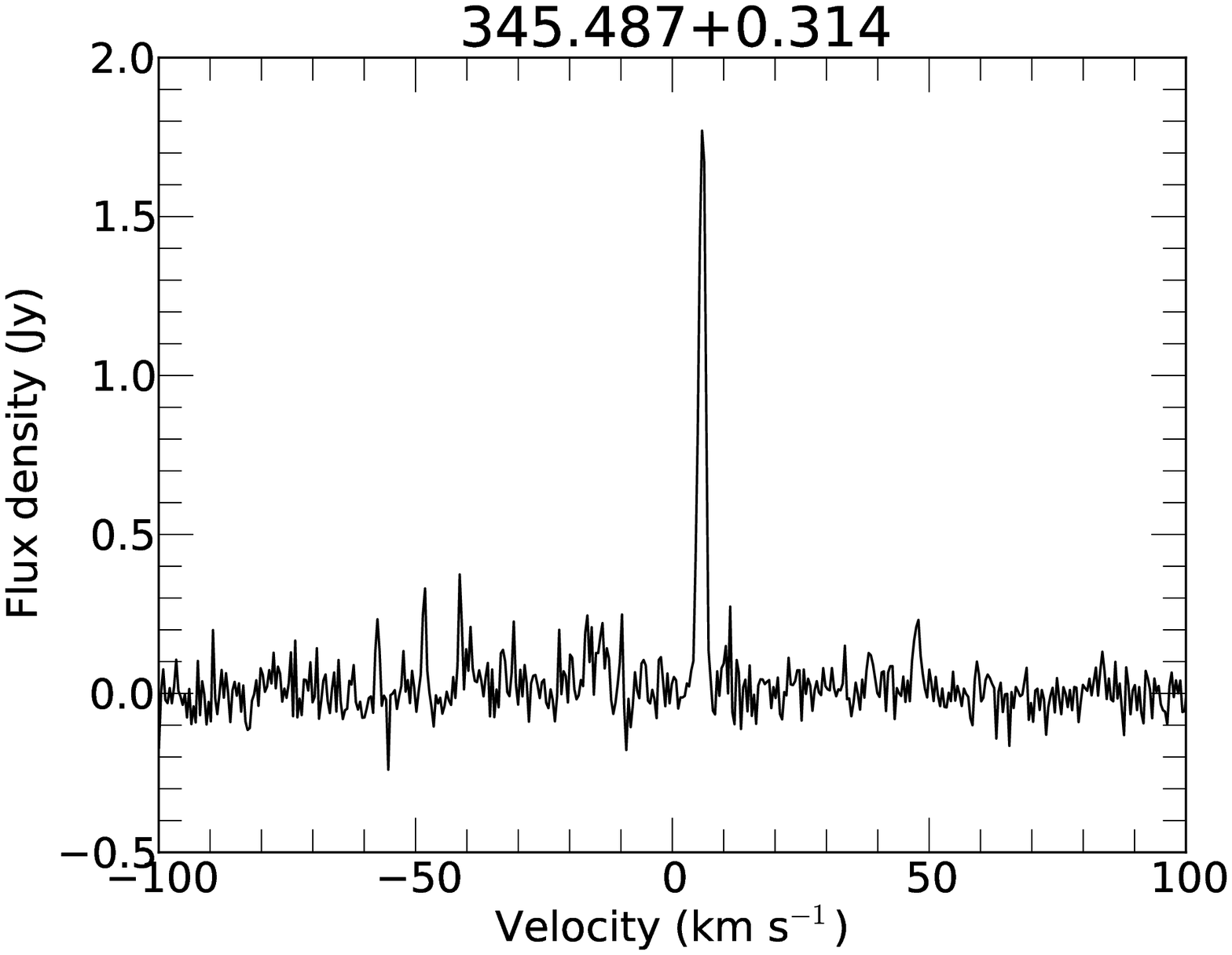}
\includegraphics[width=2.2in]{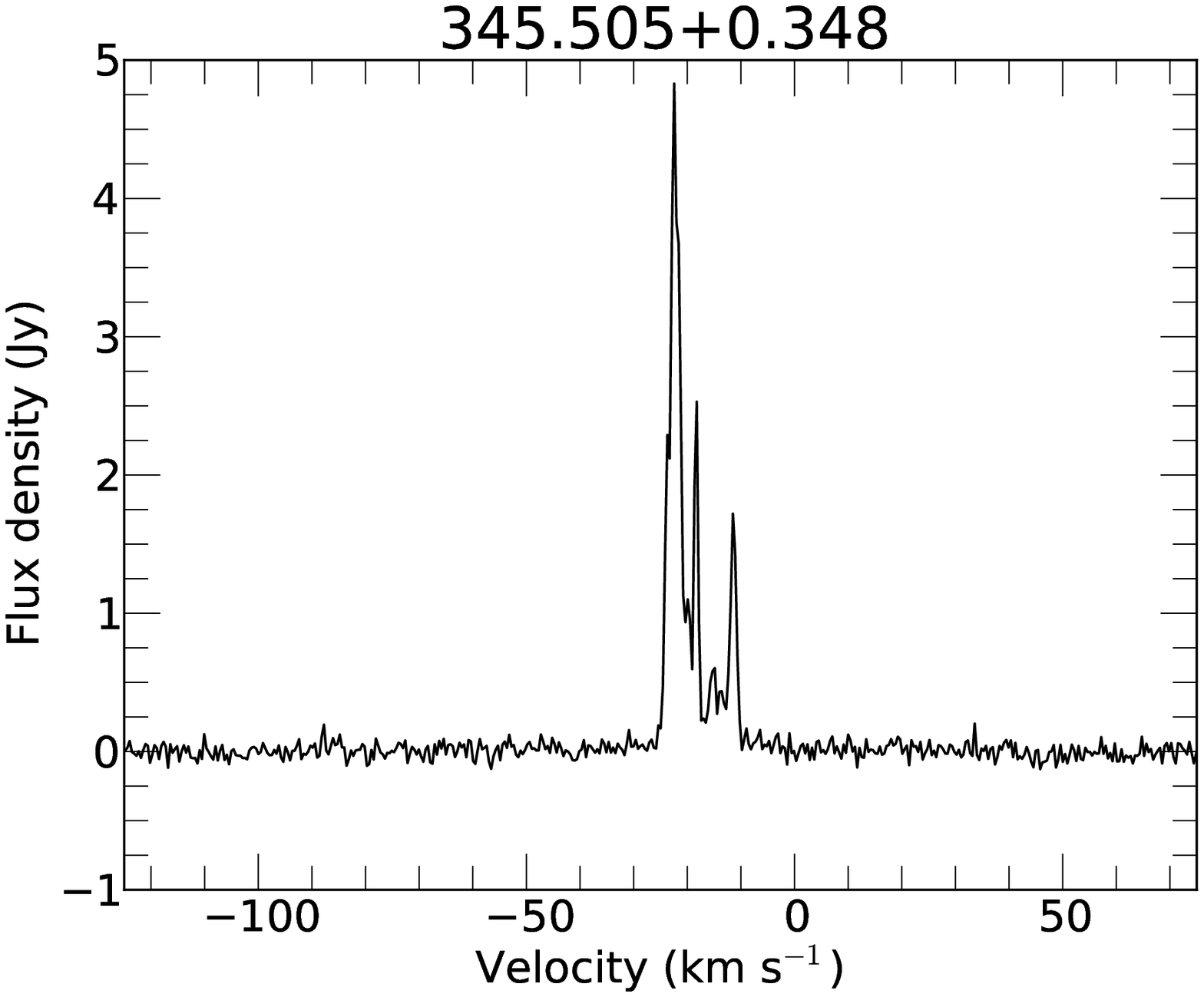}
\includegraphics[width=2.2in]{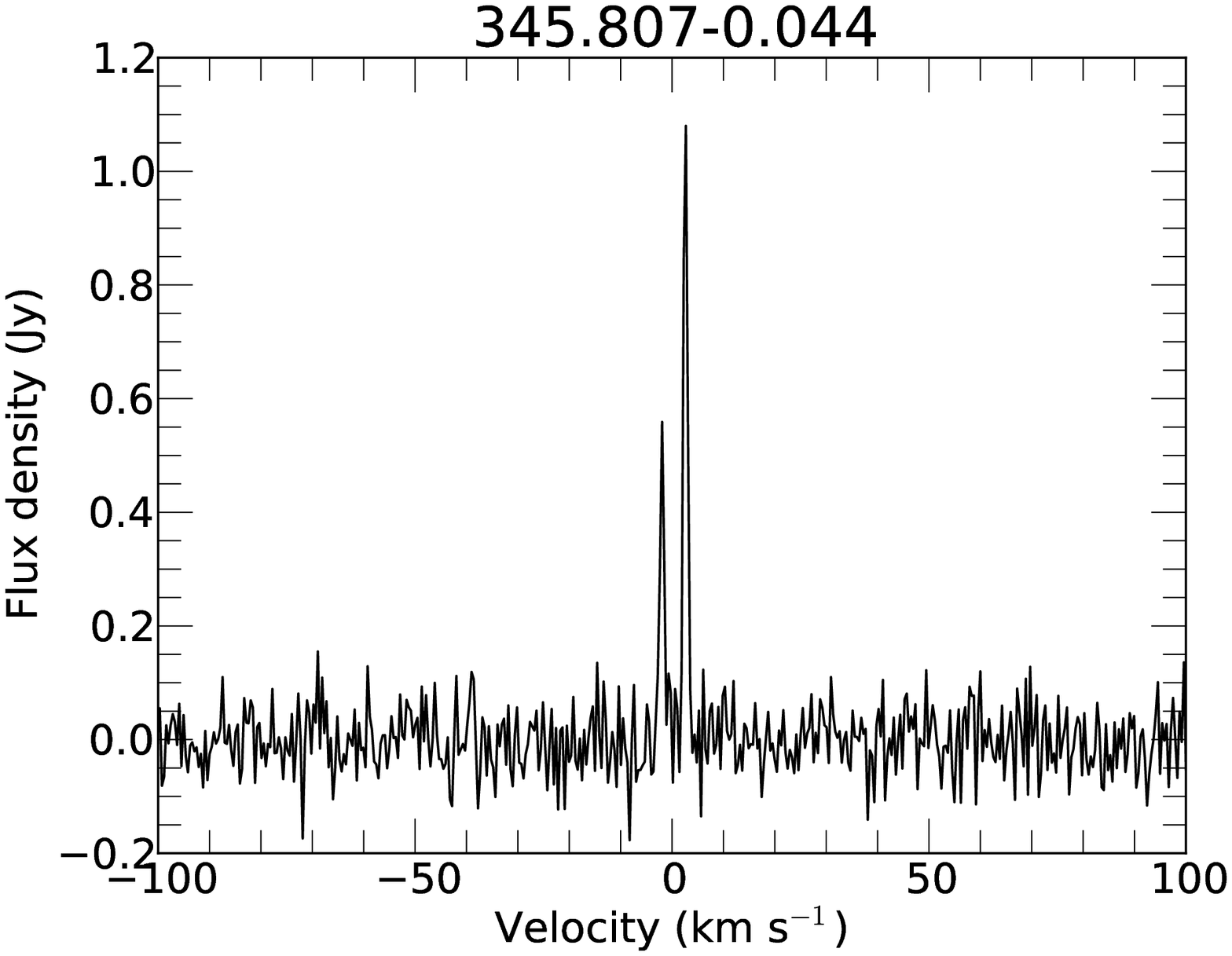}
\includegraphics[width=2.2in]{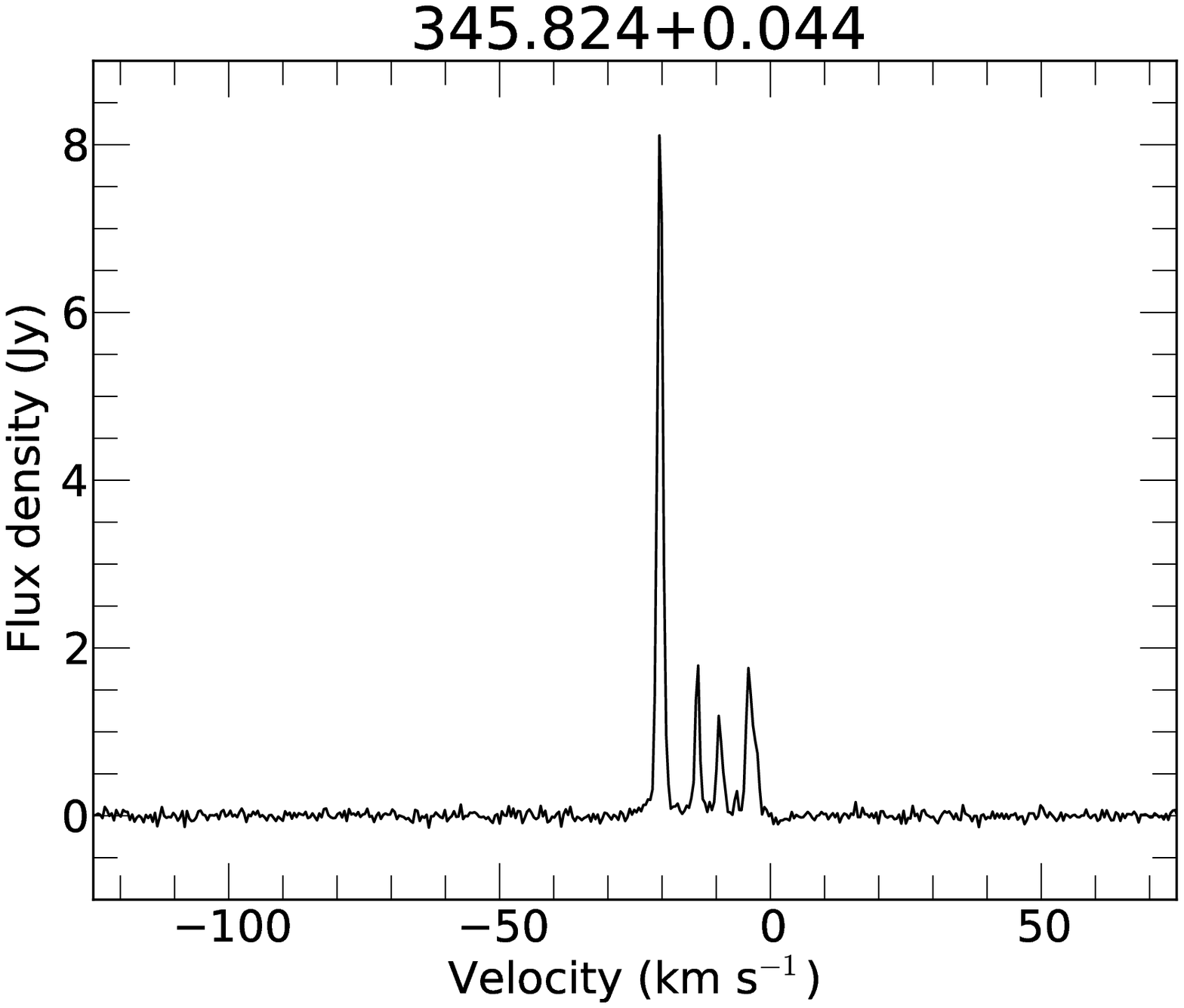}
\includegraphics[width=2.2in]{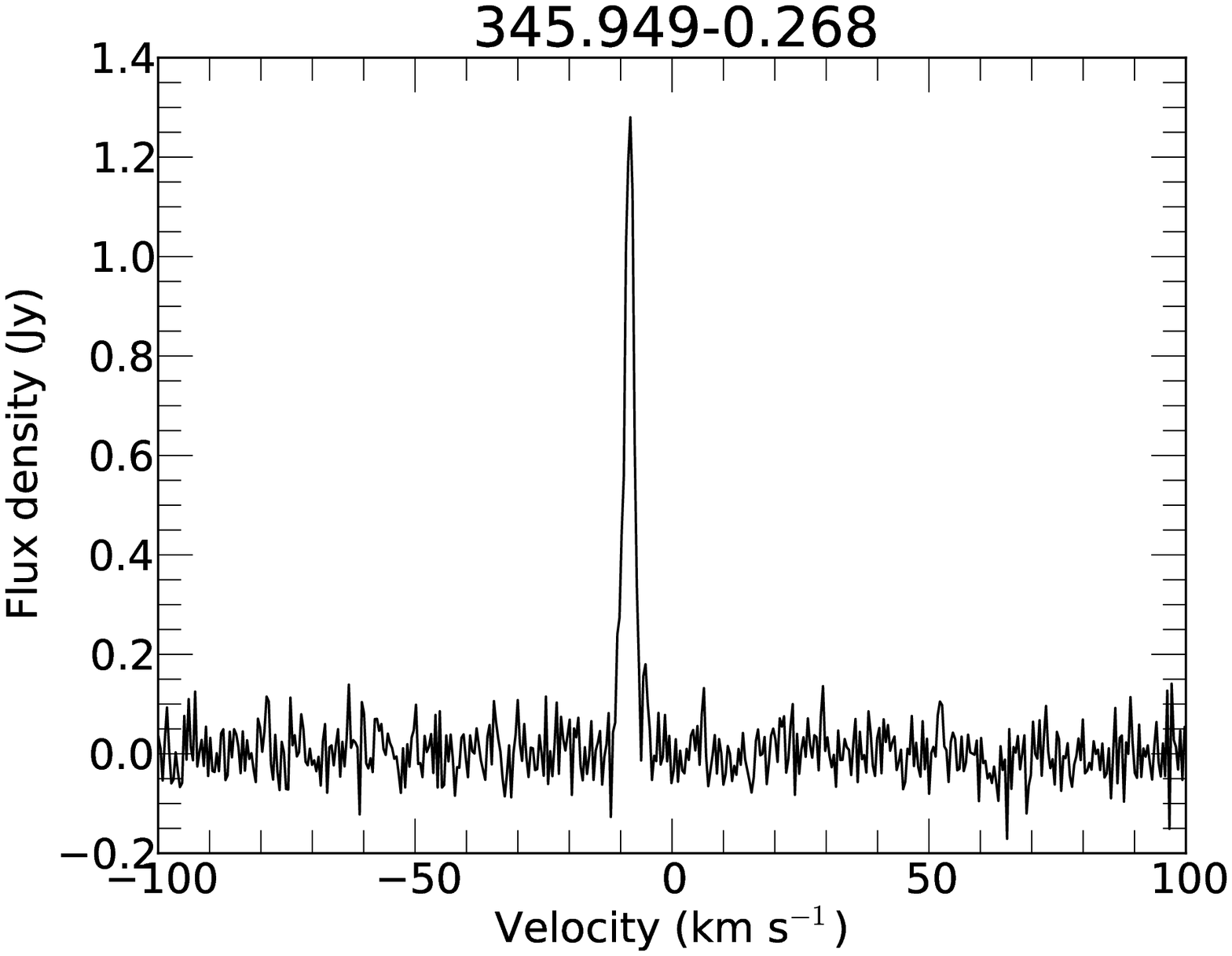}
\includegraphics[width=2.2in]{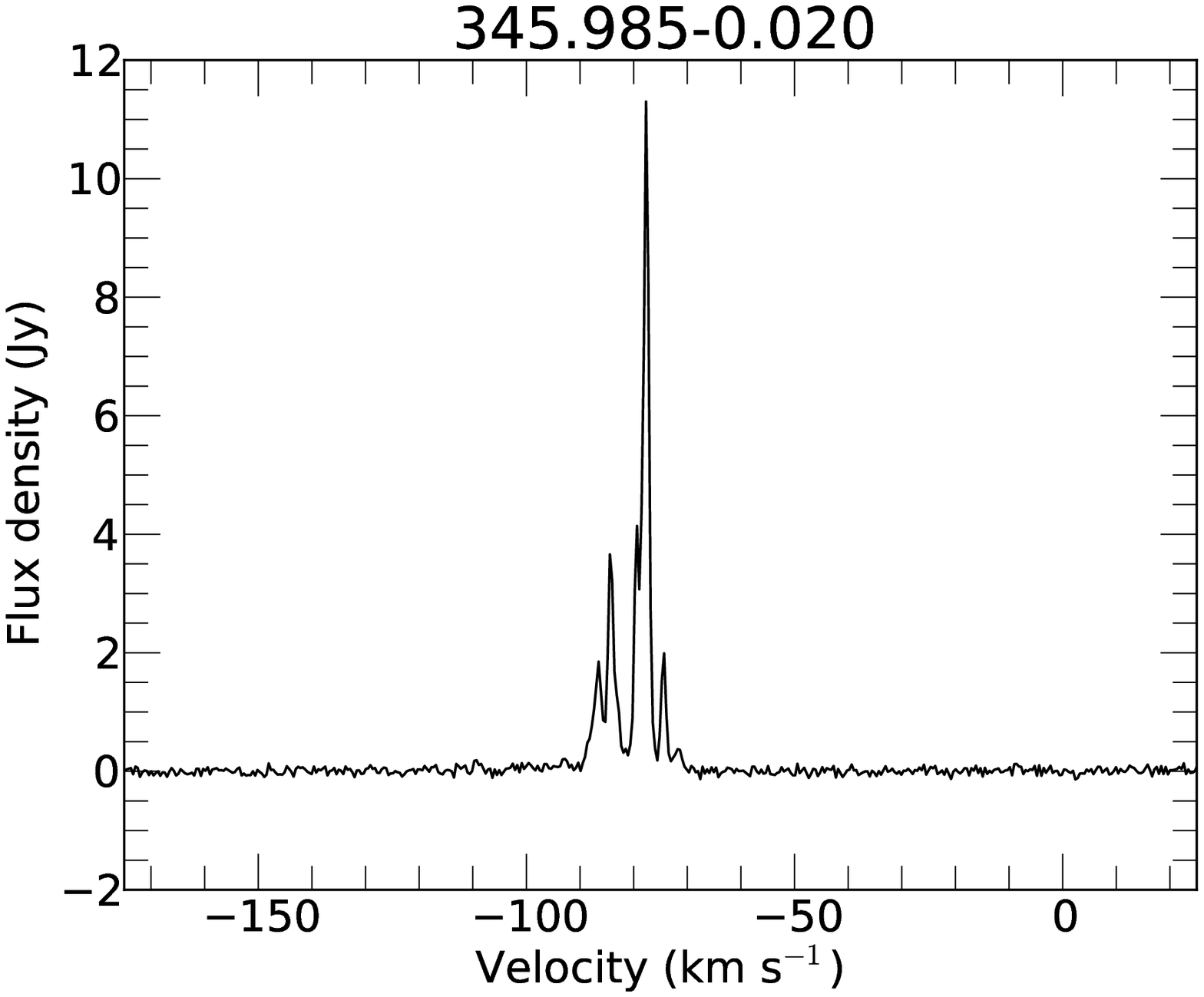}
\includegraphics[width=2.2in]{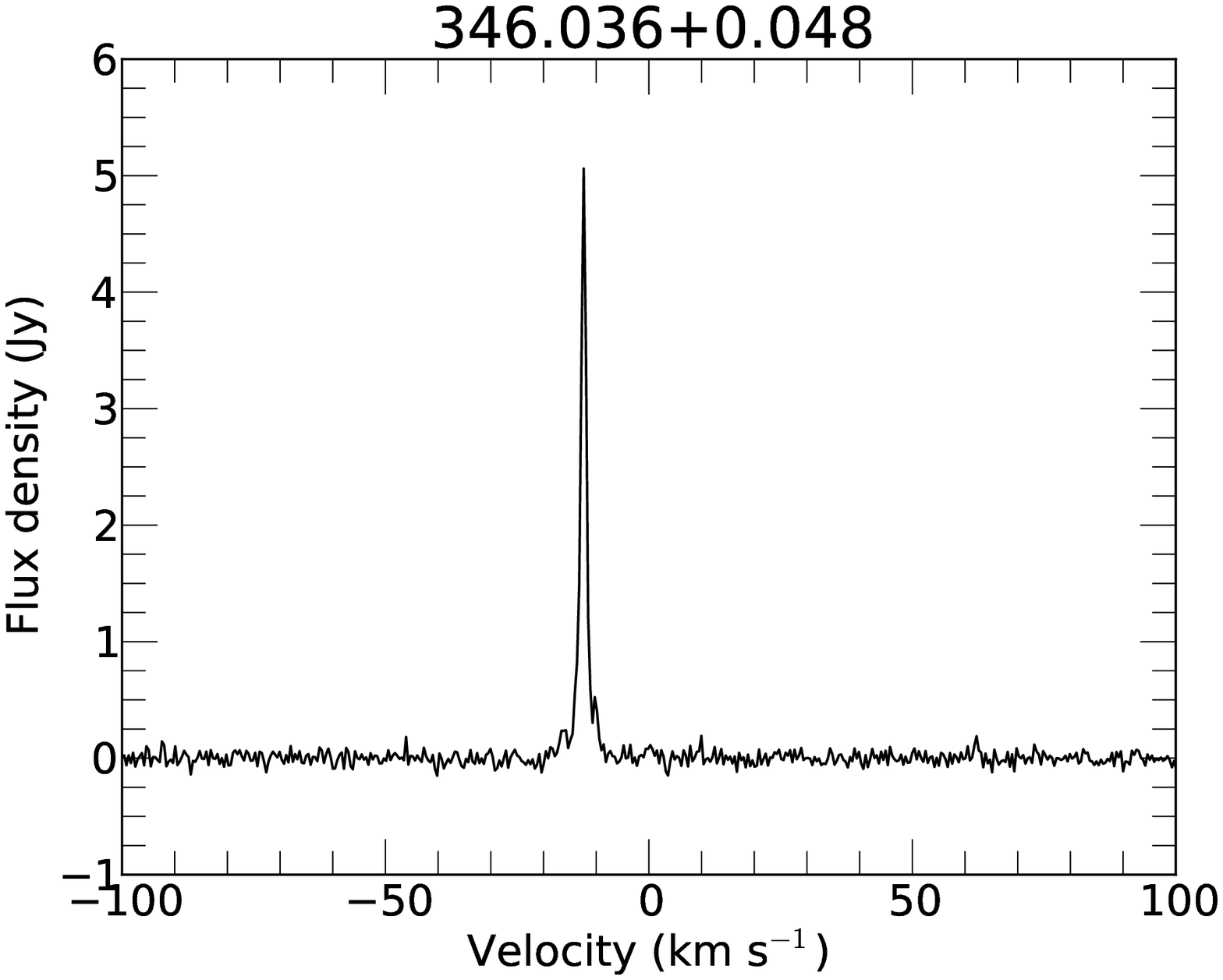}
\includegraphics[width=2.2in]{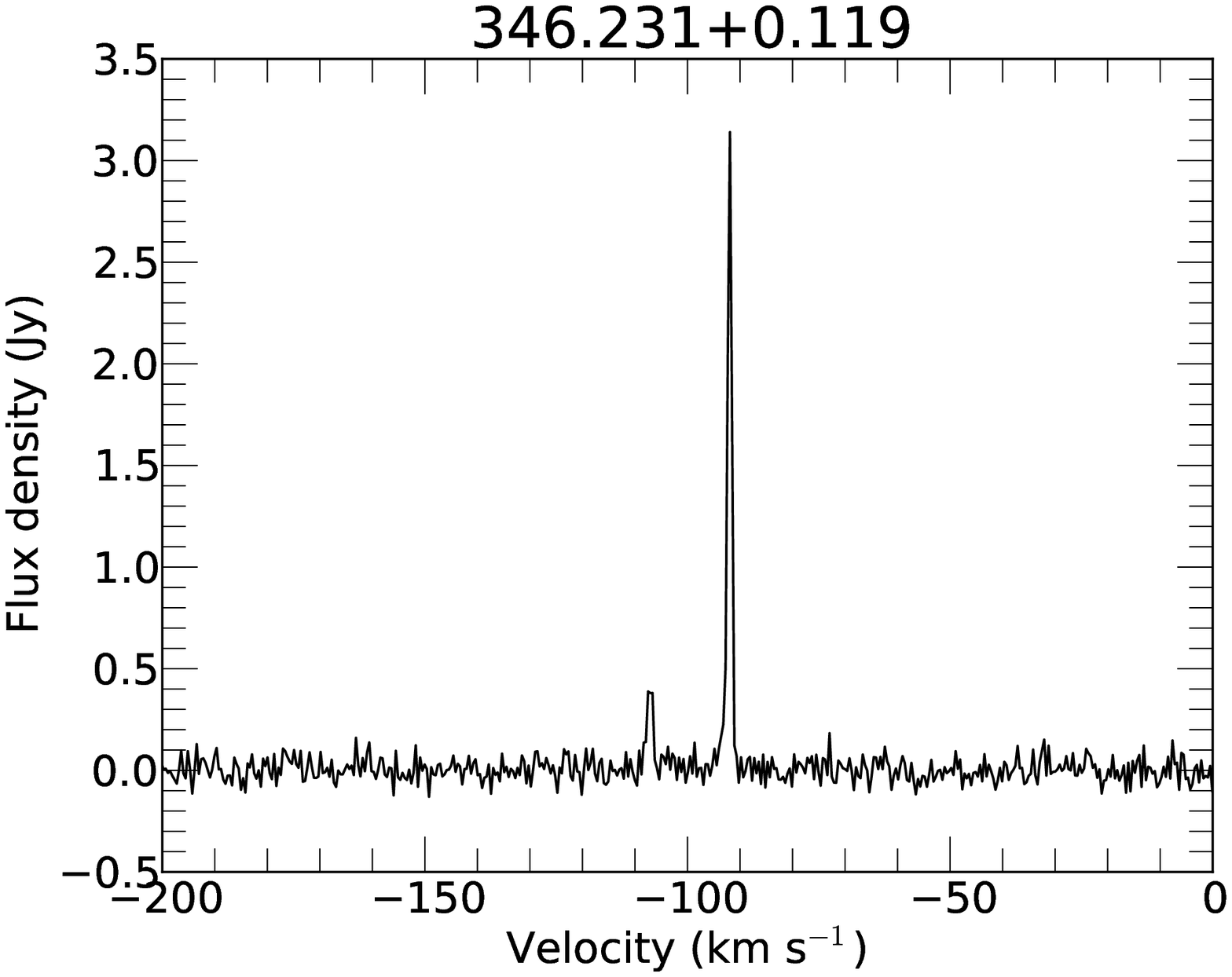}
\includegraphics[width=2.2in]{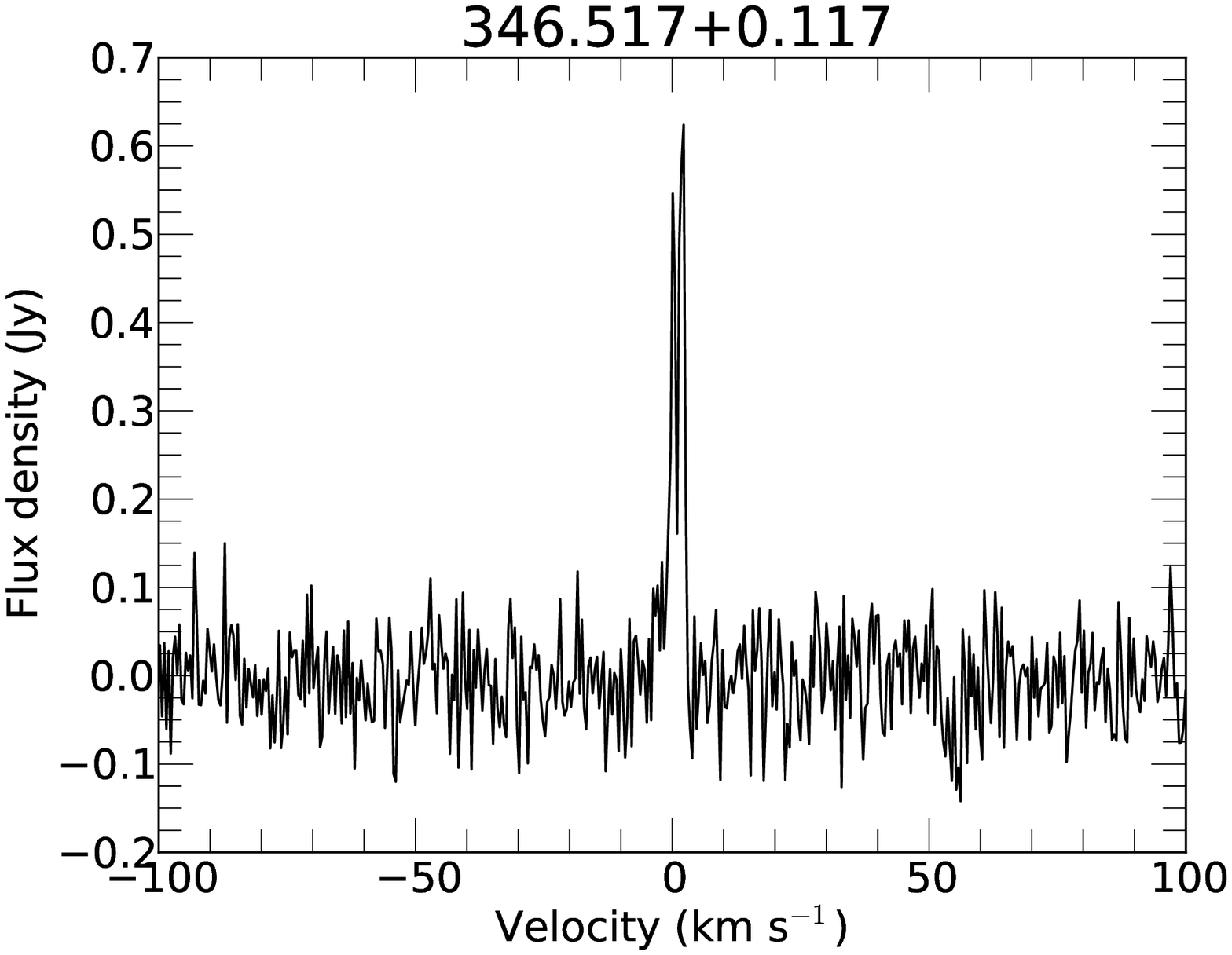}
\includegraphics[width=2.2in]{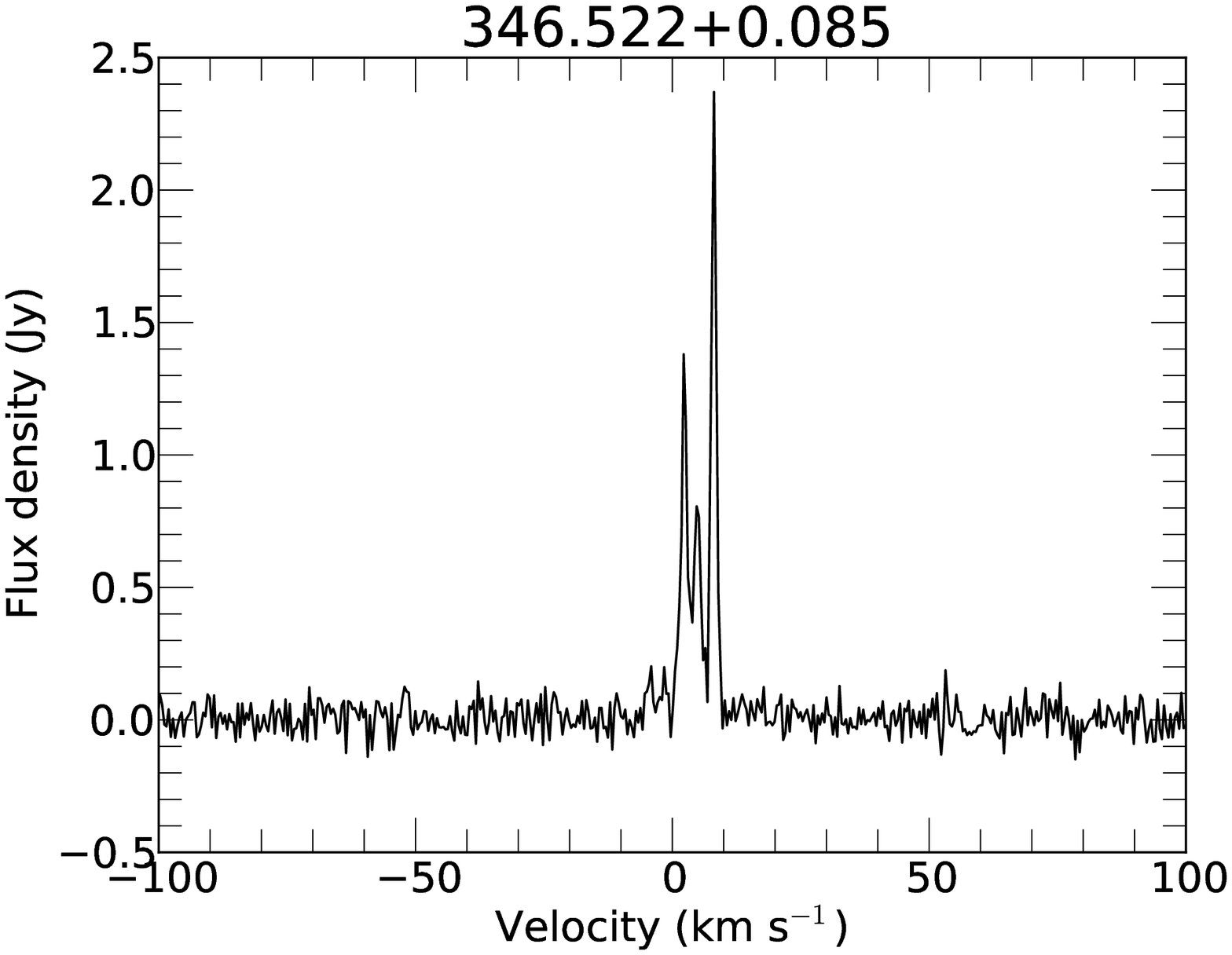}
\includegraphics[width=2.2in]{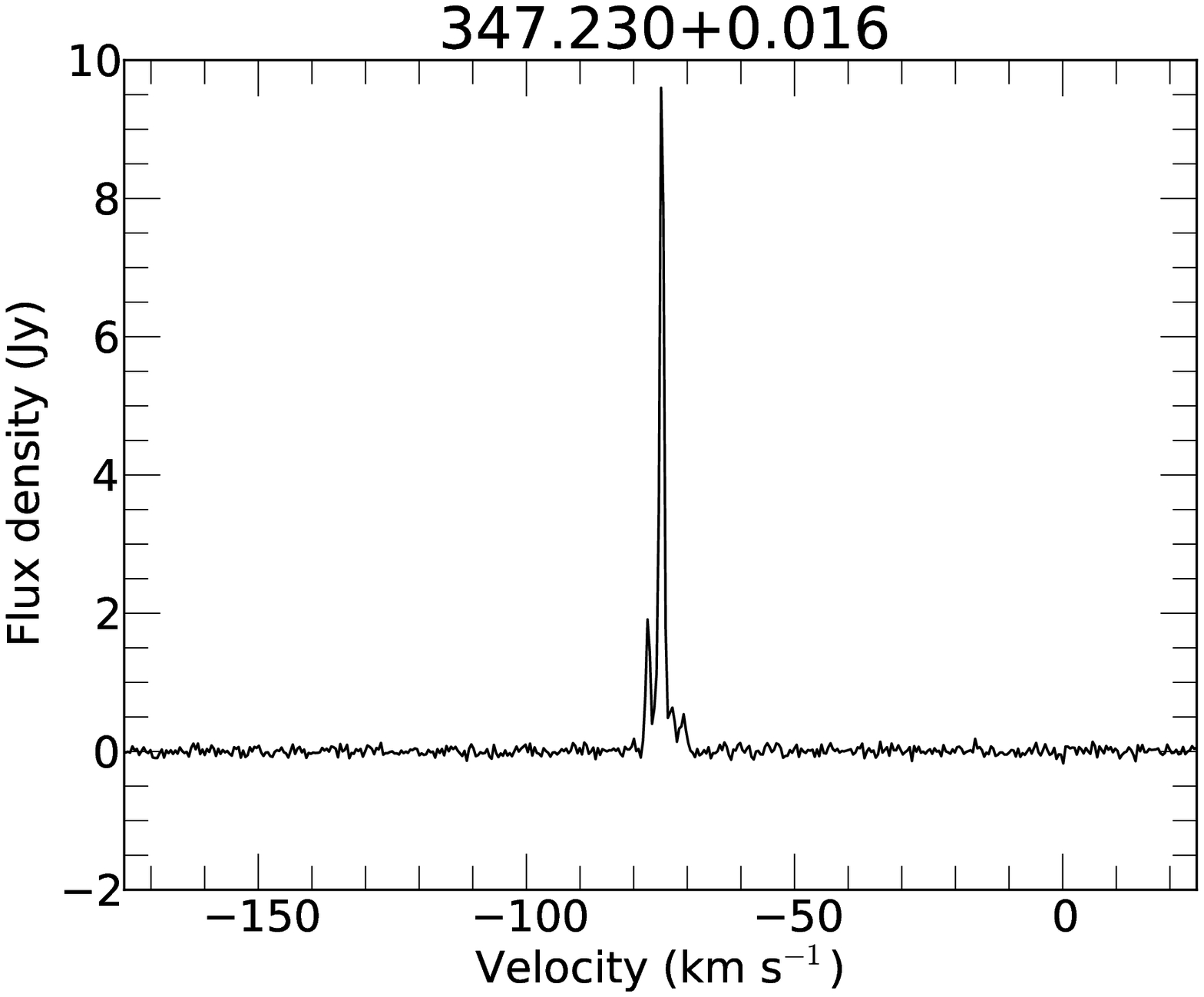}
\includegraphics[width=2.2in]{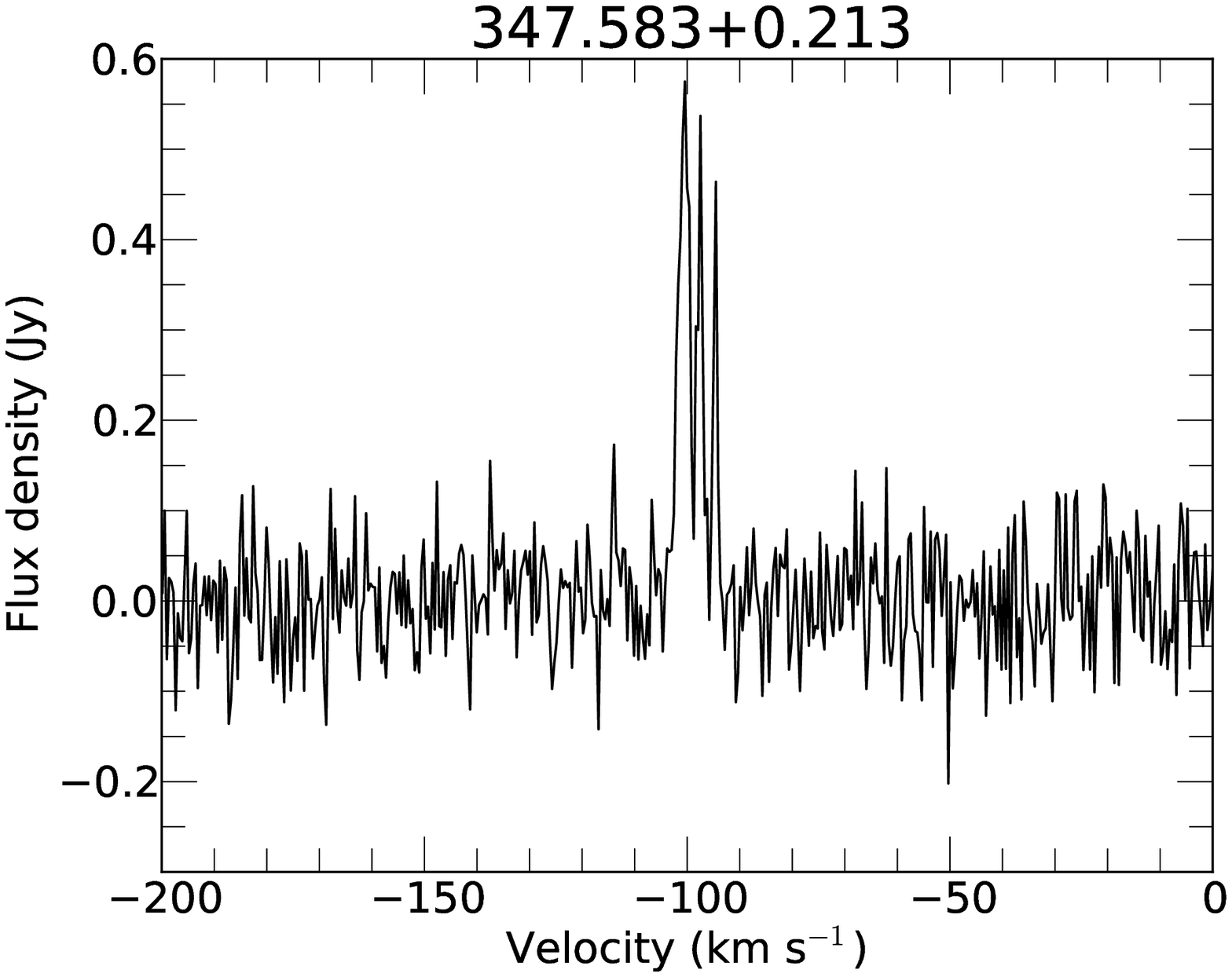}
\includegraphics[width=2.2in]{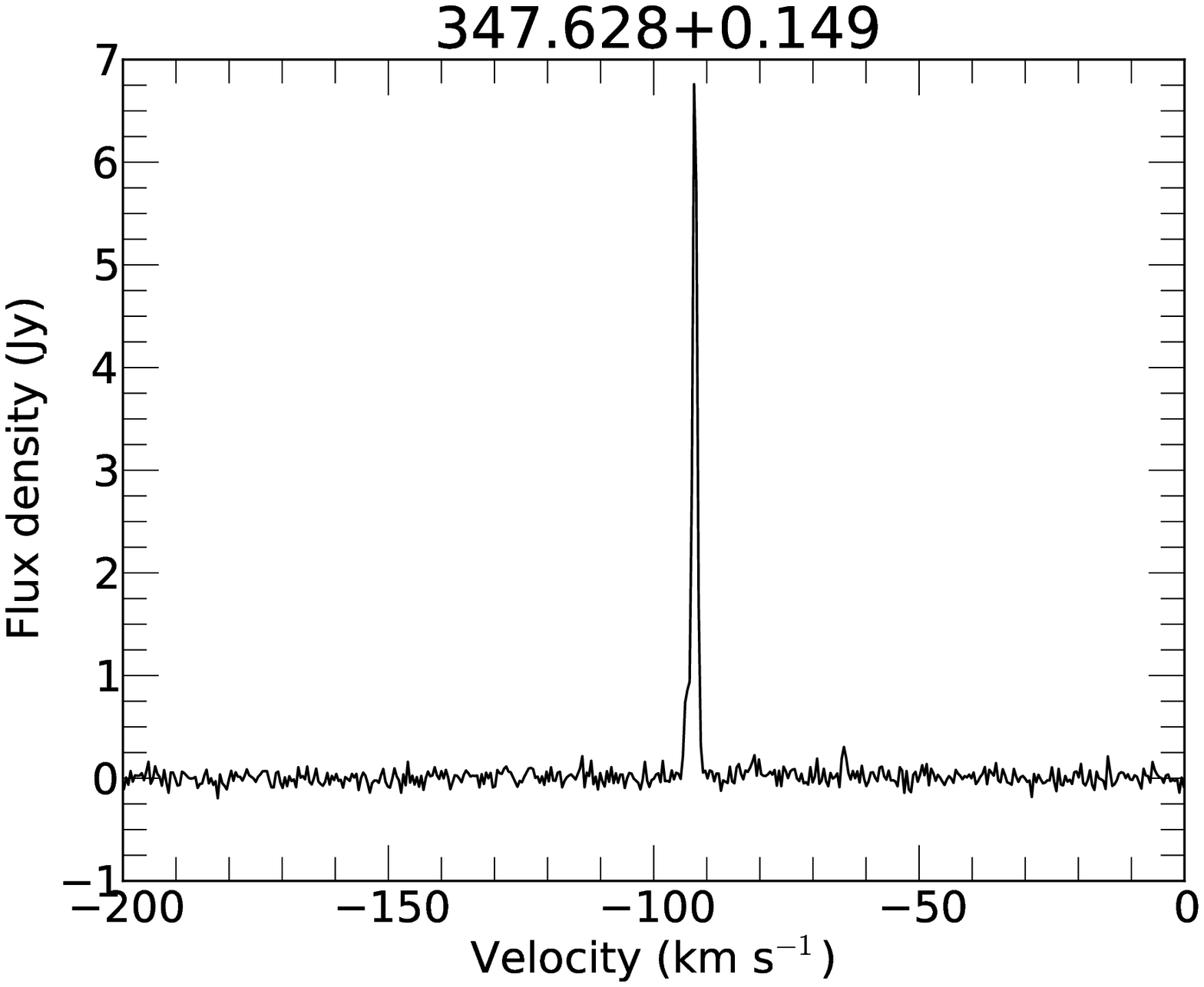}
\includegraphics[width=2.2in]{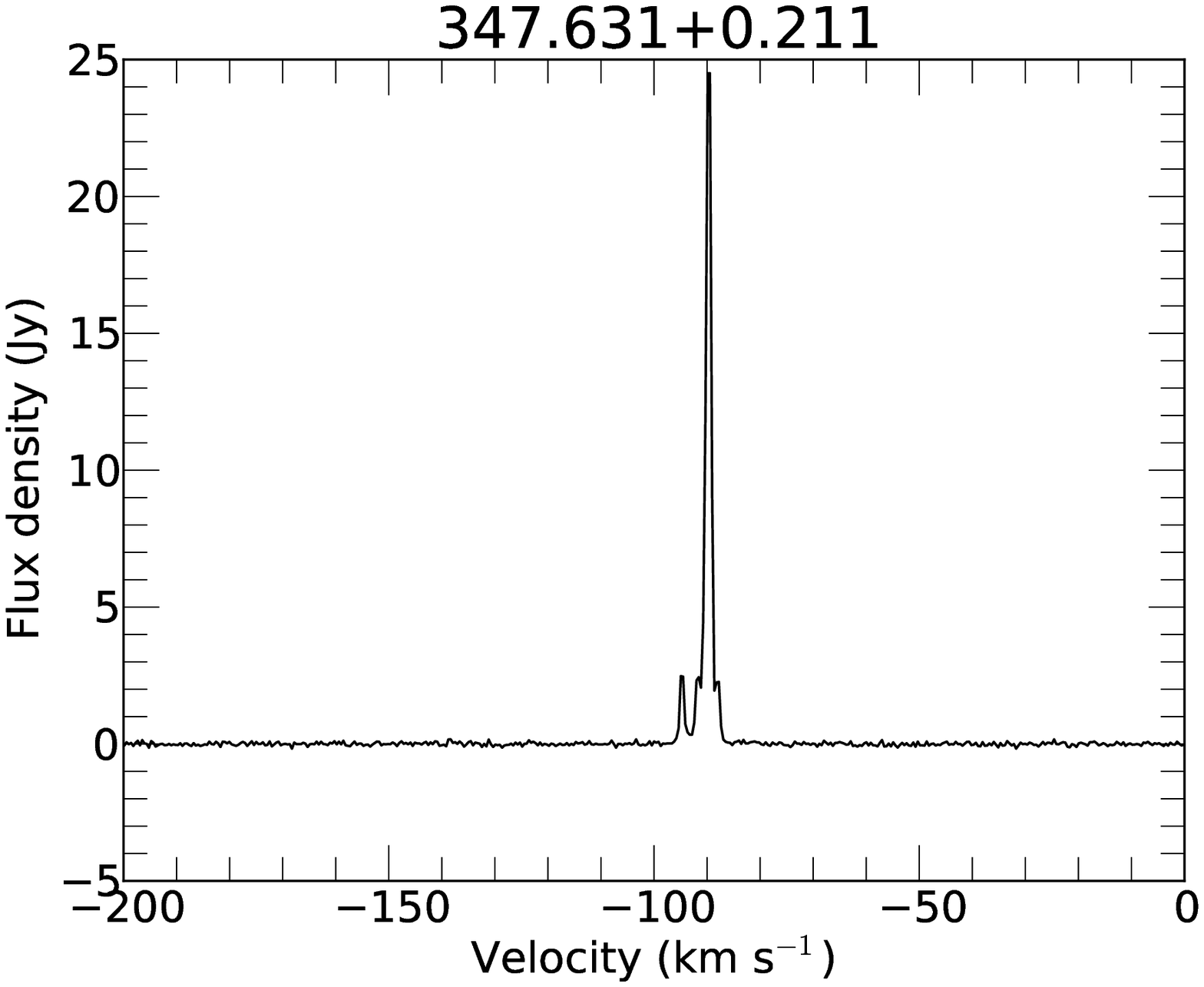}
\\
\addtocounter{figure}{-1}
  \caption{-- {\emph {continued}}}
\end{figure*}

\begin{figure*}
\includegraphics[width=2.2in]{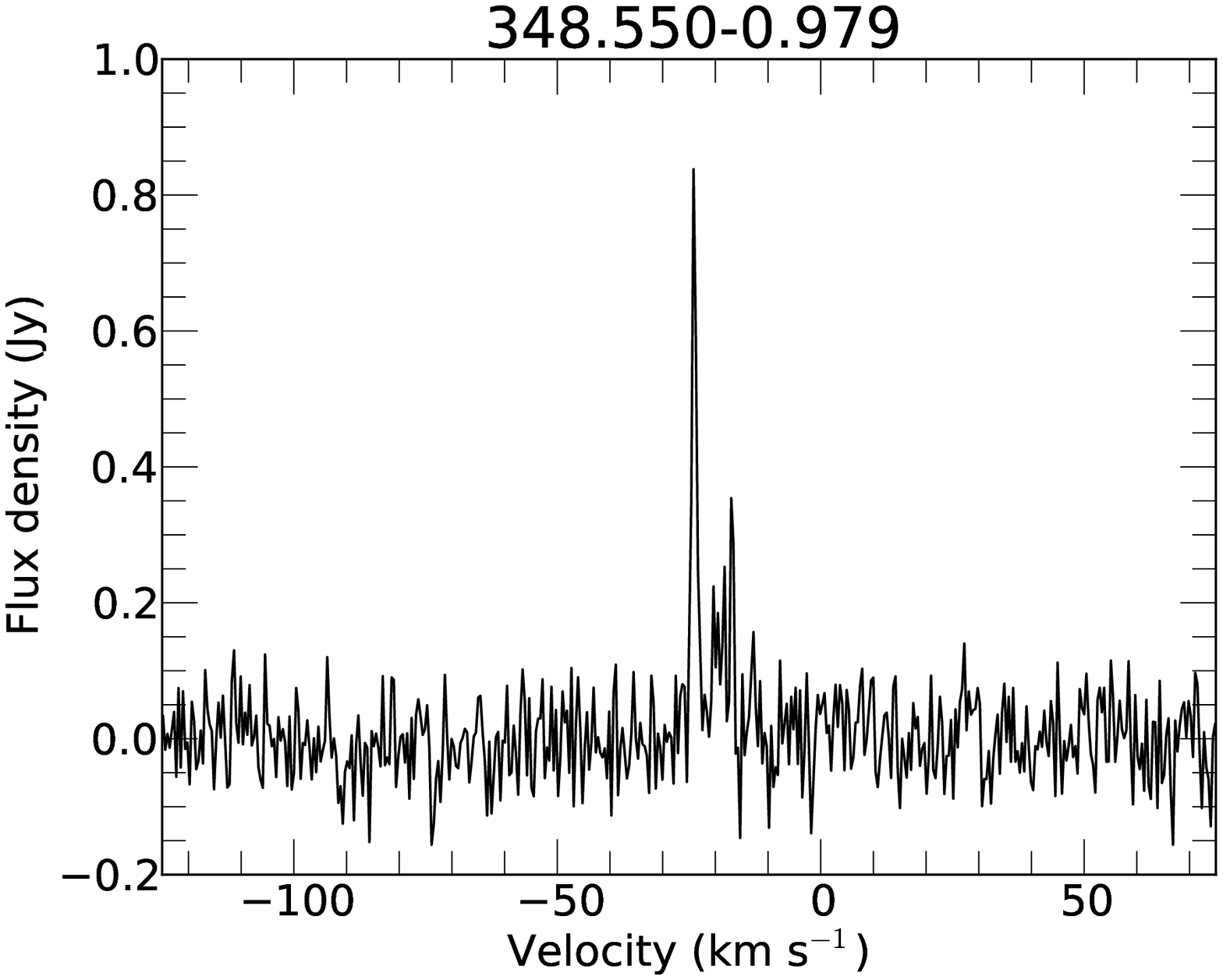}
\includegraphics[width=2.2in]{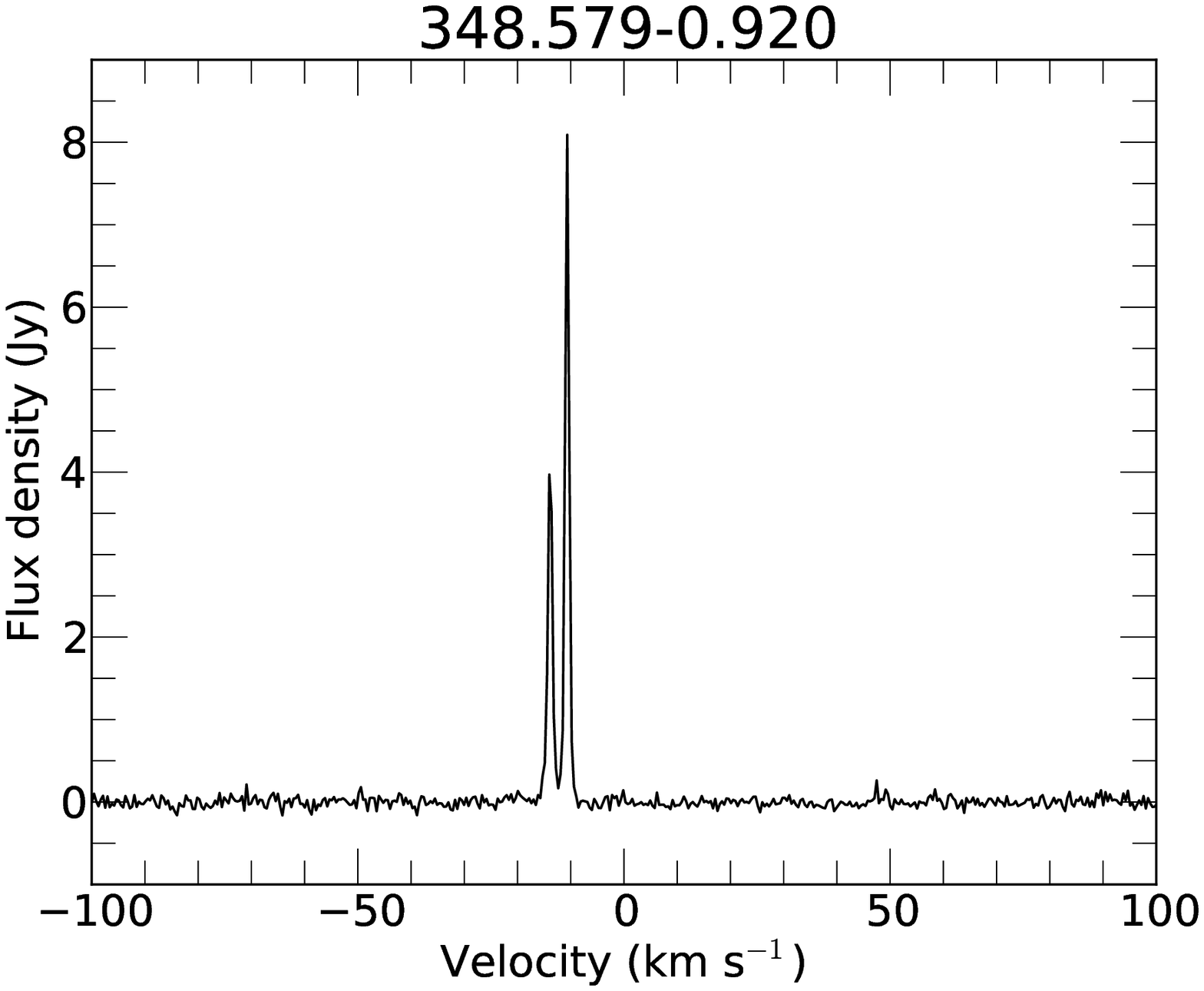}
\includegraphics[width=2.2in]{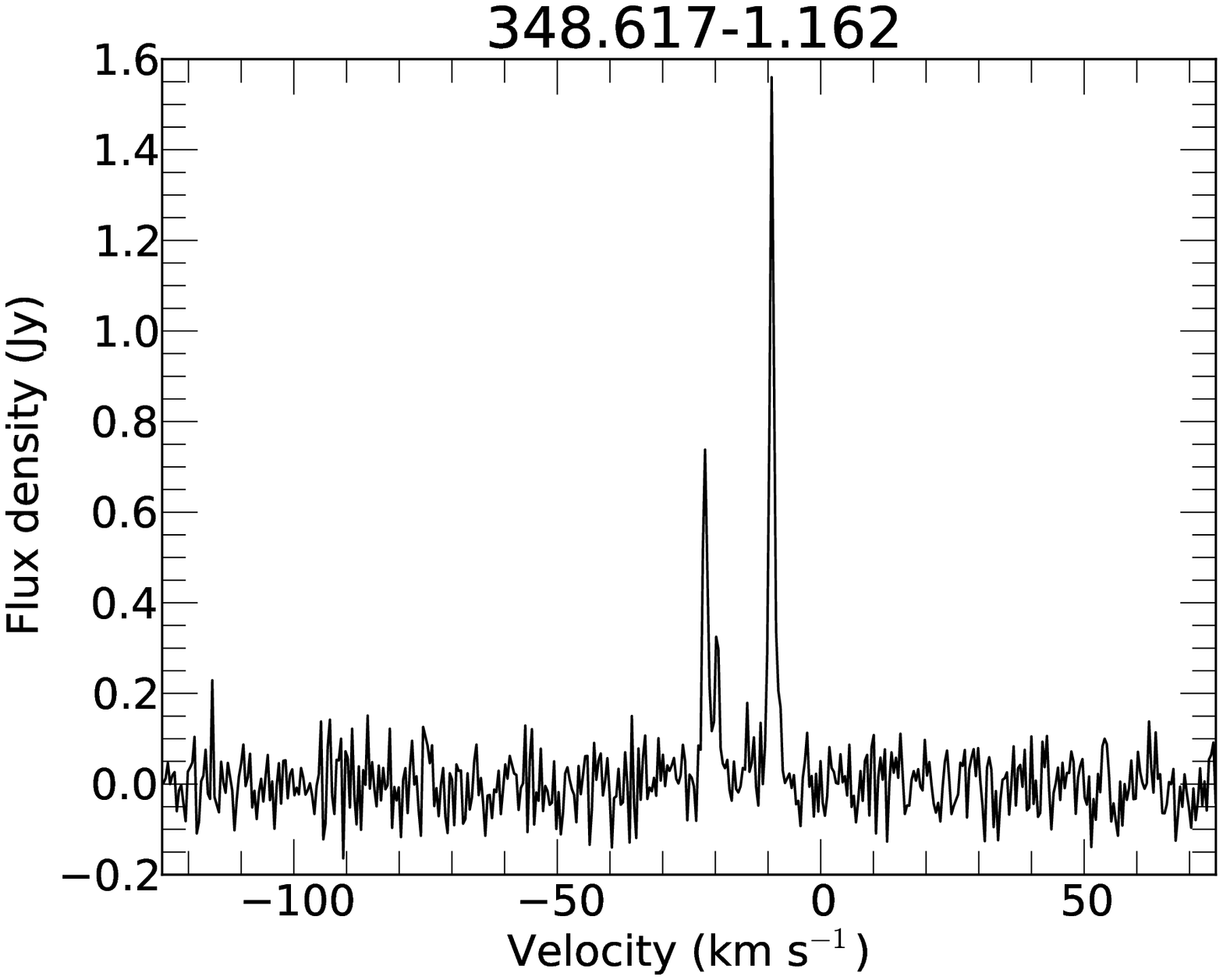}
\includegraphics[width=2.2in]{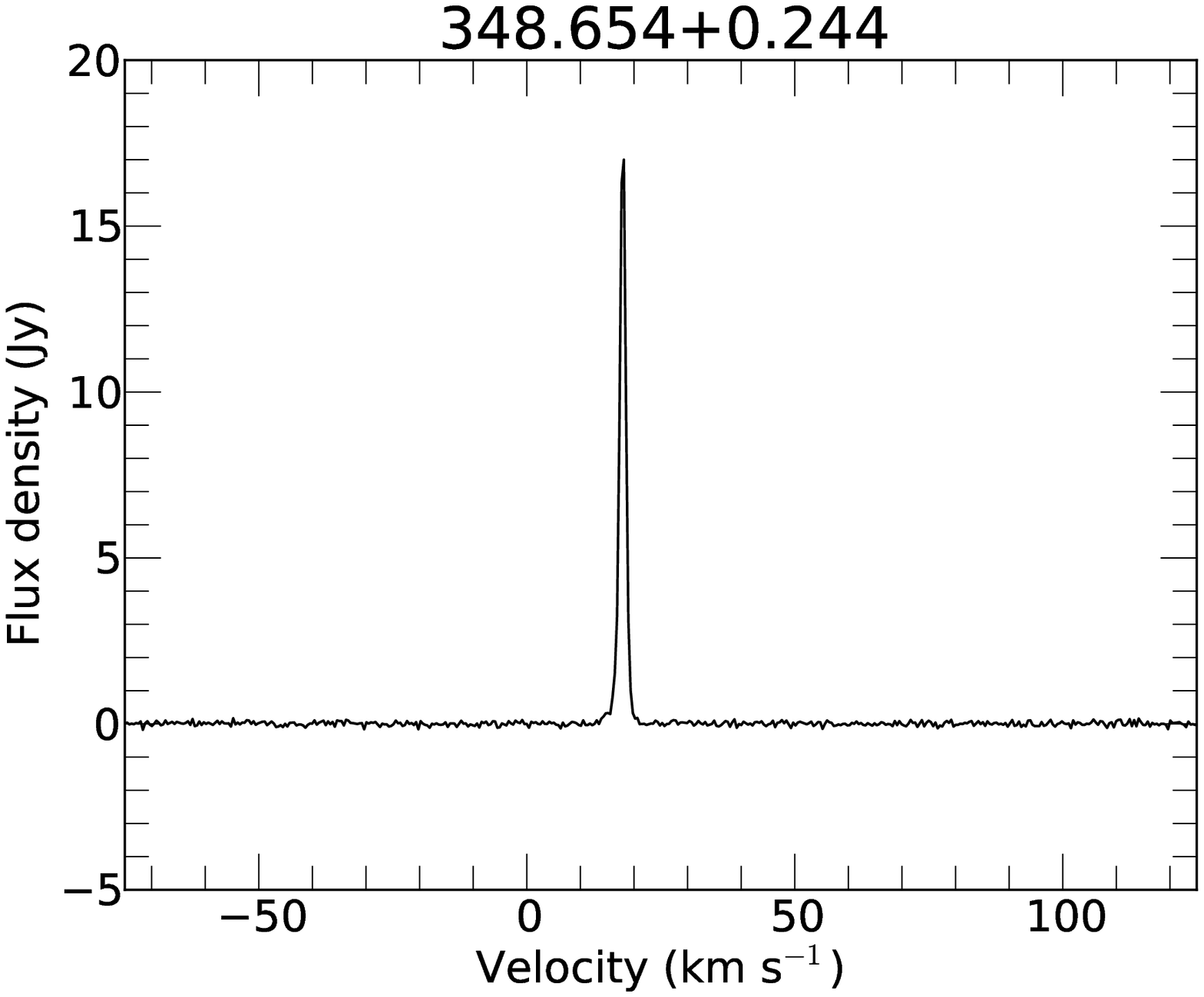}
\includegraphics[width=2.2in]{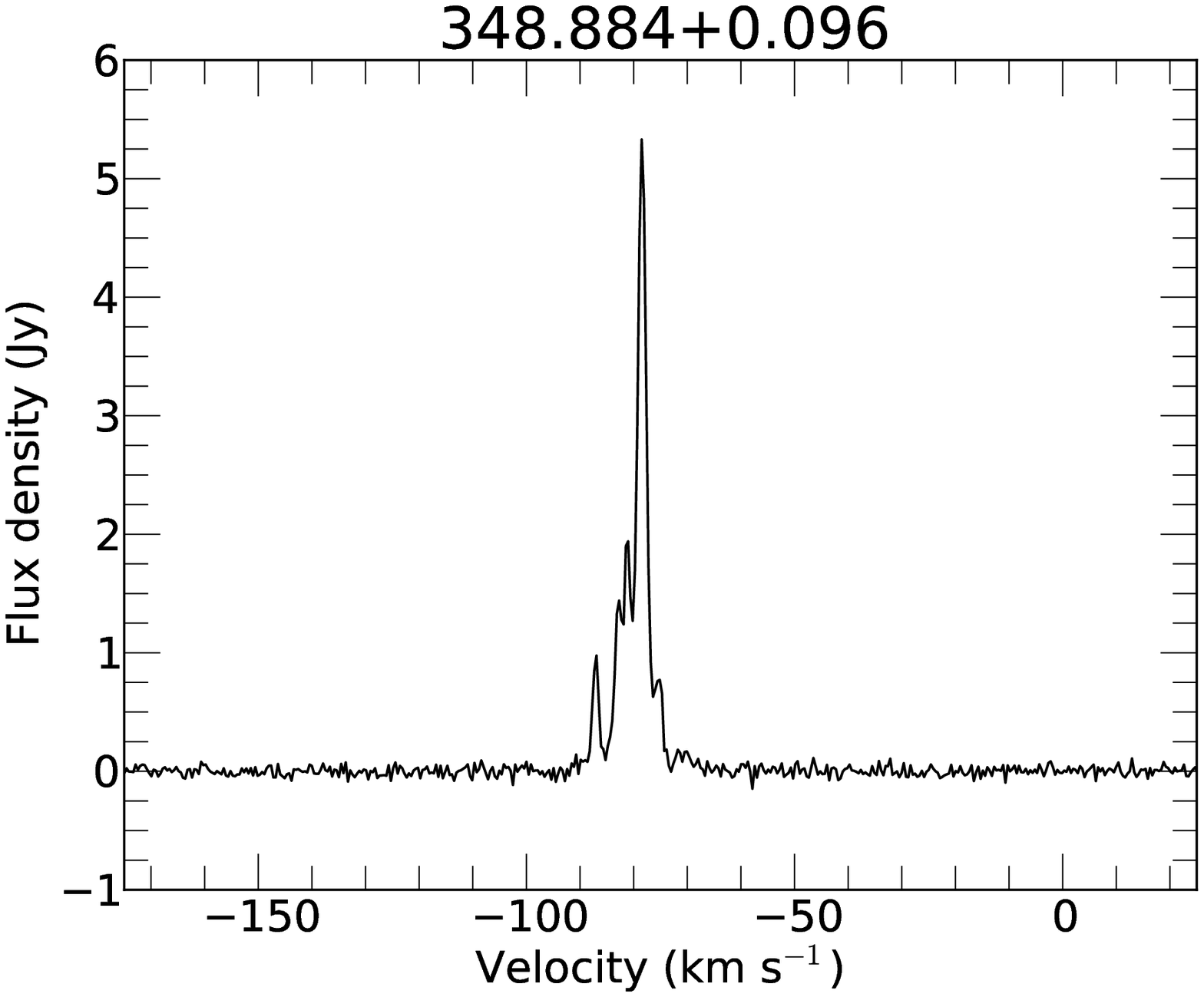}
\includegraphics[width=2.2in]{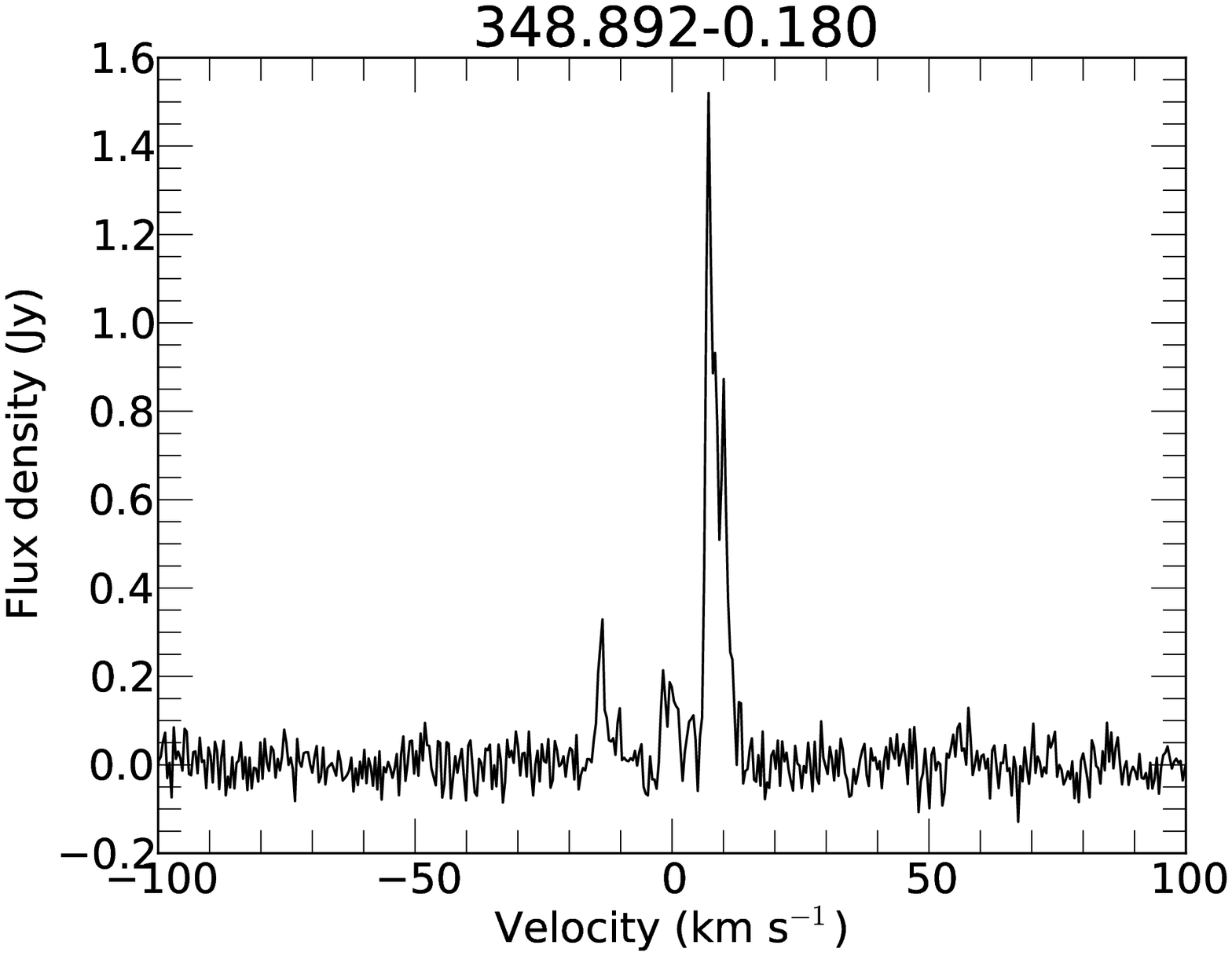}
\includegraphics[width=2.2in]{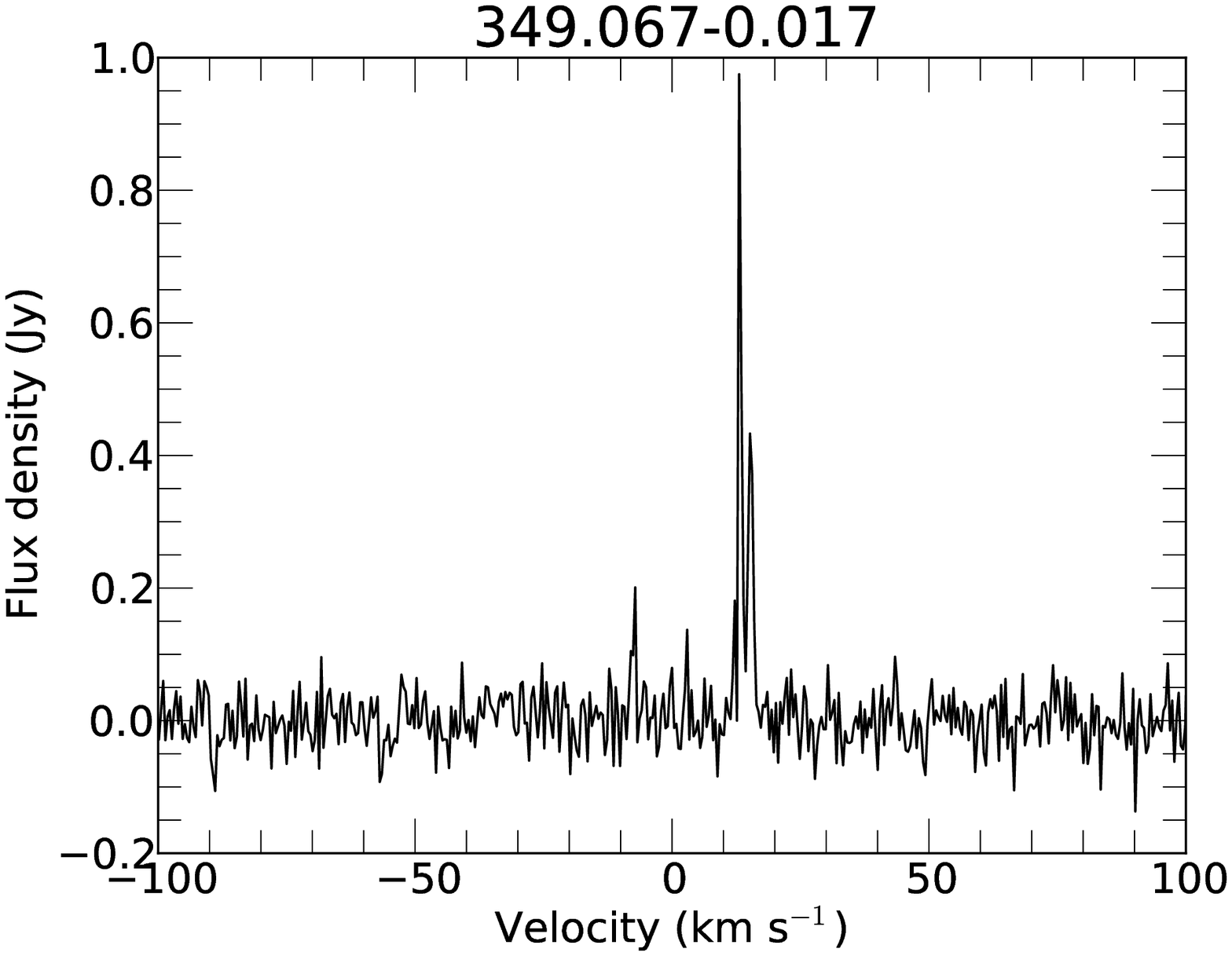}
\includegraphics[width=2.2in]{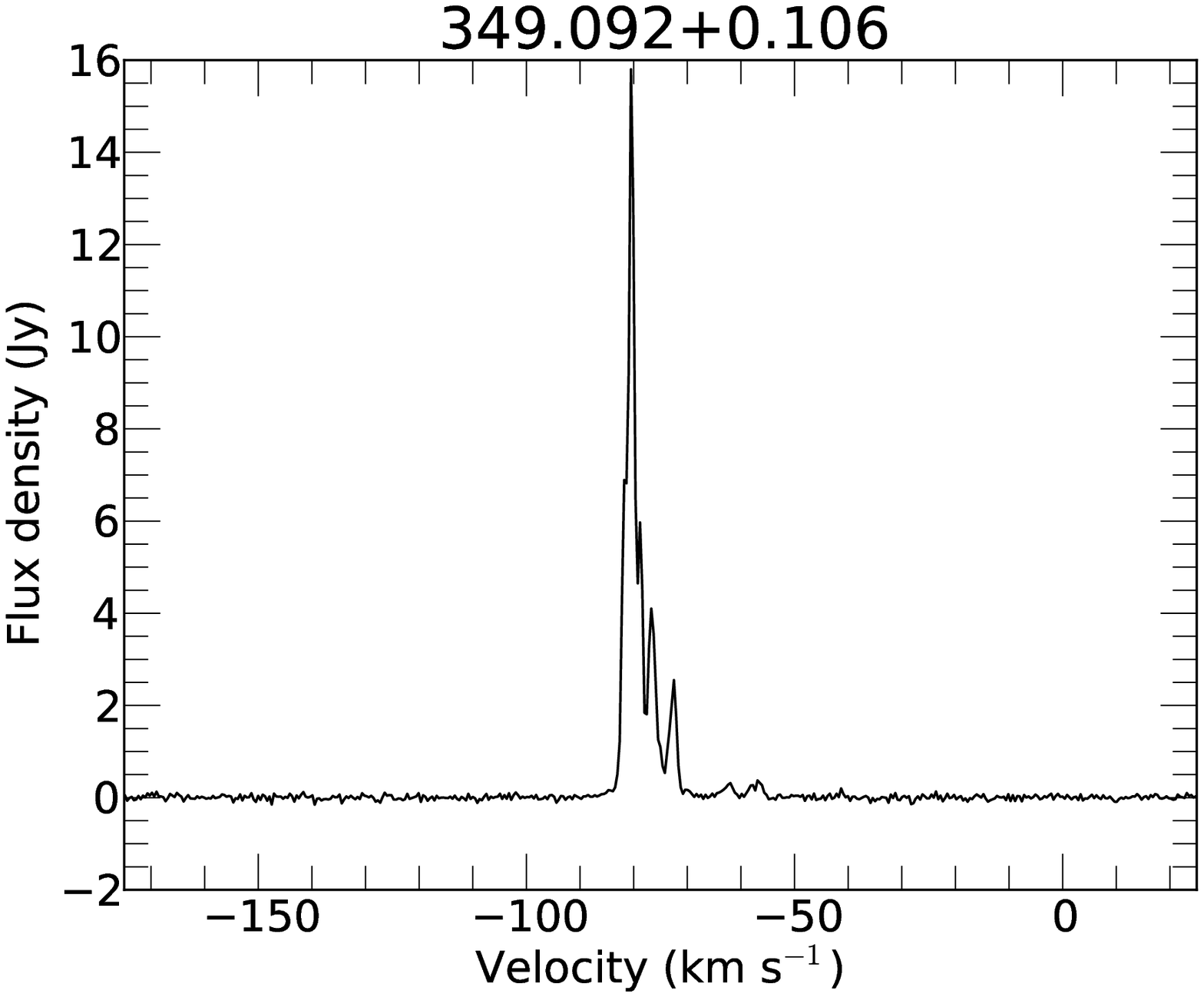}
\includegraphics[width=2.2in]{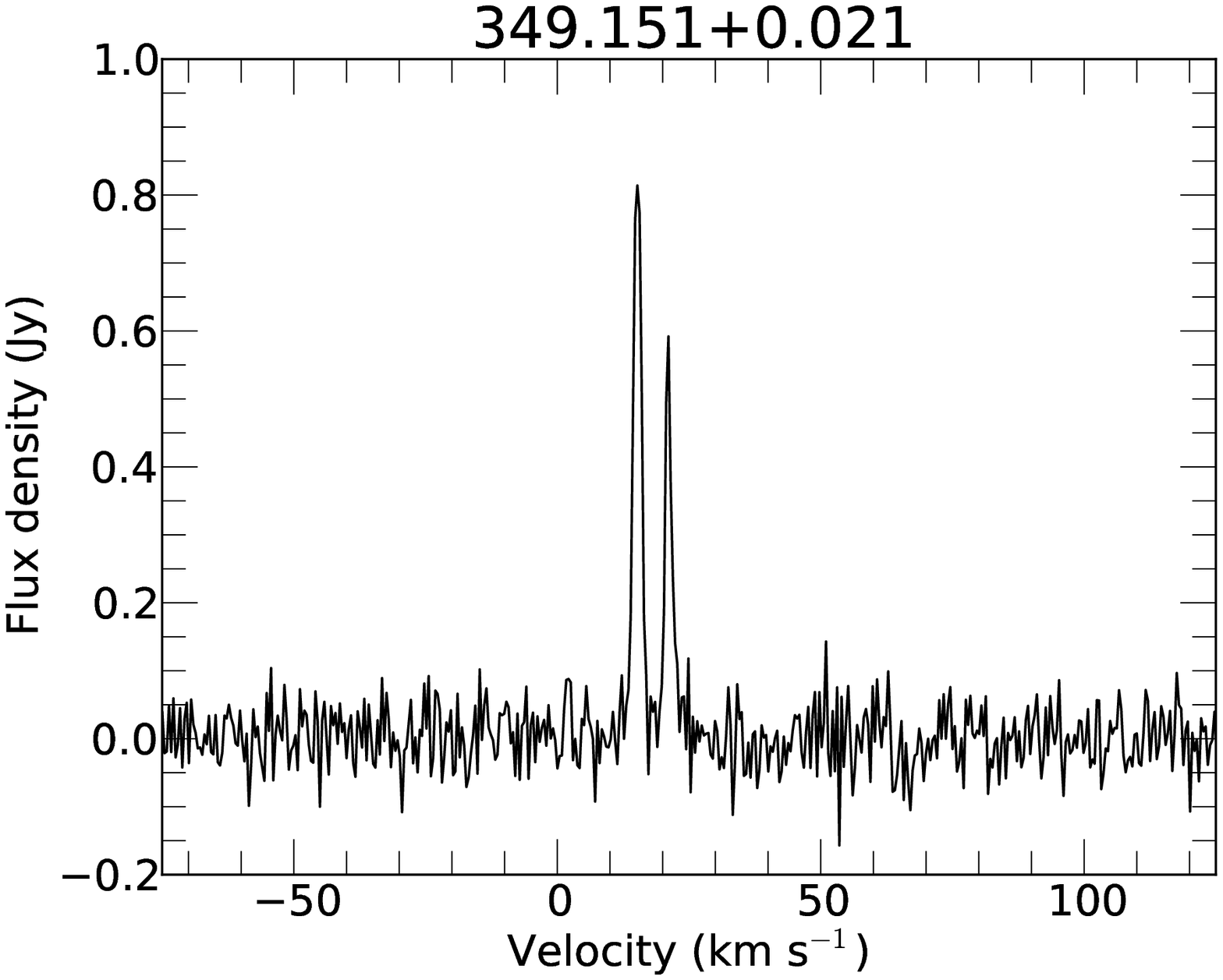}
\includegraphics[width=2.2in]{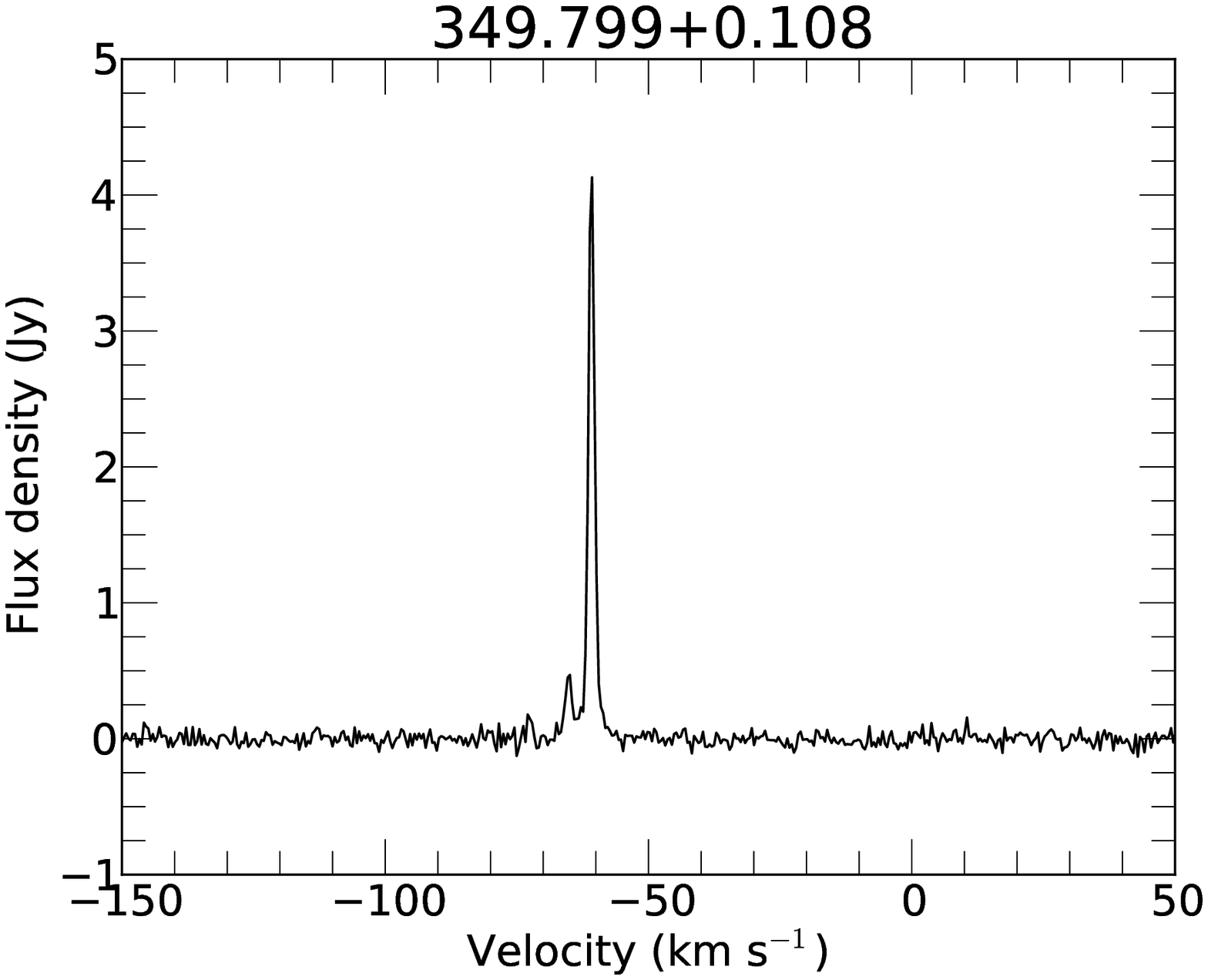}
\includegraphics[width=2.2in]{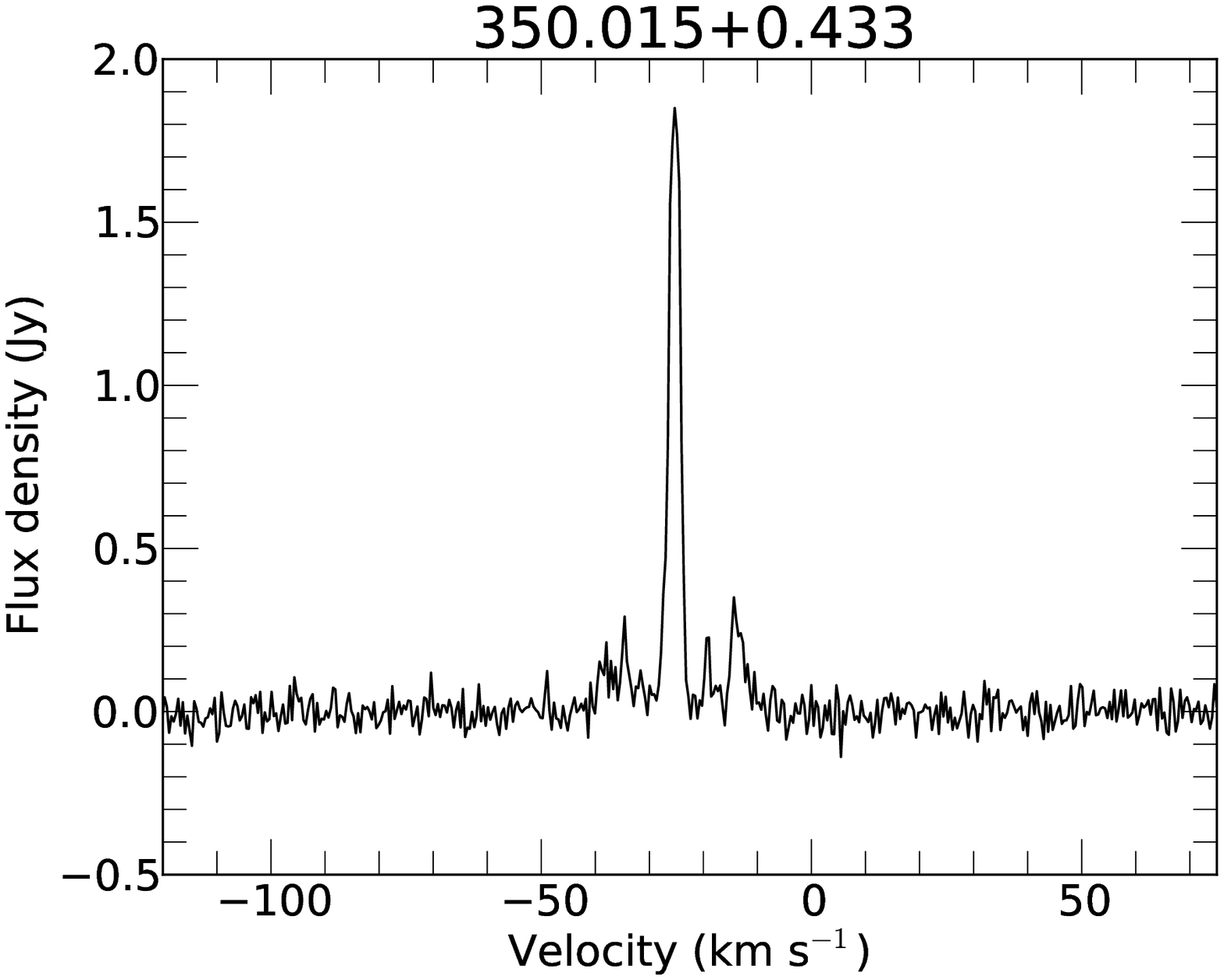}
\includegraphics[width=2.2in]{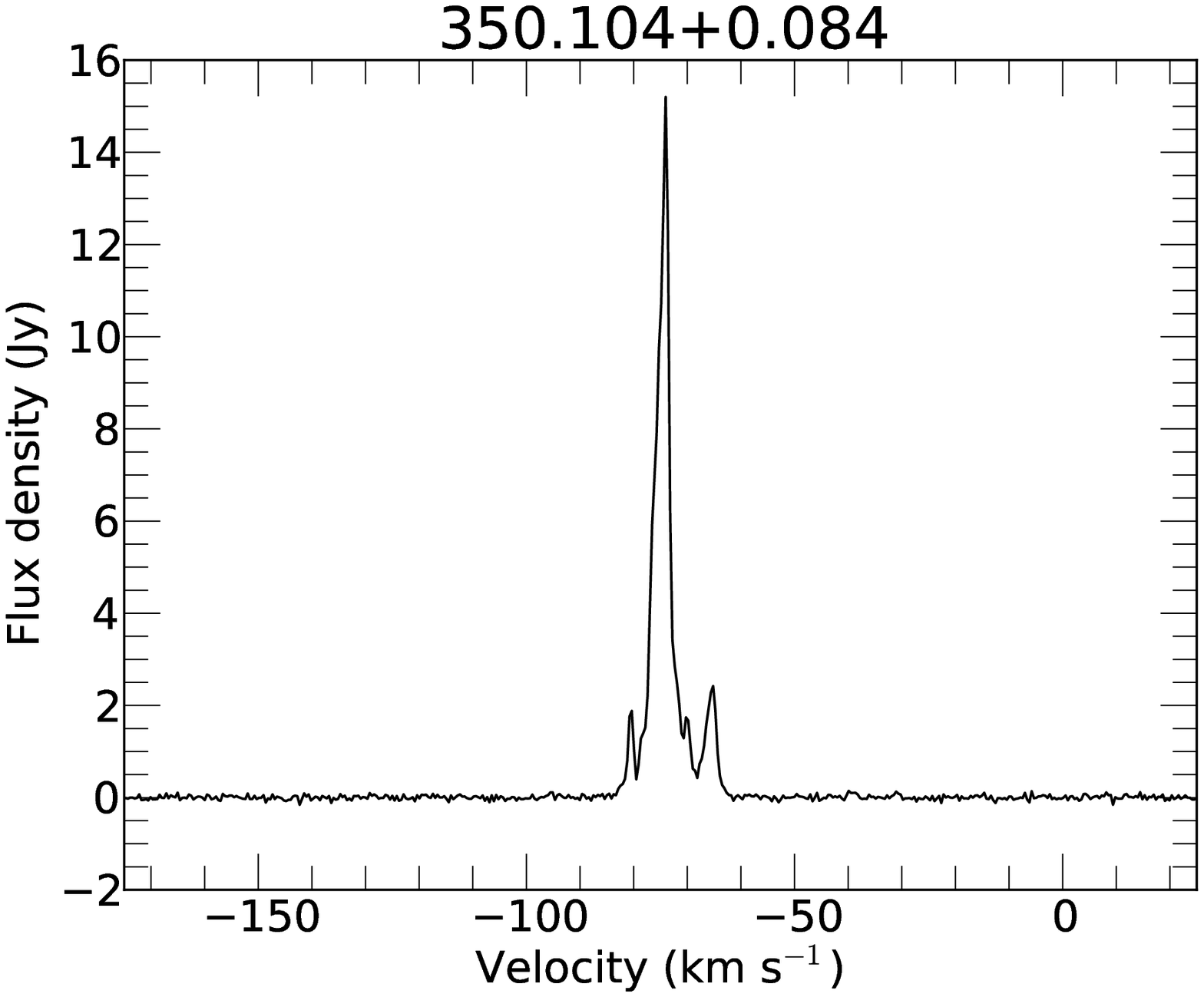}
\includegraphics[width=2.2in]{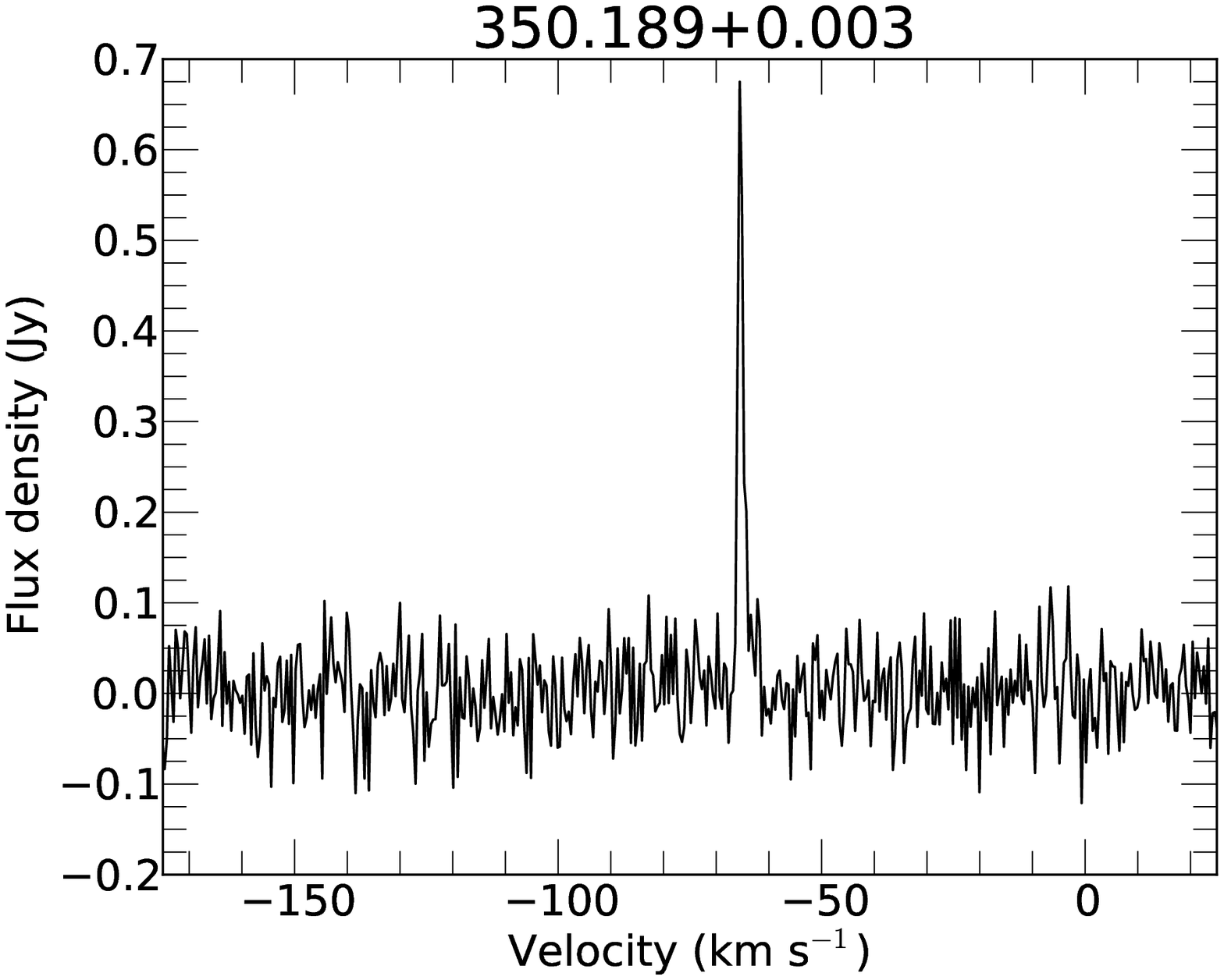}
\includegraphics[width=2.2in]{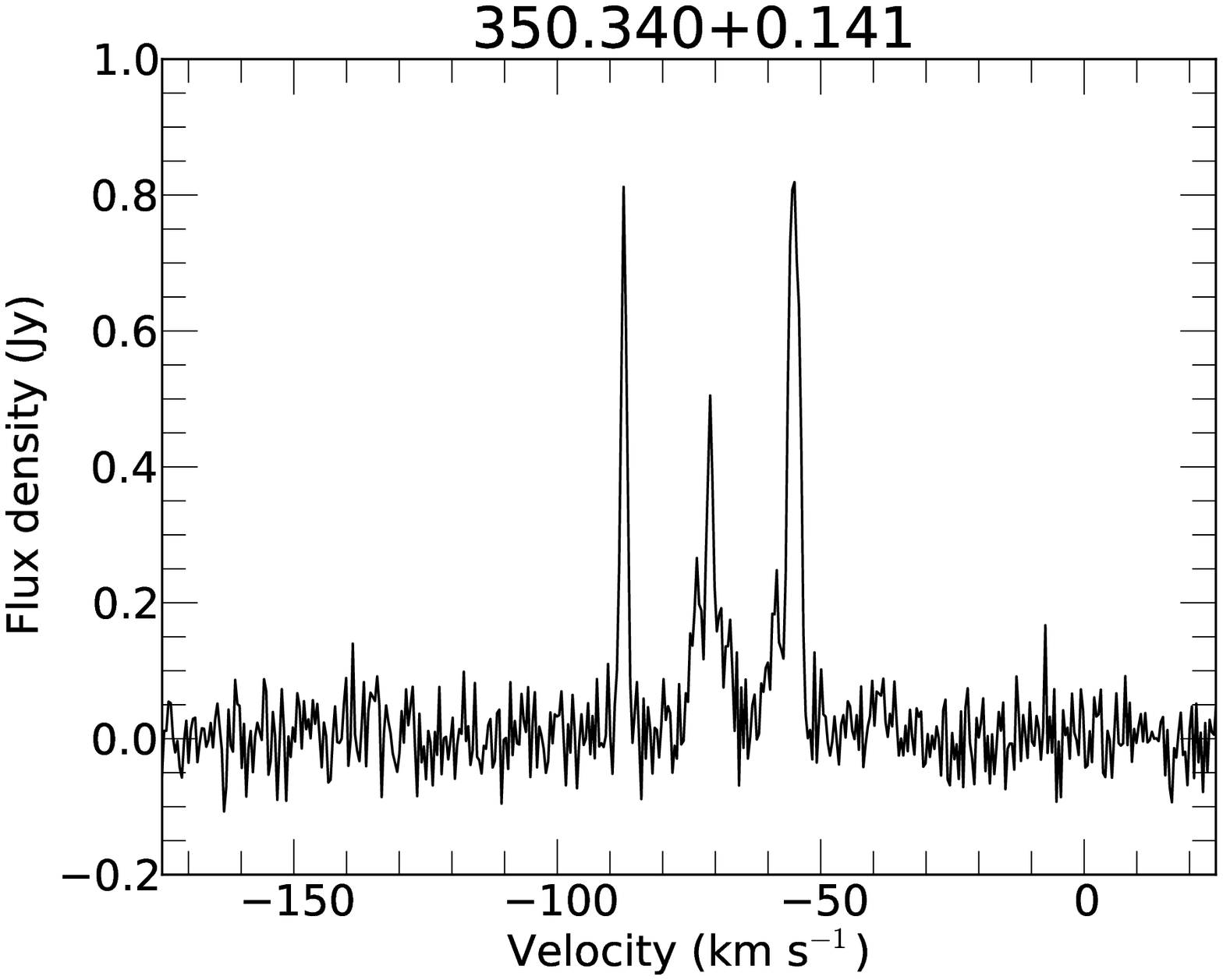}
\includegraphics[width=2.2in]{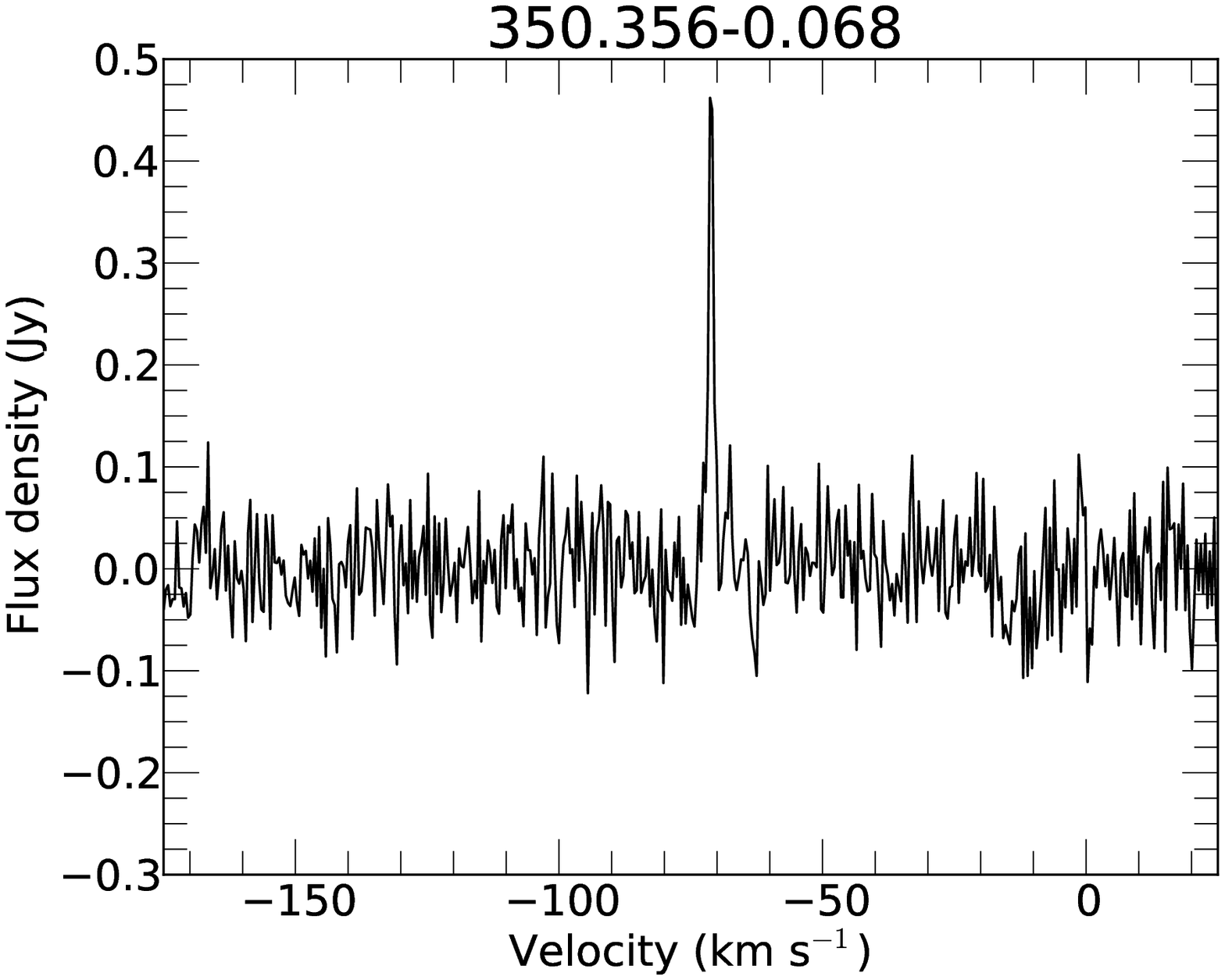}
\\
\addtocounter{figure}{-1}
  \caption{-- {\emph {continued}}}
\end{figure*}

\begin{figure*}
\includegraphics[width=2.2in]{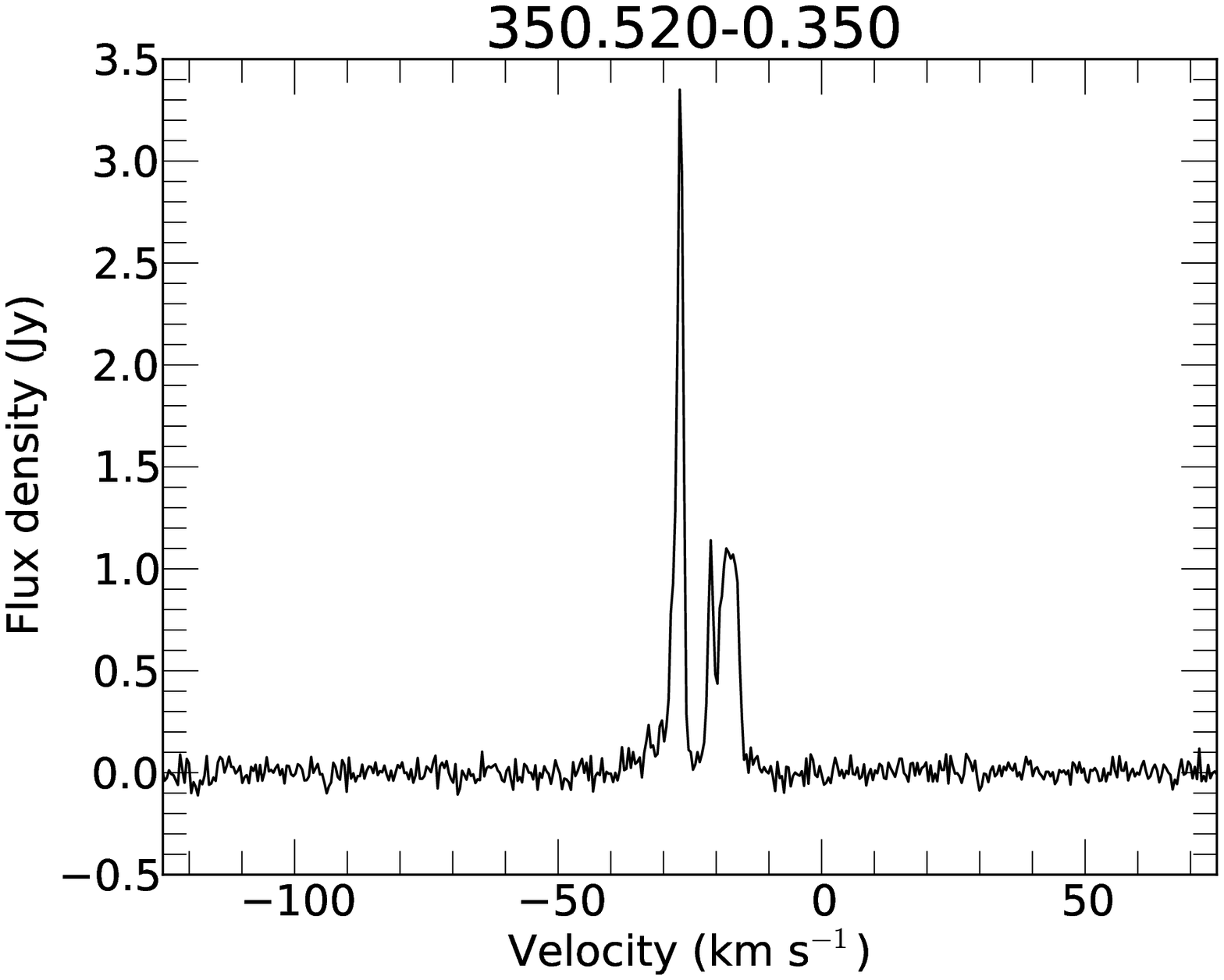}
\includegraphics[width=2.2in]{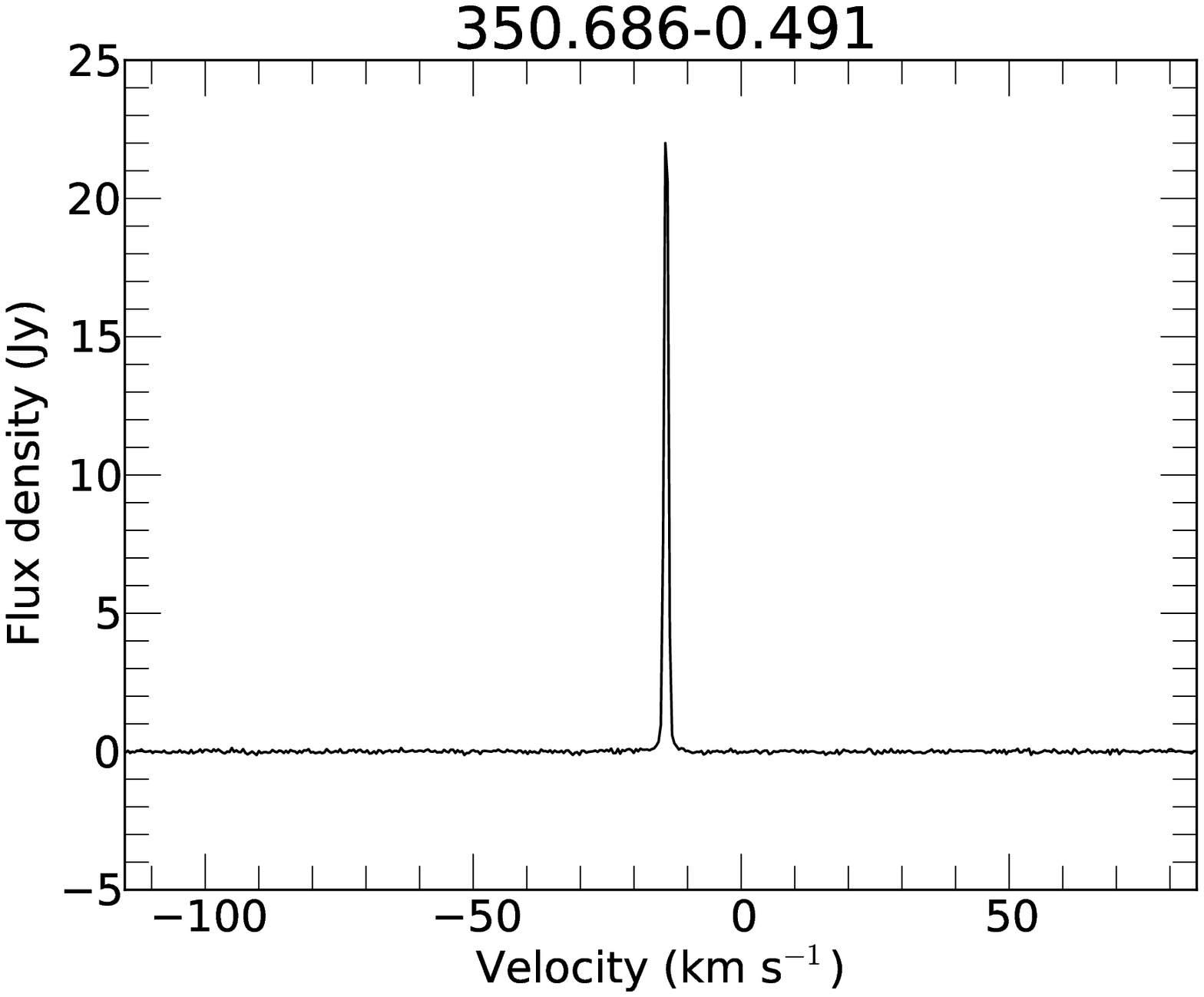}
\includegraphics[width=2.2in]{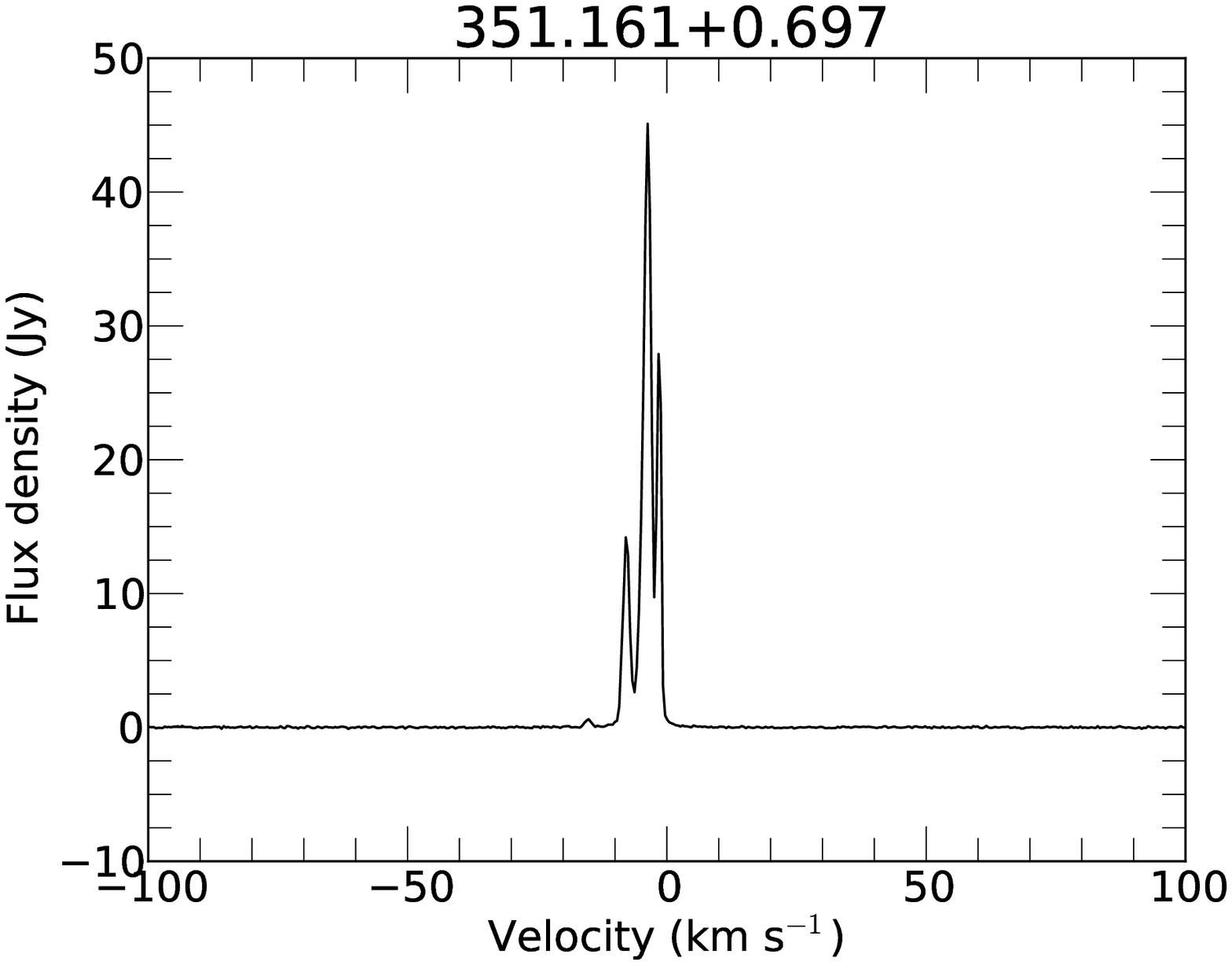}
\includegraphics[width=2.2in]{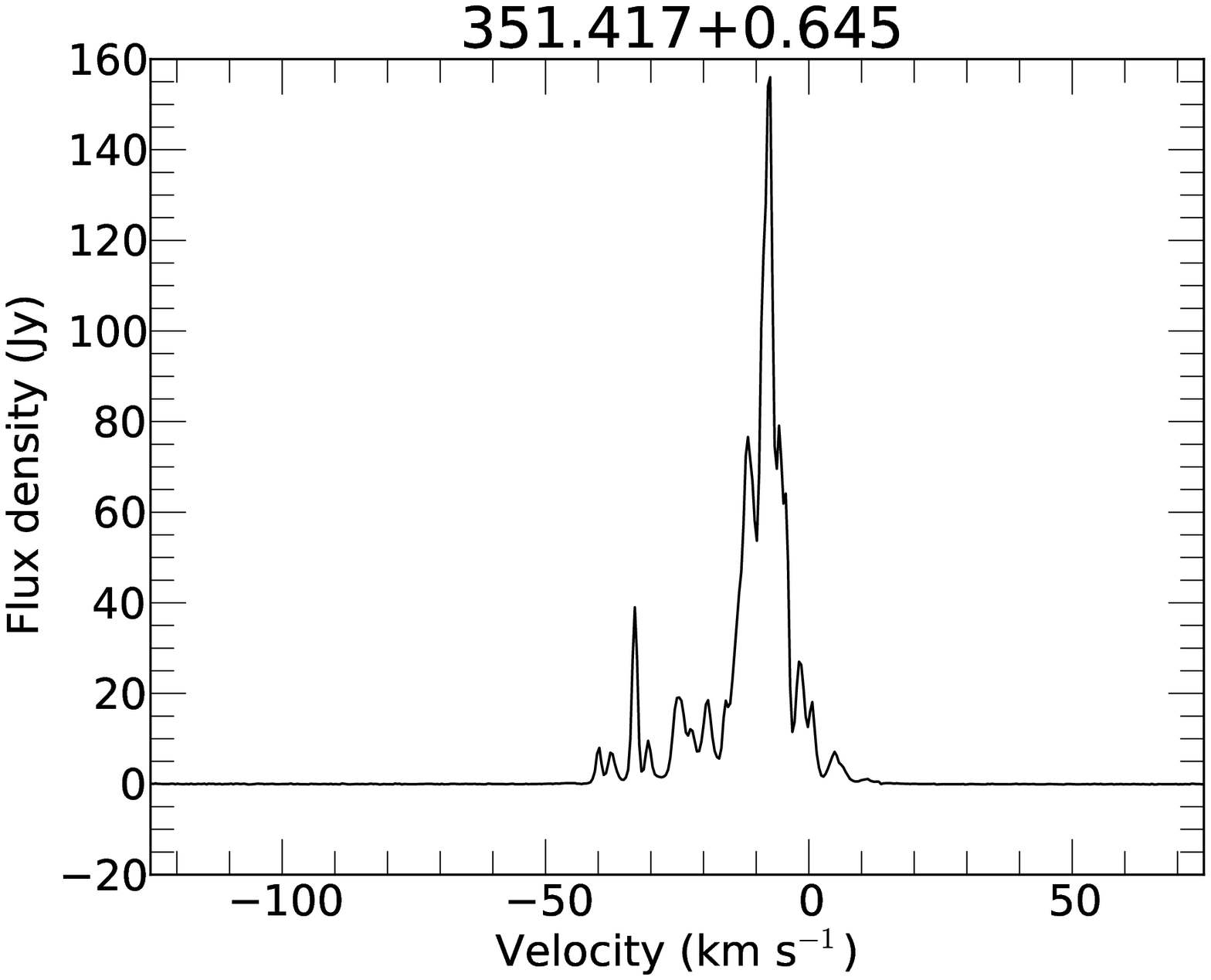}
\includegraphics[width=2.2in]{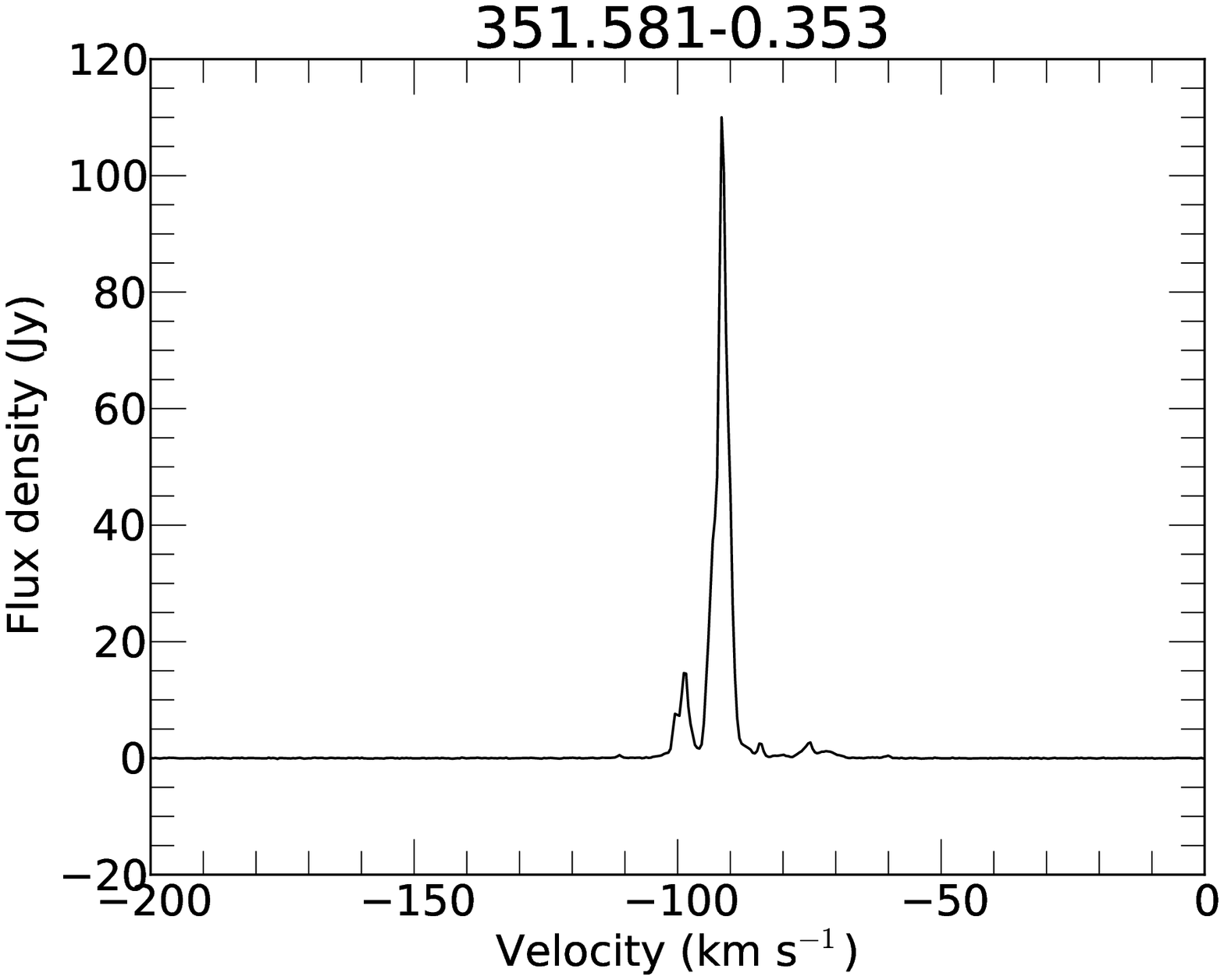}
\includegraphics[width=2.2in]{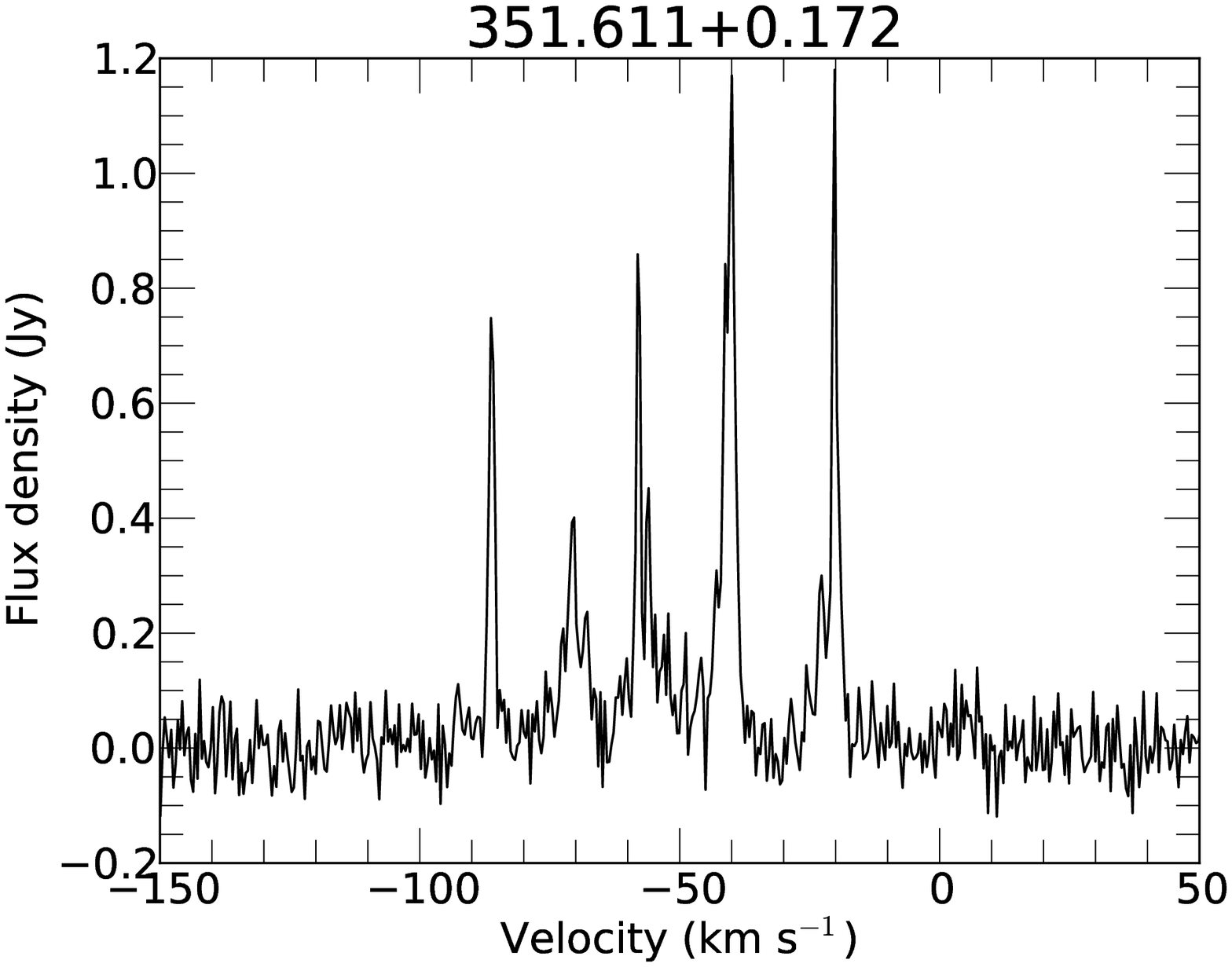}
\includegraphics[width=2.2in]{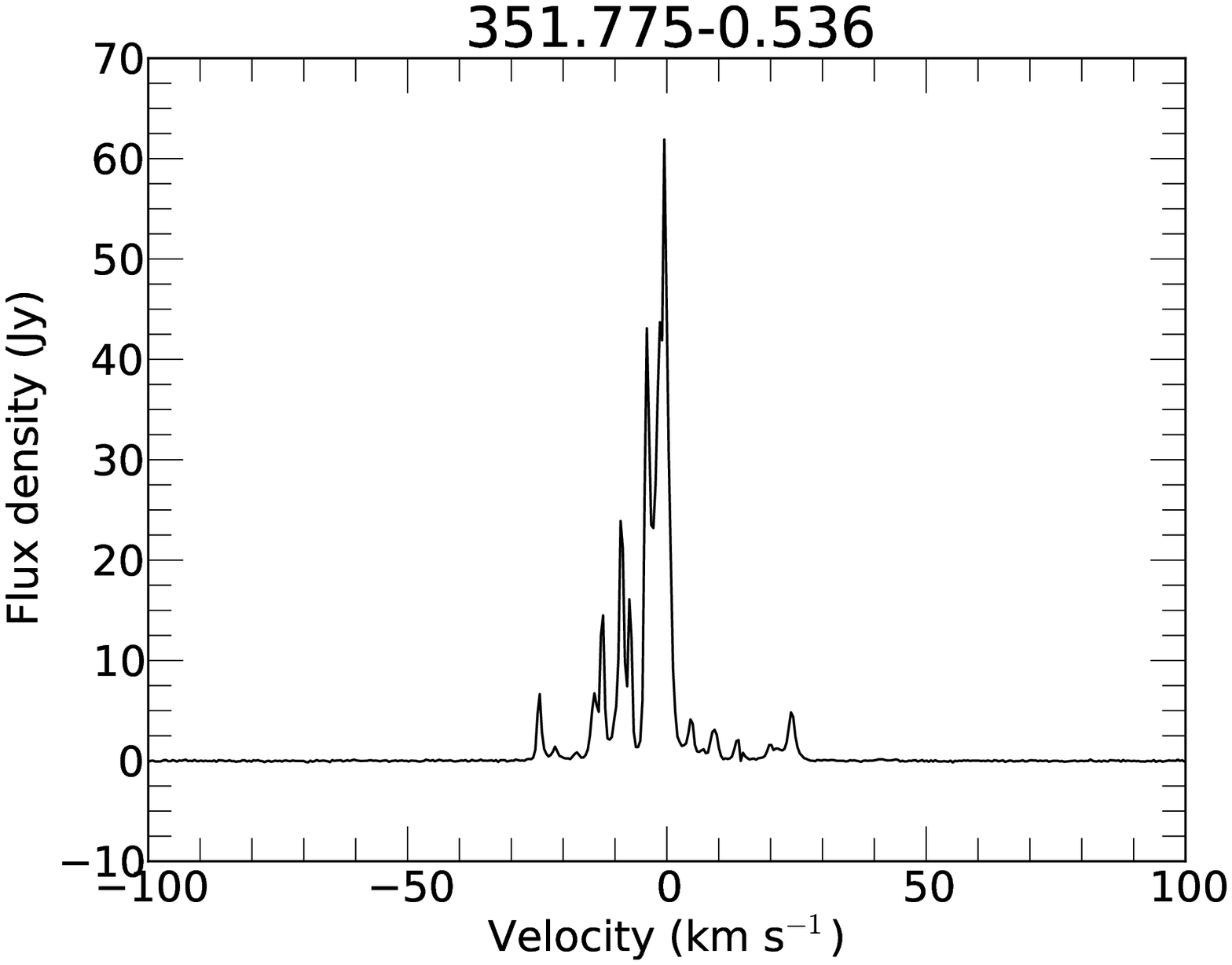}
\includegraphics[width=2.2in]{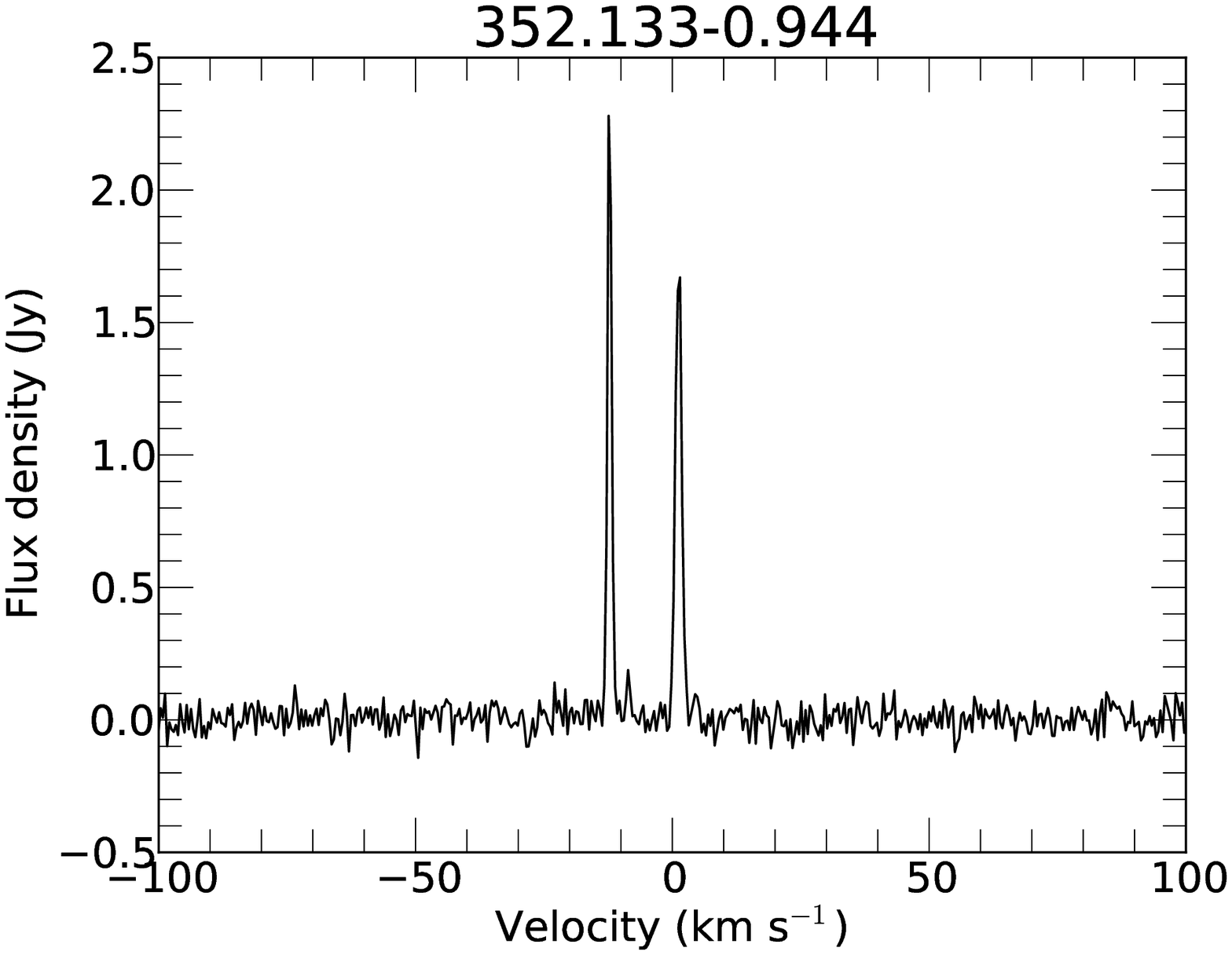}
\includegraphics[width=2.2in]{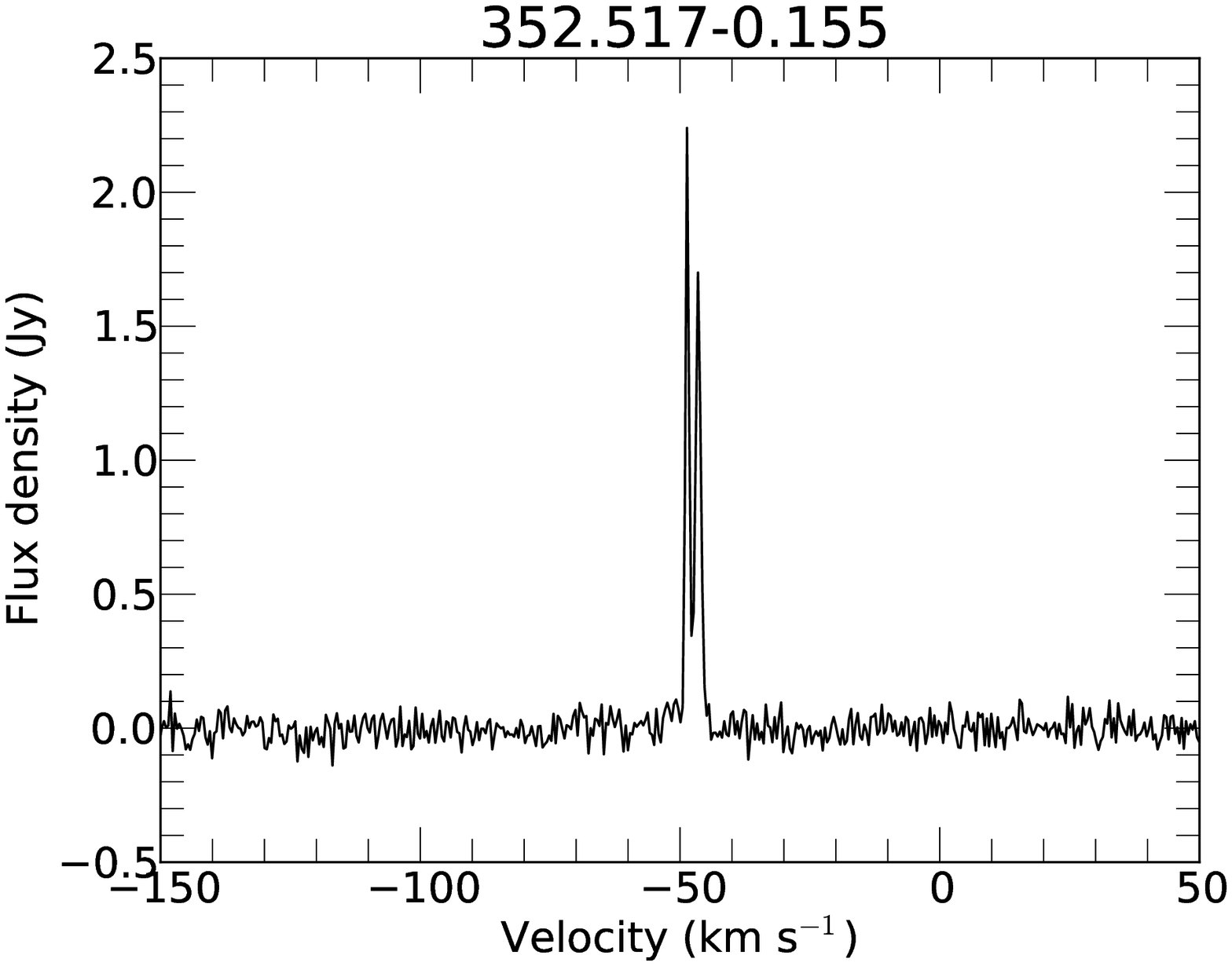}
\includegraphics[width=2.2in]{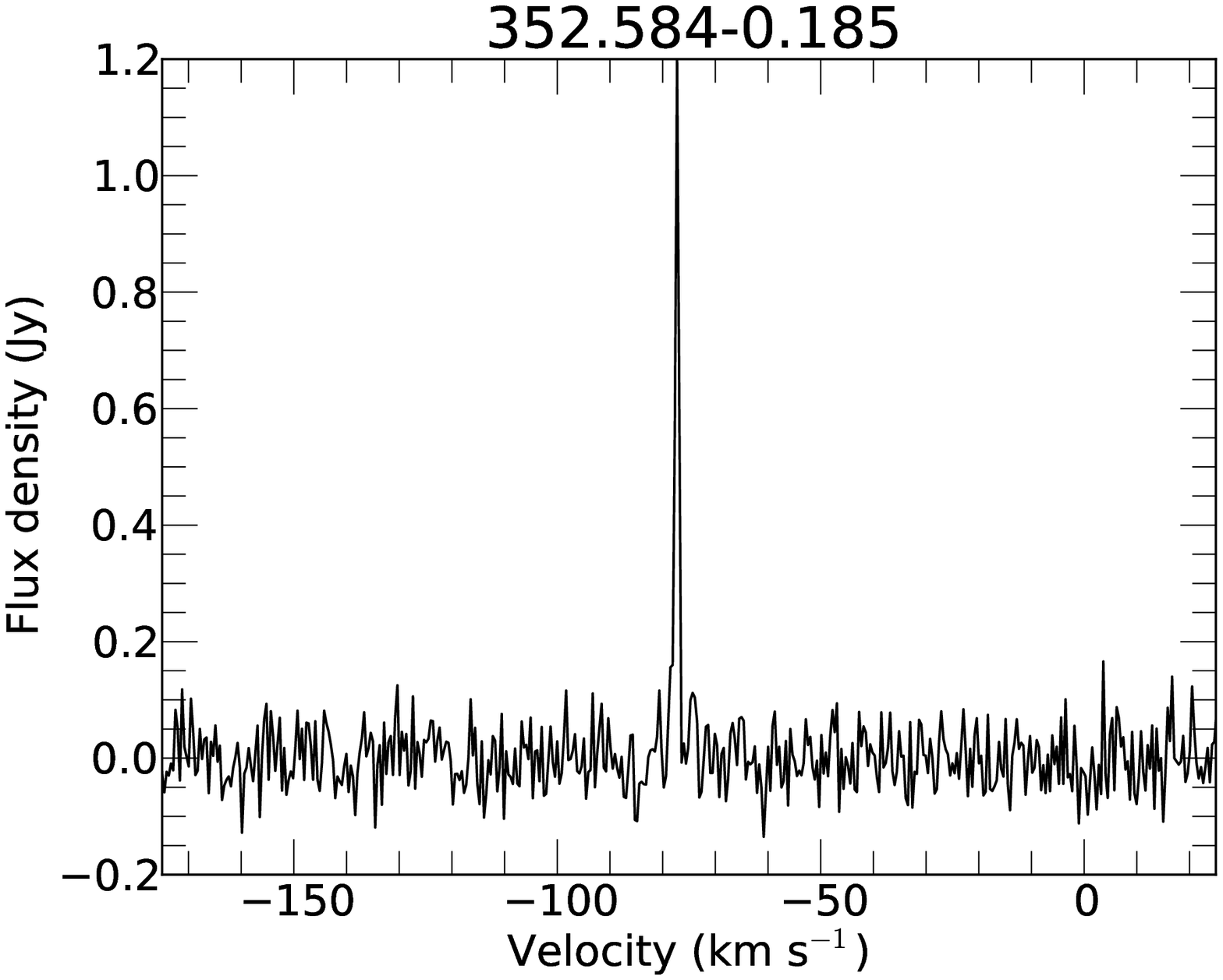}
\includegraphics[width=2.2in]{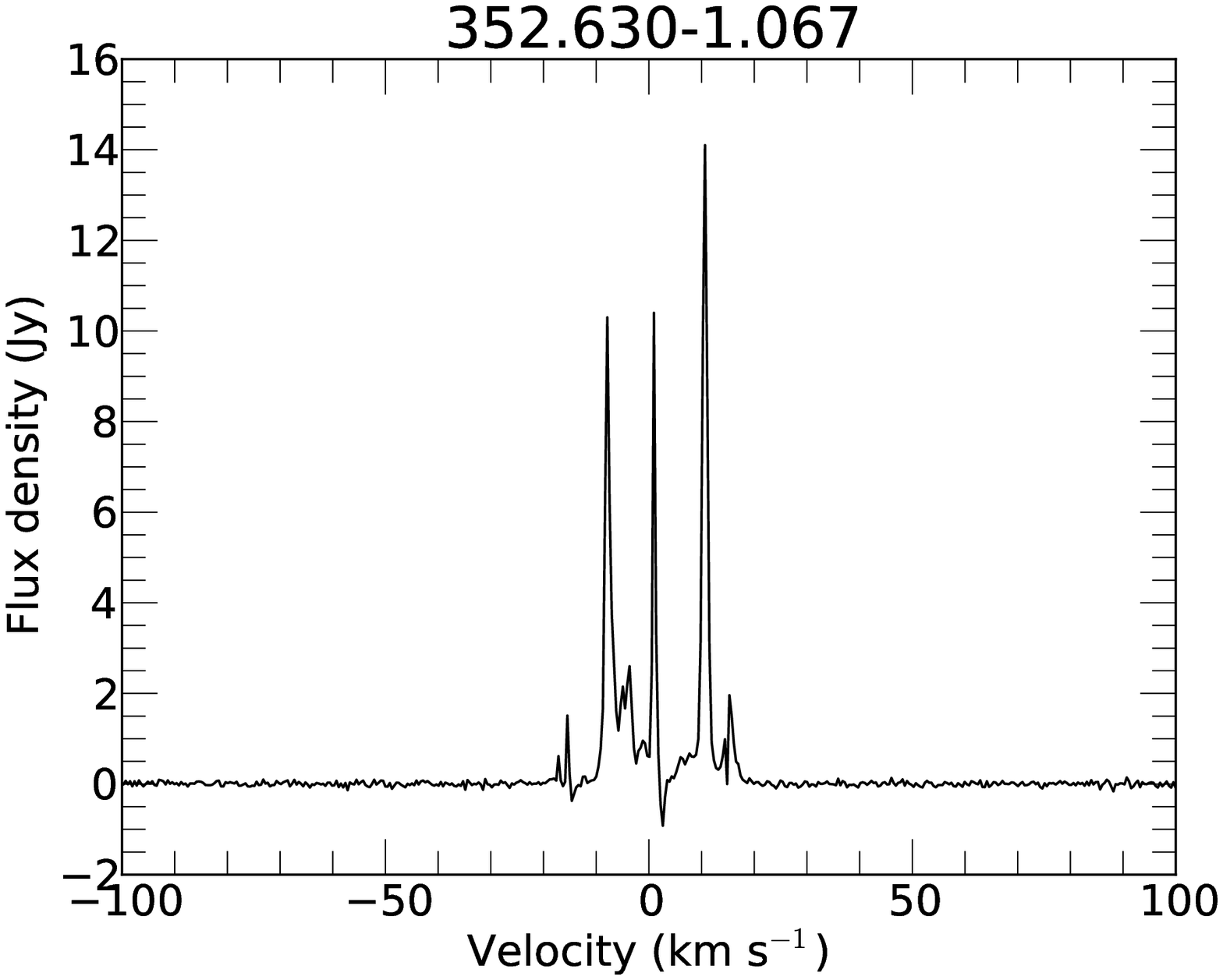}
\includegraphics[width=2.2in]{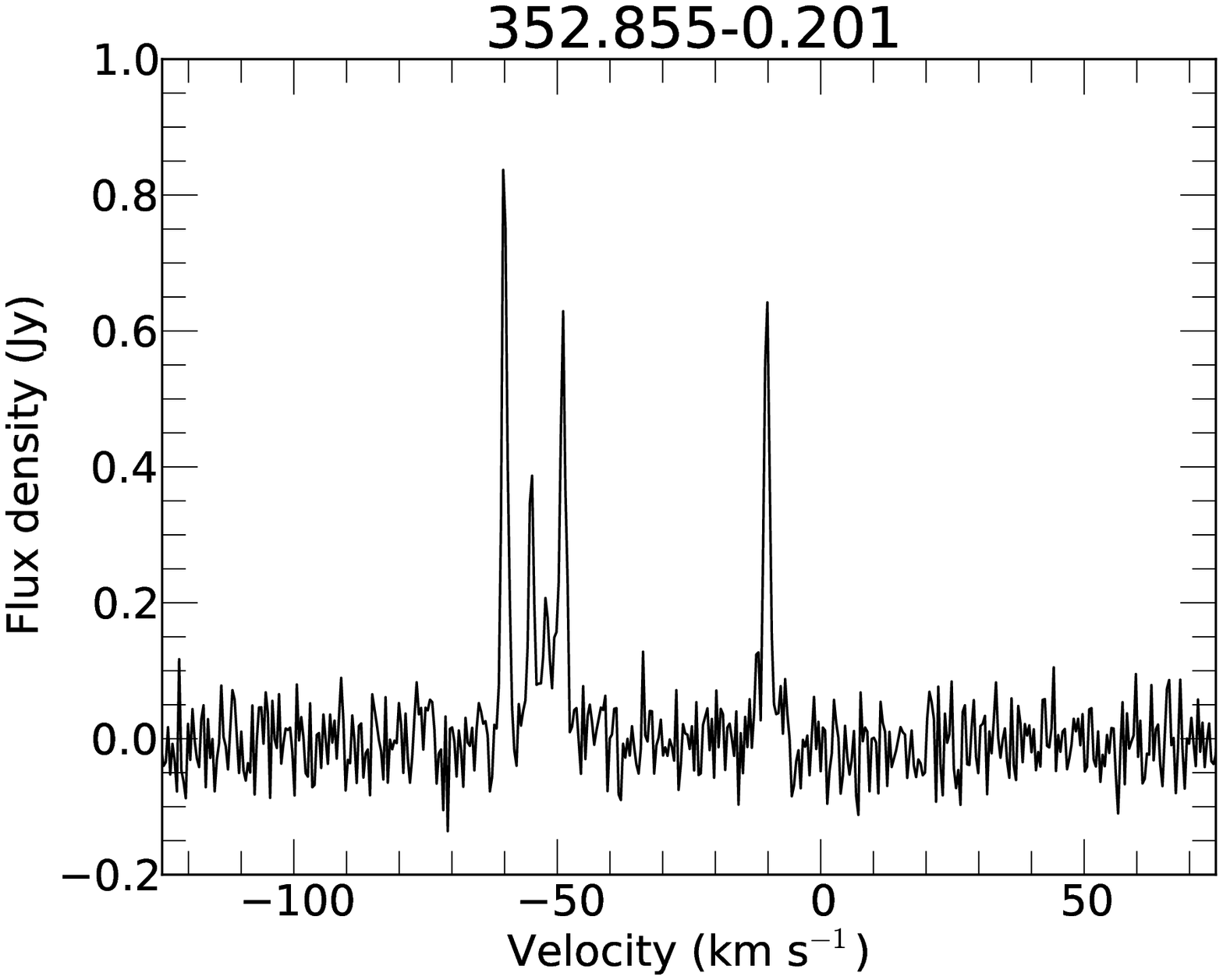}
\includegraphics[width=2.2in]{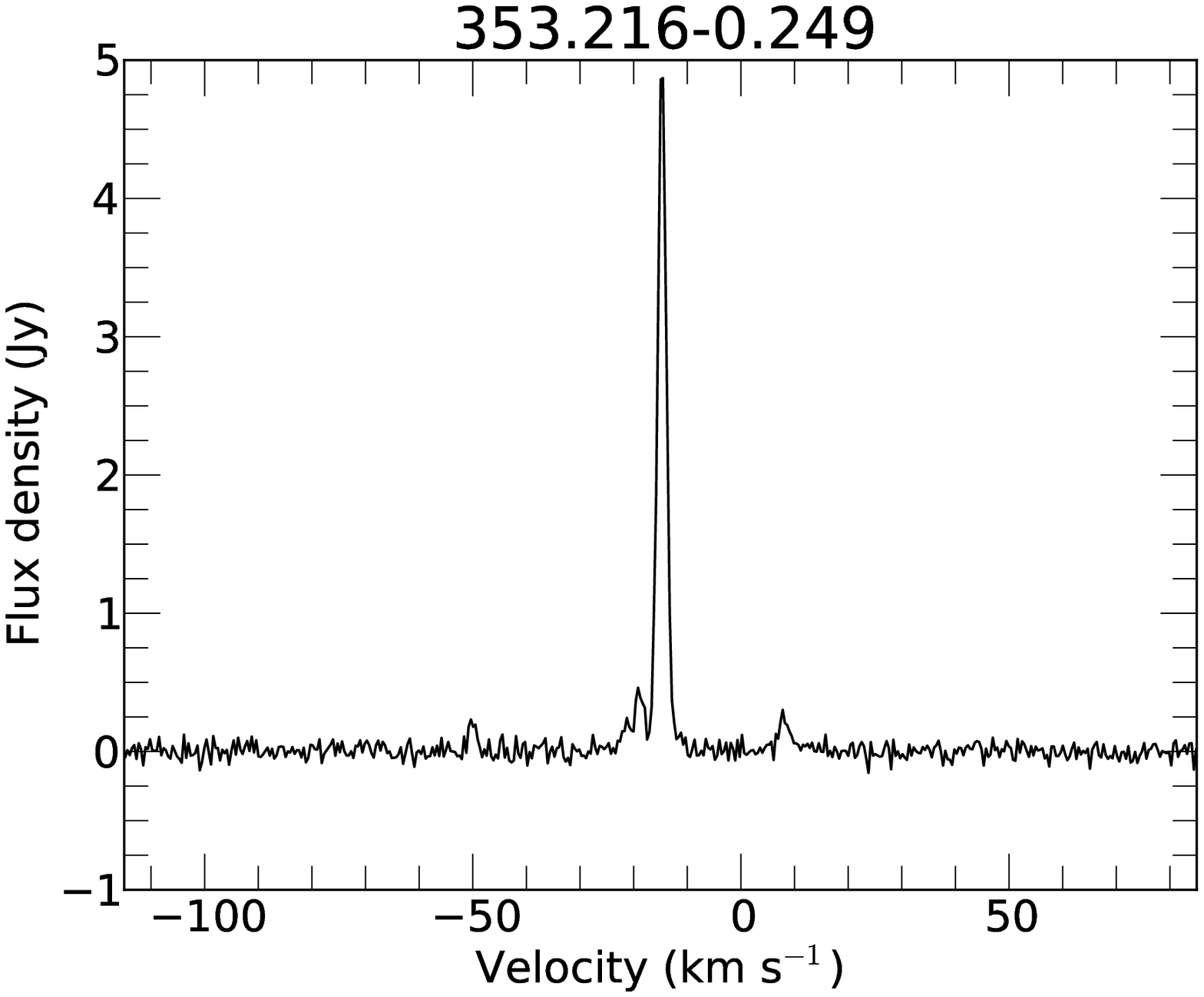}
\includegraphics[width=2.2in]{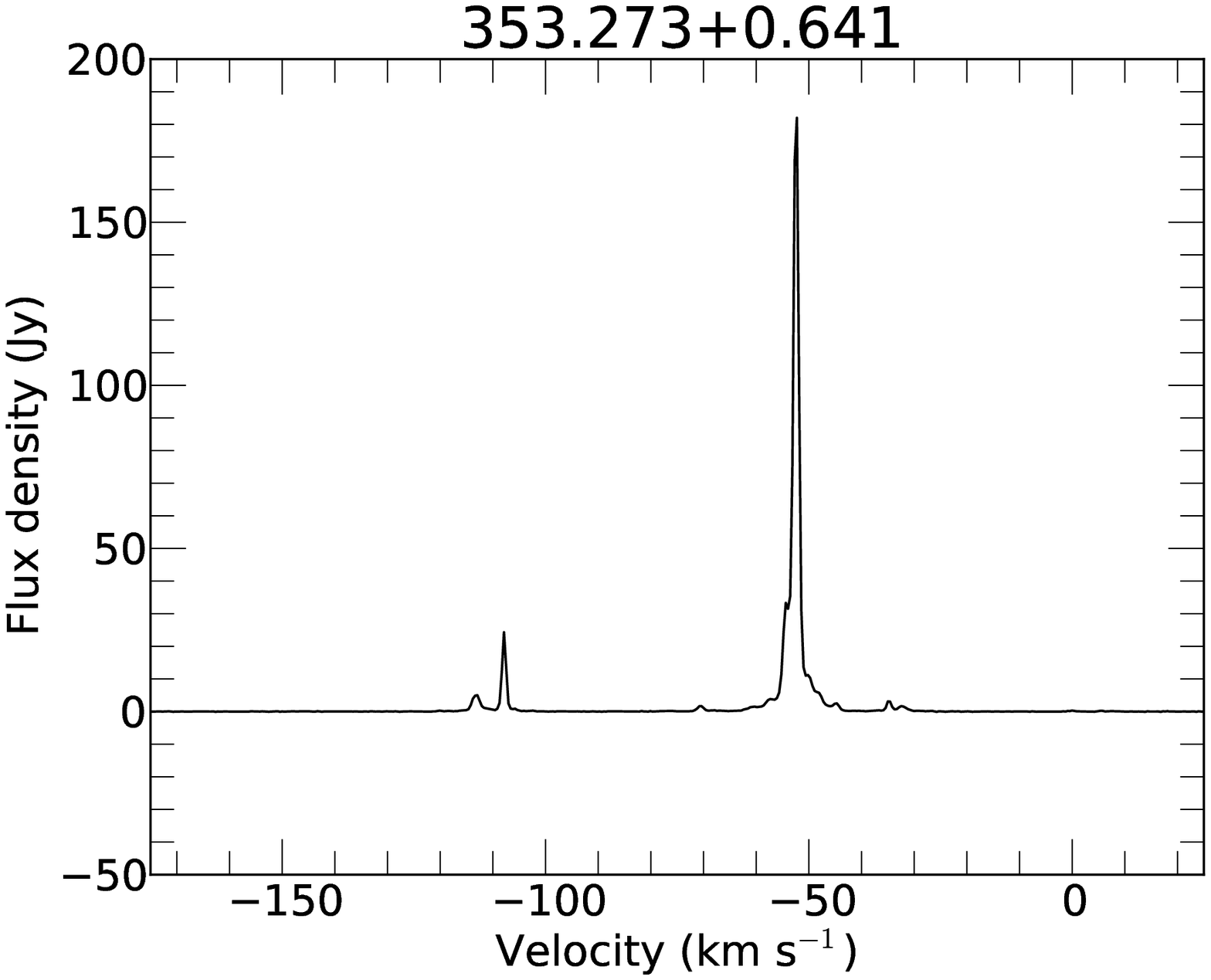}
\includegraphics[width=2.2in]{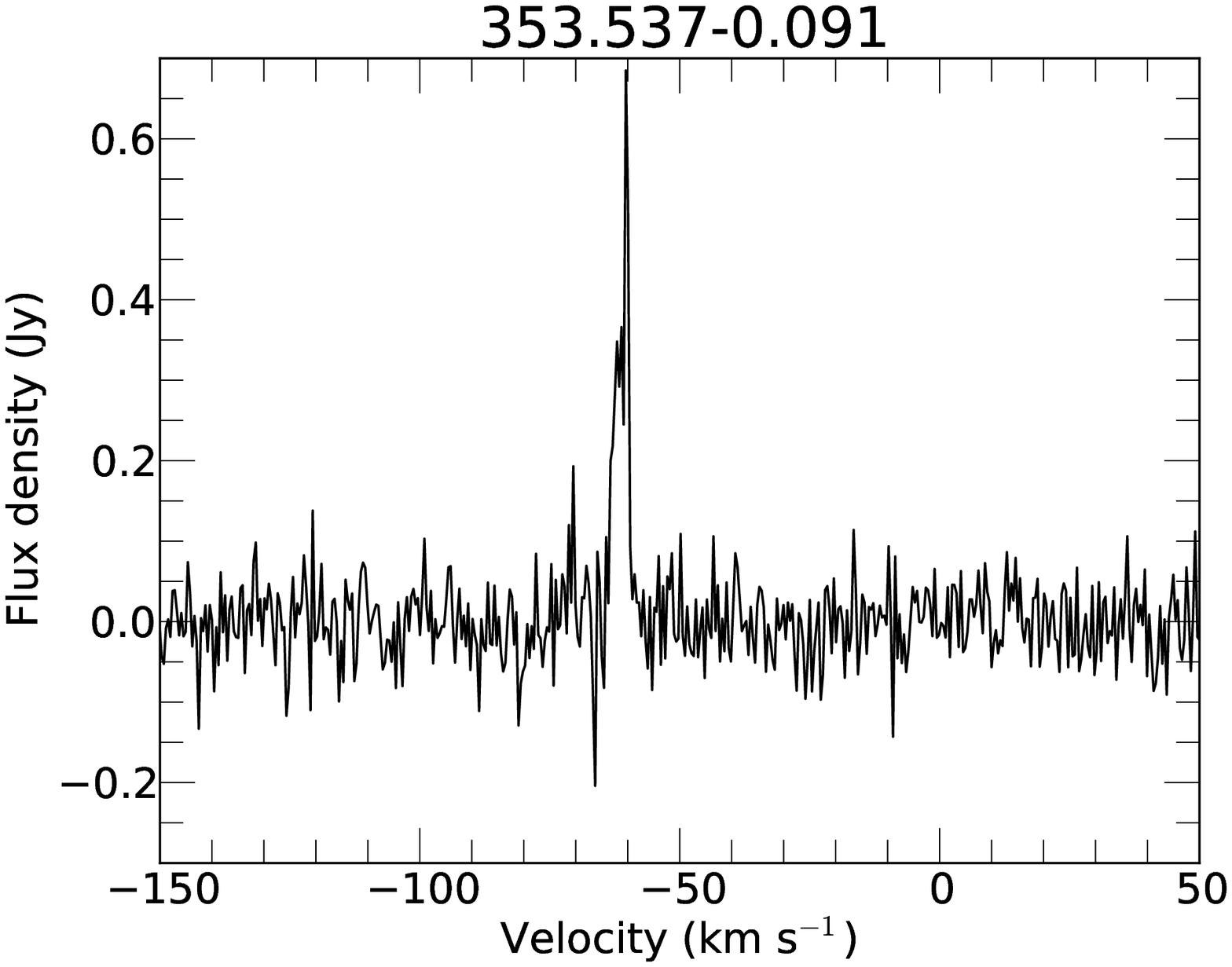}
\\
\addtocounter{figure}{-1}
  \caption{-- {\emph {continued}}}
\end{figure*}

\begin{figure*}\addtocounter{figure}{-1}
  \caption{-- {\emph {continued}}}
\includegraphics[width=2.2in]{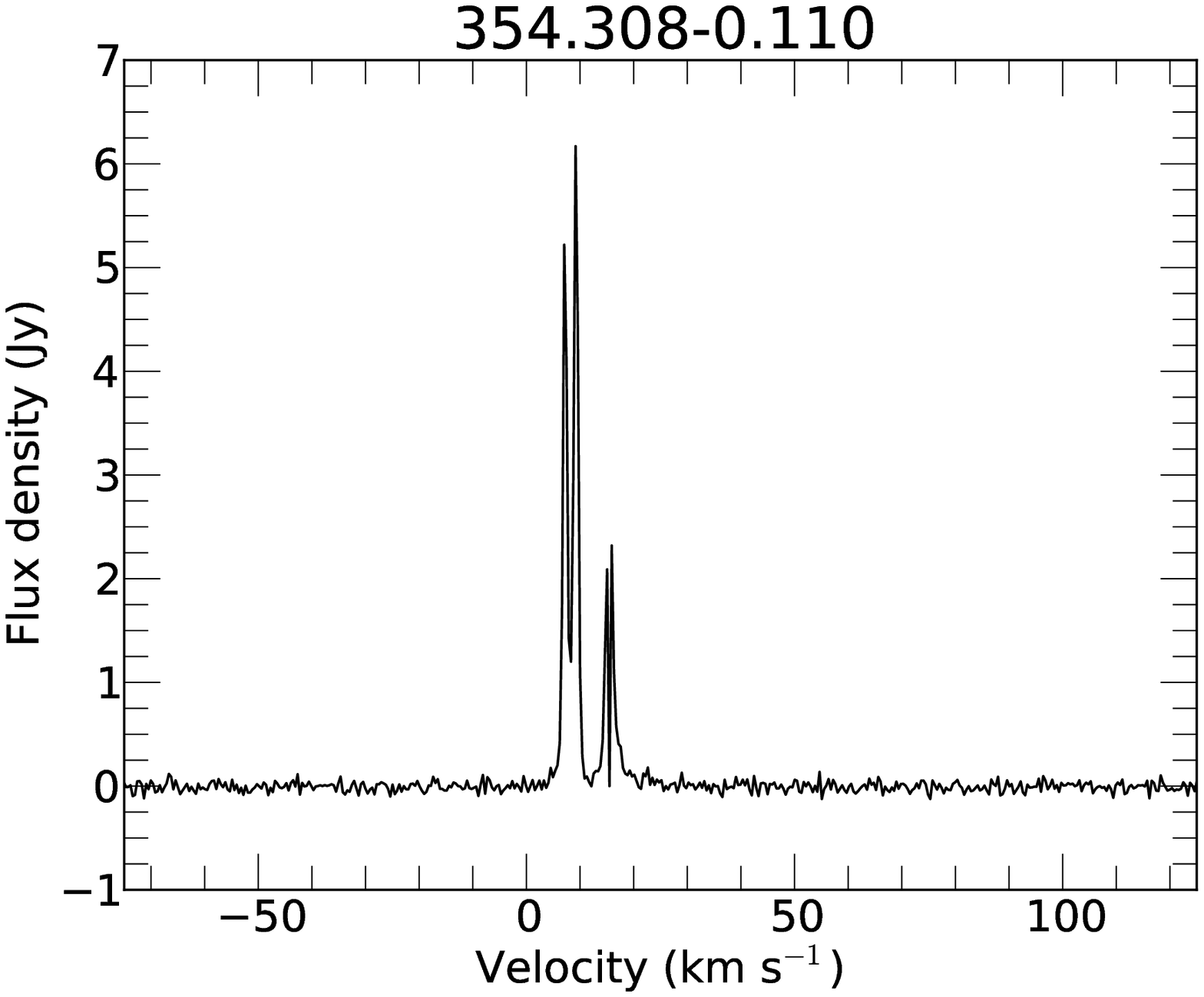}
\includegraphics[width=2.2in]{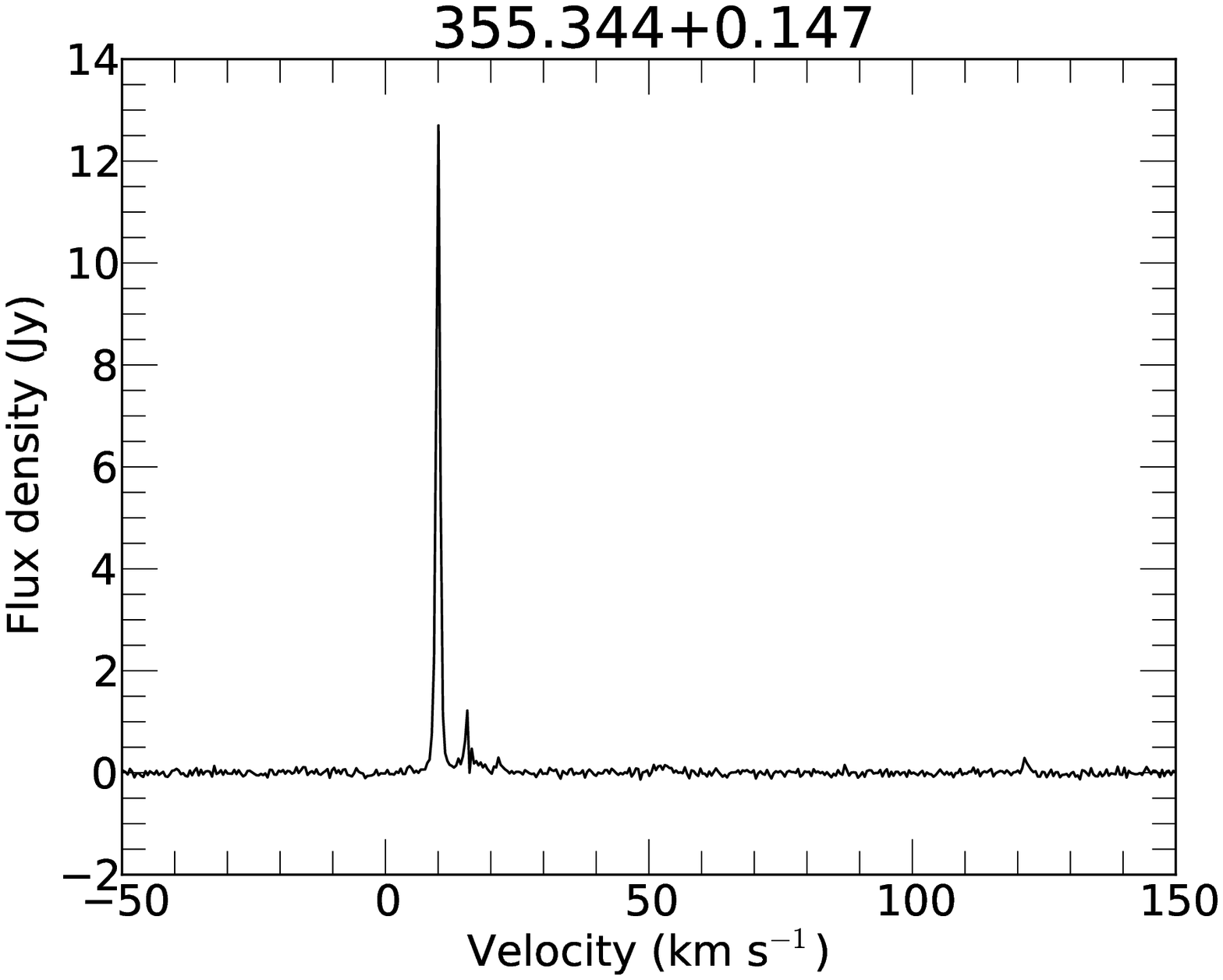}
\includegraphics[width=2.2in]{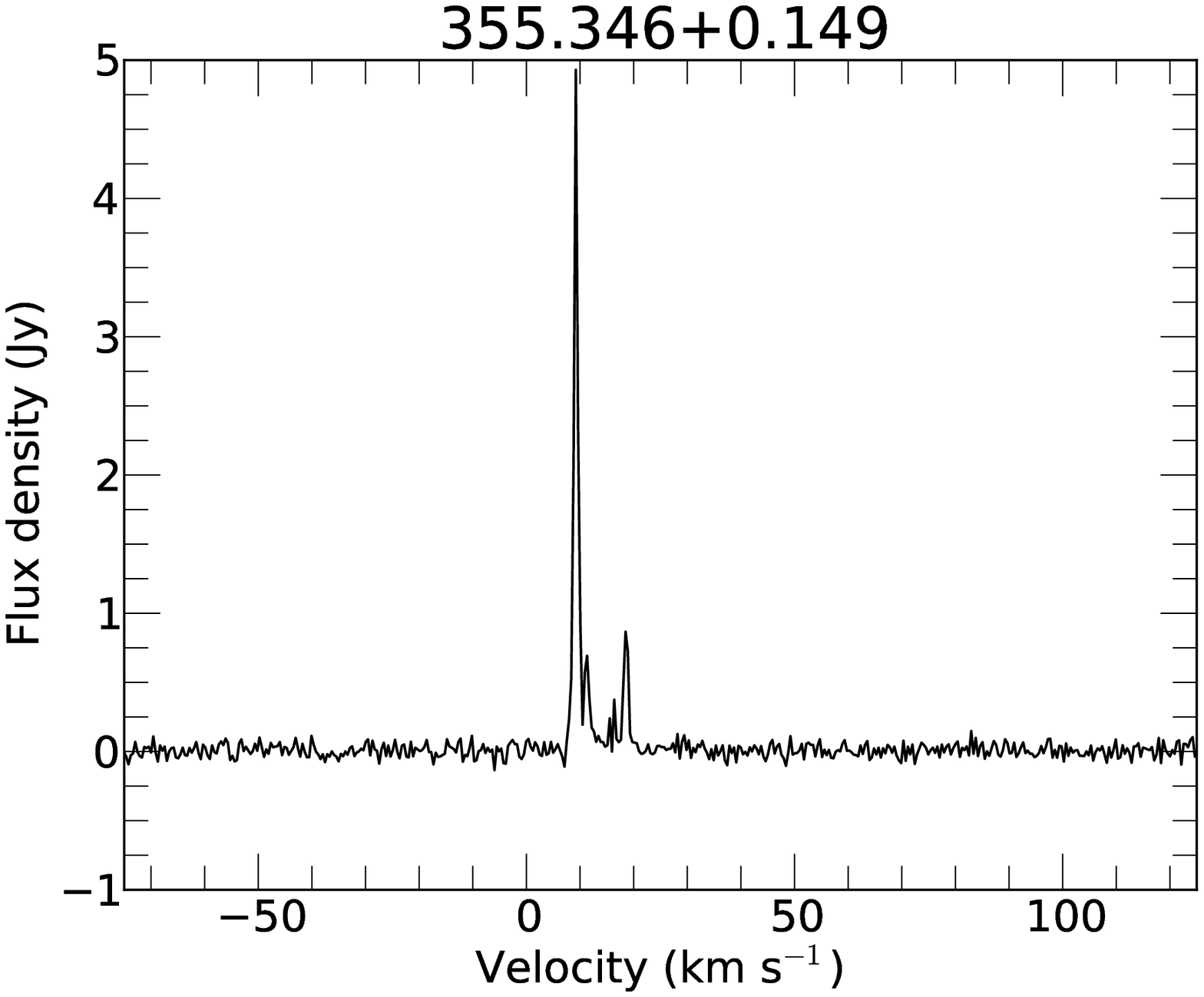}
\includegraphics[width=2.2in]{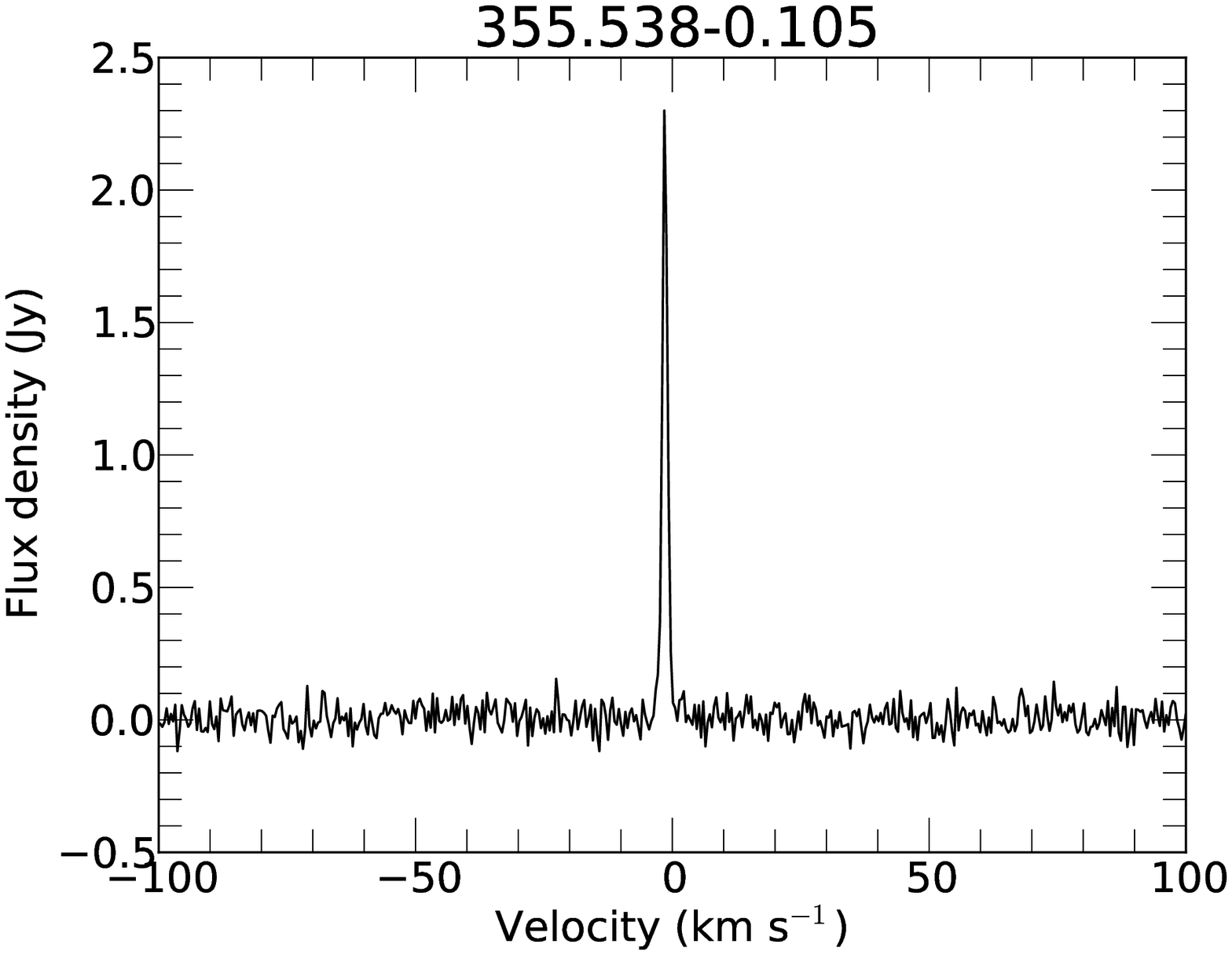}
\includegraphics[width=2.2in]{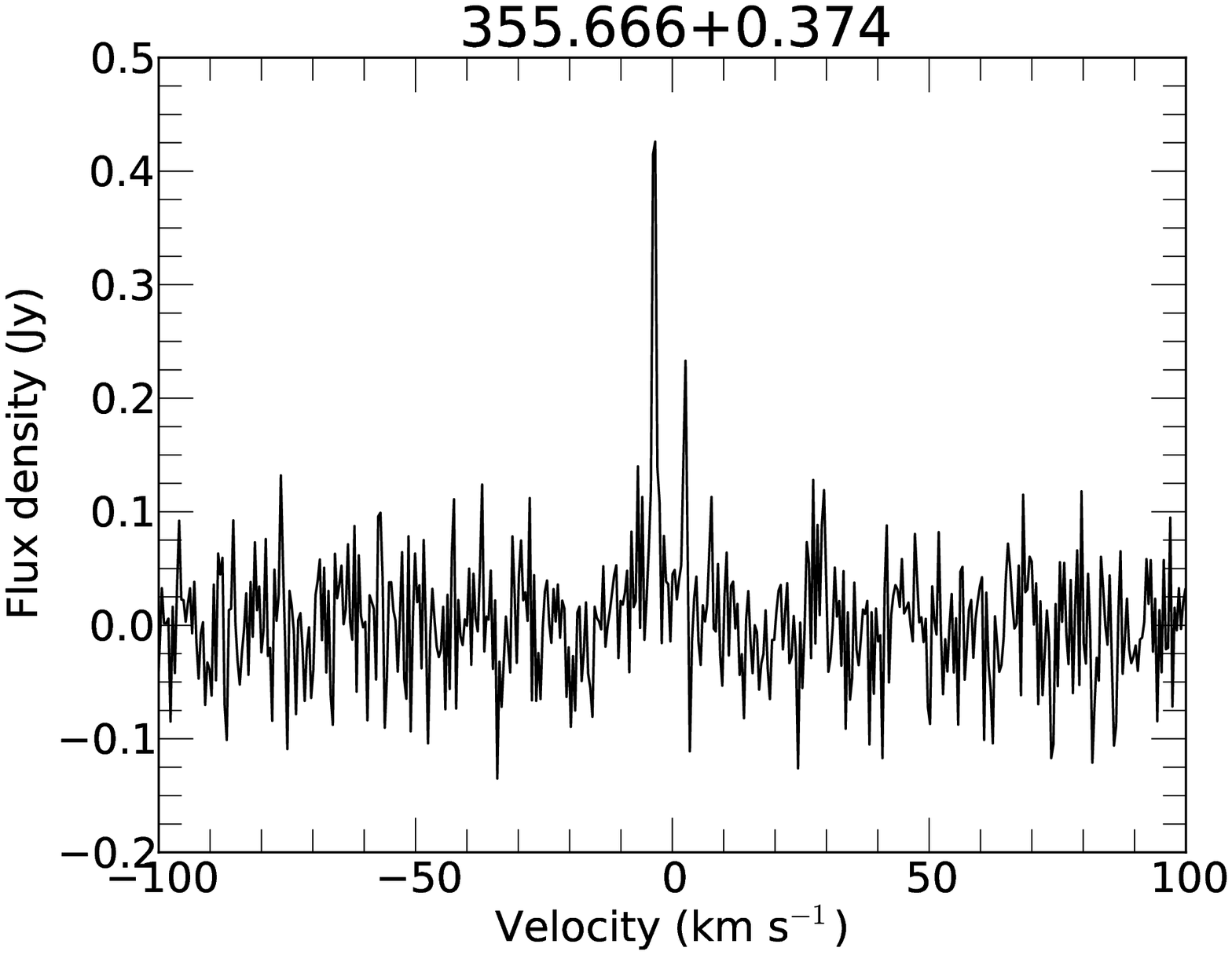}
\includegraphics[width=2.2in]{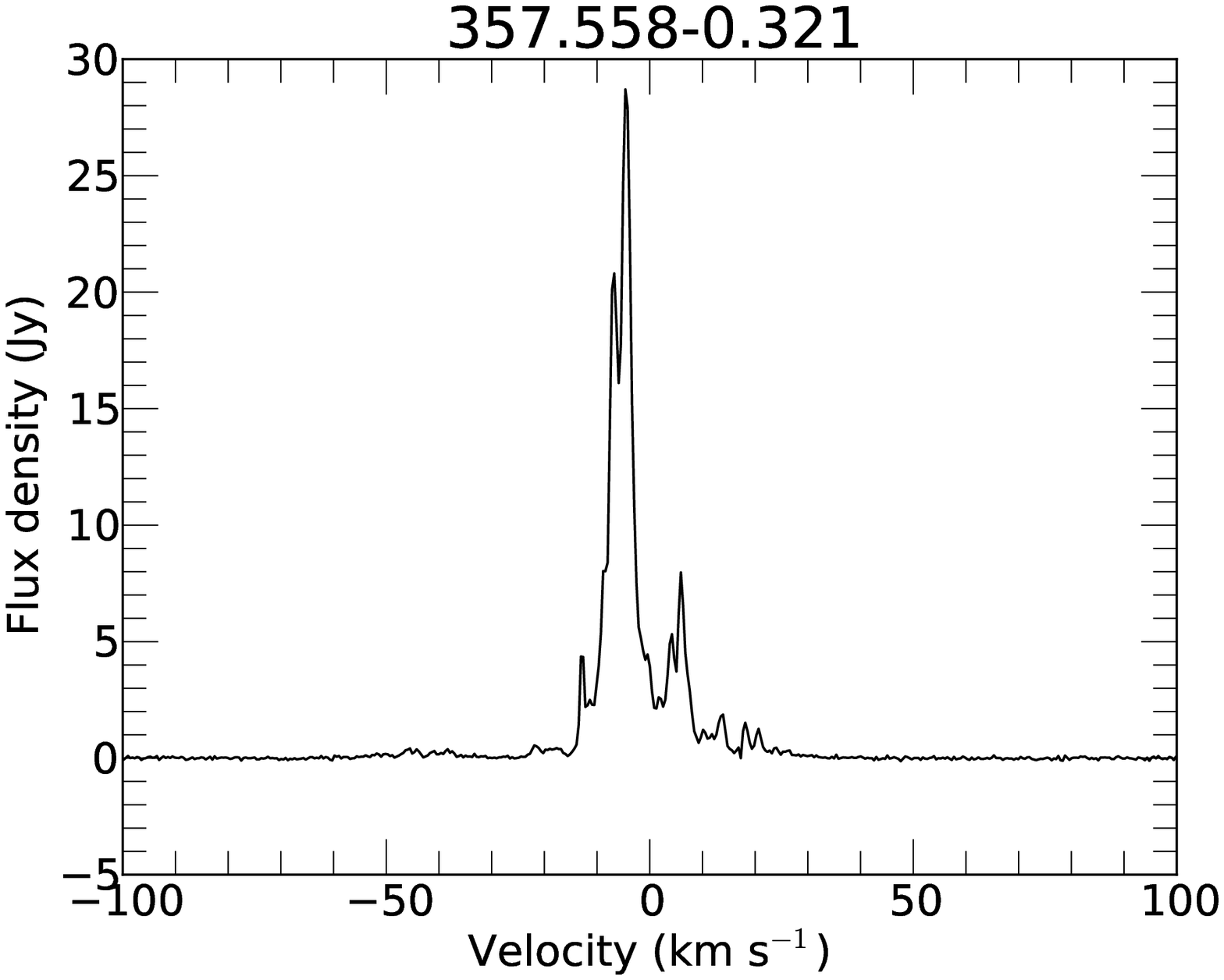}
\includegraphics[width=2.2in]{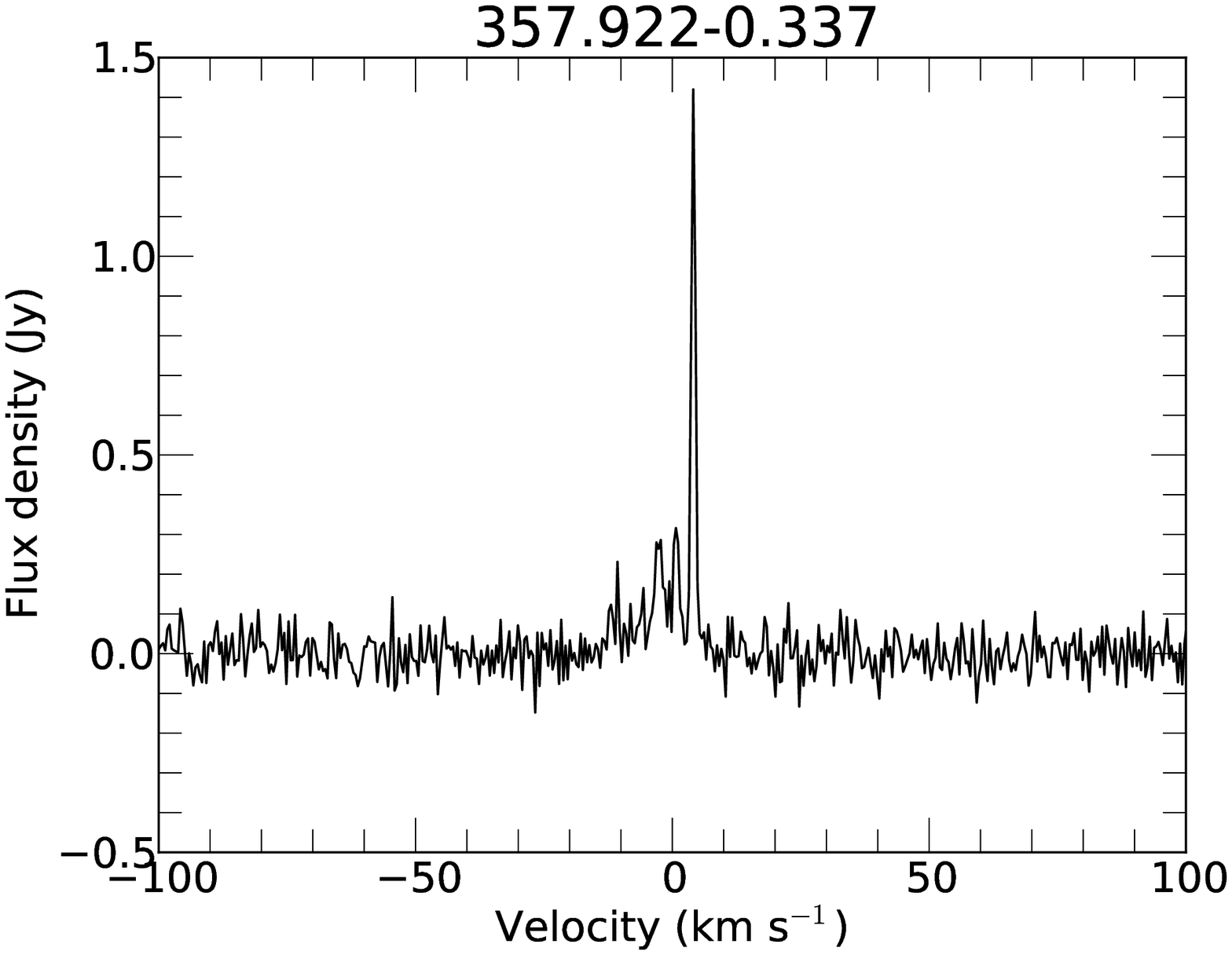}
\includegraphics[width=2.2in]{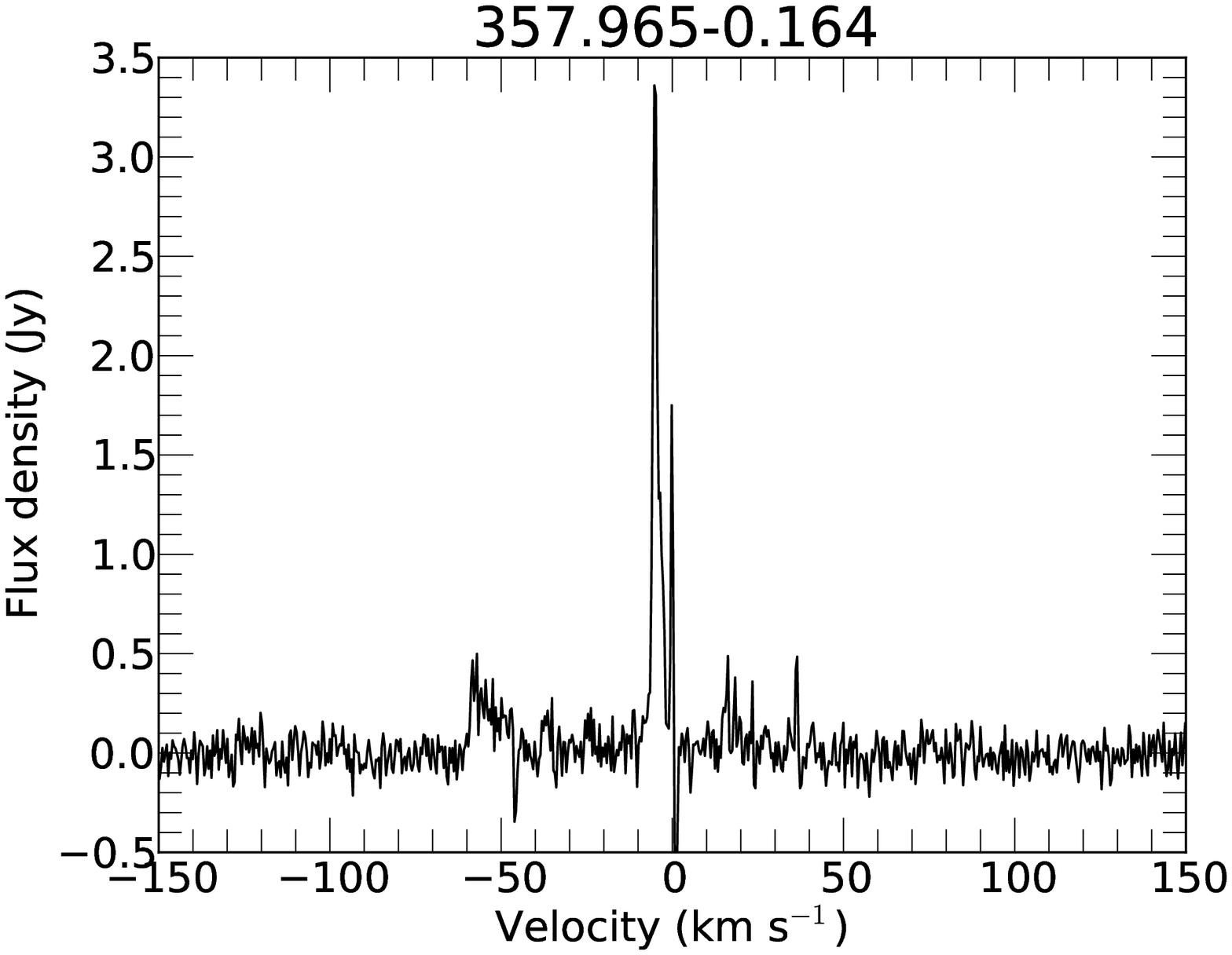}
\includegraphics[width=2.2in]{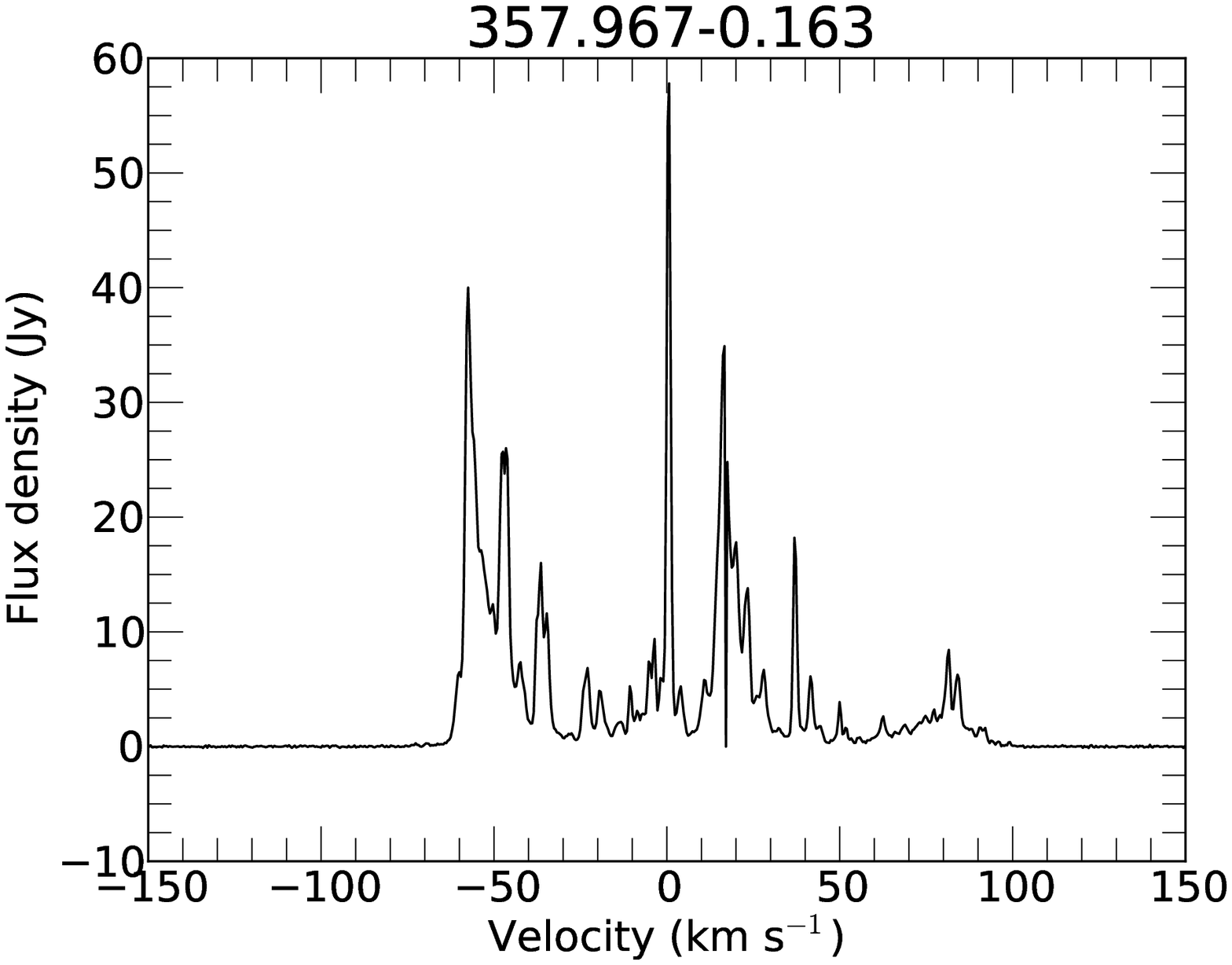}
\includegraphics[width=2.2in]{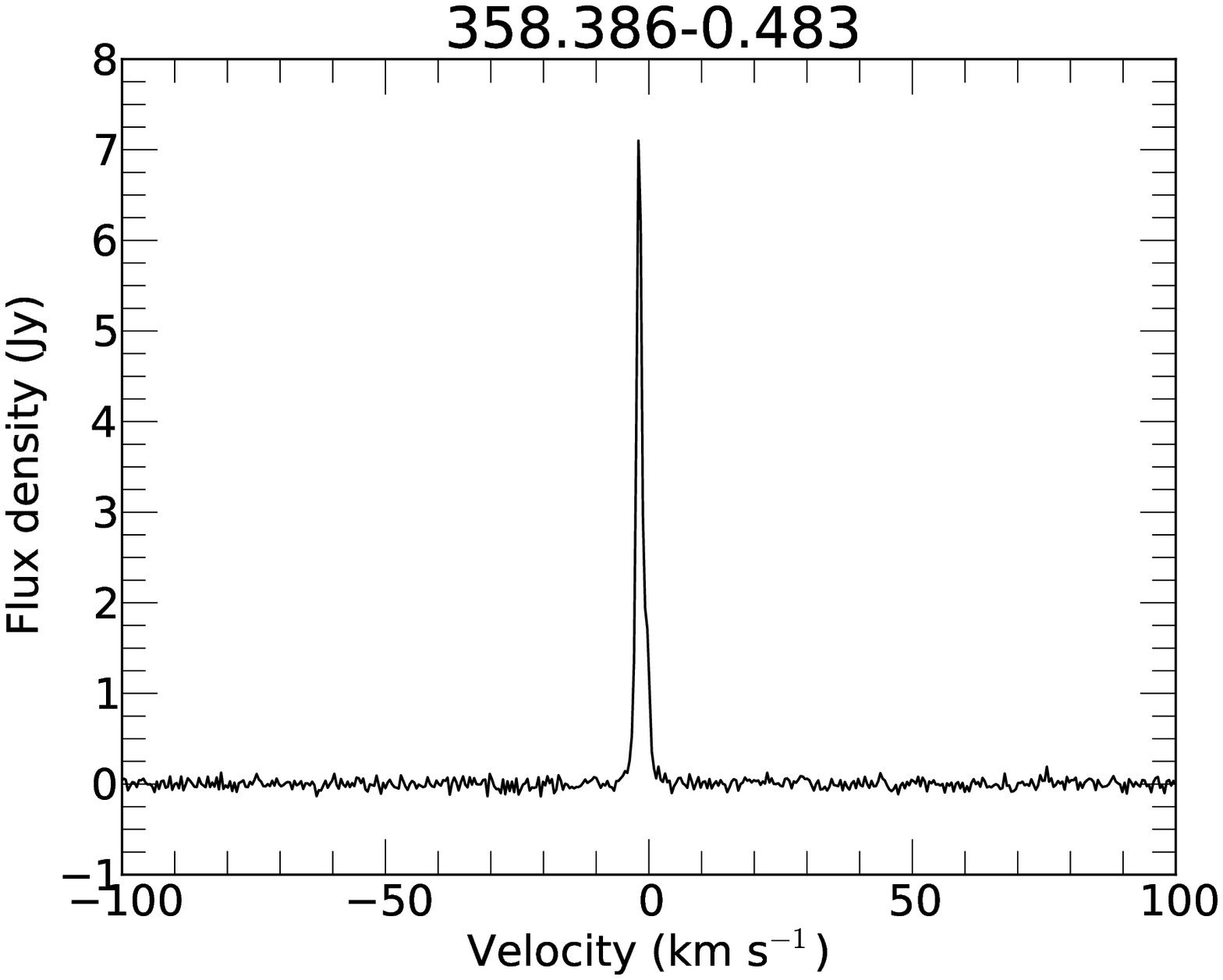}
\includegraphics[width=2.2in]{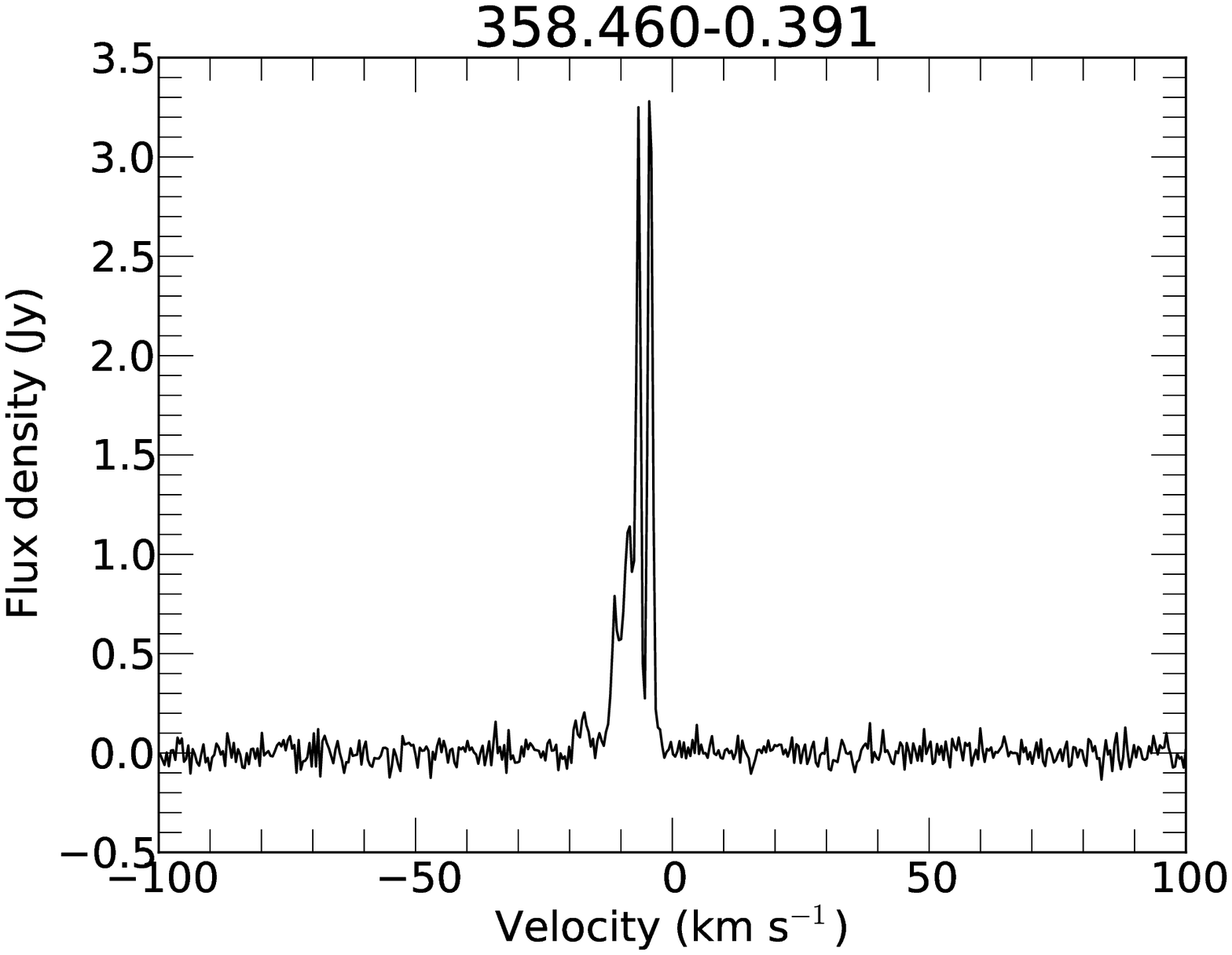}
\includegraphics[width=2.2in]{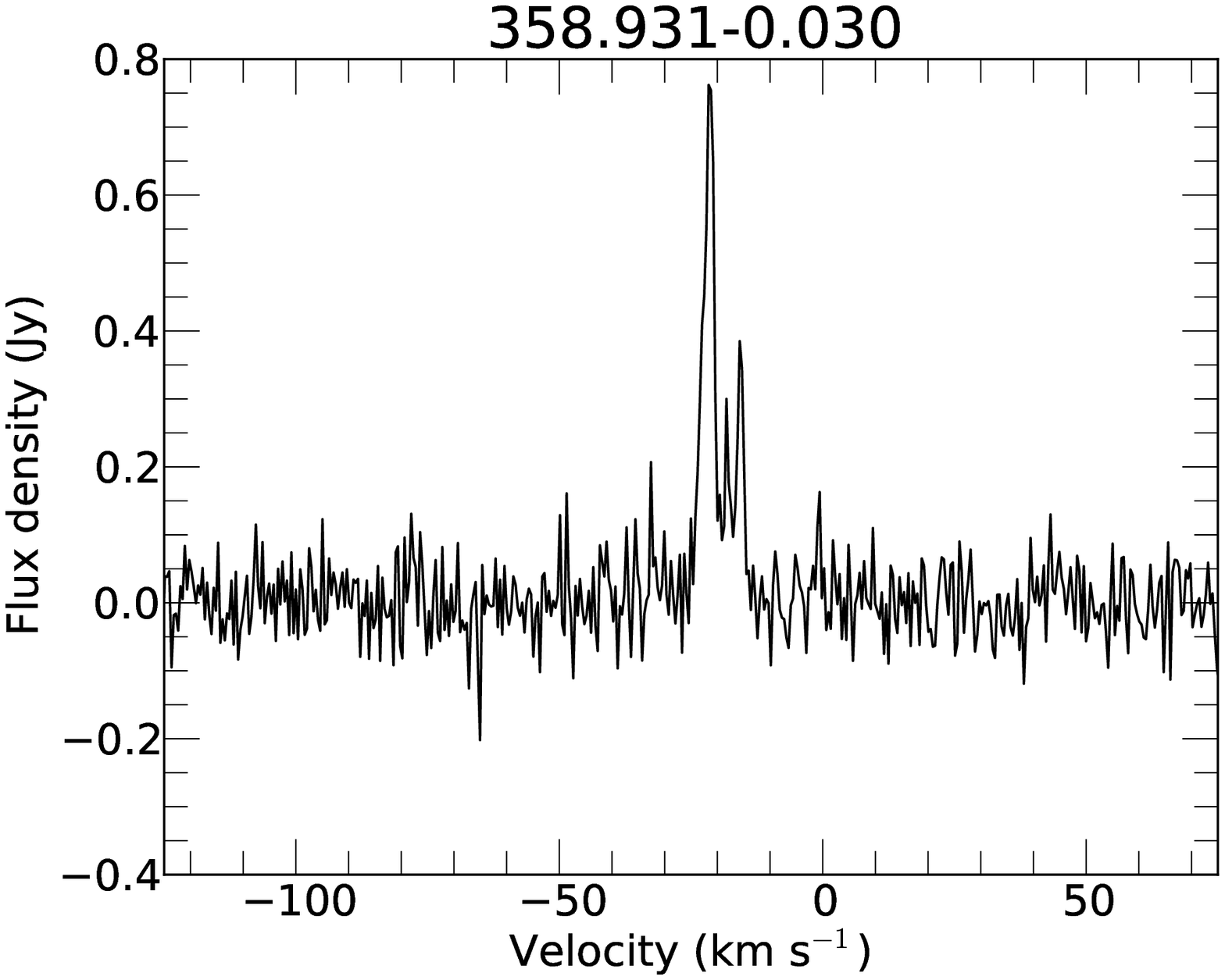}
\includegraphics[width=2.2in]{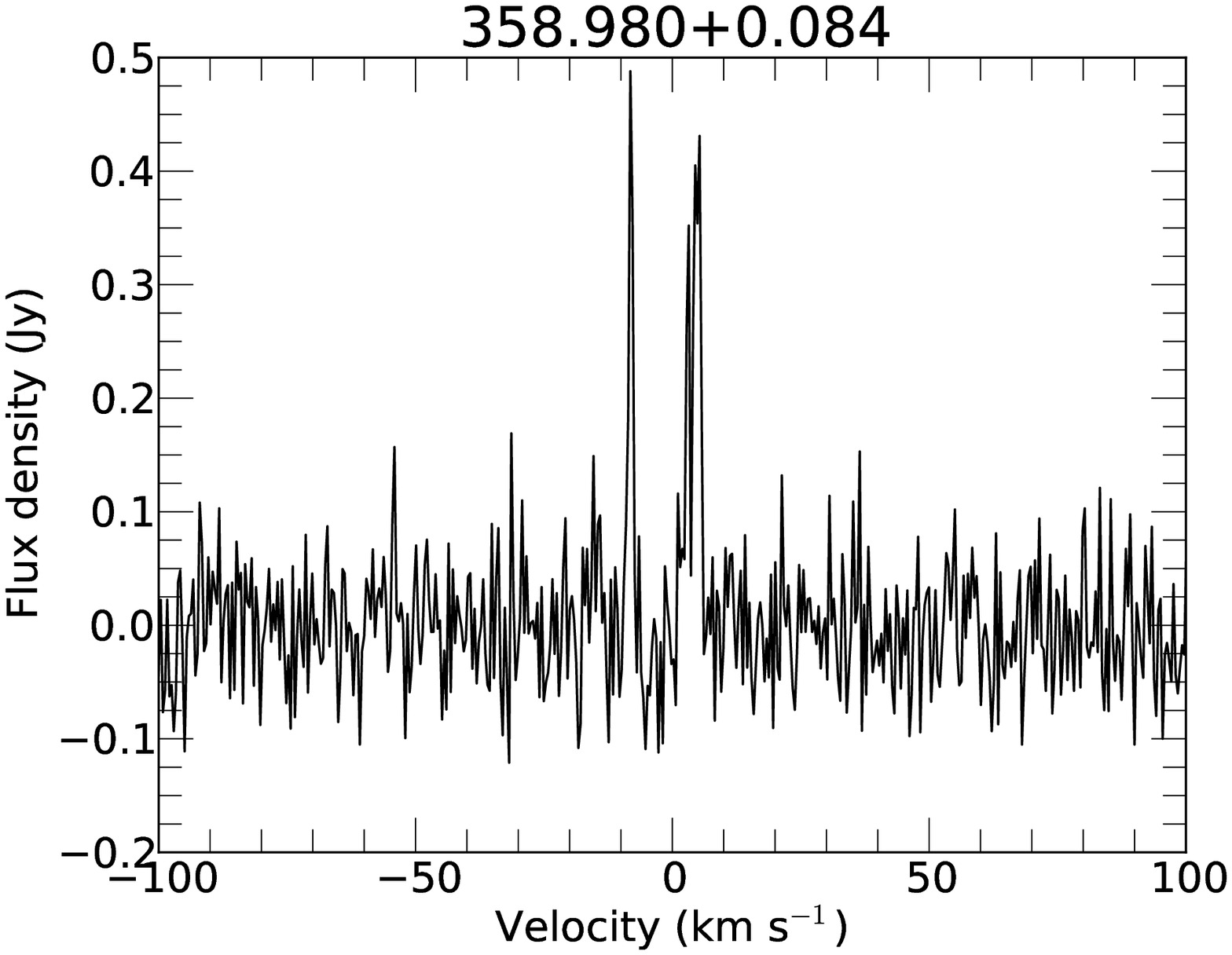}
\includegraphics[width=2.2in]{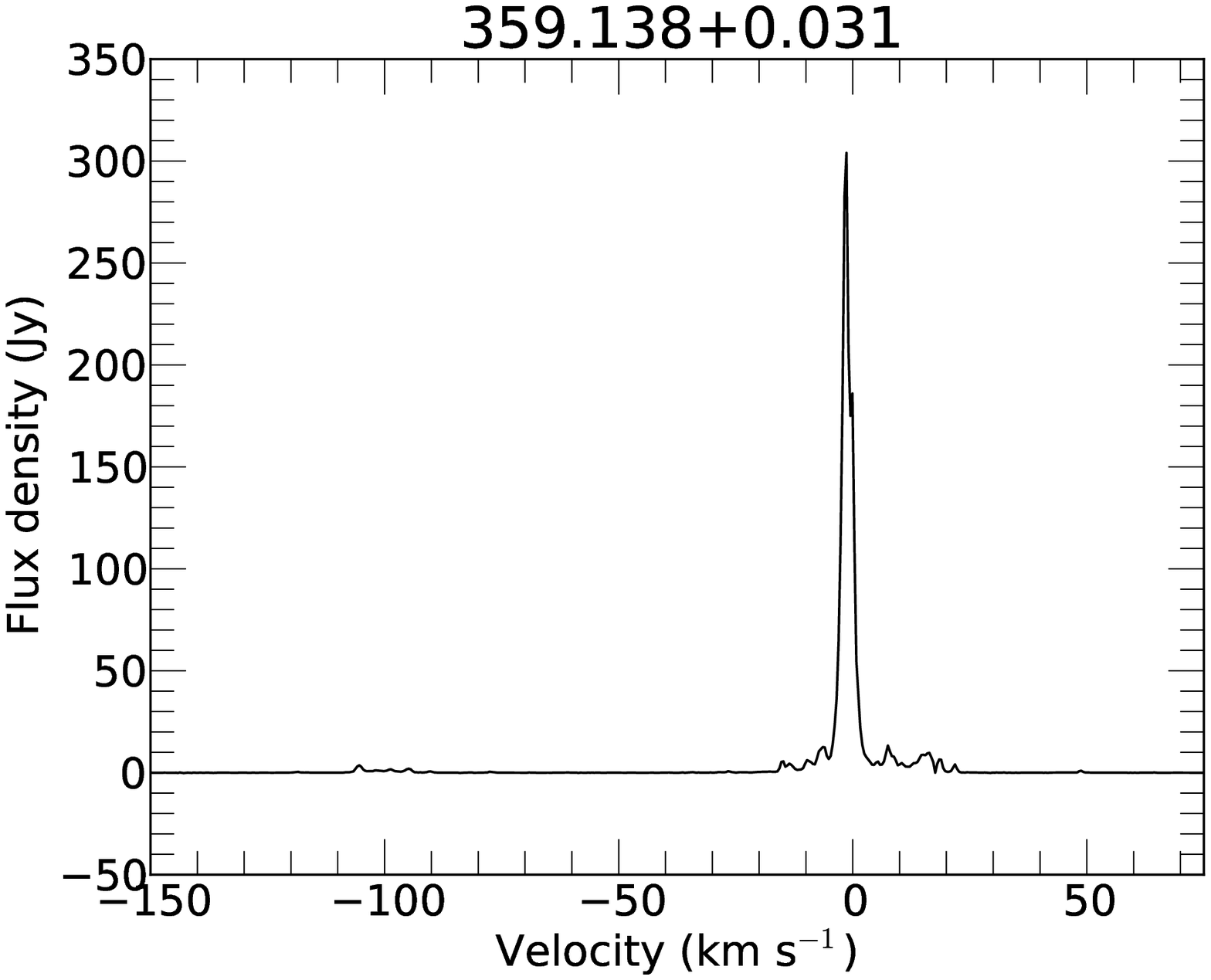}
\includegraphics[width=2.2in]{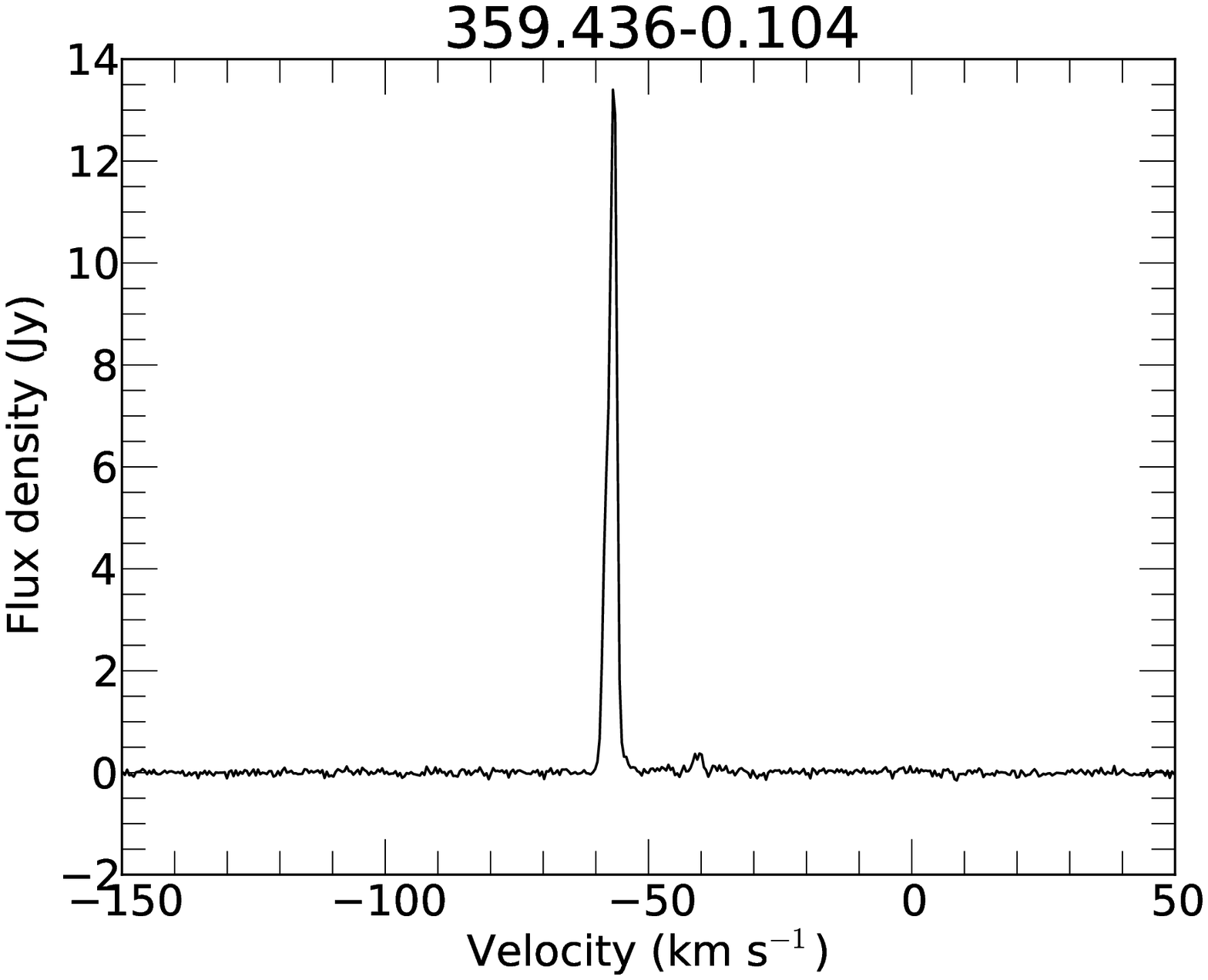}
\\
\addtocounter{figure}{-1}
  \caption{-- {\emph {continued}}}
\end{figure*}

\begin{figure*}
\includegraphics[width=2.2in]{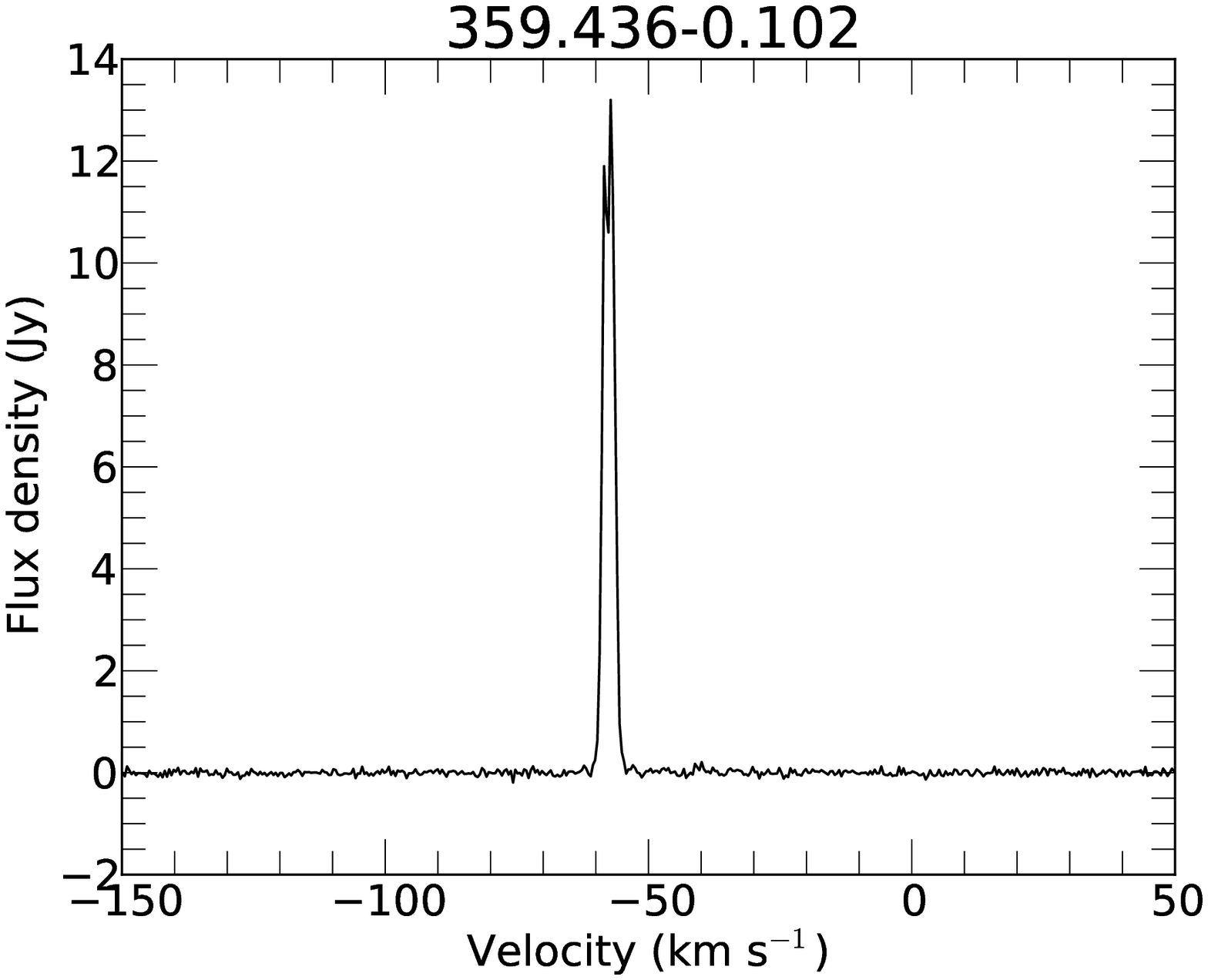}
\includegraphics[width=2.2in]{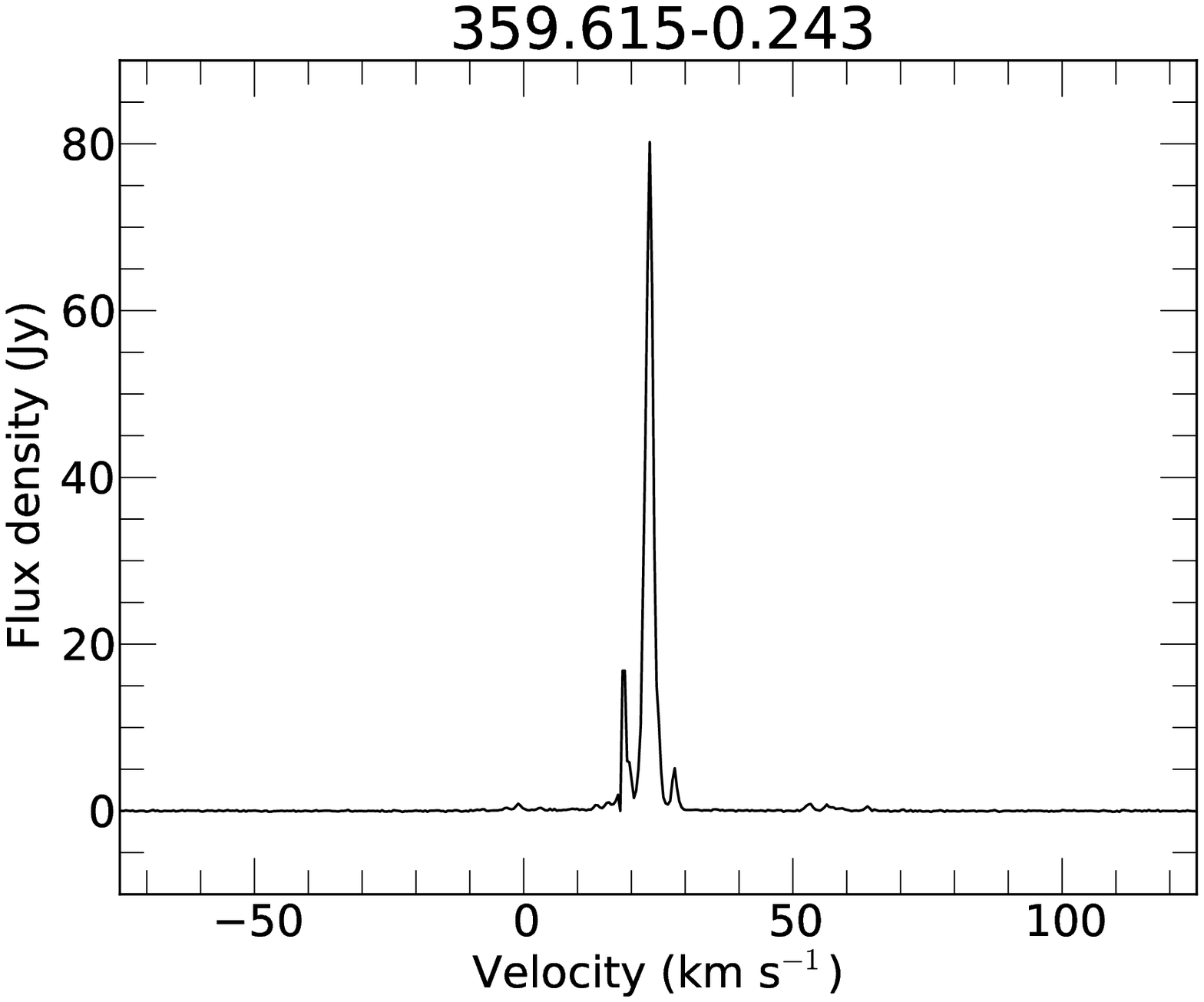}
\includegraphics[width=2.2in]{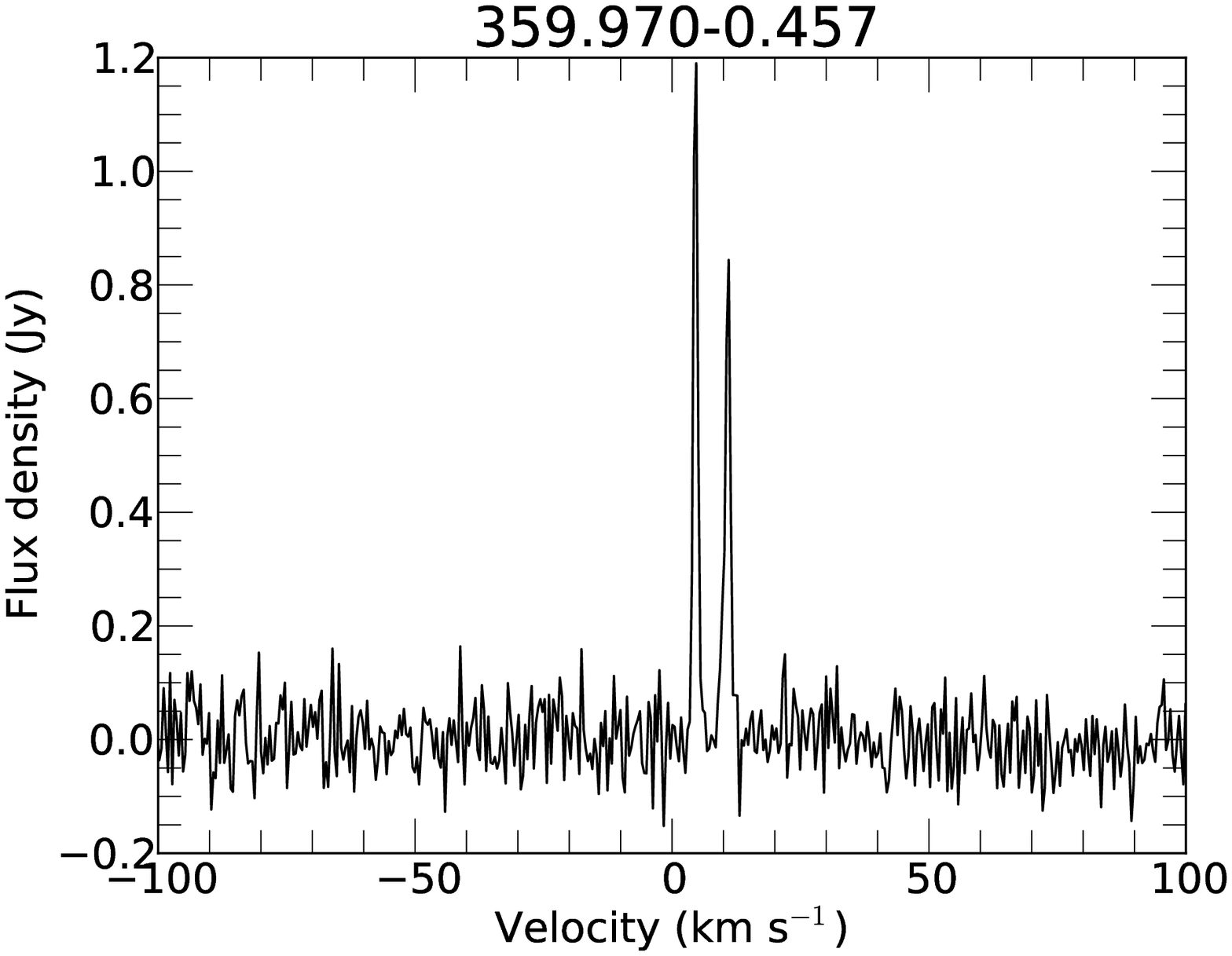}
\includegraphics[width=2.2in]{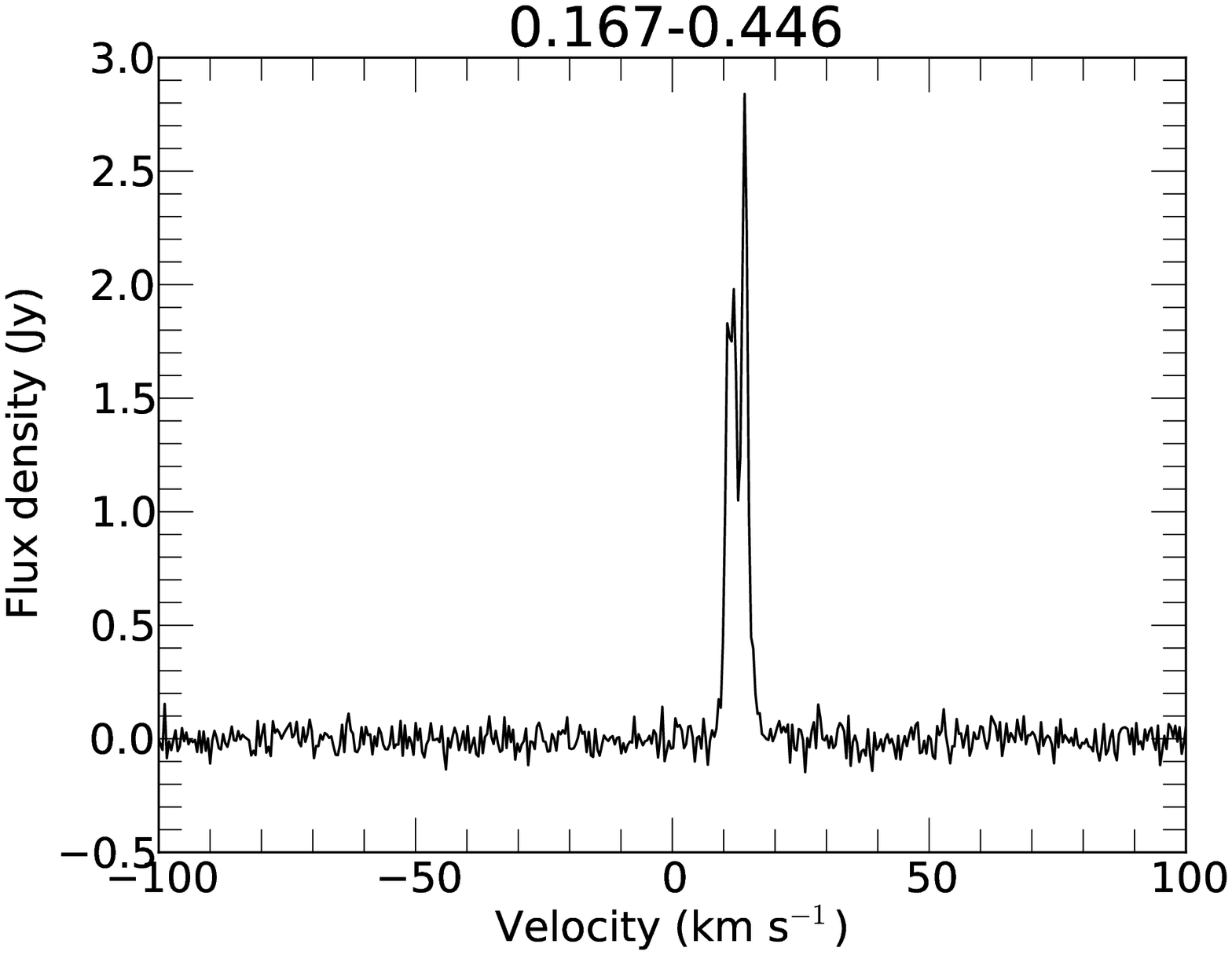}
\includegraphics[width=2.2in]{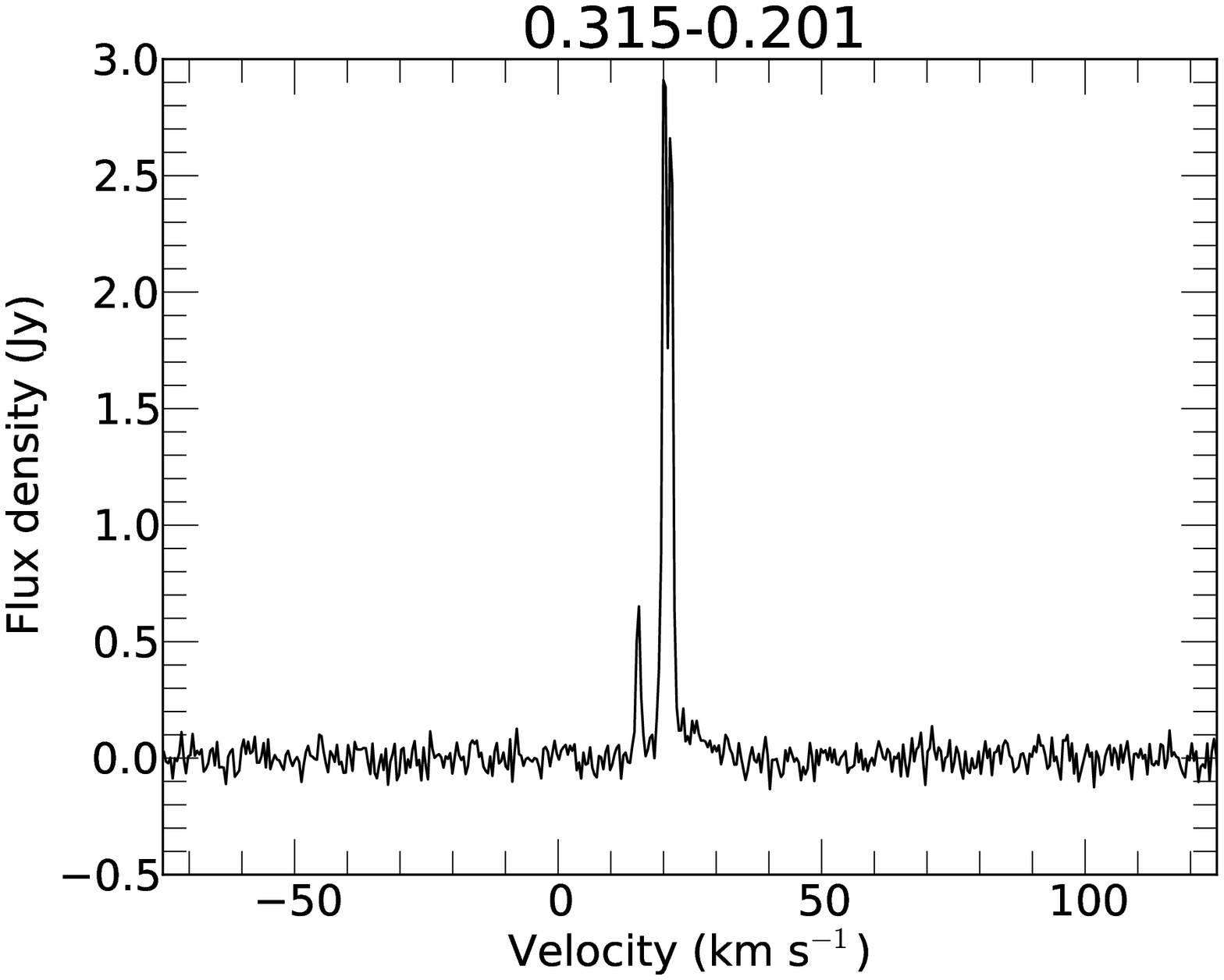}
\includegraphics[width=2.2in]{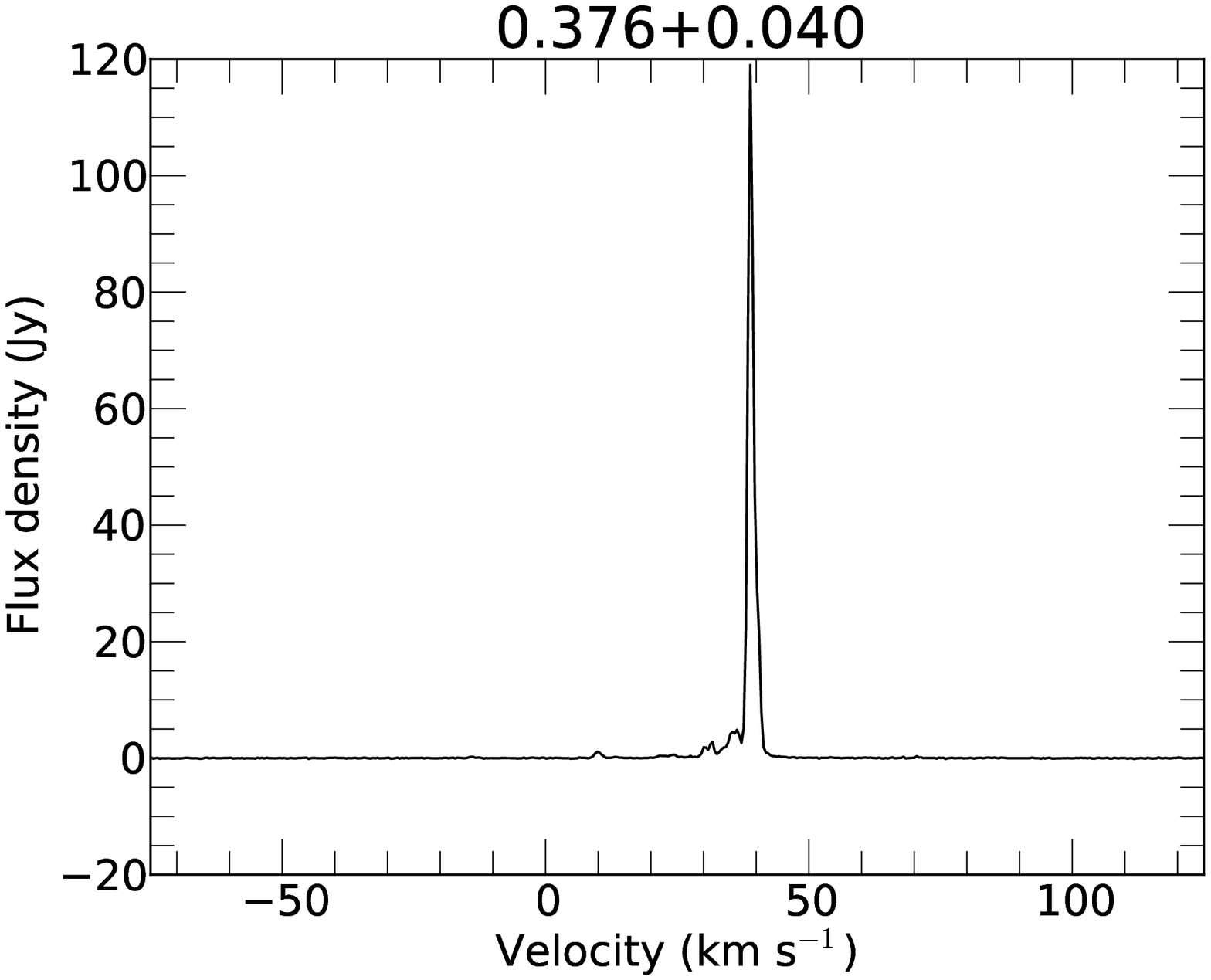}
\includegraphics[width=2.2in]{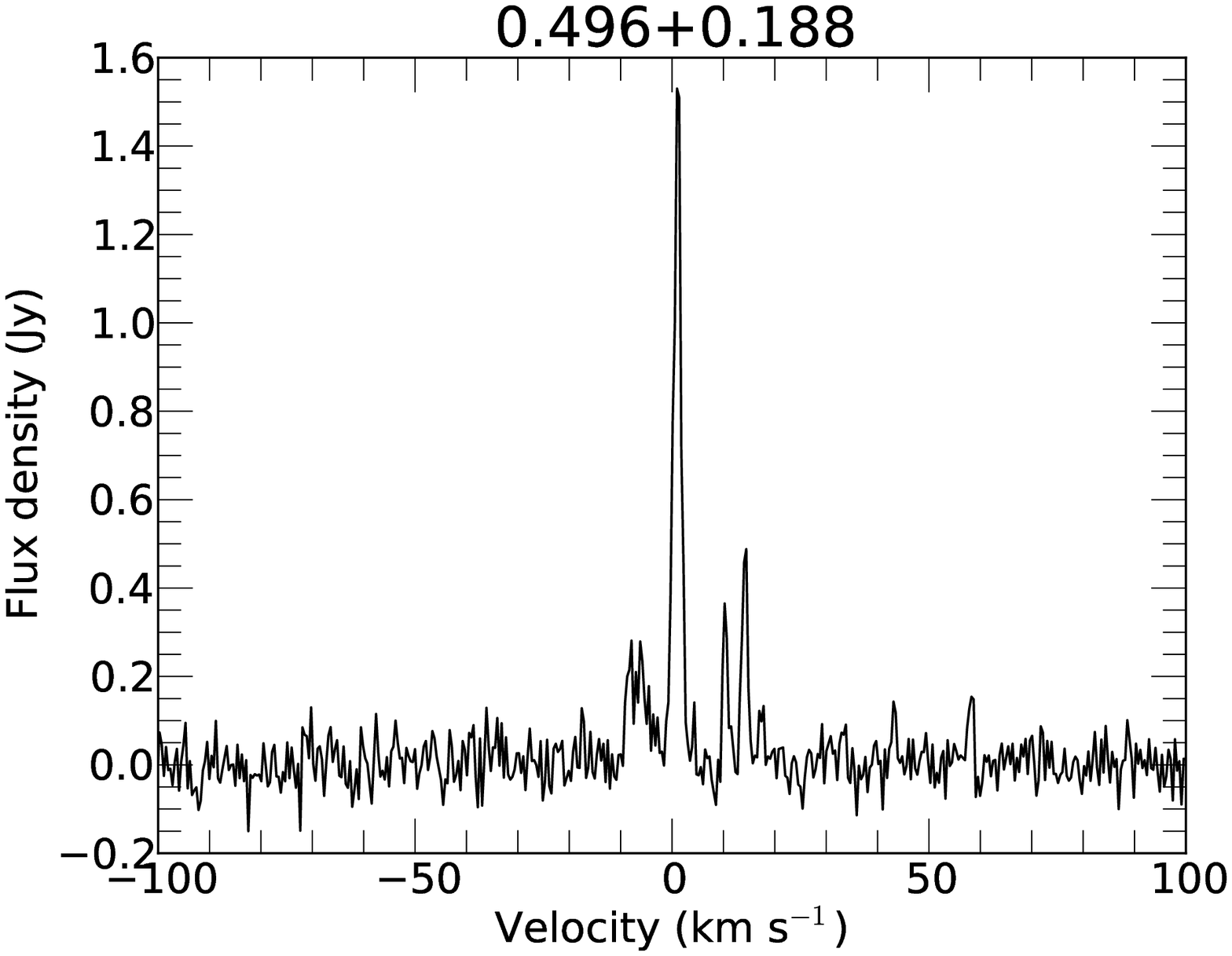}
\includegraphics[width=2.2in]{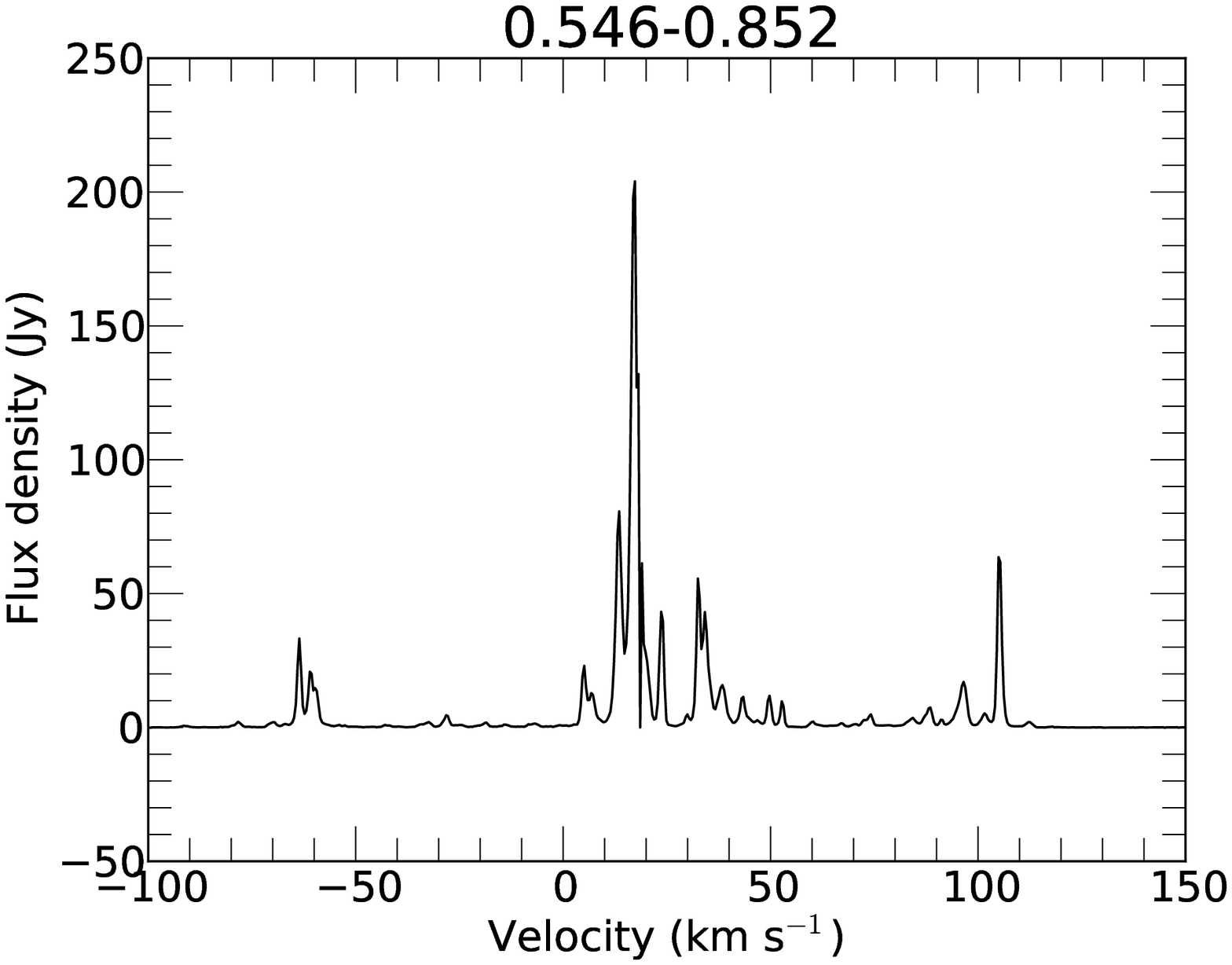}
\includegraphics[width=2.2in]{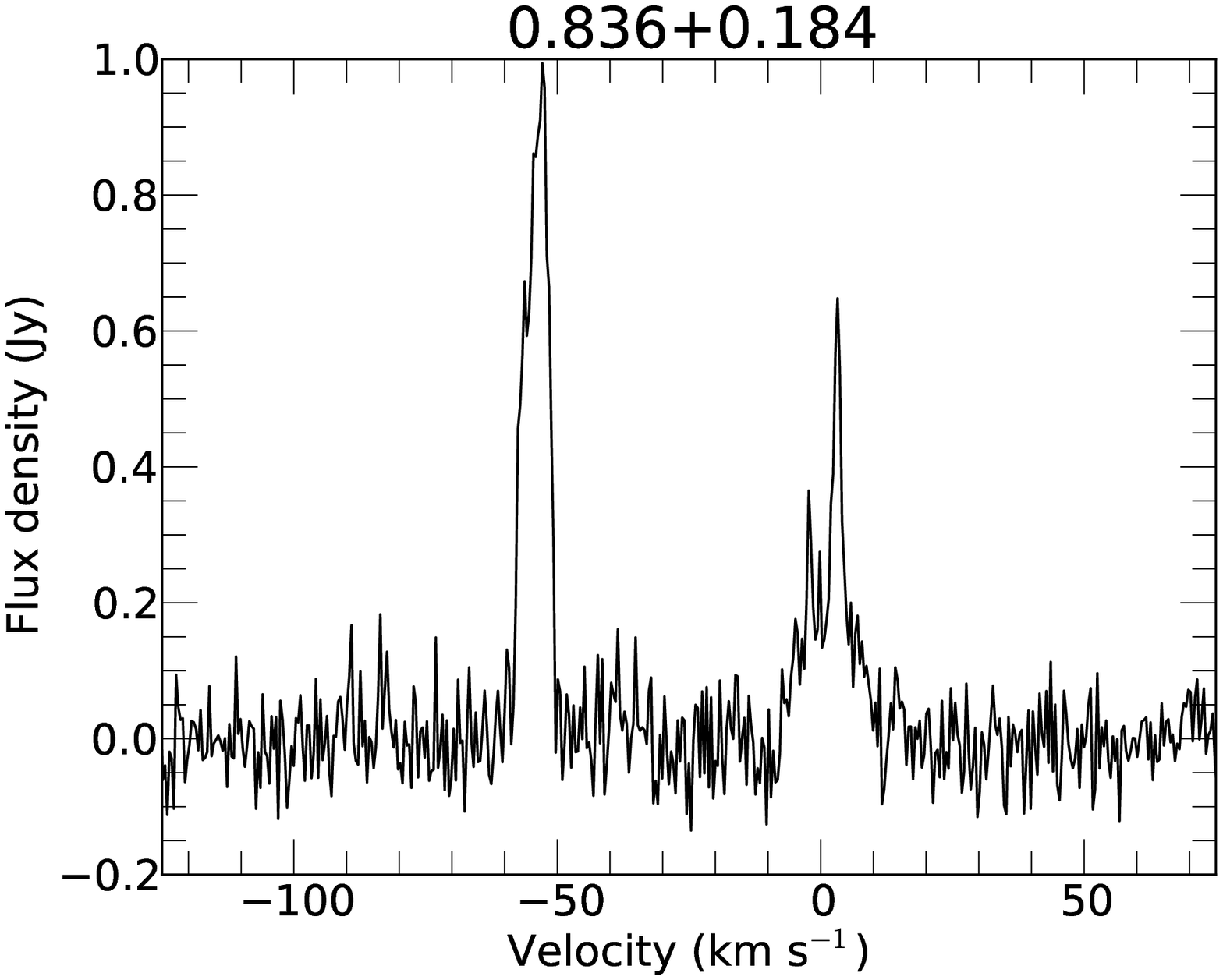}
\includegraphics[width=2.2in]{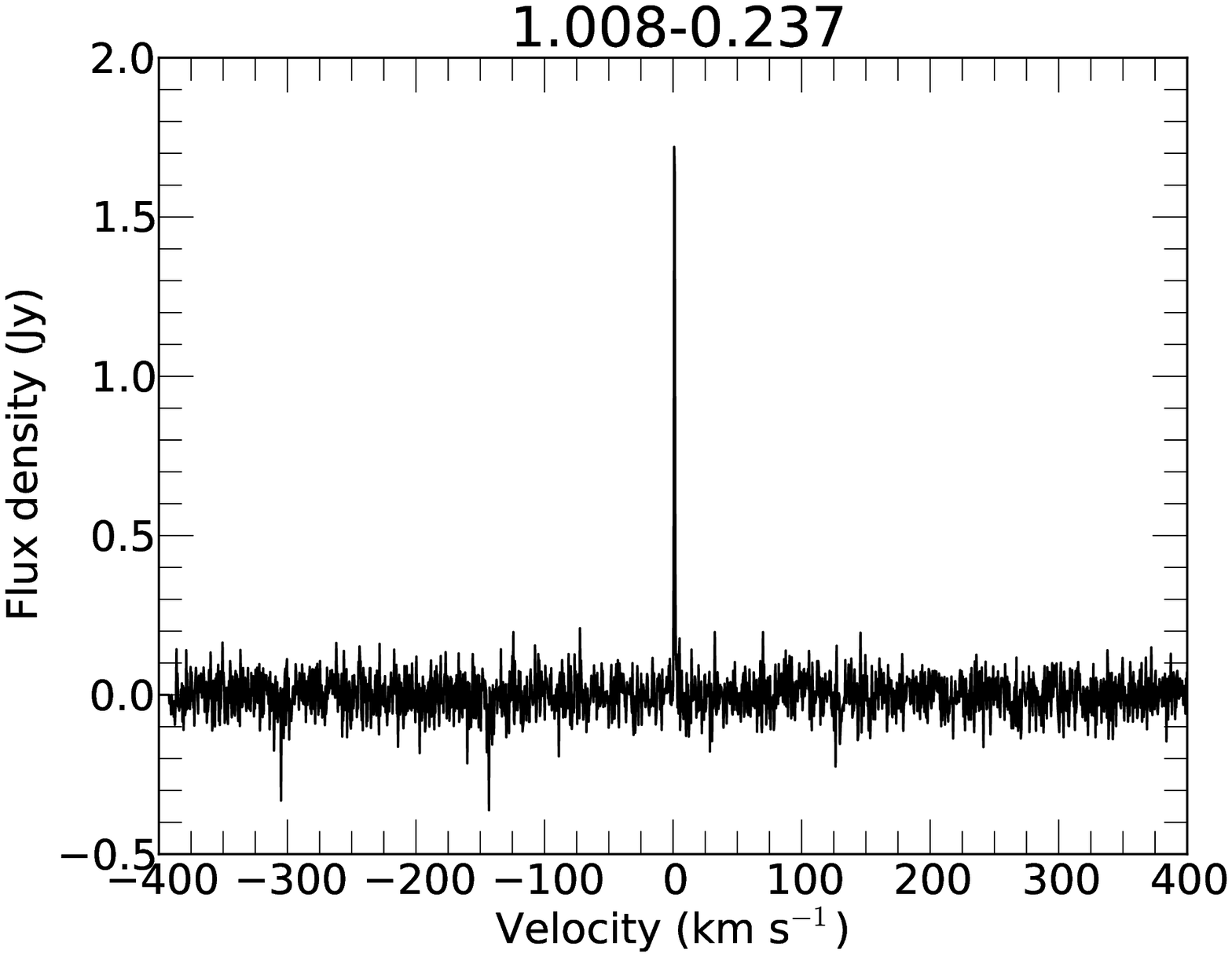}
\includegraphics[width=2.2in]{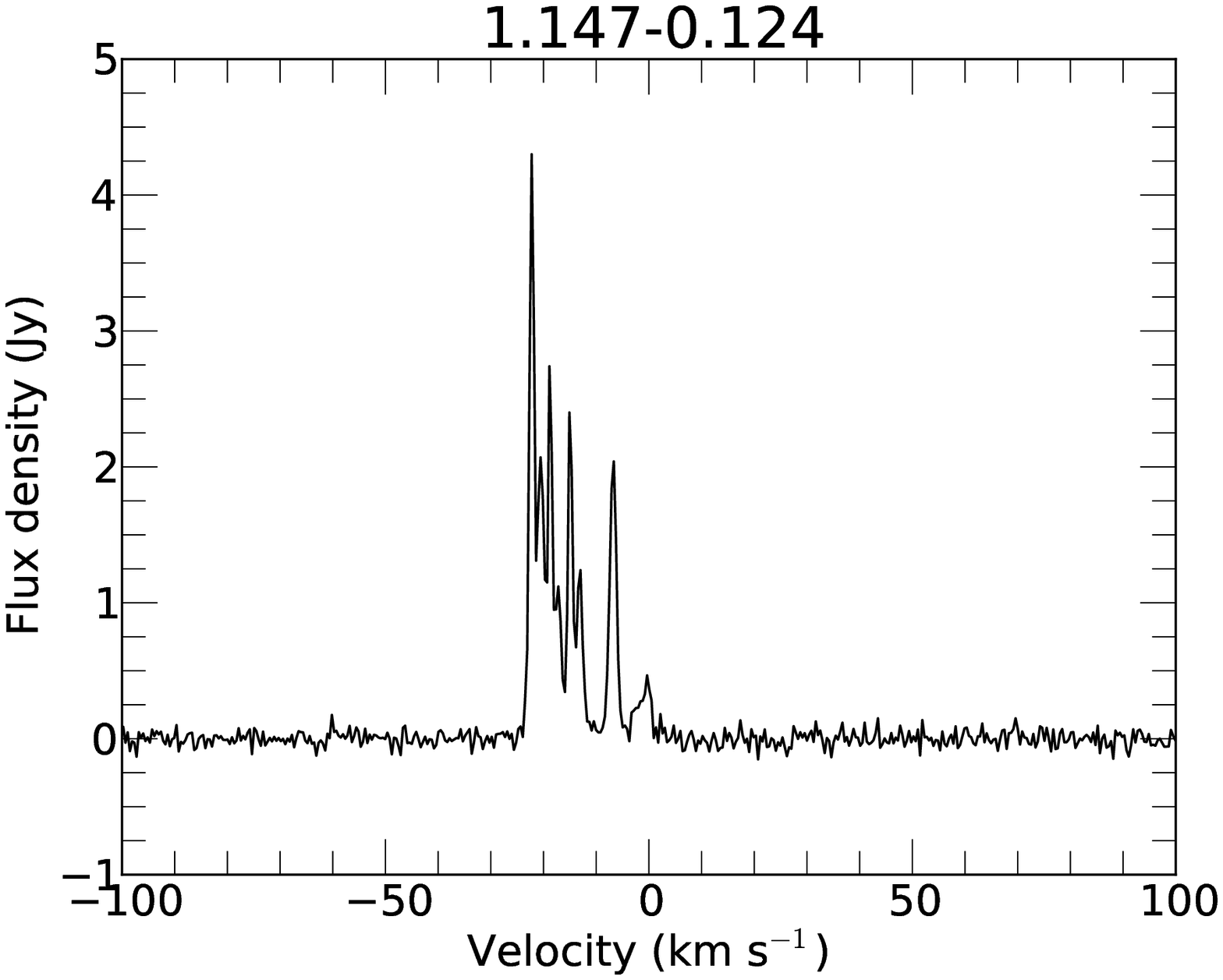}
\includegraphics[width=2.2in]{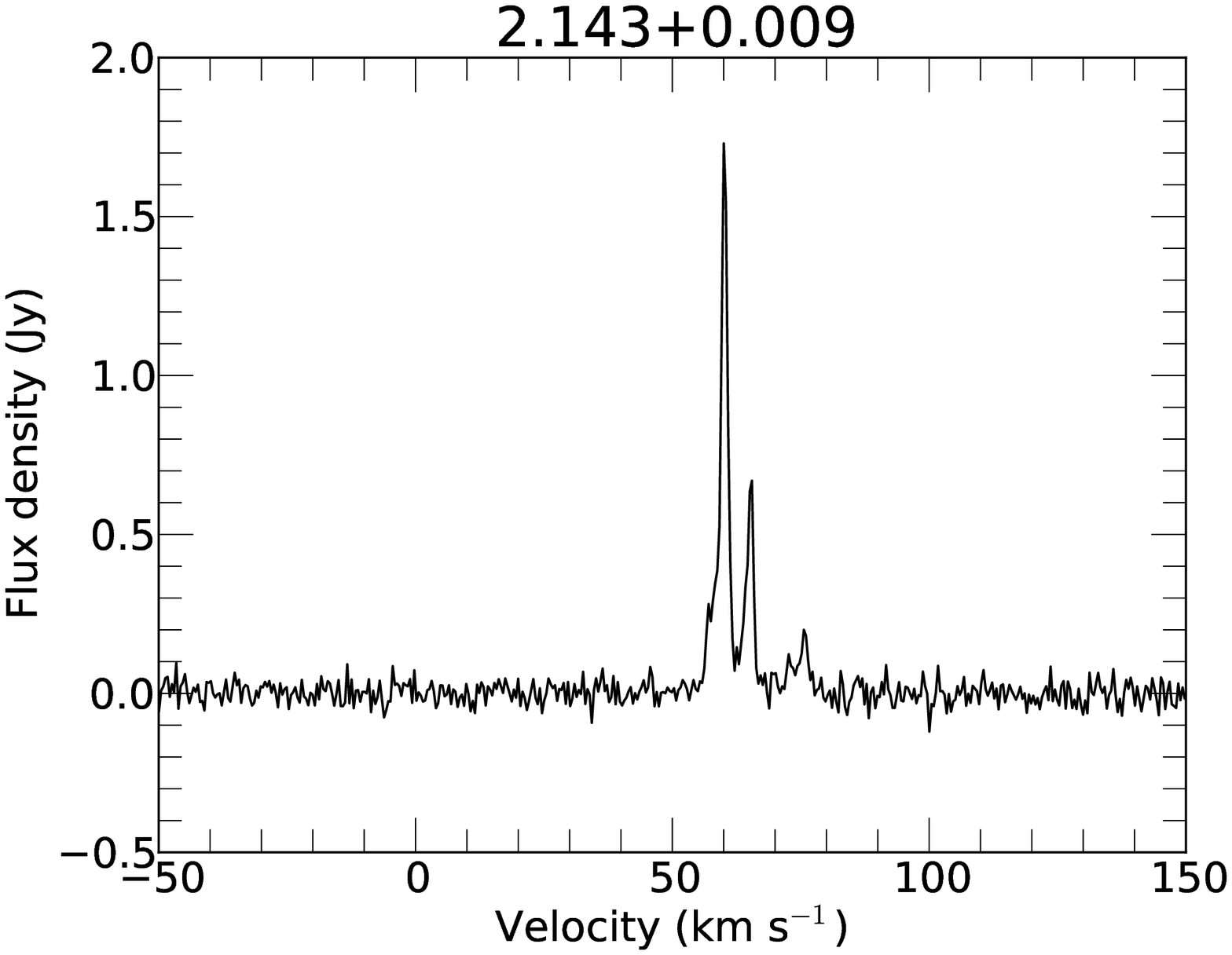}
\includegraphics[width=2.2in]{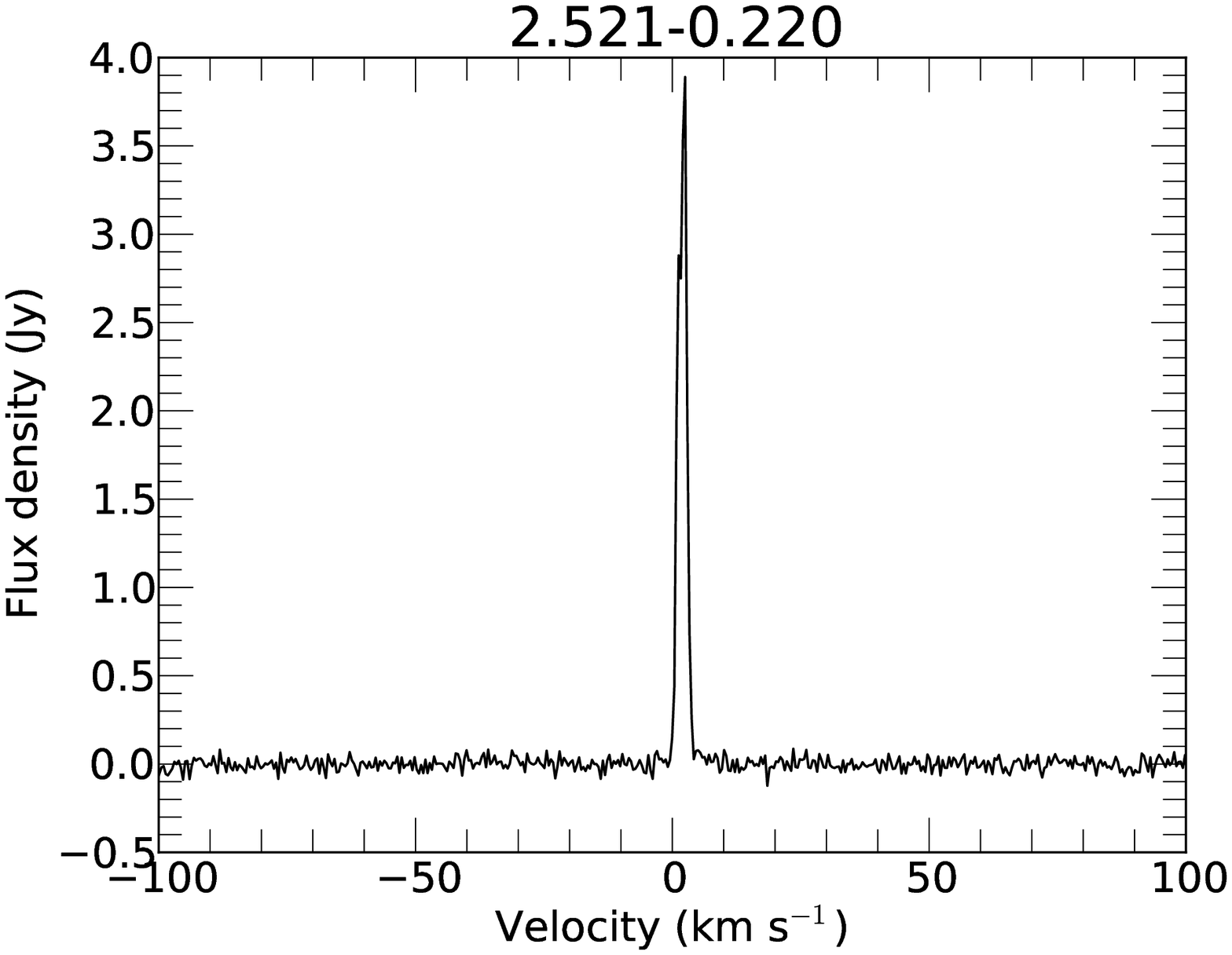}
\includegraphics[width=2.2in]{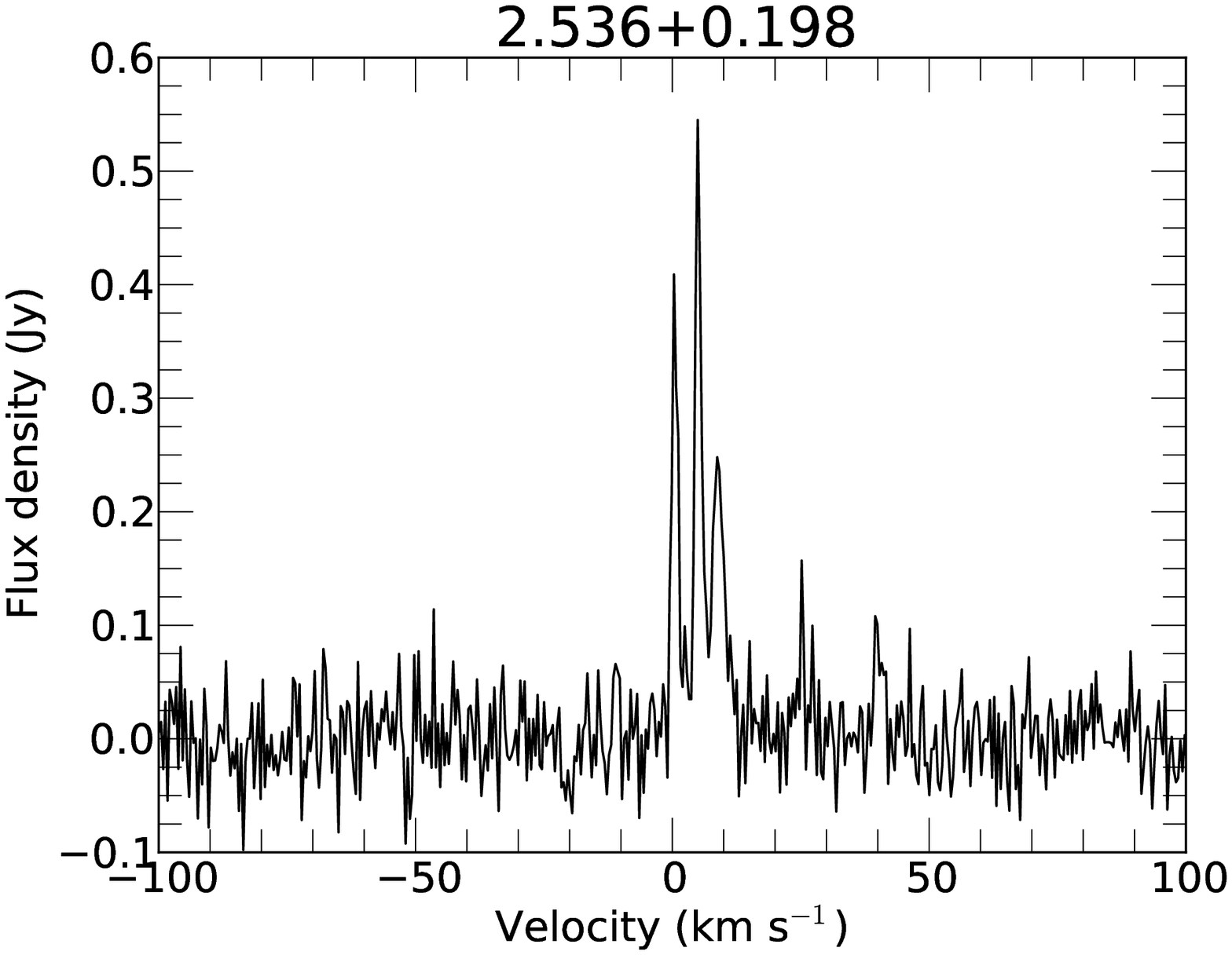}
\includegraphics[width=2.2in]{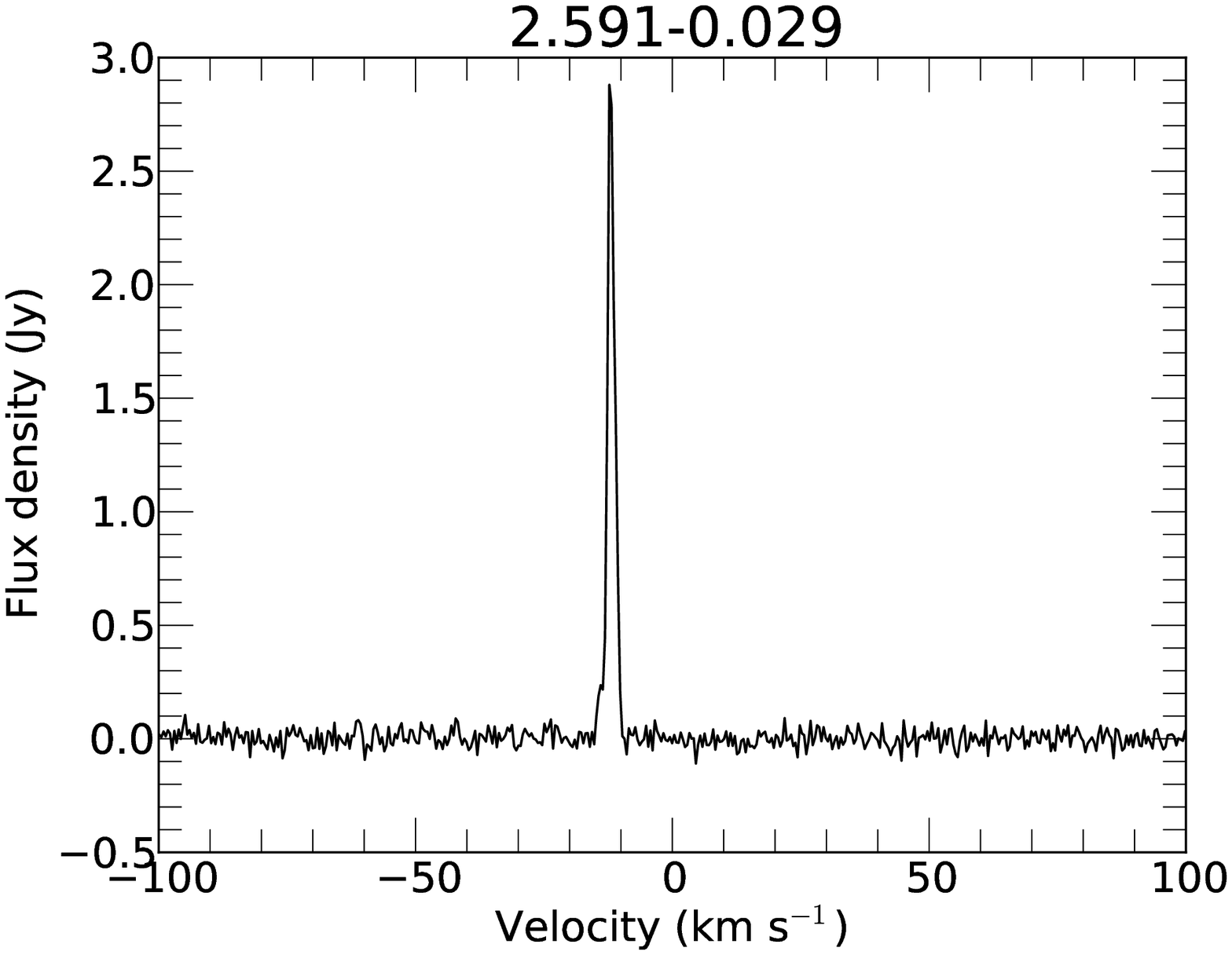}
\\
\addtocounter{figure}{-1}
  \caption{-- {\emph {continued}}}
\end{figure*}

\begin{figure*}
\includegraphics[width=2.2in]{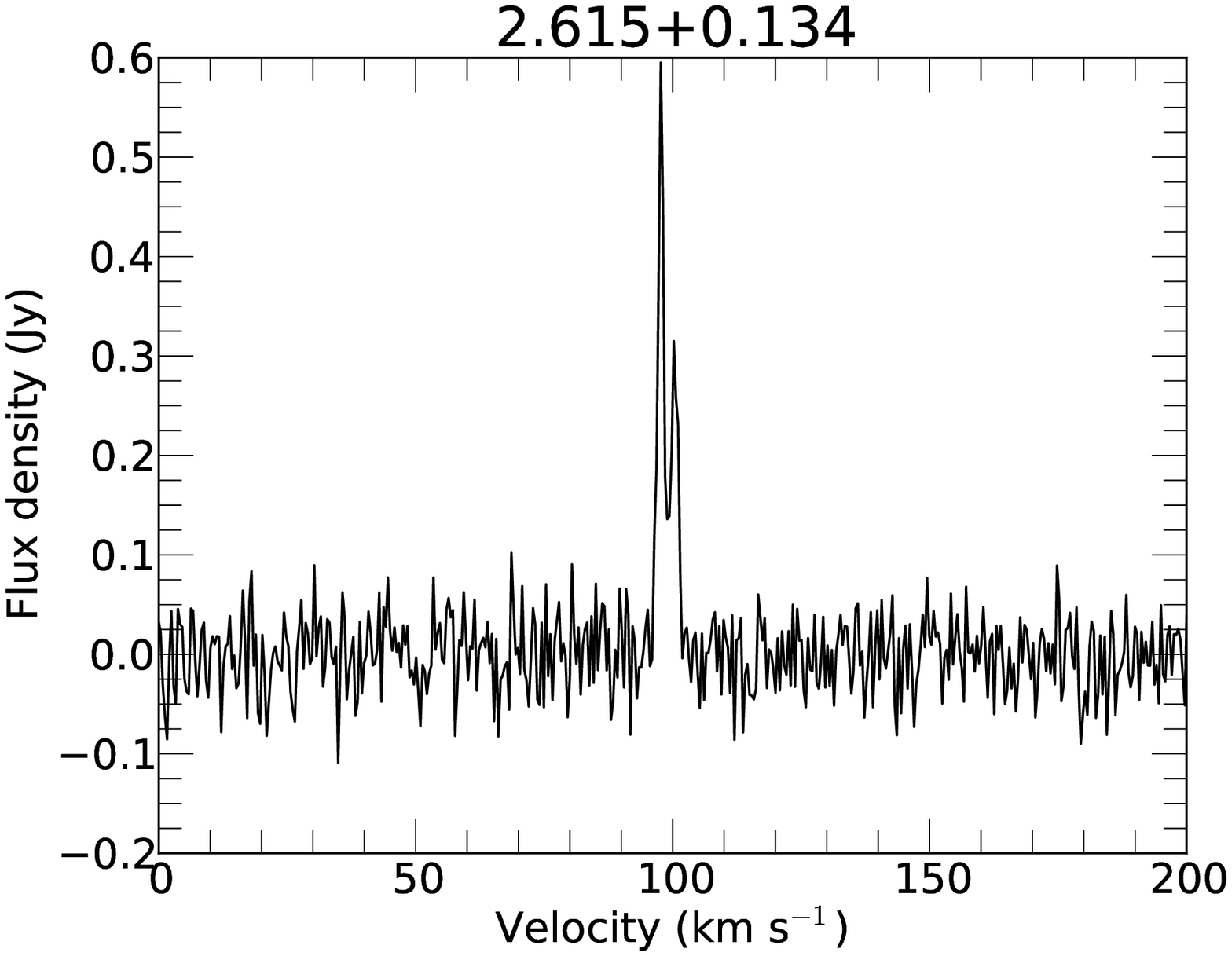}
\includegraphics[width=2.2in]{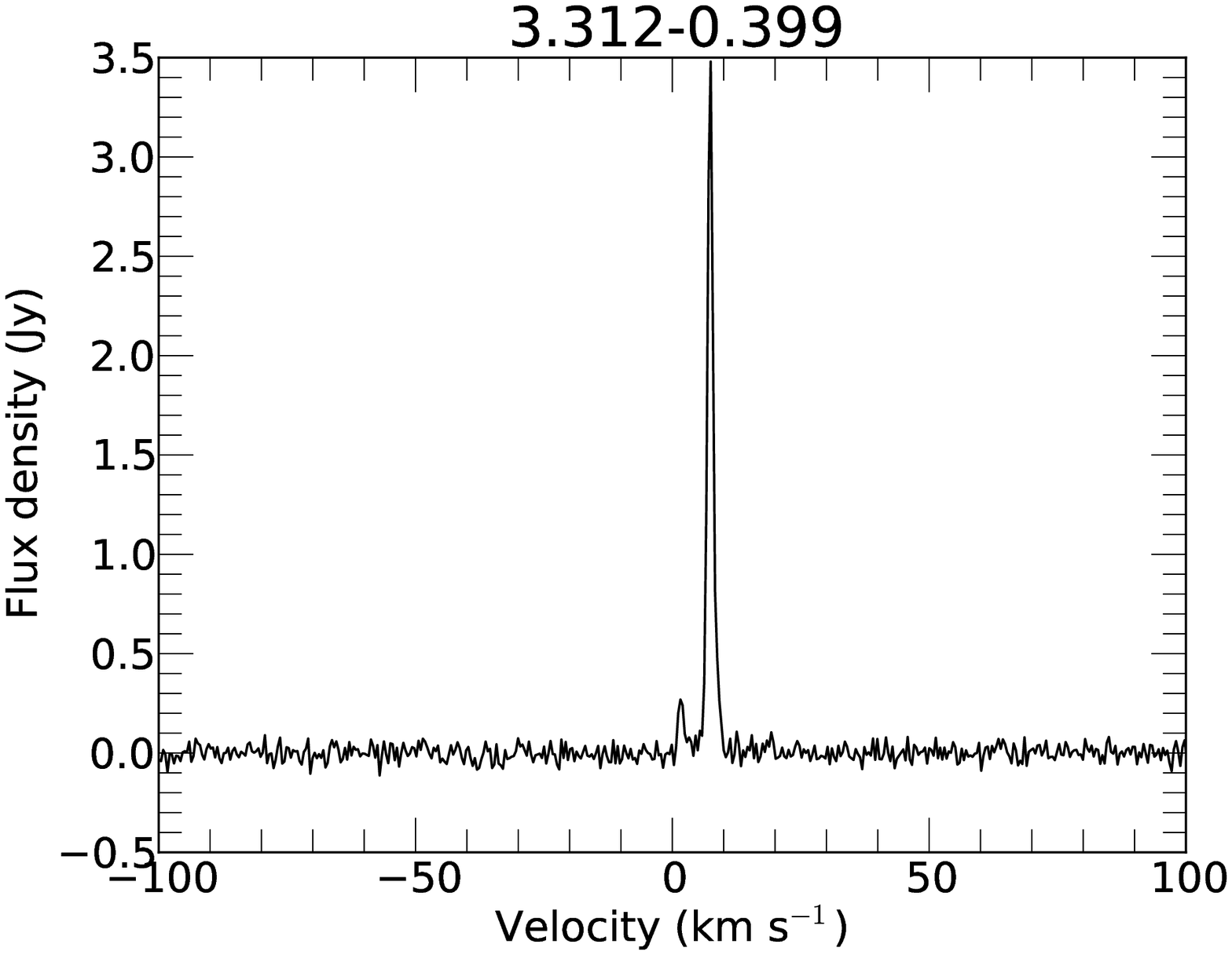}
\includegraphics[width=2.2in]{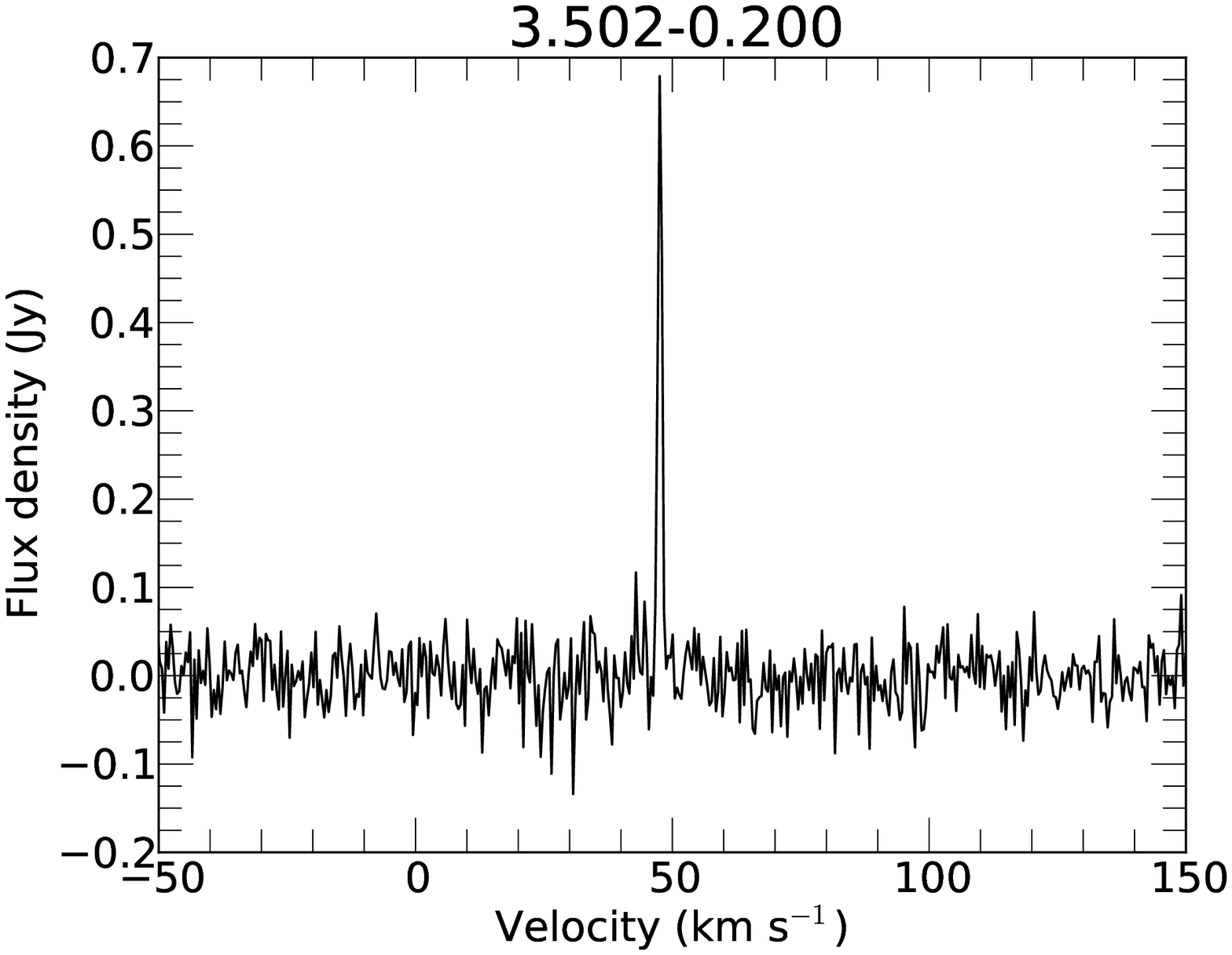}
\includegraphics[width=2.2in]{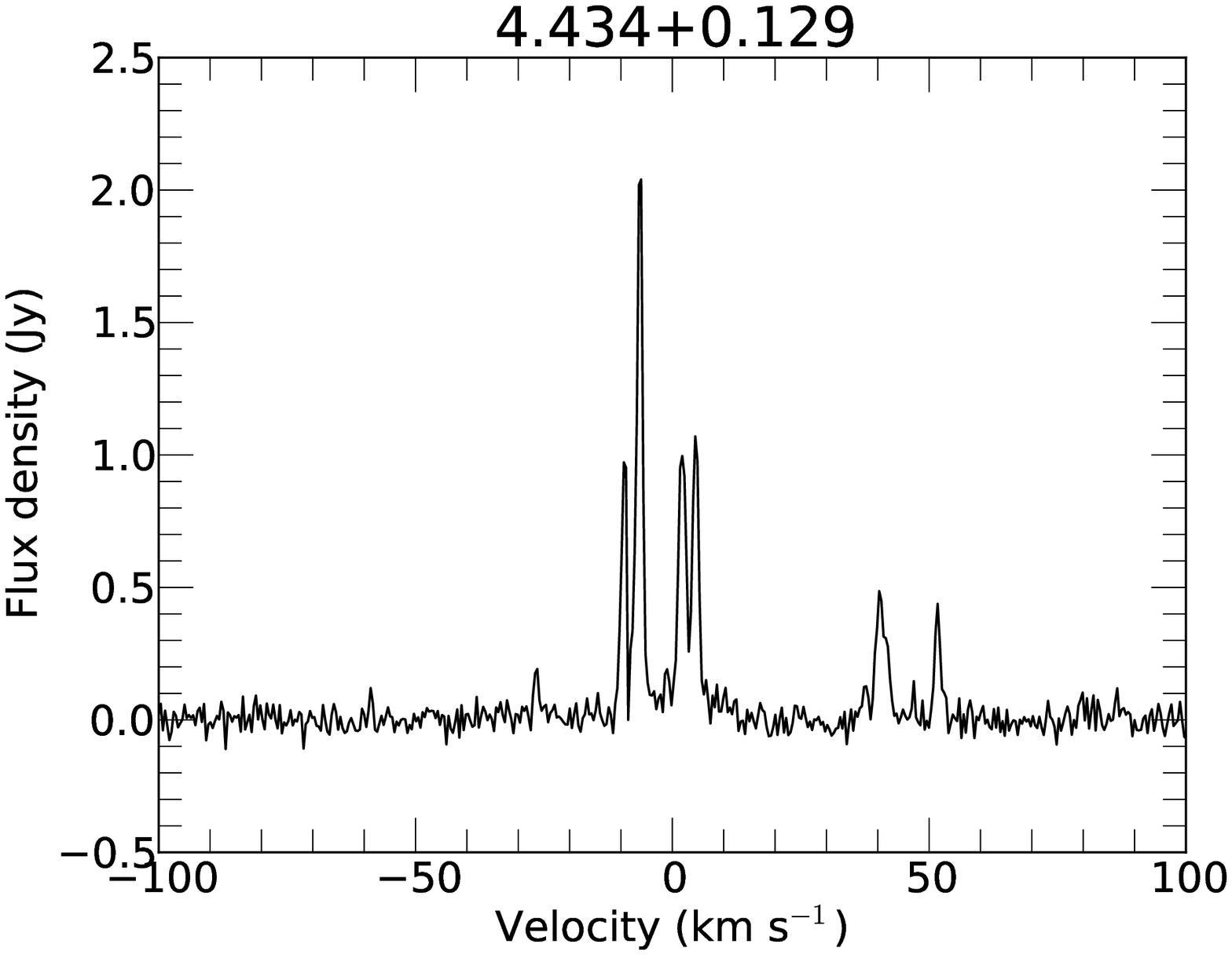}
\includegraphics[width=2.2in]{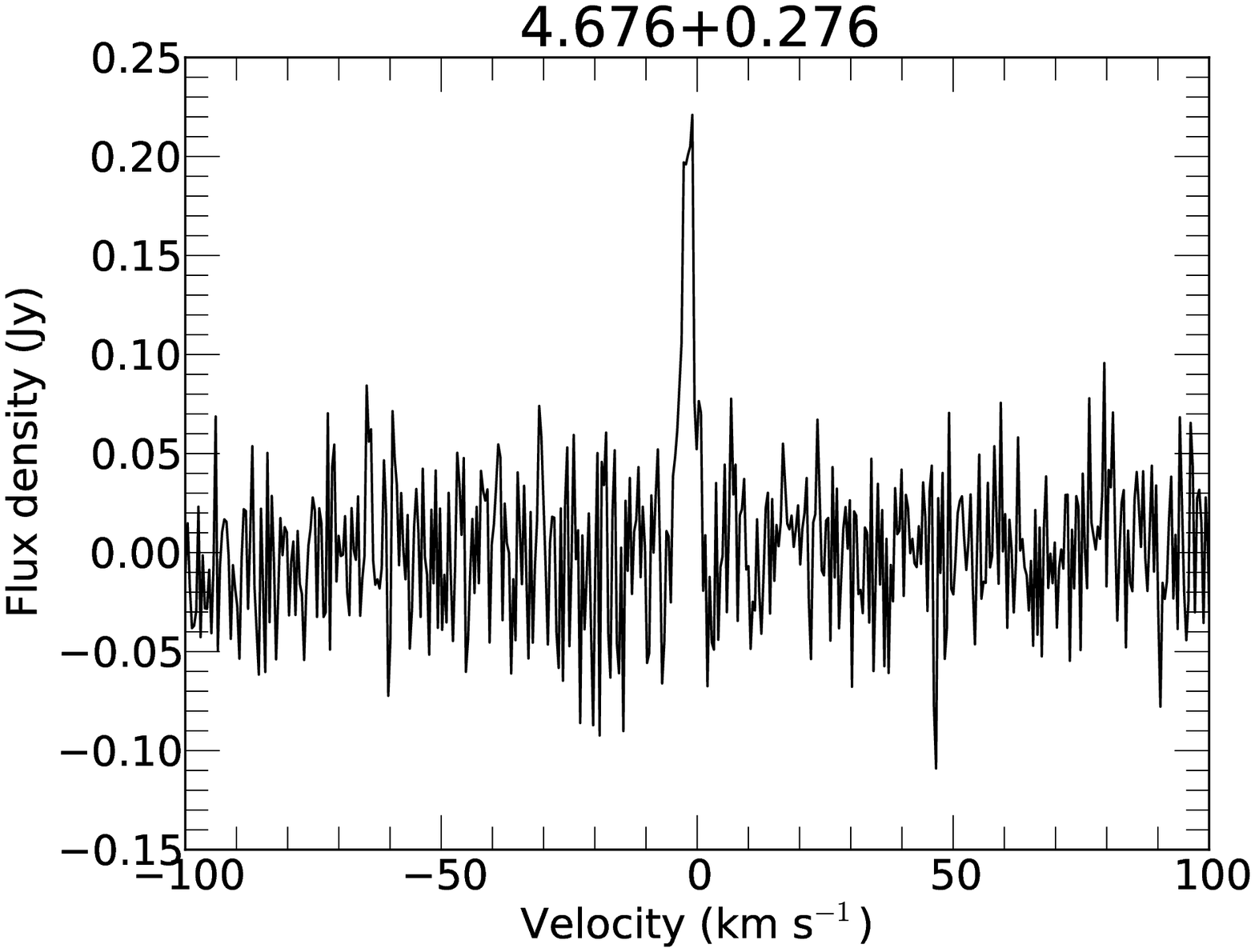}
\includegraphics[width=2.2in]{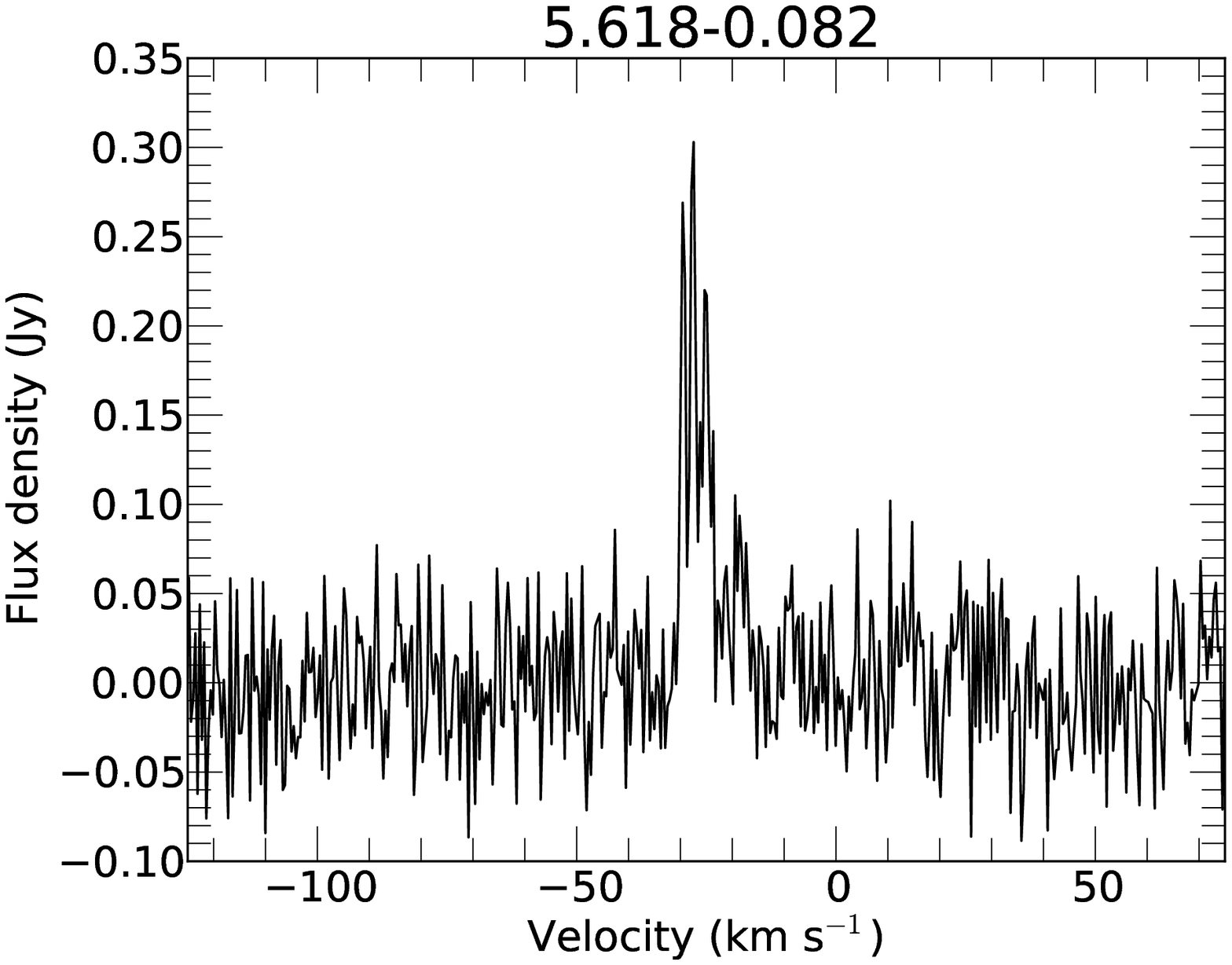}
\includegraphics[width=2.2in]{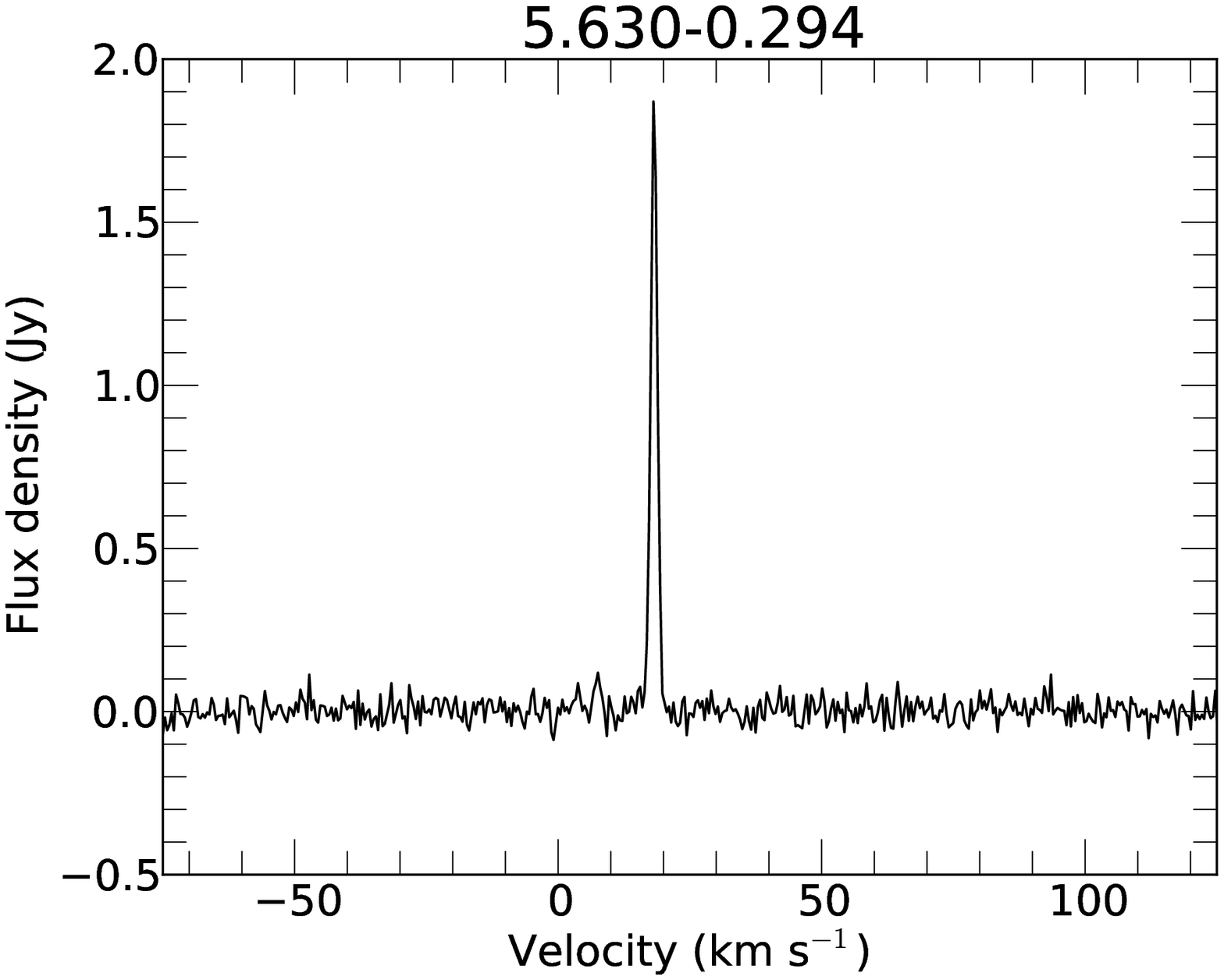}
\includegraphics[width=2.2in]{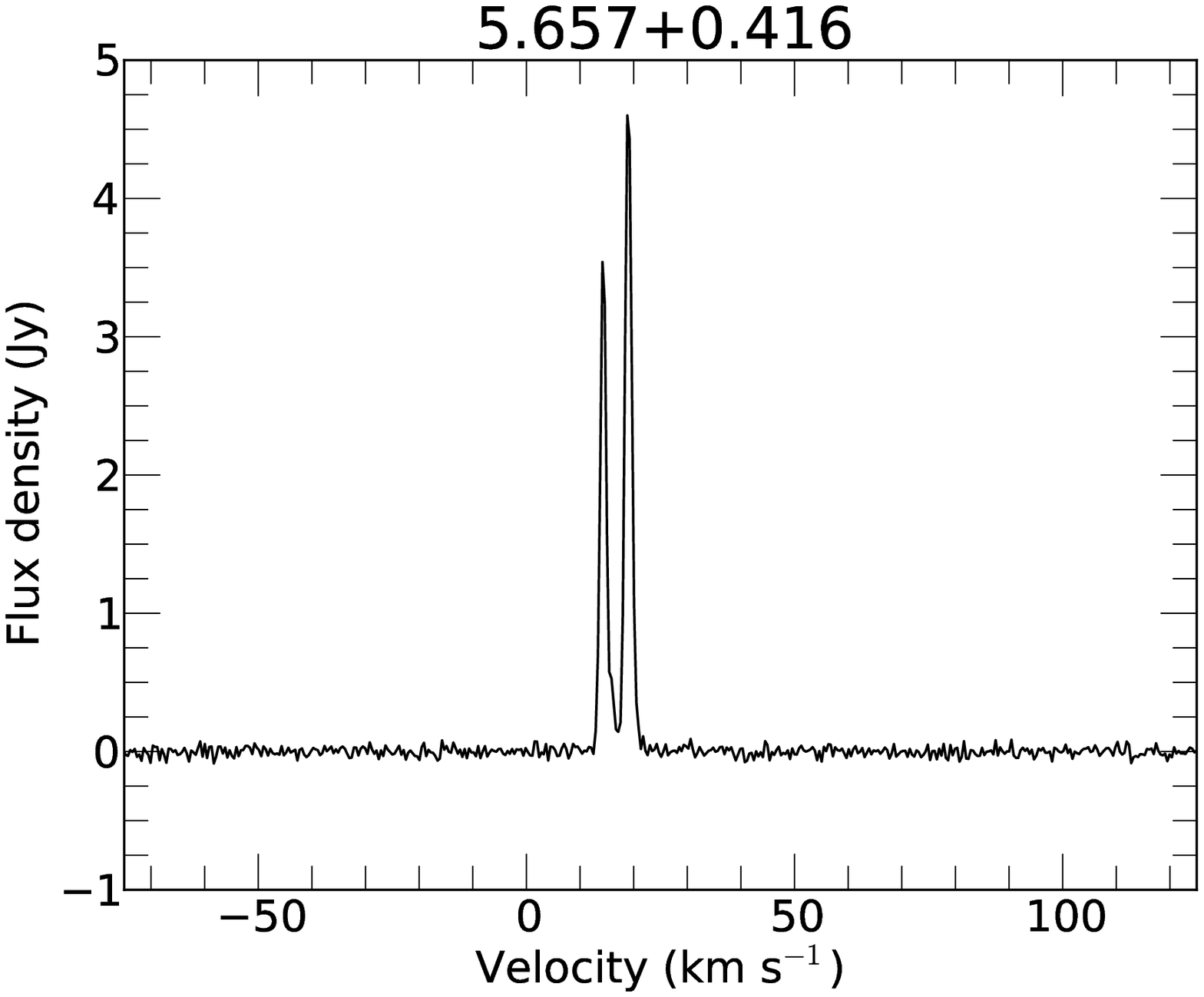}
\includegraphics[width=2.2in]{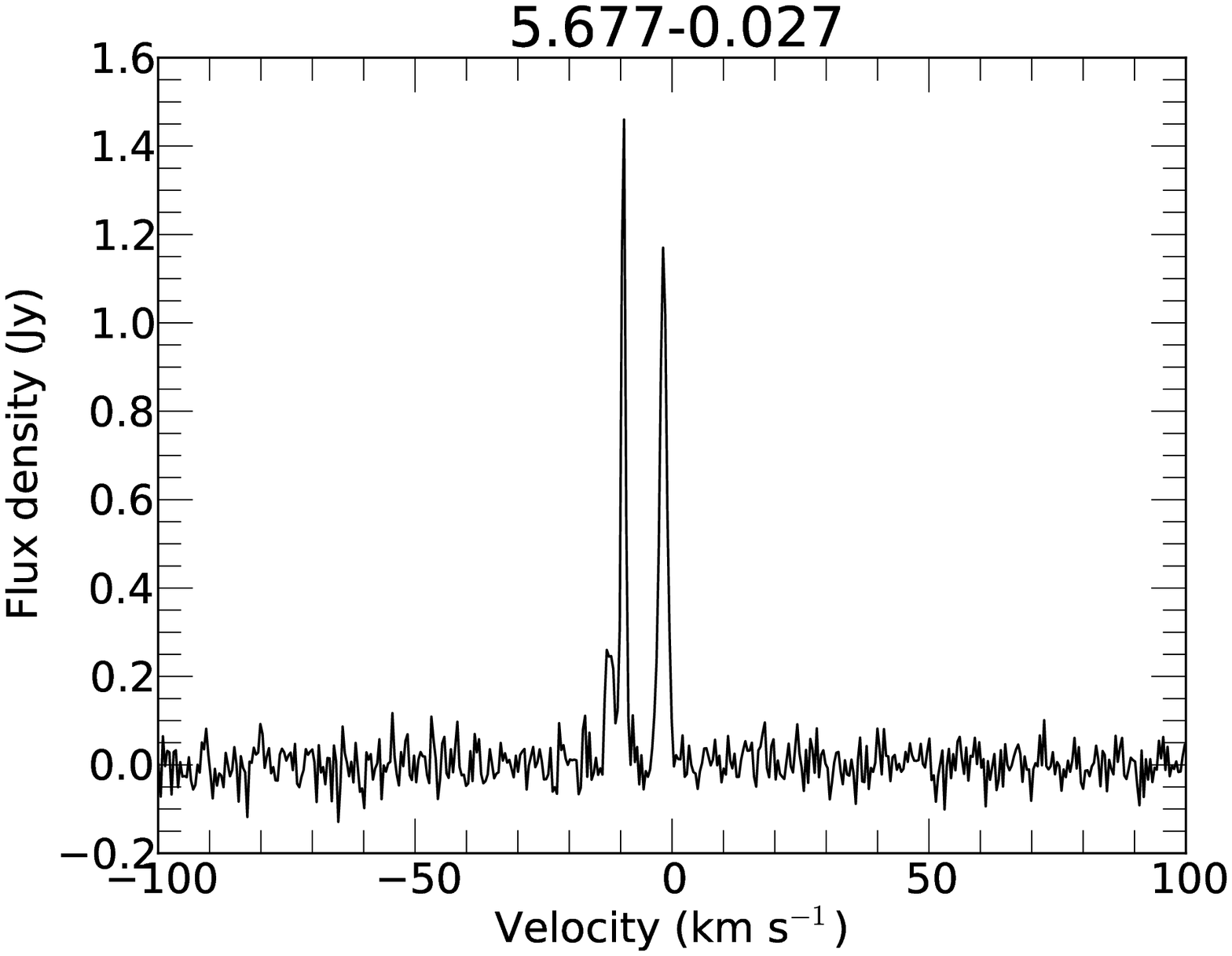}
\includegraphics[width=2.2in]{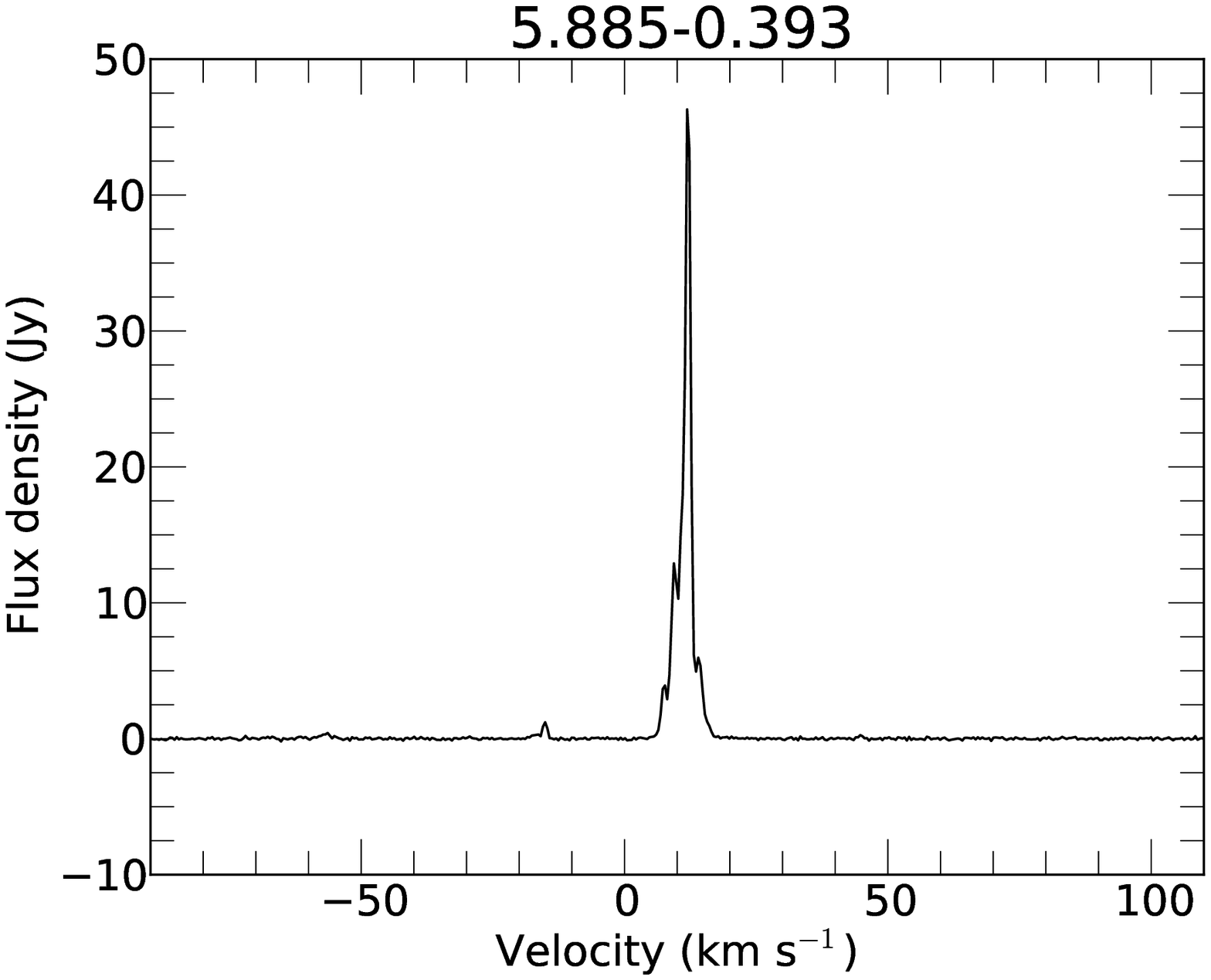}
\includegraphics[width=2.2in]{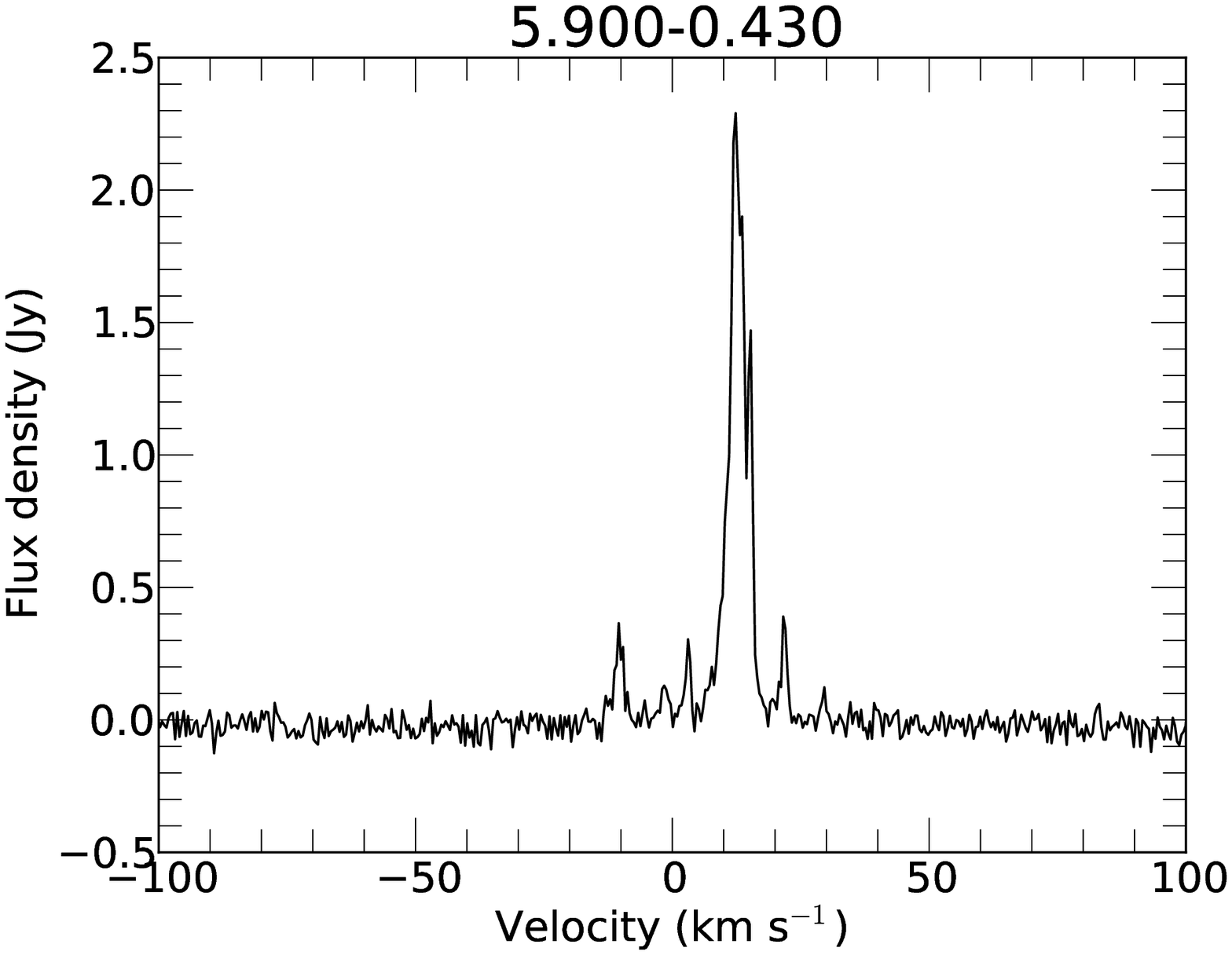}
\addtocounter{figure}{-1}
  \caption{-- {\emph {continued}}}
\end{figure*}

\end{document}